\documentclass[letterpaper,11pt]{article}
\pdfoutput=1

\usepackage{jheppub}
\usepackage{multirow}

\usepackage{subfig}
\usepackage{xspace}
\usepackage[countmax]{subfloat}

\usepackage{amssymb}
\usepackage{amsmath}
\usepackage{cancel}
\usepackage{color}
\usepackage{braket}
\usepackage{graphicx}
\usepackage{multirow}
\usepackage{verbatim}
\usepackage{amsthm}
\usepackage{slashed}
\usepackage{wasysym}
\usepackage{simplewick}
\usepackage{mathtools}
\usepackage{soul}
\usepackage{xspace}

\definecolor{dancomment}{RGB}{0,159,0}

\def\cA{\mathcal{A}}
\def\cB{\mathcal{B}}

\def\cL{\mathcal{L}}
\def\cM{\mathcal{M}}

\def\cO{\mathcal{O}}

\def\cP{\mathcal{P}}

\def\cW{\mathcal{W}}

\def\cY{\mathcal{Y}}

\def\tr{{\rm tr}}

\def\dpm{d_{+-}}
\def\dmp{d_{-+}}
\def\dpp{d_{++}}
\def\dmm{d_{--}}

\def\tO{\tilde{O}}

\def\nn{{\nonumber}}
\def\mcdot{\!\cdot\!}

\newcommand{\hard}{\mathrm{hard}}
\newcommand{\dyn}{\mathrm{dyn}}
\newcommand{\tree}{\mathrm{tree}}
\newcommand{\BPS}{\mathrm{BPS}}

\newcommand{\fd}[2]{\parbox{#1}{\includegraphics[width=#1]{#2}}}

\newcommand{\Eq}[1]{Equation~\eqref{#1}}

\DeclareRobustCommand{\Sec}[1]{Sec.~\ref{#1}}

\DeclareRobustCommand{\App}[1]{App.~\ref{#1}}
\DeclareRobustCommand{\Tab}[1]{Table~\ref{#1}}

\DeclareRobustCommand{\Fig}[1]{Fig.~\ref{#1}}

\DeclareRobustCommand{\Eq}[1]{Eq.~(\ref{#1})}
\DeclareRobustCommand{\Eqs}[2]{Eqs.~(\ref{#1}) and (\ref{#2})}
\DeclareRobustCommand{\Eqss}[3]{Eqs.~(\ref{#1}), (\ref{#2}), and (\ref{#3})}
\DeclareRobustCommand{\Ref}[1]{Ref.~\cite{#1}}
\DeclareRobustCommand{\Refs}[1]{Refs.~\cite{#1}}

\def\be{\begin{equation}}
\def\ee{\end{equation}}

\newcommand{\nbar}{{\bar n}}

\newcommand{\SCETi}{\mbox{${\rm SCET}_{\rm I}$}\xspace}
\newcommand{\SCETii}{\mbox{${\rm SCET}_{\rm II}$}\xspace}

\def\l{\langle}
\def\r{\rangle}

\def\bt{\beta}

\newcommand{\lotsdots}{{
+\cdot\cdot:\cdot\cdot 
(\cdot\cdot:\cdot\cdot\ldots\cdot\cdot:\cdot\cdot)
[\cdot\cdot:\cdot \cdot-]
}}

\newcommand{\Sl}[1]{\slashed{#1}}

\renewcommand{\arraystretch}{1.05}
\allowdisplaybreaks[3]

\newcommand{\eq}[1]{Eq.~\eqref{eq:#1}}
\newcommand{\eqs}[2]{Eqs.~\eqref{eq:#1} and \eqref{eq:#2}}
\renewcommand{\sec}[1]{Sec.~\ref{sec:#1}}
\newcommand{\secs}[2]{Secs.~\ref{sec:#1} and \ref{sec:#2}}

\newcommand{\app}[1]{App.~\ref{app:#1}}

\newcommand{\ord}[1]{\mathcal{O}(#1)}

\newcommand{\mae}[3]{\langle#1\rvert#2\rvert#3\rangle}
\newcommand{\Mae}[3]{\bigl\langle#1\bigr\rvert#2\bigr\rvert#3\bigr\rangle}

\newcommand{\ang}[1]{\langle #1 \rangle}

\newcommand{\df}{\mathrm{d}}

\newcommand{\sdt}{\!\cdot\!}

\newcommand{\al}{\alpha}

\newcommand\bn{{\bar n}}
\newcommand{\ga}{\gamma}
\newcommand{\Ga}{\Gamma}
\newcommand{\de}{\delta}

\newcommand{\ve}{\varepsilon}
\newcommand{\la}{\lambda}

\newcommand{\w}{\omega}
\newcommand{\balpha}{{\bar \alpha}}
\newcommand{\bbeta}{{\bar \beta}}
\newcommand{\bgamma}{{\bar \gamma}}
\newcommand{\bdelta}{{\bar \delta}}

\newcommand{\nslash}{\slashed{n}}
\newcommand{\bnslash}{\slashed{\bar{n}}}
\newcommand{\pslash}{\slashed{p}}

\newcommand{\qslash}{\slashed{q}}

\newcommand{\vT}{\bar{T}}

\newcommand{\vC}{\vec{C}}
\newcommand{\vO}{\vec O}

\newcommand{\lp}{\tilde p}        
\newcommand{\ldel}{\tilde \delta} 

\renewcommand{\P}{\mathrm{P}}       
\newcommand{\C}{\mathrm{C}}       
\newcommand{\bnP}{\overline {\mathcal P}}

  \newcount\hour \newcount\minute
  \hour=\time \divide \hour by 60 \minute=\time
  \newcommand{\todaytime}{\today \ -- \number\hour :\ifnum \minute<10 0\fi\number\minute}

\preprint{\begin{flushright}
MIT--CTP 4597
\end{flushright}}

\title{A Complete Basis of Helicity Operators for Subleading Factorization}

\author{Ilya Feige$^1$, Daniel W. Kolodrubetz$^2$, Ian Moult$^{3,4}$, Iain W. Stewart$^2$}

\affiliation{$^1$Center for the Fundamental Laws of Nature, Harvard University, Cambridge, MA 02138, USA}
\affiliation{$^2$Center for Theoretical Physics, Massachusetts Institute of Technology, Cambridge, MA 02139, USA}
\affiliation{$^3$Berkeley Center for Theoretical Physics, University of California, Berkeley, CA 94720, USA}
\affiliation{$^4$Theoretical Physics Group, Lawrence Berkeley National Laboratory, Berkeley, CA 94720, USA}

\emailAdd{feige@physics.harvard.edu}
\emailAdd{dkolodru@mit.edu}
\emailAdd{ianmoult@lbl.gov}
\emailAdd{iains@mit.edu}

\abstract{Factorization theorems underly our ability to make predictions for many processes involving the strong interaction. Although typically formulated at leading power, the study of factorization at subleading power is of interest both for improving the precision of calculations, as well as for understanding the all orders structure of QCD. We use the SCET helicity operator formalism to construct a complete power suppressed basis of hard scattering operators for $e^+e^-\to$ dijets, $e^- p\to e^-$ jet, and constrained Drell-Yan, including the first two subleading orders in the amplitude level power expansion. We analyze the form of the hard, jet, and soft function contributions to the power suppressed cross section for $e^+e^-\to$ dijet event shapes, and give results for the lowest order matching to the contributing operators. These results will be useful for studies of power corrections both in fixed order and resummed perturbation theory. }

\keywords{Factorization, QCD, Power Corrections}

\begin{document} 

\maketitle

\section{Introduction}\label{sec:intro}

One of the primary goals of the study of Quantum Chromodynamics (QCD) is an understanding of the all orders behavior of observables, traditionally formalized through either an operator product expansion (OPE)~\cite{Wilson:1969zs} or factorization theorems~\cite{Collins:1985ue,Collins:1988ig,Collins:1989gx}. For observables that can be handled with an OPE, a lot is known about the form of power corrections. Examples include deep inelastic scattering where the OPE has been carried out to twist-4~\cite{Jaffe:1981td,Jaffe:1982pm,Ellis:1982wd,Ellis:1982cd}, inclusive B-decays where the OPE is known to ${\cal O}(1/m_b^4)$~\cite{Mannel:1993su}, and Quarkonia production and decay, see~\cite{Brambilla:2010cs} for a review.  The description of observables with more complicated dynamics typically relies on factorization theorems and much less is known about the structure of power corrections in these cases. Power corrections have been considered for Drell-Yan~\cite{Qiu:1990xxa,Qiu:1990xy,Korchemsky:1994is,Beneke:1995pq,Korchemsky:1996iq} at ${\cal O}(\Lambda_{\rm QCD}^2/Q^2)$, for inclusive $B$ decays in the endpoint region at $\mathcal{O}((1-z)^0,(\Lambda_{\rm QCD}/m_b) ^{1,2})$~\cite{Bauer:2001mh,Leibovich:2002ys,Bauer:2002yu,Lee:2004ja,Mannel:2004as,Bosch:2004cb,Beneke:2004in,Tackmann:2005ub,Benzke:2010js}, for exclusive $B$ decays at $\mathcal{O}(\Lambda_{\rm QCD}/m_b)$~\cite{Mantry:2003uz,Blechman:2004vc,Beneke:1999br,Beneke:2000ry,Keum:2000ph,Keum:2000wi,Bauer:2004tj,Arnesen:2006vb,Arnesen:2006dc}, for event shapes $\tau$ in $e^+e^-$, $ep$, and $pp$ collisions at ${\cal O}(\Lambda_{\rm QCD}^k/(Q\tau)^k)$  \cite{Korchemsky:1994is,Dokshitzer:1995qm,Korchemsky:1995zm,Dokshitzer:1995zt,Dokshitzer:1997ew,Dokshitzer:1997iz,Korchemsky:1997sy,Korchemsky:1998ev,Korchemsky:1999kt,Korchemsky:2000kp,Belitsky:2001ij,Berger:2003pk,Lee:2006fn,Lee:2006nr,Hoang:2007vb,Gehrmann:2012sc}, and at $\mathcal{O}((1-z)^0)$ for threshold resummation \cite{Dokshitzer:2005bf,Grunberg:2007nc,Laenen:2008gt,Laenen:2008ux,Grunberg:2009yi,Laenen:2010uz,Almasy:2010wn,Bonocore:2014wua,White:2014qia,deFlorian:2014vta,Bonocore:2015esa,Bonocore:2016awd}.

A convenient formalism for studying factorization in QCD is the Soft Collinear Effective Theory (SCET) \cite{Bauer:2000ew, Bauer:2000yr, Bauer:2001ct, Bauer:2001yt}, an effective field theory describing the soft and collinear limits of QCD. SCET allows for a systematic power expansion at the level of the Lagrangian, and simplifies many aspects of factorization proofs \cite{Bauer:2002nz}. SCET has been used to study power corrections at the level of the amplitude \cite{Larkoski:2014bxa} and to derive factorization theorems  for $B$ decays using subleading power operators (eg.~\cite{Bauer:2002aj,Beneke:2003pa,Mantry:2003uz,Blechman:2004vc,Bauer:2004tj,Lee:2004ja,Hill:2004if,Bosch:2004cb,Beneke:2004in,Benzke:2010js}), where many interesting processes only start at subleading power. More recently, progress has been made towards understanding the subleading factorization and resummation of the event shape thrust in $e^+e^-$ \cite{Freedman:2013vya,Freedman:2014uta}. Such subleading factorization theorems are technically cumbersome, and significant work is still required to gain a simplified and more complete understanding. 


In this paper we consider the formalism required to study subleading factorization theorems in SCET, focusing in particular on subleading hard scattering operators.  Using the results of \Ref{Kolodrubetz:2016uim}, we further develop and explore a set of SCET helicity operator building blocks which is valid for constructing operator bases at any order in the power expansion. These operators extend the leading power basis of  \Ref{Moult:2015aoa}, where it was shown that the use of helicity operators greatly simplifies the construction of operator bases, as well as matching calculations, for processes with many final state jets (see also \cite{Moult:2014pja} for an application of helicity operators to Higgs processes, where they were used to simplify the matching to fixed order helicity amplitudes).  As we will see, helicity operators can also be used to simplify the construction of subleading power bases of hard scattering operators, where multiple fields can appear in the same collinear sector. After reviewing the helicity operator building blocks, we will focus on the case of hard scattering operators involving two back-to-back collinear sectors, which is relevant for proving subleading factorization theorems for a number of phenomenologically important process, namely $e^+e^-\to$ dijets, $e^- p\to e^-$ jet, and threshold Drell-Yan or Drell-Yan with an inclusive jet veto (which we refer to as constrained Drell-Yan for short).

\begin{figure}[t!]
%
%
\begin{center}
\includegraphics[width=0.23\columnwidth]{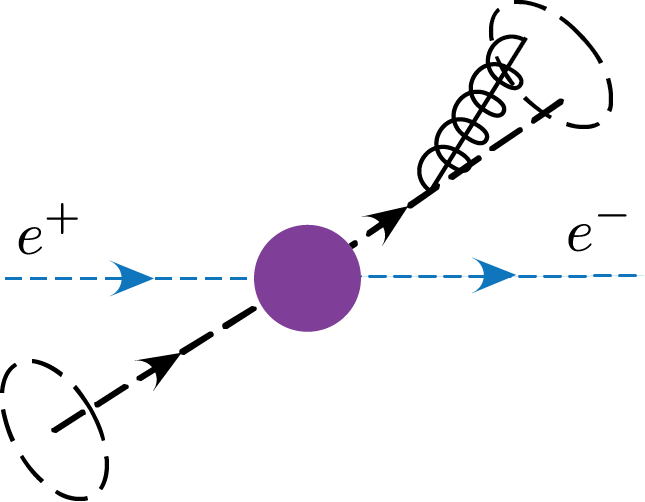} 
\hspace{0.1cm}
\includegraphics[width=0.23\columnwidth]{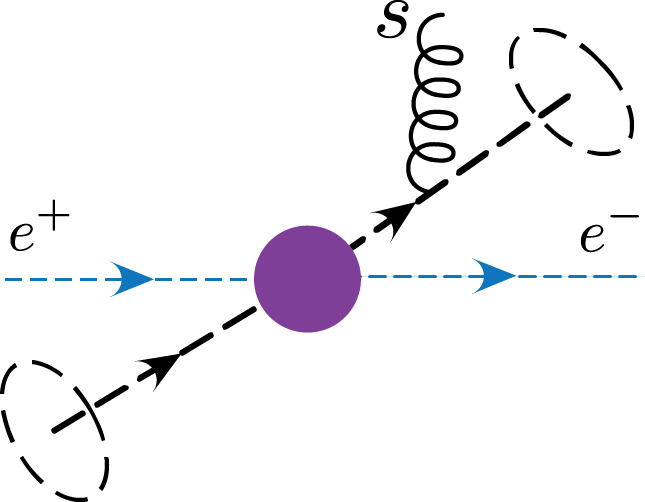} 
\hspace{0.1cm}
\includegraphics[width=0.23\columnwidth]{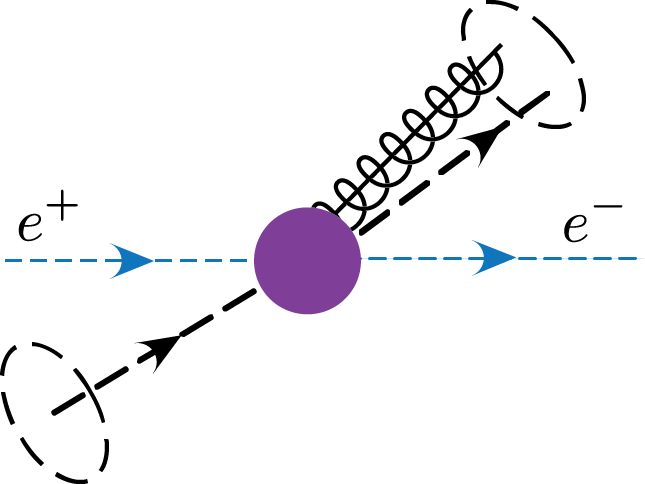} 
\hspace{0.1cm}
\includegraphics[width=0.23\columnwidth]{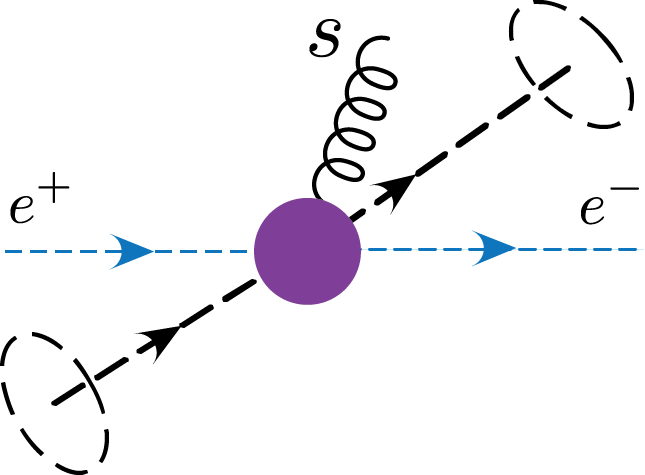} 
\raisebox{0cm}{ \hspace{-0.2cm} 
  $a$)\hspace{3.4cm}
  $b$)\hspace{3.4cm} 
  $c$)\hspace{3.4cm}
  $d$)\hspace{4cm} } 
\\[-25pt]
\end{center}
\vspace{-0.4cm}
\caption{ 
Examples of the contributions to thrust, $\tau$, in the dijet limit, at leading power in a) and b), and subleading power in c) and d).  There is an extra collinear gluon in a) from splitting, and in b) there is an extra gluon from soft emission. In c) the extra energetic gluon is collinear with the quark, but occurs without a nearly onshell parent propagator. Likewise in d) the extra soft emission amplitude is subleading. } 
\label{fig:subleadingamp}
\end{figure}

As an example to illustrate the expected form of subleading factorization theorems for collider observables, consider the $e^+e^-$ event shape thrust $T=1-\tau$~\cite{Farhi:1977sg} at center-of-mass energy $Q$. In the $Q\gg\Lambda_{\rm QCD}$ and  $\tau\ll 1$ limit \cite{Korchemsky:1999kt,Fleming:2007qr,Schwartz:2007ib,Abbate:2010xh} where the events are dominated by back-to-back jets, one can derive a leading power factorization formula for this process, given by 
\begin{align} \label{eq:fact0}
  \frac{\df\sigma}{\df\tau} 
   &= Q\sigma_0\, H^{(0)}(Q,\mu) \int\!\! \df s\: J^{(0)}_\tau(s,\mu) 
   \ S^{(0)}_\tau \Big(Q\tau -\frac{s}{Q},\mu \Big)  
  \ +\  {\cal O}\Big(\tau^0 , \frac{\Lambda_{\rm QCD}}{Q\tau }\Big) \,.
\end{align}
This factorization formula involves hard contributions from the scale $Q$ in a leading power hard function $H^{(0)}$, collinear contributions from the scale $Q\tau^{1/2}$ in the leading power thrust jet function obtained by combining two standard jet functions, $J_\tau^{(0)} = [ J^{(0)}\otimes J^{(0)}]$, and ultrasoft contributions from the scales $Q\tau$ and $\Lambda_{\rm QCD}$ in the leading power thrust soft function $S_\tau^{(0)}$. Here $\sigma_0$ is the $e^+e^-\to q\bar q$ Born cross section and $\mu$ is the factorization scale. For simplicity, we take $Q\tau \sim \Lambda_{\rm QCD}$ so that we do not have to elaborate further on the factorization of the soft function into perturbative and non-perturbative parts. The factorized product $H^{(0)} \times J_\tau^{(0)} \otimes S_\tau^{(0)}$ in \eq{fact0} includes contributions at all orders in $\alpha_s$ and powers of $\ln\tau$, which have the power law scaling of ${\cal O}(\tau^{-1})$, including $\delta(\tau)$ terms. As indicated, there are perturbative and nonperturbative  power corrections to this formula starting at ${\cal O}(\tau^0)$.  The soft function $S_\tau^{(0)}$ contains all nonperturbative corrections of ${\cal O}\Big( \frac{\Lambda_{\rm QCD}^k}{\tau (Q\tau)^k} \Big)$, so the first missing power corrections are ${\cal O}(\frac{\Lambda_{\rm QCD}}{Q\tau})$, which is also ${\cal O}(\tau^0)$ in our simplified counting.

In general the cross section can be expanded in powers of $\tau$,
\begin{align}
\frac{\df\sigma}{\df\tau} &=\frac{\df\sigma^{(0)}}{\df\tau} +\frac{\df\sigma^{(1)}}{\df\tau} +\frac{\df\sigma^{(2)}}{\df\tau}+\frac{\df\sigma^{(3)}}{\df\tau} +{\cal O}(\tau)\,,
\end{align}
where $\df\sigma^{(n)}/\df\tau\sim \tau^{-1+n/2}$ denotes the suppression relative to the leading term in powers of $\sqrt{\tau}$. Although for most observables the odd terms vanish, this convention gives a simple correspondence with the amplitude level power expansion. The explicit expression for $\df\sigma^{(0)}/\df\tau$ is given by the first term, $H^{(0)} \times J_\tau^{(0)} \otimes S_\tau^{(0)}$, shown in \eq{fact0}.  In SCET the hard scattering operators and Lagrangians governing the soft and collinear dynamics can be factorized from each other at any order in the power expansion.\footnote{This assumes that leading power Lagrangian interactions that can couple soft and collinear modes through Glauber exchange operators~\cite{Rothstein:2016bsq} that involve $1/\cP_\perp^2$ potential can be ignored at the active parton level.  It is known that this is the case for the full $e^+e^-\to $2-jet event shapes at leading power and for inclusive Drell-Yan at leading power \cite{Collins:1988ig}, and that this is not the case for spectator effects and ${\cal O}(\alpha_s^4)$ perturbative corrections in certain Drell-Yan event shapes~\cite{Gaunt:2014ska,Zeng:2015iba}. } Therefore factorization theorems can also be derived for the power suppressed contributions $\df\sigma^{(i)}/\df\tau$, for $i>0$,  corresponding to the power corrections in \eq{fact0}. A prototypical example of this is the factorization theorem for the decay rate of $b\to s\gamma$ at large $E_\gamma$, which has the same form as \eq{fact0} at leading power, and where a factorization theorem for the ${\cal O}(\tau^0)$ terms has been derived~\cite{Lee:2004ja}. It involves subleading hard, jet, and soft functions, since power corrections can not change the relevant degrees of freedom. Based on this we expect that the higher order power corrections to the thrust cross section will obey schematic factorization theorems of the form
\begin{align} \label{eq:sigma}
&\hspace{-0.25cm}\frac{\df\sigma^{(n)}}{\df\tau} =
 Q\sigma_0 
\sum_{j}  H^{(n_{Hj})}_{j} \otimes \Big[ J^{(n_{Jj})}_{j} J^{(n'_{Jj})}_{j}\Big]  \otimes S_j^{(n_{Sj})}  
,\end{align}
where $j$ sums over the multiple contributions that appear at each order, $n_{Hj}+n_{Jj}+n_{Jj}'+n_{Sj}=n$, and $\otimes$ denotes a set of convolutions. In this formula, the cross section at each power has been factorized into hard functions $H_j^{(n_{Hj})}$, jet functions $J_j^{(n_{Jj})}$, and soft functions $S_j^{(n_{Sj})}$ which may be leading or subleading power depending on the value of the $n_{Xj}$ indices.  The hard function contains the dependence on the underlying hard partonic process, but can be chosen to be independent of the particular event shape that is being measured. The jet functions (which describe the collinear radiation along the jet directions) as well as the soft function (which describes the soft radiation in the event) depend on the particular measurement function.

Deriving a subleading power factorization theorem using SCET, like that in \eq{sigma},  consists of several steps. First, one must demonstrate the existence of a finite basis of hard scattering operators in SCET at the appropriate order in the power expansion, and determine an explicit basis of such operators. Matching calculations from QCD to SCET are required to determine the Wilson coefficients of these operators and the structure of their collinear Wilson lines. The soft and collinear dynamics are entirely described by the Lagrangian of the effective theory, and the subleading power SCET Lagrangian is also required at the same order in the power expansion. At leading power, the BPS field redefinition \cite{Bauer:2002nz} can be used to factorize the Lagrangian into non-interacting pieces describing each collinear sector, as well as the soft sector. The only exception to this is the leading power Glauber Lagrangian~\cite{Rothstein:2016bsq} which couples together soft and collinear fields, and which will violate factorization if it can not be shown to give canceling contributions or that it is irrelevant. Power suppressed Lagrangians have been analyzed in the literature~\cite{Beneke:2002ni,Chay:2002vy,Manohar:2002fd,Pirjol:2002km,Beneke:2002ph,Bauer:2003mga}, and the \SCETi Lagrangian is currently known to ${\cal O}(\lambda^2)$ \cite{Bauer:2003mga} (excluding power suppressed Glauber exchange operators).  Beyond leading power and after the BPS field redefinition, the subleading Lagrangians (including power suppressed Glauber operators) will involve factorized products of soft and collinear fields. Here factorization at the Lagrangian level again only requires showing that the leading power Glauber Lagrangian is not required. Next, one must consider factorization of the observable, and demonstrate that one can define suitable measurements that are separately made in the soft and collinear matrix elements. Finally, starting with the full QCD expression for the appropriate observable, one must go through a number of expansions and algebraic manipulations to factorize the cross section into a product of squared matrix elements, each involving only collinear or soft fields.  This step leads to field theoretic definitions of the subleading jet, soft and hard functions appearing in the factorization theorem of \eq{sigma}. The degree to which these steps require lengthy and tedious calculations is determined by the complexity of the operator basis.

Traditionally, an operator basis is constructed by enumerating all possible operators consistent with symmetry constraints. These operators are formed from the SCET fields, along with Lorentz, Dirac and color structures. Beyond leading power, the determination of a minimal operator basis becomes complicated, even for processes with a limited number of collinear sectors, such as $pp\to \mu^+\mu^-$ (constrained Drell-Yan) or $e^+e^- \to$ dijets. The algebraic manipulations in SCET required to achieve factorization are similarly complicated, making subleading factorization laborious. In this paper, we show that by working with operators of definite helicity, the operator basis is easy to construct, and does not involve complicated Lorentz or Dirac structures, simplifying the algebraic manipulations required for factorization. Many symmetry properties are also made manifest in the helicity basis. We demonstrate how these provide simplifications both at the level of the hard scattering operator basis as well as in factorized matrix elements.

An outline of this paper is as follows. In \sec{scet} we provide a brief review of SCET with an emphasis on the field content, power counting, and construction of SCET operators. In \sec{helops} we describe our basis of helicity operators and discuss their symmetry properties. We focus in particular on the treatment of operators involving ultrasoft fields, and use the BPS field redefinition to define collinear and ultrasoft gauge invariant operator building blocks. Additionally, selection rules on the hard scattering operators due to angular momentum conservation are described \cite{Kolodrubetz:2016uim}. Extensions of the formalism to SCET$_\text{II}$ as well as SCET with massive collinear quarks are also discussed, as are complications associated with evanescent operators. We then demonstrate the utility of our helicity basis in \sec{eeJets} by constructing an $\mathcal{O}(\lambda^2)$ basis of hard scattering operators with two back-to-back collinear sectors,  as relevant for $e^+e^-\to$ dijets, $e^- p\to e^-$+jet, and constrained Drell-Yan. Using the symmetry properties of the operators, we enumerate those which can contribute to the factorized cross section at $\mathcal{O}(\lambda^2)$, and in \Sec{sec:matching}, we perform the tree level matching to these operators. We conclude in \sec{conclusions}.

\section{Review of SCET}\label{sec:scet}

SCET is an effective field theory of QCD describing the interactions of collinear and soft particles in the presence of a hard interaction \cite{Bauer:2000ew, Bauer:2000yr, Bauer:2001ct, Bauer:2001yt, Bauer:2002nz}. Since SCET describes collinear particles (which are characterized by a large momentum along a particular light-like direction), as well as soft particles, it is natural to use light-cone coordinates. For each jet direction we define two light-like reference vectors $n_i^\mu$ and $\bn_i^\mu$ such that $n_i^2 = \bn_i^2 = 0$ and $n_i\sdt\bn_i = 2$. One typical choice for these quantities is
\begin{equation}
n_i^\mu = (1, \vec{n}_i)
\,,\qquad
\bn_i^\mu = (1, -\vec{n}_i)
\,,\end{equation}
where $\vec{n}_i$ is a unit three-vector. Given a choice for $n_i^\mu$ and $\bn_i^\mu$, any four-momentum $p$ can then be written as
\begin{equation} \label{eq:lightcone_dec}
p^\mu = \bn_i\sdt p\,\frac{n_i^\mu}{2} + n_i\sdt p\,\frac{\bn_i^\mu}{2} + p^\mu_{n_i\perp}\
\,.\end{equation}
A particle with momentum $p$ close to the $\vec{n}_i$ direction, such that the components of $p$ scale as $(n_i\!\cdot\! p, \bn_i \!\cdot\! p, p_{n_i\perp}) \sim \bn_i\!\cdot\! p$ $\,(\la^2,1,\la)$, where $\la \ll 1$ is a small formal power counting parameter, are referred to as $n_i$ collinear. The formal scaling of $\lambda$ is determined by the form of measurements or kinematic restrictions on the QCD radiation. To ensure that $n_i$ and $n_j$ refer to distinct collinear directions, we must have
\begin{equation} \label{eq:nijsep}
  n_i\sdt n_j  \gg \la^2 \qquad\text{for}\qquad i\neq j
\,.\end{equation}
Since distinct reference vectors, $n_i$ and $n_i'$, with $n_i\cdot n_i' \sim \ord{\lambda^2}$ both describe the same collinear physics, one can label a collinear sector by any member of a set of equivalent vectors, $\{n_i\}$. This freedom is manifest as a symmetry of the effective theory known as reparametrization invariance (RPI) \cite{Manohar:2002fd,Chay:2002vy}. Specifically, the three classes of RPI transformations are
\begin{alignat}{3}\label{eq:RPI_def}
&\text{RPI-I} &\qquad &  \text{RPI-II}   &\qquad &  \text{RPI-III} \nn \\
&n_{i \mu} \to n_{i \mu} +\Delta_\mu^\perp &\qquad &  n_{i \mu} \to n_{i \mu}   &\qquad & n_{i \mu} \to e^\alpha n_{i \mu} \nn \\
&\bar n_{i \mu} \to \bar n_{i \mu}  &\qquad &  \bar n_{i \mu} \to \bar n_{i \mu} +\epsilon_\mu^\perp  &\qquad & \bar n_{i \mu} \to e^{-\alpha} \bar n_{i \mu}\,.
\end{alignat}
Here, we have $\Delta^\perp \sim \lambda$, $\epsilon^\perp \sim \lambda^0$, and $\alpha\sim \lambda^0$. The parameters $\Delta^\perp$ and $\epsilon^\perp$ are infinitesimal, and satisfy $n_i\cdot \Delta^\perp=\bar n_i\cdot \Delta^\perp=n_i \cdot \epsilon^\perp=\bar n_i \cdot \epsilon^\perp=0$. RPI will be exploited to simplify the structure of the subleading power operator basis in \sec{eeJets}.

The effective theory is constructed by expanding momenta into label and residual components
\begin{equation} \label{eq:label_dec}
p^\mu = \lp^\mu + k^\mu = \bn_i \sdt\lp\, \frac{n_i^\mu}{2} + \lp_{n_i\perp}^\mu + k^\mu\,.
\,\end{equation}
Here, $\bn_i \cdot\lp \sim Q$ and $\lp_{n_i\perp} \sim \la Q$ are the large label momentum components, where $Q$ is the scale of the hard interaction, while $k\sim \la^2 Q$ is a small residual momentum describing fluctuations about the label momentum. A multipole expansion is then performed to obtain fields with momenta of definite scaling, namely collinear quark and gluon fields for each collinear direction, as well as soft quark and gluon fields. Independent gauge symmetries are enforced for each set of fields.

Due to the multipole expansion, the SCET fields for $n_i$-collinear quarks and gluons, $\xi_{n_i,\lp}(x)$ and $A_{n_i,\lp}(x)$,  are written in position space with respect to the residual momentum and in momentum space with respect to the large momentum components. They are labeled by their collinear direction $n_i$ and their large momentum $\lp$. The label momentum operator $\cP_{n_i}^\mu$, gives the large label component of the momentum, $\cP_{n_i}^\mu\, \xi_{n_i,\lp} = \lp^\mu\, \xi_{n_i,\lp}$, while derivatives give the residual momentum dependence, $i \partial^\mu \sim k \sim \la^2 Q$. The label momentum operator is defined such that when acting on a product of fields, $\cP_{n_i}$ gives the sum of the label momenta of all $n_i$-collinear fields. We will often use the shorthand notation $\bnP_{n_i} = \bn\sdt\cP_{n_i}$ for the large label momentum component.  

Soft degrees of freedom are described in the effective theory by separate quark and gluon fields. We will assume that we are working in the SCET$_\text{I}$ theory where these soft degrees of freedom are referred to as ultrasoft so as to distinguish them from the soft modes of SCET$_\text{II}$ \cite{Bauer:2002aj}. Extensions of our formalism to treat SCET$_\text{II}$ problems will be discussed in \Sec{sec:scet_II}. In SCET$_\text{I}$, the ultrasoft modes do not carry label momenta, but have residual momentum dependence with $i \partial^\mu \sim \la^2Q$, and are able to exchange residual momenta between different collinear sectors. They are therefore described by fields $q_{us}(x)$ and $A_{us}(x)$ without label momenta, and without a collinear sector label. 

SCET is formulated as an expansion in powers of $\la$, constructed so that manifest power counting is maintained at all stages of a calculation. As a consequence of the multipole expansion, all fields and derivatives acquire a definite power counting \cite{Bauer:2001ct}, shown in \Tab{tab:PC}. The SCET Lagrangian is also expanded as a power series in $\lambda$
\begin{align} \label{eq:SCETLagExpand}
\cL_{\text{SCET}}=\cL_\hard+\cL_\dyn= \sum_{i\geq0} \cL_\hard^{(i)}+\cL_G^{(0)}+\sum_{i\geq0} \cL^{(i)} \,,
\end{align}
where $(i)$ denotes objects at ${\cal O}(\lambda^i)$ in the power counting. The Lagrangians $ \cL_\hard^{(i)}$ contain the hard scattering operators $O^{(i)}$, whose structure is determined by the matching process, as described in \Sec{sec:matching}. The leading power Glauber Lagrangian~\cite{Rothstein:2016bsq}, $\cL_G^{(0)}$, describes leading power interactions between soft and collinear modes in the form of potentials. It breaks factorization unless it can be shown to cancel out or absorbed into other interactions such as Wilson line directions. The $\cL^{(i)}$ describe the dynamics of ultrasoft and collinear modes in the effective theory, including subleading power corrections to the Glauber Lagrangian. The subleading Lagrangians (excluding subleading power corrections to the Glauber Lagrangian) are explicitly known to $\mathcal{O}(\lambda^2)$, and the relevant ones for our analysis can be found, along with their Feynman rules in \cite{iain_notes}.

Factorization theorems used in jet physics are typically derived at leading power in $\lambda$. In this case, interactions involving hard processes in QCD are matched to a basis of leading power SCET hard scattering operators $O^{(0)}$, the dynamics in the effective theory are described by the leading power Lagrangian, $\cL^{(0)}$, and the measurement function, which defines the action of the observable, is expanded to leading power. Higher power terms in the $\lambda$ expansion, known as power corrections, arise from three sources: subleading power hard scattering operators $O^{(i)}$, subleading Lagrangian insertions, and subleading terms in the expansion of the measurement functions which act on soft and collinear radiation. The first two sources are independent of the details of the particular measurement, only requiring that it is an SCET$_\text{I}$ dijet observable, while the third depends on its precise definition. Although we will not discuss subleading measurement functions in this paper, an example, for the case of thrust, is given in \App{app:meas}.

\begin{table}
\begin{center}
\begin{tabular}{| l | c | c |c |c|c| r| }
  \hline                       
  Operator & $\cB_{n_i\perp}^\mu$ & $\chi_{n_i}$& $\cP_\perp^\mu$&$q_{us}$&$D_{us}^\mu$ \\
  Power Counting & $\lambda$ &  $\lambda$& $\lambda$& $\lambda^3$& $\lambda^2$ \\
  \hline  
\end{tabular}
\end{center}
\caption{
Power counting for building block operators in $\text{SCET}_\text{I}$.
}
\label{tab:PC}
\end{table}

Gauge invariant collinear operators in the effective theory are constructed out of products of gauge invariant building blocks. These building blocks are formed from gauge invariant combinations of fields and Wilson lines~\cite{Bauer:2000yr,Bauer:2001ct}.  The collinearly gauge-invariant quark and gluon fields are defined as
\begin{align} \label{eq:chiB}
\chi_{{n_i},\w}(x) &= \Bigl[\delta(\w - \bnP_{n_i})\, W_{n_i}^\dagger(x)\, \xi_{n_i}(x) \Bigr]
\,,\\
\cB_{{n_i}\perp,\w}^\mu(x)
&= \frac{1}{g}\Bigl[\delta(\w + \bnP_{n_i})\, W_{n_i}^\dagger(x)\,i  D_{{n_i}\perp}^\mu W_{n_i}(x)\Bigr]
 \,. \nn
\end{align}
Here we have chosen a convention such that for $\chi_{{n_i},\w}$, we have $\w > 0$ for an incoming quark and $\w < 0$ for an outgoing antiquark. For $\cB_{{n_i},\w\perp}$, $\w > 0$ ($\w < 0$) corresponds to outgoing (incoming) gluons. The covariant derivative which appears in \eq{chiB} is defined as,
\begin{equation}
i  D_{{n_i}\perp}^\mu = \cP^\mu_{{n_i}\perp} + g A^\mu_{{n_i}\perp}\,,
\end{equation}
and $W_{n_i}$ are Wilson lines of ${n_i}$-collinear gluons in label momentum space defined as
\begin{equation} \label{eq:Wn}
W_{n_i}(x) = \biggl[~\sum_\text{perms} \exp\Bigl(-\frac{g}{\bnP_{n_i}}\,\bn\sdt A_{n_i}(x)\Bigr)~\biggr]\,,
\end{equation}
In general the structure of Wilson lines must be derived by a matching calculation from QCD. These Wilson lines sum up arbitrary emissions of ${n_i}$-collinear gluons off of particles from other sectors, which due to the power expansion always appear in the ${\bar{n}_i}$ direction. The gluon emissions summed in the Wilson lines are $\ord{\lambda^0}$ in the power counting. In \eqs{chiB}{Wn} the label momentum operators act only on the fields inside the square brackets. The Wilson line $W_{n_i}(x)$ is localized with respect to the residual position $x$, and we can therefore treat
$\chi_{{n_i},\w}(x)$ and $\cB_{{n_i},\w}^\mu(x)$ as local quark and gluon fields from the perspective of ultrasoft derivatives $\partial^\mu$ that act on $x$. 

The complete set of collinear and ultrasoft building blocks for constructing hard scattering operators or subleading Lagrangians at any order in the power counting is given in \Tab{tab:PC}. All other field and derivative combinations can be reduced to this set by the use of equations of motion and operator relations~\cite{Marcantonini:2008qn}. Since these building blocks carry vector or spinor Lorentz indices they must be contracted to form scalar operators, which also involves the use of objects like $\{n_i^\mu, \bn_i^\mu, \gamma^\mu, g^{\mu\nu}, \epsilon^{\mu\nu\sigma\tau}\}$. A key advantage of the helicity operator approach discussed below is that this is no longer the case; all the building blocks will be scalars. 

As shown in \Tab{tab:PC}, both the collinear quark and collinear gluon building block fields scale as ${\cal O}(\lambda)$. For the majority of jet processes there is a single collinear field operator for each collinear sector at leading power.  (Although for fully exclusive processes that directly produce hadrons there will be multiple building blocks from the same sector in the leading power operators since they form color singlets in each sector.) Also, since $\cP_\perp\sim \lambda$, this operator will not typically be present at leading power (exceptions could occur, for example, in processes picking out P-wave quantum numbers). At subleading power, operators for all processes can involve multiple collinear fields in the same collinear sector, as well as $\cP_\perp$ operator insertions. The power counting for an operator is obtained by simply adding up the powers for the building blocks it contains. To ensure consistency under renormalization group evolution the operator basis in SCET must be complete, namely all operators consistent with the symmetries of the problem must be included.

Dependence on the ultrasoft degrees of freedom enters the operators through the ultrasoft quark field $q_{us}$, and the ultrasoft covariant derivative $D_{us}$, defined as 
\begin{equation}
i  D_{us}^\mu = i  \partial^\mu + g A_{us}^\mu\,,
\end{equation}
from which we can construct other operators including the ultrasoft gluon field strength. All operators in the theory must be invariant under ultrasoft gauge transformations. Collinear fields transform under ultrasoft gauge transformations as background fields of the appropriate representation. The power counting for these operators is shown in \Tab{tab:PC}. Since they are suppressed relative to collinear fields, ultrasoft fields typically do not enter factorization theorems in jet physics at leading power. An example where ultrasoft fields enter at leading power is $B \to X_s \gamma$ in the photon endpoint region, which is described at leading power by a single collinear sector, and an ultrasoft quark field for the b quark~\cite{Bauer:2000ew}.

\section{Helicity Operators}\label{sec:helops}

The use of on-shell helicity amplitudes has been fruitful for the study of scattering amplitudes in gauge theories and gravity (see e.g. \cite{Dixon:1996wi,Elvang:2013cua,Dixon:2013uaa,Henn:2014yza} for pedagogical reviews).  By focusing on amplitudes for external states with definite helicity and color configurations many simplifications arise. The helicity approach to SCET operators of \Ref{Moult:2015aoa} takes advantage of the fact that collinear SCET fields are themselves gauge invariant, and are each associated with a fixed external label direction with respect to which helicities can naturally be defined. Instead of considering operators formed from Lorentz and Dirac structures (each of which contributes to multiple states with different helicity combinations) helicity operators can be associated with external states of definite helicity. This approach greatly simplifies the construction of a minimal operator bases for processes with many active partons, and facilitates the matching to fixed order calculations which are often performed using spinor helicity techniques.

We now briefly summarize our spinor helicity conventions. Further identities, as well as our phase conventions, can be found in \app{helicity}. To simplify our discussion we take all momenta and polarization vectors as outgoing, and label all fields and operators by their outgoing helicity and momenta. We use the standard spinor helicity notation 
\begin{align} \label{eq:braket_def}
|p\rangle\equiv \ket{p+} &= \frac{1 + \ga_5}{2}\, u(p)
  \,,
 & |p] & \equiv \ket{p-} = \frac{1 - \ga_5}{2}\, u(p)
  \,, \\
\bra{p} \equiv \bra{p-} &= \mathrm{sgn}(p^0)\, \bar{u}(p)\,\frac{1 + \ga_5}{2}
  \,, 
 & [p| & \equiv \bra{p+} = \mathrm{sgn}(p^0)\, \bar{u}(p)\,\frac{1 - \ga_5}{2}
  \,, \nn 
\end{align}
with $p$ lightlike. The polarization vector of an outgoing gluon with momentum $p$ can be written
\begin{equation}
 \ve_+^\mu(p,k) = \frac{\mae{p+}{\ga^\mu}{k+}}{\sqrt{2} \langle kp \rangle}
\,,\qquad
 \ve_-^\mu(p,k) = - \frac{\mae{p-}{\ga^\mu}{k-}}{\sqrt{2} [kp]}
\,,\end{equation}
where $k\neq p$ is an arbitrary lightlike reference vector.

The polarization vectors and spinors satisfy the standard identities.
\begin{align}
&p\cdot \epsilon_{\pm}(p,k)=k \cdot\epsilon_{\pm}(p,k) =0\,, \qquad  \epsilon_{\pm}(p,k)\cdot \epsilon_{\pm}(p,k)=0\,, \qquad \epsilon_{\pm}(p,k)\cdot \epsilon_{\mp}(p,k)=-1\,, \nn\\
&\Sl{\epsilon}_+(p,k)|k\rangle=\Sl{\epsilon}_-(p,k)|k]=0\,, \qquad
[k|\Sl{\epsilon}_-(p,k)=\langle k|\Sl{\epsilon}_+(p,k)=0\,.
\end{align}
Additional identities can be found in \App{app:helicity}.

In this section we discuss the extension of the helicity operator approach of \cite{Moult:2015aoa} to subleading powers. We review the full set of subleading power building block operators introduced in \cite{Kolodrubetz:2016uim}, and provide more details about them.  In \sec{coll} we describe operators involving only collinear fields, and in \sec{pperp} we describe operators involving insertions of the $\cP_\perp$ operator. We also give all the Feynman rules for these operators. The organization of color bases is discussed in \sec{color}. The inclusion of ultrasoft fields in the hard scattering operators is more involved, since the usual SCET building blocks are gauge covariant rather than gauge invariant. In \sec{BPS} we discuss the BPS field redefinition, and how it can be used to define ultrasoft gauge invariant helicity operators to be used as basis elements. In \sec{constr_subops}, we give the complete list of scalar building blocks needed to construct an operator basis at arbitrary power in $\lambda$. Next, in \sec{extensions} we examine the extension of this formalism to \SCETii, SCET with massive quarks and evanescent operators. We briefly discuss how to carry out matching calculations at subleading power in \sec{matching}, and the C and P properties of the operators in \sec{PandC}. In \Sec{sec:ang_cons} we discuss interesting constraints from angular momentum conservation which first appear at subleading power when there are multiple fields in the same collinear sector.

\subsection{Collinear Gauge Invariant Helicity Building Blocks}\label{sec:coll}

We define a collinear gluon field of definite helicity as
\begin{equation} \label{eq:cBpm_def}
\cB^a_{i\pm} = -\ve_{\mp\mu}(n_i, \bn_i)\,\cB^{a\mu}_{n_i\perp,\w_i}
\,,\end{equation}
where $a$ is an adjoint color index. This is sufficient for the treatment of collinear gluons even at subleading power. With this definition, for an outgoing gluon with polarization $\pm$, momentum $p$, $p^0>0$ (or an incoming gluon with polarization $\mp$, momentum $-p$, $p_0<0$), and color $a$, the nonzero tree-level Feynman rules are\footnote{The precise definition of this delta function and measure are
\begin{align} \label{eq:labelsums}
\ldel(\lp_i - p) &\equiv \delta_{\{n_i\},p}\,\delta(\w_i - \bn_i\cdot p)\,,\nn\\
\int\!\df\lp &\equiv \sum_{\{n_i\}} \int\!\df\w_i \nn
\,,\end{align}
where 
\begin{equation}
\delta_{\{n_i\},p} =
\begin{cases}
   1 &\quad n_i\cdot p = \ord{\lambda^2}
   \,,\\
   0 &\quad \text{otherwise}\nn
\,.\end{cases}
\end{equation}
The Kronecker delta is nonzero if the collinear momentum $p$ is in the $\{ n_i\}$
equivalence class, i.e. $p$ should be considered as collinear with
the $i$th jet.  The sum in the second line runs over the different
equivalence classes. }
\begin{align} \label{eq:gluonbaseFR}
\l g_\pm^a(p) | \cB_{i \pm}^b |0 \r & = \delta^{ab} \ldel (\lp_i -p)\,,  \\
\l 0 | \cB_{i \pm}^b |g_\mp^a(-p) \r & = \delta^{ab} \ldel (\lp_i -p), \nn
\end{align}
where we have followed \Ref{Moult:2015aoa} in using $k = \bn$ as our reference vector. Despite the fact that $\cB_{i \pm}^a=\cB_{i \pm}^a(x)$, our external gluon state has zero residual momentum, so we do not get an additional phase.
We also define quark fields with definite helicity, given by
\begin{align} \label{eq:definitehelicityquarkdef}
\chi_{i \pm}^\alpha &= \frac{1\,\pm\, \gamma_5}{2} \chi_{n_i, - \omega_i}^\alpha\,,\qquad \bar{\chi}_{i \pm}^\balpha =  \bar{\chi}_{n_i, - \omega_i}^\balpha \frac{1\,\mp\, \gamma_5}{2}\,.
\end{align}
For external quarks of definite helicity, with $n_i$-collinear momentum $p$, the spinor appearing in SCET Feynman rules is,
\begin{align}
\frac{1\,\pm\, \gamma_5}{2} \frac{\slashed{n}_i \slashed{\bar{n}}_i}{4} u(p) = \frac{\slashed{n}_i \slashed{\bar{n}}_i}{4} | p \pm \r \equiv |p \pm \r_{n_i},
\end{align}
where $|p \pm \r_{n_i}$ is a convenient short-hand notation for the projected spinor, and is proportional to $|n_i \pm \r$ (see \ref{eq:ketn}). Using this, we get the nonzero tree-level Feynman rules for incoming ($p^0<0$) and outgoing ($p^0>0$) quarks with definite helicity $\pm$ and color $\alpha$ (or $\balpha$),
\begin{align} \label{eq:quarkbaseFR}
\Mae{0}{\chi^\beta_{i\pm}}{q_\pm^\balpha(-p)}
&= \delta^{\beta\balpha}\,\ldel(\lp_i - p)\, \ket{(-p_i)\pm}_{n_i} \,,\\
\Mae{q_\pm^\alpha(p)}{\bar\chi^\bbeta_{i\pm}}{0}
&= \delta^{\alpha\bbeta}\,\ldel(\lp_i - p)\, {}_{n_i\!}\bra{p_i\pm}
\,, \nn \\
\Mae{0}{\bar\chi^\bbeta_{i\pm}}{\bar{q}_\mp^\alpha(-p)}
&= \delta^{\alpha\bbeta}\,\ldel(\lp_i - p)\,{}_{n_i\!}\bra{(-p_i)\pm}
\,, \nn \\
\Mae{\bar{q}_\mp^\balpha(p)}{\chi^\beta_{i\pm}}{0}
&= \delta^{\beta\balpha}\,\ldel(\lp_i - p)\, \ket{p_i\pm}_{n_i} \,.\nn
\end{align}
We wish to take advantage of the fact that fermions come in pairs, to simplify the treatment of Dirac structures when constructing an operator basis. \Ref{Moult:2015aoa} therefore defined the currents
\begin{align} \label{eq:jpm_def}
 J_{ij\pm}^{\balpha\beta}
& = \mp\, \sqrt{\frac{2}{\omega_i\, \omega_j}}\, \frac{   \ve_\mp^\mu(n_i, n_j) }{\langle n_j\mp | n_i\pm\rangle}   \, \bar{\chi}^\balpha_{i\pm}\, \gamma_\mu \chi^\beta_{j\pm}
\,, \\
 J_{ij0}^{\balpha\beta}
& =\frac{2}{\sqrt{\vphantom{2} \omega_i \,\omega_j}\,  [n_i n_j] } \bar \chi^\balpha_{i+}\chi^\beta_{j-}
\,, \qquad
(J^\dagger)_{ij0}^{\balpha\beta}=\frac{2}{\sqrt{ \vphantom{2} \omega_i \, \omega_j}  \langle n_i  n_j \rangle  } \bar \chi^\balpha_{i-}\chi^\beta_{j+}
. \nn
\end{align}
These currents have been defined such that they are invariant under an RPI-III transformation, which can be easily from the fact that $\omega_i$ scales as $\bn_i$ and the $| n_i \rangle$ scale as $\sqrt{n_i}$. In \secs{eeJets}{matching}, we will focus on the case of two back-to-back jet directions, so it is worth writing down the currents in that case. The tree-level Feynman rules for the currents with general sectors are given by
\begin{align}\label{eq:tree_feyn}
\l q_{+}^{\alpha_1}(p_1) \bar{q}_{-}^{\balpha_2}(p_2) |  J_{12 +}^{\bbeta_1 \beta_2} | 0 \r  
&= e^{i\Phi(J_{12+})}\  \delta^{\alpha_1 \bbeta_1} \delta^{\beta_2 \balpha_2} 
  \tilde{\delta} (\lp_1 - p_1) \tilde{\delta} (\lp_2 - p_2)
 \,,  \\
\l q_{-}^{\alpha_1}(p_1) \bar{q}_{+}^{\balpha_2}(p_2) |  J_{12 -}^{\bbeta_1 \beta_2} | 0 \r  
 &= e^{i\Phi(J_{12-})}\  \delta^{\alpha_1 \bbeta_1} \delta^{\beta_2 \balpha_2} 
 \tilde{\delta} (\lp_1 - p_1) \tilde{\delta} (\lp_2 - p_2)
 \,, \nn  \\
\l q_{+}^{\alpha_1}(p_1) \bar{q}_{+}^{\balpha_2}(p_2) |  J_{12\, 0}^{\bbeta_1 \beta_2} | 0 \r  
 &= e^{i\Phi(J_{12\,0})}\ \delta^{\alpha_1 \bbeta_1} \delta^{\beta_2 \balpha_2}  
   \tilde{\delta} (\lp_1 - p_1) \tilde{\delta} (\lp_2 - p_2)
 \,, \nn \\
\l q_{-}^{\alpha_1}(p_1) \bar{q}_{-}^{\balpha_2}(p_2) |  (J^\dagger)_{12 \,0}^{\bbeta_1 \beta_2} | 0 \r 
 &= e^{i\Phi(J^\dagger_{12\,0})}\ \delta^{\alpha_1 \bbeta_1} \delta^{\beta_2 \balpha_2} 
   \tilde{\delta} (\lp_1 - p_1) \tilde{\delta} (\lp_2 - p_2)
 \,, \nn 
\end{align}
where the phases appearing here are given by
\begin{align} \label{eq:phases1}
e^{i\Phi(J_{12\pm})} &= \l n_1 \mp | \bar{n}_1 \pm \r \l n_2 \pm | \bar{n}_2 \mp \r 
  \frac{\l \bar{n}_1 \pm | p_1 \mp \r \l \bar{n}_2 \mp | p_2 \pm \r}
   {8\,\sqrt{\vphantom{2}\omega_1 \, \omega_2}}
  \,,\\
e^{i\Phi(J_{12\, 0})} &= \l \bar{n}_1 n_1 \r \l n_2 \bar{n}_2 \r \frac{[p_1 \bar{n}_1] [\bar{n}_2 p_2] }{8\,\sqrt{\vphantom{2}\omega_1\,\omega_2}}
  \,,\nn \\
e^{i\Phi(J^\dagger_{12\, 0})} &= [ \bar{n}_1 n_1 ] [ n_2 \bar{n}_2 ] \frac{\l p_1 \bar{n}_1 \r \l \bar{n}_2 p_2 \r }{8\,\sqrt{\vphantom{2}\omega_1\, \omega_2}}
  \,.\nn
\end{align}
Note that the spinor products $\l \bar{n}_1 n_1 \r$, $\l n_2 \bar{n}_2 \r $, etc., depend on the choice of quantization axis for the spinors, and hence are not all trivial even though $n_1\cdot \bn_1=n_2\cdot\bn_2=2$.

If we consider two back-to-back collinear directions given by $n$ and $\bn$, our currents have definite helicity, given by
 \begin{align} \label{eq:jpm_back_to_bacjdef}
 & \text{h=$\pm$ 1:}
 & J_{n \bn \pm}^{\balpha\beta}
 & = \mp\, \sqrt{\frac{2}{\omega_n\, \omega_\bn}}\, \frac{   \ve_\mp^\mu(n, \bn) }{\langle \bn \mp | n \pm\rangle}   \, \bar{\chi}^\balpha_{n\pm}\, \gamma_\mu \chi^\beta_{\bn \pm}
 \,, \\
 & \text{h=0:}
 & J_{n \bn 0}^{\balpha\beta}
 & =\frac{2}{\sqrt{\vphantom{2} \omega_n \,\omega_\bn}\,  [n \bn] } \bar \chi^\balpha_{n+}\chi^\beta_{\bn-}
 \,, \qquad
 (J^\dagger)_{n \bn 0}^{\balpha\beta}=\frac{2}{\sqrt{ \vphantom{2} \omega_n \, \omega_\bn}  \langle n  \bn \rangle  } \bar \chi^\balpha_{n-}\chi^\beta_{\bn+}
 . \nn
 \end{align}
The current $J_{n \bn 0}^{\balpha \beta}$ transforms as a scalar under rotations about the $n$ axis, i.e. has helicity zero.\footnote{In Ref.~\cite{Moult:2015aoa} the $J_{ij0}^{\balpha\beta}$ current was denoted as $J_{ijS}^{\balpha\alpha}$. We choose to use the $0$ subscript to emphasize the helicity in the back-to-back case and conform with our notation for subleading currents below.}  Similarly, the currents $J_{n \bn \pm}^{\balpha \beta}$ have helicity $h=\pm1$.

Our notation above for the $\cB^a_{i\pm}$ and $\chi_{i \pm}^\alpha$ fields and the currents $J_{ij\pm}^{\bar\alpha\beta}$, $J_{ij\,0}^{\bar\alpha\beta}$, and $(J^\dagger)_{ij\, 0}^{\bar\alpha\beta}$ follows Ref.~\cite{Moult:2015aoa}, and these objects suffice for the construction of leading power operators. The $\Phi$ phases in \eq{phases1} were set to zero in Ref.~\cite{Moult:2015aoa}, since with only one particle in each collinear sector we are free to choose the $\tilde p_i$ to have zero label $\perp$-momenta, and with this choice all the phases vanish.  However, at subleading power, multiple collinear fields can be present in the same collinear sector and the phases can not a priori be set to zero. Note that the phases for each current are given by a product of phases,  one from each collinear sector. 

We now look at how to treat the sectors at subleading power that contain multiple collinear fields, as was discussed in \cite{Kolodrubetz:2016uim}.  Collinear gluons appear in gauge invariant building blocks of definite helicity, and therefore operators with multiple collinear gluons in the same sector can simply be obtained by multiplying copies of the gluon building blocks, such as $\cB^a_{i +}\cB^b_{i+}$. However, for quarks we must introduce new helicity currents. For $i=j$ the products of quark building blocks in \eq{jpm_def} all vanish ($\bar\chi_{i\pm}^{\bar\alpha} \gamma_\mu \chi_{i\pm}^\beta=0$, $\bar \chi_{i+}^{\bar\alpha} \chi_{i-}^\beta=0$, and $\bar\chi_{i-}^{\bar\alpha} \chi_{i+}^\beta=0$) and hence are not suitable for handling quarks in the same collinear sector. This follows from the SCET projection relations
\begin{align} \label{eq:proj}
\frac{\Sl n_i \Sl {\bar n}_i}{4}  \chi_{n_i}=\chi_{n_i}, \qquad \Sl n_i \chi_{n_i}=0 
  \,,
\end{align}
which enforce that a quark anti-quark pair of the same chirality, in the same sector, must have zero helicity, while a quark anti-quark pair of opposite chirality must have helicity $\pm 1$. 
Indeed, the scalar current $\bar \chi_{n_i} \chi_{n_i}=0$, vanishes, as do the plus and minus helicity components of the vector current $\bar \chi_{n_i} \gamma_\perp^{\pm} \chi_{n_i}=0$.

We therefore define the helicity operators involving two collinear quarks in the same sector as \cite{Kolodrubetz:2016uim}
\begin{align}\label{eq:coll_subl}
 & \text{h=0:}
 & J_{i0}^{\balpha \beta} 
  &= \frac{1}{2 \sqrt{\vphantom{2} \omega_{\bar \chi} \, \omega_\chi}}
  \: \bar \chi^\balpha_{i+}\, \Sl{\bar n}_i\, \chi^\beta_{i+}
   \,,\qquad
   J_{i\bar 0}^{\balpha \beta} 
  = \frac{1}{2 \sqrt{\vphantom{2} \omega_{\bar \chi} \, \omega_\chi}}
  \: \bar \chi^\balpha_{i-}\, \Sl {\bar n}_i\, \chi^\beta_{i-}
 \,, \\[5pt]
  & \text{h=$\pm 1$:}
 & J_{i\pm}^{\balpha \beta}
  &= \mp  \sqrt{\frac{2}{ \omega_{\bar \chi} \, \omega_\chi}}  \frac{\epsilon_{\mp}^{\mu}(n_i,\bar n_i)}{ \big(\l n_i \mp | \bar{n}_i \pm \r \big)^2}\: 
   \bar \chi_{i\pm}^\balpha\, \gamma_\mu \Sl{\bar n}_i\, \chi_{i\mp}^\beta
 \,. \nn
\end{align}
Note that these currents are only labeled by a single collinear sector $i$. Once again, we can easily see the RPI-III invariance, as the scaling power of $n_i$ is the same as the scaling power of $\bn_i$ in each of these currents. The  $J_{i0}^{\balpha \beta}$ and $J_{i\bar 0}^{\balpha \beta}$ transform as a scalar under rotations about the $n_i$ axis, i.e. have helicity zero. Similarly, the operators $J_{i\pm}^{\balpha \beta}$ have helicity $h=\pm1$.  These currents use only the reference vector associated with the particular jet in question for their construction. The Feynman rules for these currents with external quark states are
\begin{align}\label{eq:same_sector_feynrules}
& \l q_{+}^{\alpha_1}(p_1) \bar{q}_{-}^{\balpha_2}(p_2) |  J_{i0}^{\bbeta_1 \beta_2} | 0 \r  =  e^{i\Phi(J_{i0})}\: \delta^{\alpha_1 \bbeta_1} \delta^{\beta_2 \balpha_2}  \tilde{\delta} (\lp_1 - p_1) \tilde{\delta} (\lp_2 - p_2)\,,  \\ 
& \l q_{-}^{\alpha_1}(p_1) \bar{q}_{+}^{\balpha_2}(p_2) |  J_{i\bar 0}^{\bbeta_1 \beta_2} | 0 \r  =  e^{i\Phi(J_{i\bar 0})}\: \delta^{\alpha_1 \bbeta_1} \delta^{\beta_2 \balpha_2}  \tilde{\delta} (\lp_1 - p_1) \tilde{\delta} (\lp_2 - p_2)\,, \nn \\ 
& \l q_{+}^{\alpha_1}(p_1) \bar{q}_{+}^{\balpha_2}(p_2) |  J_{i +}^{\bbeta_1 \beta_2} | 0 \r =  e^{i\Phi(J_{i+})}\: \delta^{\alpha_1 \bbeta_1} \delta^{\beta_2 \balpha_2}  \tilde{\delta} (\lp_1 - p_1) \tilde{\delta} (\lp_2 - p_2)\,,\nn \\
& \l q_{-}^{\alpha_1}(p_1) \bar{q}_{-}^{\balpha_2}(p_2) |  J_{i -}^{\bbeta_1 \beta_2} | 0 \r =  e^{i\Phi(J_{i-})}\: \delta^{\alpha_1 \bbeta_1} \delta^{\beta_2 \balpha_2}  \tilde{\delta} (\lp_1 - p_1) \tilde{\delta} (\lp_2 - p_2)\,, \nn
\end{align}
where the phases here are given by
\begin{align}
 e^{i\Phi(J_{i0})} &= \frac{1}{2} \frac{ [p_1  \bar{n}_i ]  \l \bar{n}_i   p_2  \r}
  {\sqrt{\vphantom{2} \omega_1 \, \omega_2} }\,, \qquad
   e^{i\Phi(J_{i\bar0})} = \frac{1}{2} \frac{\l p_1  \bar{n}_i  \r   [ \bar{n}_i  p_2 ]}
  {\sqrt{\vphantom{2} \omega_1 \, \omega_2} }\,,\\
  e^{i\Phi(J_{i\pm})} &= \frac{1}{2} \frac{\l p_1 \pm | \bar{n}_i \mp \r \l \bar{n}_i \pm | p_2 \mp \r}{\sqrt{\vphantom{2} \omega_1 \, \omega_2}} 
    \,.\nn  
\end{align}
Together, the currents in \Eq{eq:jpm_def}, as well as \Eq{eq:coll_subl} are sufficient to describe collinear sectors with multiple quark fields. The complete set of quark currents will also include those with ultrasoft quark building blocks, which we will consider below in \Sec{sec:BPS}.

\subsection{$\cP_\perp$ Operators}\label{sec:pperp}

Along with multiple collinear fields in the same sector, subleading power operators can involve explicit insertions of the $\cP_{i\perp}^\mu$ operator, where $i$ denotes a particular collinear sector.  $\cP_{i\perp}^\mu$ is included as part of the operator to ensure that the Wilson coefficient includes only the dependence on the hard kinematics and has a uniform power counting. Since the $\cP_{i\perp}^\mu$ operator acts on the perpendicular subspace defined by the vectors $n_i, \bar n_i$, which is spanned by the polarization vectors $\epsilon(n_i, \bar n_i)$, it naturally decomposes as 
\begin{align} \label{eq:Pperppm}
\cP_{i\perp}^{+}(n_i,\bar n_i)=-\epsilon^-(n_i,\bar n_i) \cdot \cP_{i\perp}\,, \qquad \cP_{i\perp}^{-}(n_i,\bar n_i)=-\epsilon^+(n_i,\bar n_i) \cdot \cP_{i\perp}\,.
\end{align} 
It is important to emphasize that the subscript $\pm$ refers to the helicity about the $n_i$ axis, and not the lightcone components of the momenta. This decomposition is performed for the $\cP_{i\perp}$ operator in each sector. Note that it suffices to allow the operator $\cP_{i\perp}$ to act only on fields with collinear label $i$.\footnote{The $i$-sector operator $\cP_{i\perp}^\pm$ does not in general have a well defined power counting when acting on the field $\cB^\lambda_{j\perp}$ for $j \ne i$. Instead, when we carry out the multipole expansion we decompose any derivative acting on $\cB^\lambda_j$ using vectors for the $j$-sector, as $i\partial_{\rm tot}^\mu = (n_j^\mu/2) \bn_j\cdot\cP_j + \cP_{j\perp}^\mu + i\partial_{\rm us}^\mu$. Therefore only the ${\cal O}(\lambda)$ $\perp$-operator $\cP_{j\perp}^\mu$ acts on $\cB^\lambda_{j\perp}$.} Therefore, when acting on a field, we will drop the collinear sector label on $\cP_{i\perp}$, as it is determined by the label of the field. For example, we will simply write 
\begin{align}
 \cP_{i \perp} \cdot \cB_{i \perp} = \cP_{\perp} \cdot \cB_{i \perp} \,.
\end{align}

To see how this decomposition applies to operators written in more familiar notation, we consider the example operator $ \cP_\perp \cdot \cB_{i\perp}$. Using the completeness relation
\begin{align}\label{eq:completeness}
\sum\limits_{\lambda=\pm} \epsilon^\lambda_\mu(n_i,\bar n_i) \left (\epsilon^\lambda_\nu(n_i,\bar n_i) \right )^*=-g_{\mu \nu}+\frac{n_{i\mu} \bar n_{i\nu}+n_{i\nu} \bar n_{i\mu}}{n_i\cdot \bar n_i} = - g^\perp_{\mu \nu} ( n_i, \bn_i)\,,
\end{align}
the decomposition into our basis is given by\footnote{The sign convention in \Eq{eq:Pperppm} is made so that dot products, as in \Eq{eq:pdotbexample}, agree with using a $(+,-,-,-)$ metric for the contraction.}
\begin{align}\label{eq:pdotbexample}
\cP_\perp \cdot \cB_{i\perp}=-\cP_{\perp}^{+} \cB_{i-}-\cP_{\perp}^{-} \cB_{i+}\,.
\end{align}
When acting within an operator containing multiple fields, square brackets are used to denote which fields are acted upon by the $\cP_{\perp}^{\pm}$ operator. For example
\begin{align}
\cB_{i+} \left [ \cP_{\perp}^{\pm}  \cB_{i-}  \right]  \cB_{i+}\,,
\end{align}
indicates that the $\cP_{\perp}^{+}$ or $\cP_{\perp}^{-}$ operator acts only on the middle field.

In general, we can decompose the action of the $\cP_{\perp}^{\pm}$ operators into a superposition of terms where $\cP_{\perp}^{\pm}$ will act only on a single field within a quark current. To define a general notation for these currents, we will follow \cite{Kolodrubetz:2016uim} and use curly braces and write $\cP_{\perp}^{\pm}$ to the left of the current if it acts on only the first field in the current and write $(\cP_{\perp}^{\pm})^\dagger$ to the right of the current if it acts on only the second field. As an example, we can look at the helicity currents with two quarks in the same sector with a single $\cP_{\perp}^{\pm}$ insertion acting either on the first or second field, we write
\begin{align}\label{eq:p_perp_notation}
  \big\{ \cP_{\perp}^\lambda J_{i 0 }^{\balpha \beta} \big\}  
  & = \frac{1}{2 \sqrt{\vphantom{2}\omega_{\bar \chi} \, \omega_\chi }} \:
   \Big[  \cP_{\perp}^{\lambda}  \bar \chi^\balpha_{i +}\Big] \Sl {\bar n}_i \chi^\beta_{i+}
  \,, \\
 \big\{ J_{i0 }^{\balpha \beta} (\cP_{\perp}^{\lambda})^\dagger \big\}
  &=  \frac{1}{2\sqrt{\vphantom{2}\omega_{\bar \chi} \, \omega_\chi}} \:
  \bar \chi^\balpha_{i+} \Sl {\bar n}_i \Big[   \chi^\beta_{i+} (\cP_{\perp}^{\lambda})^\dagger \Big]
  \,. \nn
\end{align}
Following the notation defined above, the use of the use of curly brackets $\{ \cP_\perp^\lambda \cdots \}$ and $\{  \cdots (\cP_\perp^\lambda)^\dagger \}$ indicate that the $\cP_\perp$ operators act on only one of quark fields in the current, in the manner shown. 
Also note for \eq{p_perp_notation} that the choice of $\lambda=+$ or $-$ for $\cP_\perp^\lambda$ is independent of the $\pm$ choice for the quark building block fields. 
The same notation will be used for $\cP_{\perp}^{\pm}$ insertions into other currents. If we wish to instead indicate a $\cP_\perp$ operator that acts on both building blocks in a current then we use the standard square bracket notation, for example, $\big[\cP_\perp^\lambda J_{i \pm }^{\balpha \beta}\big]$.

The Feynman rules for collinear operators involving insertions of the $\cP_\perp$ operator follow straightforwardly from the corresponding Feynman rules without the  $\cP_\perp$ insertion, as given in Eqs. (\ref{eq:cBpm_def}), (\ref{eq:tree_feyn}), and (\ref{eq:same_sector_feynrules}). For example, using the $h=+1$ current of \Eq{eq:tree_feyn} as an example, we have
\begin{align}\label{eq:tree_feyn_perp}
\l q_{+}^{\alpha_1}(p_1) \bar{q}_{-}^{\balpha_2}(p_2) |  \big\{\cP_\perp^+ J_{12 +}^{\bbeta_1 \beta_2}  \big\}  | 0 \r  
&=-\epsilon^-(n_i,\bar n_i)\cdot \tilde p_{1\perp}\, e^{i\Phi(J_{12+})}\  \delta^{\alpha_1 \bbeta_1} \delta^{\beta_2 \balpha_2} 
  \tilde{\delta} (\lp_1 - p_1) \tilde{\delta} (\lp_2 - p_2)
 \,, \nn \\
 \l q_{+}^{\alpha_1}(p_1) \bar{q}_{-}^{\balpha_2}(p_2) | \big\{ \cP_\perp^-  J_{12 +}^{\bbeta_1 \beta_2} \big\}| 0 \r  
&=-\epsilon^+(n_i,\bar n_i)\cdot \tilde p_{1\perp}\, e^{i\Phi(J_{12+})}\  \delta^{\alpha_1 \bbeta_1} \delta^{\beta_2 \balpha_2} 
  \tilde{\delta} (\lp_1 - p_1) \tilde{\delta} (\lp_2 - p_2)
 \,, \nn \\
 \l q_{+}^{\alpha_1}(p_1) \bar{q}_{-}^{\balpha_2}(p_2) | \big\{ J_{12 +}^{\bbeta_1 \beta_2} (\cP_\perp^+)^\dagger   \big\} | 0 \r  
&=-\epsilon^-(n_i,\bar n_i)\cdot \tilde p_{2\perp}\, e^{i\Phi(J_{12+})}\  \delta^{\alpha_1 \bbeta_1} \delta^{\beta_2 \balpha_2} 
  \tilde{\delta} (\lp_1 - p_1) \tilde{\delta} (\lp_2 - p_2)
 \,, \nn  \\
 \l q_{+}^{\alpha_1}(p_1) \bar{q}_{-}^{\balpha_2}(p_2) | \big\{ J_{12 +}^{\bbeta_1 \beta_2} (\cP_\perp^-)^\dagger   \big\} | 0 \r  
&=-\epsilon^+(n_i,\bar n_i)\cdot \tilde p_{2\perp}\, e^{i\Phi(J_{12+})}\  \delta^{\alpha_1 \bbeta_1} \delta^{\beta_2 \balpha_2} 
  \tilde{\delta} (\lp_1 - p_1) \tilde{\delta} (\lp_2 - p_2)
 \,,  
\end{align}
where the phase $\Phi(J_{12+})$ was defined in \Eq{eq:phases1}.

\subsection{Color Bases}\label{sec:color}

To this point we have focused on the helicity structure of the operators, with color indices left free. Consider an operator formed from a product of the currents of collinear helicity fields, $O^{a_1\dotsb \alpha_n}$. An important feature of each of the collinear helicity fields is that they are collinear gauge invariant. Furthermore, all the operators, including $\cP_\perp$, behave as local operators with respect to ultrasoft gauge transformations, transforming like background fields. This implies that for the collinear operators, the constraints of gauge invariance are equivalent to that of a global color. It is therefore straightforward to write down a color basis for these operators. Following and generalizing the notation of \Ref{Moult:2015aoa}, we write
\begin{equation} \label{eq:Opm_color}
\vO^\dagger_\lotsdots  = O_\lotsdots^{a_1\dotsb \alpha_n}\, \vT^{\, a_1\dotsb \alpha_n}
 \,.
\end{equation}
Here $\vT^{\, a_1\dotsb\alpha_n}$ is a row vector of color structures that spans the color conserving subspace. The $a_i$ are adjoint indices and the $\alpha_i$ are fundamental indices.  The color structures do not necessarily have to be independent, but must be complete. Color structures which do not appear in the matching at a particular order will be generated by renormalization group evolution. Subtleties associated with the use of non-orthogonal color bases were discussed in detail in \Ref{Moult:2015aoa}, but they will not play a role here because we do not explicitly carry out the full factorization. Note that a decomposition as in \Eq{eq:Opm_color} is not possible in a gauge invariant manner in the full theory due to the covariant derivative, $D^\mu=\partial^\mu +ig A^\mu$, which does not transform uniformly under color. In other words, the full theory gauge invariance relates different possible color structures. 

The goal of the subscripts on the $O$ in \eq{Opm_color} is to enumerate the helicities of the gluon, quark, and derivative building blocks in the operator. We have introduced it in a manner that is general enough to account for the presence of the ultrasoft building blocks that will be discussed in later sections. Outside of all parentheses the ultrasoft gluon helicities are listed first, followed by a colon and the enumeration of the collinear gluon helicities (note that in the absence of any ultrasoft gluons we omit the colon entirely). The helicities of the various types of quark currents are listed inside the round parentheses and are separated by colons. (In addition we use a semicolon to distinguish quark currents involving different flavors, though this notation is not made explicit in \eq{Opm_color}.) Finally in the square brackets we first list the $\cP_\perp$ helicities, followed by a colon and then entries $\pm$ or $0$ to indicate the presence of ultrasoft derivatives $\partial_{us(i)\pm}$ and $\partial_{us(i)0}$ to be discussed below. Explicit examples that fully exploit this notation will be given in \sec{constr_subops}. When working at leading power, it is usually true that all possible combinations of helicity labels need to be included in the basis. As will be reviewed in \Sec{sec:ang_cons}, angular momentum conservation of the hard scattering process places many constraints on subleading power operators \cite{Kolodrubetz:2016uim}, so that various helicity combinations can be eliminated. Operators with $\cP_\perp$s acting on collinear gauge invariant objects can also often be eliminated by RPI and momentum conservation considerations.

We can demonstrate how the helicity and color decomposition works with two leading power examples, which only involve operators formed from collinear fields. We consider $pp \to 2$ jets, and for simplicity, we restrict to the $q \bar q\, q' \bar q'$ channel with distinct quark flavors to demonstrate the use of the collinear quark fields, and the $gggg$ channel to demonstrate the use of the collinear gluon fields. For the quark channel we take $q$ to be $n_1$ collinear,  $\bar{q}$ to be $n_2$ collinear,  $q'$ to be $n_3$ collinear, and  $\bar{q}'$ to be $n_4$ collinear. For the gluon channel we take the collinear gluon fields to lie in four distinct collinear sectors, labelled $n_1$ through $n_4$.

Using the notation of the traditional SCET building blocks in \Tab{tab:PC}, such operators are given by
\begin{align}
O^{\, \balpha \beta\bgamma\delta} = \bar \chi_{n_1}^\balpha \Gamma_1 \chi^\beta_{n_2}  \bar \chi^\bgamma_{n_3} \Gamma_2 \chi^\delta_{n_4}\,,
\end{align}
for the four quark case, and
\begin{align}\label{eq:oldgggg}
O^{a b c d} = \cB_{n_1 \perp}^{ \mu a} \cB_{n_2 \perp }^{\nu b} \cB_{n_3 \perp}^{ \sigma c} \cB_{n_4 \perp}^{ \delta d} \Gamma_{\mu \nu \sigma \delta},
\end{align}
for the four gluon case. Here $\Gamma_{\mu \nu \sigma \delta}$ is a shorthand for all allowed Lorentz structures, while $\Gamma_1$ and $\Gamma_2$ are shorthand for all possible Lorentz and Dirac structures, including contractions. Actually enumerating a minimal basis of these structures is a nontrivial task. On the other hand, using the helicity basis described in this section, we find \cite{Moult:2015aoa} that there are four independent helicity operators for the quark process,
\begin{alignat}{2} \label{eq:qqQQ_basis}
&O_{(+;+)}^{\balpha\bt\bgamma\delta}
= J_{12+}^{\balpha\bt}\, J_{34+}^{\bgamma\delta}
\,, \qquad &
&O_{(+;-)}^{\balpha\bt\bgamma\delta}
= J_{12+}^{\balpha\bt}\, J_{34-}^{\bgamma\delta}
\,,\\
&O_{(-;+)}^{\balpha\bt\bgamma\delta}
= J_{12-}^{\balpha\bt}\, J_{34+}^{\bgamma\delta}
\,, \qquad&
&O_{(-;-)}^{\balpha\bt\bgamma\delta}
= J_{12-}^{\balpha\bt}\, J_{34-}^{\bgamma\delta}
\,,\nn
\end{alignat}
with a color basis given by
\begin{equation} \label{eq:qqqq_color}
 \vT^{\, \al\bbeta\ga\bdelta} =
\Bigl(
  \de_{\al\bdelta}\, \de_{\ga\bbeta}\,,\, \delta_{\al\bbeta}\, \de_{\ga\bdelta}
\Bigr)
\,.\end{equation}
Similarly, we can immediately write down a basis of helicity operators for the gluon process
\begin{align} \label{eq:helicitygggg}
 \mathcal{O}^{a b c d}_{++++} 
 &= \frac{1}{4!}  \cB^a_{1 +} \cB^b_{2 +} \cB^c_{3 +} \cB^d_{4 +} \,\,,   &\mathcal{O}^{a b c d}_{+++-} &= \frac{1}{3!}  \cB^a_{1 +} \cB^b_{2 +} \cB^c_{3 +} \cB^d_{4 -}\,,
  \\
  \mathcal{O}^{a b c d}_{++--} 
  &= \frac{1}{4}  \cB^a_{1 +} \cB^b_{2 +} \cB^c_{3 -} \cB^d_{4 -}  \,\,,   &\mathcal{O}^{a b c d}_{+---} &= \frac{1}{3!}  \cB^a_{1 -} \cB^b_{2 -} \cB^c_{3 -} \cB^d_{4 +}\,,
   \nn \\
   \mathcal{O}^{a b c d}_{----} 
   &= \frac{1}{4!}  \cB^a_{1 -} \cB^b_{2 -} \cB^c_{3 -} \cB^d_{4 -}\,, \nn
\end{align}
with a color basis 
\begin{equation} \label{eq:color_gggg}
\vT^{ abcd} =
\frac{1}{2}\begin{pmatrix}
\tr[abcd] + \tr[dcba] \\ \tr[acdb] + \tr[bdca] \\ \tr[adbc] + \tr[cbda] \\
2\tr[ab]\, \tr[cd] \\ 2\tr[ac]\, \tr[db] \\ 2\tr[ad]\, \tr[bc]
\end{pmatrix}^{\!\!\!T}
.\end{equation}
(See \Ref{Moult:2015aoa} for a more detailed discussion. For a pedagogical review of color bases for QCD amplitudes see \Refs{Dixon:1996wi,Dixon:2013uaa}.) Note that there is no complication of dealing with Dirac structures, or using equations of motion to determine the minimal operator basis. The operators only encode relevant information on the helicities and color configuration of the particles. By using the helicity building blocks, which behave like scalar fields, we reduce the process of constructing an operator basis to simply enumerating unique combinations of the scalar objects. The hard kinematics is then described by the Wilson coefficients of these operators. 

The simplicity of the color bases for the collinear operators does not, however, naively extend to operators involving ultrasoft fields. Indeed, the ultrasoft derivatives are local at the ultrasoft scale, requiring the use of the ultrasoft covariant derivative $D_{us}$, and reintroducing the problem of the color decomposition that is present in the full theory, namely that the constraints of gauge invariance must be implemented.  In the next section we show how this issue can be overcome by using objects that account for the action of the BPS field redefinition. By working in terms of the resulting more non-local operators, we can reduce the constraints of gauge invariance to global color, enabling us to extend the simple color decomposition discussed in this section for the collinear building blocks to ultrasoft building blocks.

\subsection{Ultrasoft Gauge Invariant Helicity Building Blocks}\label{sec:BPS}

The BPS field redefinition is defined by \cite{Bauer:2002nz}
\be \label{eq:BPSfieldredefinition}
\cB^{a\mu}_{n\perp}\to \cY_n^{ab} \cB^{b\mu}_{n\perp} , \qquad \chi_n^\alpha \to Y_n^{\alpha \bar \beta} \chi_n^\beta,
\ee
and is performed in each collinear sector. Here $Y_n$, $\cY_n$ are fundamental and adjoint ultrasoft Wilson lines, respectively, and we note that
\be \label{eq:adjointtofundamental}
Y_n T^a Y_n^\dagger =  T^b {\cal Y}_n^{ba}\,.
\ee
  For a general representation, r, the ultrasoft Wilson line is defined by
\be
Y^{(r)}_n(x)=\bold{P} \exp \left [ ig \int\limits_{-\infty}^0 ds\, n\cdot A^a_{us}(x+sn)  T_{(r)}^{a}\right]\,,
\ee
where $\bold P$ denotes path ordering.  The BPS field redefinition has the effect of decoupling the ultrasoft degrees of freedom from the leading power collinear Lagrangian \cite{Bauer:2002nz}. When this is done consistently for S-matrix elements it accounts for the full physical path of ultrasoft Wilson lines~\cite{Chay:2004zn,Arnesen:2005nk}, so $Y_n^{(r)}$ may also occur with a path from $(0,\infty)$. Indeed for $e^+e^-\to$ dijets all the ultrasoft Wilson lines occur with paths from $(0,\infty)$ (see eg.~\cite{Bauer:2002ie}).

After the BPS field redefinition, the fields $\cB_{n\perp}$, and $\chi_n$ are ultrasoft gauge singlets, but still carry a global color index. We can use the BPS field redefinition to define ultrasoft quark and gluon fields that are ultrasoft gauge invariant. These operators are non-local at the ultrasoft scale, and involve the ultrasoft Wilson lines. For their construction, it is essential that the non-locality is dictated by the form of the BPS decoupling. In particular, the matching is first done onto the SCET Lagrangian pre-BPS field redefinition, which is local at the hard scale, and then the BPS decoupling is performed.

We begin by defining an ultrasoft gauge invariant quark building block field
\begin{align} \label{eq:usgaugeinvdef}
\psi_{us(i)}=Y^\dagger_{n_i} q_{us}\,,
\end{align}
where the direction of the Wilson line $n_i$ is a label for a collinear sector. Since the ultrasoft quarks are not naturally associated with an external label direction, $n_i$ can be chosen arbitrarily. However, there is often a convenient or obvious choice. The definition in \Eq{eq:usgaugeinvdef} straightforwardly generalizes to matter in an arbitrary representation.  We also perform the following decomposition of the gauge covariant derivative in an arbitrary representation, $r$,
\begin{align}\label{eq:soft_gluon}
Y^{(r)\,\dagger}_{n_i} i D^{(r)\,\mu}_{us} Y^{(r)}_{n_i }=i \partial^\mu_{us} + [Y_{n_i}^{(r)\,\dagger} i D^{(r)\,\mu}_{us} Y^{(r)}_{n_i}]=i\partial^\mu_{us}+T_{(r)}^{a} g \cB^{a\mu}_{us(i)}\,,
\end{align}
where we have defined the ultrasoft gauge invariant gluon building block field by
\begin{align} \label{eq:softgluondef}
g \cB^{a\mu}_{us(i)}= \left [   \frac{1}{in_i\cdot \partial_{us}} n_{i\nu} i G_{us}^{b\nu \mu} \cY^{ba}_{n_i}  \right] \,.
\end{align}
In the above equations the derivatives act only within the square brackets. Again, the choice of collinear sector label $n_i$ here is arbitrary. This is the ultrasoft analogue of the gauge invariant collinear gluon field, which can be written
\begin{align}  
g\cB_{n_i\perp}^{A\mu} =\left [ \frac{1}{\bar \cP}    \bar n_{i\nu} i G_{n_i}^{B\nu \mu \perp} \cW^{BA}_{n_i}         \right]\,.
\end{align}
From the expression for the gauge invariant ultrasoft gluon field of \Eq{eq:softgluondef} we see the price we have paid for working with ultrasoft gauge invariant operators. Unlike the ultrasoft fields, the building block field $\cB^{A\mu}_{us(i)}$ is non-local at the scale $\lambda^2$, and depends on the choice of a collinear direction $n_i$.  Note that in the case that the derivative operator acts in the opposite direction, we have
\begin{align}\label{eq:soft_gluon_reverse}
Y^{(r)\,\dagger}_{n_i} i \overleftarrow{D}^{(r)\,\mu}_{us} Y^{(r)}_{n_i }=i \overleftarrow \partial^\mu_{us} + [Y_{n_i}^{(r)\,\dagger} i \overleftarrow{D}^{(r)\,\mu}_{us} Y^{(r)}_{n_i}]=i \overleftarrow\partial^\mu_{us}-T_{(r)}^{a} g \cB^{a\mu}_{us(i)}\,,
\end{align}

The subleading Lagrangians  can also be written after BPS field redefinition in terms of the gauge invariant ultrasoft gluon field, as was done in Ref.~\cite{Larkoski:2014bxa},
\begin{align} \label{eq:LKB}
{\cal L}_{n_i}^{(1)\text{BPS}} &=   \hat K_{n_i}^{(1)} + \hat K_{\cB n_i\mu}^{(1) a}  g \cB_{us(i)}^{a\mu} +\cL_{\xi_n \psi_{us}}^{(1)\text{BPS}}
 \,, \\
 {\cal L}_{n_i}^{(2)\text{BPS}} &=   \hat K_{n_i}^{(2)} + \hat K_{\cB n_i\mu}^{(2) a}  g \cB_{us(i)}^{a\mu}  
   + \hat K_{\cB \cB n_i\mu\nu}^{(2) ab}  g \cB_{us(i)}^{a\mu} g \cB_{us(i)}^{b\nu} +\cL_{\chi_n \psi_{us}}^{(2)\text{BPS}}
   \,. \nn
\end{align} 
Here $\hat K_{n_i}^{(1)}$ and $\hat K_{n_i}^{(2)}$ contain only collinear fields and $i\partial_{us}^\mu$ derivatives. Their particular form is not relevant for the current discussion, we merely want to emphasize that the decomposition into gauge invariant building blocks is also convenient at the level of the Lagrangian, allowing it to be written in a factorized form. The terms $\cL_{\xi_n \psi_{us}}^{(1)\text{BPS}}$ and $\cL_{\xi_n \psi_{us}}^{(2)\text{BPS}}$ involve ultrasoft quark fields. Note that the superscript $i$ for $\hat K^{(i)}$ indicates the Lagrangian that these terms contribute to and not their individual power counting.

With the ultrasoft gauge invariant operators defined, we can now introduce ultrasoft fields and currents of definite helicity, following closely the collinear operators, but with some important differences. Throughout this section we implicitly work post BPS field redefinition. In \Sec{sec:constr_subops} we will show how an operator basis can be constructed prior and post BPS field redefinition, and how to easily treat the corresponding color bases.

We begin by defining ultrasoft gluon helicity fields which are ultrasoft gauge invariant
\begin{equation} \label{eq:Bus}
\cB^a_{us(i)\pm} = -\ve_{\mp\mu}(n_i, \bn_i)\,\cB^{a\mu}_{us(i)},\qquad  \cB^a_{us(i)0} =\bar n_\mu  \cB^{a \mu}_{us(i)}   
\,.\end{equation}
From \eq{softgluondef}, we can see that the ultrasoft gluon field satisfies the relation
\begin{align}\label{eq:ndotBiszero}
 n_i\cdot \cB^{a}_{us(i)}= 0\,.
\end{align}
For the collinear gauge invariant gluon field there are only two building block fields, which correspond to the two physical helicities of the gluon. On the other hand, for the ultrasoft gauge invariant gluon field we use three building block fields to describe the two physical degrees of freedom. This occurs because the ultrasoft gluons are homogeneous and not fundamentally associated with any direction. Therefore, without making a further gauge choice their polarization vectors do not lie in the perpendicular space of any fixed external reference vector.  Note that if we use the ultrasoft gauge freedom to choose $\cB^a_{us(j)0}=0$, then we will still have $\cB^a_{us(i)\pm}\ne 0$ and $\cB^a_{us(i)0} \ne 0$ for $i\ne j$. We could remove $\cB^a_{us(i)0}$ using the ultrasoft gluon equation of motion, in a manner analogous to how $[W_n^\dagger i n\cdot D_n W_n]$ is removed for the collinear building blocks. However this would come at the expense of needing to allow inverse ultrasoft derivatives, $1/(in\cdot\partial_{us})$, to appear when building operators. In the collinear case these $1/\bnP$ factors are ${\cal O}(\lambda^0)$ and can be absorbed into the Wilson coefficients, but this is not possible for the ultrasoft case. Therefore we choose to forbid inverse ultrasoft derivatives and allow $\cB^a_{us(i)0}$ to appear.

When writing down the Feynman rules for external ultrasoft gluons, we have the freedom to choose the reference vector for their polarizations. Choosing a general reference vector $k$, the tree level Feynman rules for the ultrsoft gluon field are
\begin{align} \label{eq:usgluonbaseFR}
\l g_{us}^a(p) | \cB_{us(i) \pm}^b(x) |0 \r & = - \varepsilon_{\mp \mu}(n_i \bn_i) \big[ \varepsilon^\mu(p,k) - \frac{p^\mu}{n \cdot p} n \cdot \varepsilon(p,k)\big] \delta^{ab}e^{ip\cdot x}\,, 
 \\
\l g_{us}^a(p) | \cB_{us(i) 0}^b(x) |0 \r & =  \big[ \bn \cdot \varepsilon^\mu(p,k) - \frac{\bn \cdot p}{n \cdot p} n \cdot \varepsilon(p,k)\big] \delta^{ab}e^{ip\cdot x}\,, 
\nn \\
\l 0 | \cB_{us(i) \pm}^b(x) |g_{us}^a(-p) \r & =  - \varepsilon_{\mp \mu}(n_i \bn_i) \big[ \varepsilon^{* \mu}(p,k) - \frac{p^\mu}{n \cdot p} n \cdot \varepsilon^*(p,k)\big] \delta^{ab}e^{ip\cdot x}\,,
\nn \\ 
\l 0 | \cB_{us(i) 0}^b(x) |g_{us}^a(-p) \r & = \big[ \bn \cdot \varepsilon^{*\mu}(p,k) - \frac{\bn \cdot p}{n \cdot p} n \cdot \varepsilon^*(p,k)\big] \delta^{ab}e^{ip\cdot x}\,. \nn
\end{align}

We also decompose the ultrasoft partial derivative operator $\partial_{us}^\mu$ into lightcone components,
\begin{equation}  \label{eq:partialus}
\partial_{us(i)\pm} = -\ve_{\mp\mu}(n_i, \bn_i)\,\partial^{\mu}_{us},\qquad   \partial_{us(i)0} =\bar n_{i\mu} \partial^{\mu}_{us}, \qquad \partial_{us(i)\bar 0} = n_{i \mu} \partial^{\mu}_{us}
\,.\end{equation}
If the $\partial_{us(i)\bar 0}$ operator acts on an $n$-collinear field, then it can be eliminated using the equations of motion \cite{Marcantonini:2008qn}, and therefore such a combination does not need to be included in our basis.
In contrast with the collinear case, we cannot eliminate the $\bn_i \cdot \partial_{us}$ using the equations of motion without introducing inverse ultrasoft derivative (e.g. $1/(\bn_i \cdot \partial_{us})$), so we instead keep these operators explicitly in our basis. When inserting ultrasoft derivatives into operators we will use the same curly bracket notation defined for the $\cP_\perp$ operators in \Eq{eq:p_perp_notation}. In other words, $\{i \partial_{us(i)\lambda} J\}$ indicates that the ultrasoft derivative acts from the left on the first field in $J$ and $\{  J (i\partial_{us(i) \lambda} )^\dagger \}$ indicates that it acts from the right on the second field in $J$. Note that the appearance of $\partial_{us(i) 0}$ and $\partial_{us(i)\bar 0}$ is also constrained by RPI-III invariance.

Gauge invariant ultrasoft quark fields also appear explicitly in the operator basis at subleading powers. Due to fermion number conservation they are conveniently organized into scalar currents. From \Tab{tab:PC}, we see that ultrasoft quark fields power count like $\lambda^3$. However, for factorization theorems involving a single collinear sector, as appear in factorization theorems describing a variety of both inclusive, and exclusive $B$ decays, 
operators involving ultrasoft quarks appear at leading power. The currents involving ultrasoft quarks that are necessary to define subleading power operators at any desired order are 
\begin{align} \label{eq:Jus}
 J_{i(us)\pm}^{\balpha\beta}
  &= \mp
  \frac{\ve_\mp^\mu(n_i, \bar n_i)}{ \l \bn_i \mp | n_i \pm \r}\: 
  \bar{\chi}^\balpha_{i\pm}\,  \gamma_\mu \psi^\beta_{us(i)\pm}
 \,,  \\
 J_{i(\overline{us})\pm}^{\balpha\beta}
  &=\mp  \frac{\ve_\mp^\mu(\bar{n}_i, n_i)}{ \l n_i \mp | \bar{n}_i \pm \r}\: \bar{\psi}_{us(i) \pm}^\balpha\, \gamma_\mu \chi^\beta_{i\pm} 
 \, , \nn \\
 J_{i(us)0}^{\balpha\beta}
  &= \bar \chi^\balpha_{i+}\psi^\beta_{us(i)-}
  \,, \qquad\qquad\qquad
  (J^\dagger)_{i(us)0}^{\balpha\beta}
  =   \bar \psi^\balpha_{us(i)-} \chi^\beta_{i+}
  \,, \nn\\
 J_{i(\overline{us})0}^{\balpha\beta}
  &=  \bar\psi^\balpha_{us(i)+} \chi^\beta_{i-}
  \,, \qquad\qquad\qquad
  (J^\dagger)_{i(\overline{us})0}^{\balpha\beta}
  =   \bar \chi^\balpha_{i-}\psi^\beta_{us(i)+}
  \,, \nn\\
 J_{(us)^2 ij\pm}^{\balpha\beta}
  &= \mp\, \frac{\ve_\mp^\mu(n_i, n_j)}{\langle n_j\mp | n_i\pm\rangle}\: 
  \bar{\psi}^\balpha_{us(i)\pm} \gamma_\mu \psi^\beta_{us(j)\pm} 
  \,,\nn \\
 J_{(us)^2 ij 0}^{\balpha\beta}
  &=
  \bar \psi^\balpha_{us(i)+}\psi^\beta_{us(j)-}
  \,,\qquad\qquad\quad
 (J^\dagger)_{(us)^2 ij 0}^{\balpha\beta}
  =
  \bar \psi^\balpha_{us(i)-}\psi^\beta_{us(j)+}
  \,. \nn
\end{align}
For the mixed collinear-ultrasoft currents we choose the collinear sector label $i$ in order to specify the ultrasoft quark building block field.  For the ultrasoft-ultrasoft currents, there is freedom to choose two sectors to use to construct the collinear gauge invariant ultrasoft fields, and we leave the choices arbitrary, $i$ and $j$, which appear as subscripts after the $(us)^2$ on the currents. Although the ultrasoft quark carries a label, this label is only associated with the ultrasoft Wilson line structure used to define the building block field. In particular, it is important to realize that the ultrasoft quark field does not satisfy the projection relations of \eq{proj}, that is satisfied by the collinear quarks. Note that all the ultrasoft quark currents are already RPI-III invariant except for $J_{(us)^2 ij 0}^{\balpha\beta}$ for generic $i$ and $j$. 

The Feynman rules for the collinear-ultrasoft currents to external states are 
\begin{align}\label{eq:feyn_col_us}
\l q_{+}^{\alpha_1}(p_1) \bar{q}_{us-}^{\balpha_2}(p_2) |  J_{1(us)+}^{\bbeta_1 \beta_2}(x) | 0 \r &= \frac{\sqrt{\omega_1}}{2}e^{i\Phi(J_{1(us)+})} \l n_1 p_2 \r\, \delta^{\alpha_1 \bbeta_1} \delta^{\beta_2 \balpha_2}  \tilde{\delta} (\lp_1 - p_1)  e^{ip_2\cdot x}\,,  \\
\l q_{-}^{\alpha_1}(p_1) \bar{q}_{us+}^{\balpha_2}(p_2) |  J_{1(us)-}^{\bbeta_1 \beta_2}(x) | 0 \r &=\frac{\sqrt{\omega_1}}{2} e^{i\Phi(J_{1(us)-})} [ n_1 p_2 ] \, \, \delta^{\alpha_1 \bbeta_1} \delta^{\beta_2 \balpha_2}  \tilde{\delta} (\lp_1 - p_1) e^{ip_2\cdot x}\,,  \nn \\
\l q_{us+}^{\alpha_1}(p_1) \bar{q}_{-}^{\balpha_2}(p_2) |  J_{2(\overline{us})+}^{\bbeta_1 \beta_2}(x) | 0 \r &=\sqrt{\frac{\omega_2}{2}} e^{i\Phi(J_{2(\overline{us})+})} [n_2 p_1] \, \delta^{\alpha_1 \bbeta_1} \delta^{\beta_2 \balpha_2}  \tilde{\delta} (\lp_2 - p_2) e^{ip_1\cdot x}\,, \nn  \\
\l q_{us-}^{\alpha_1}(p_1) \bar{q}_{+}^{\balpha_2}(p_2) |  J_{2(\overline{us})-}^{\bbeta_1 \beta_2}(x) | 0 \r &=\sqrt{\frac{\omega_2}{2}} e^{i\Phi(J_{2(\overline{us})+})} \l n_2 p_1\r \, \delta^{\alpha_1 \bbeta_1} \delta^{\beta_2 \balpha_2}  \tilde{\delta} (\lp_2 - p_2) e^{ip_1\cdot x}\,,  \nn \\
\l q_{+}^{\alpha_1}(p_1) \bar{q}_{us+}^{\balpha_2}(p_2) |  J_{1(us)0}^{\bbeta_1 \beta_2}(x) | 0 \r &= \sqrt{\frac{\omega_1}{2}} e^{i\Phi(J_{1(us)0})} [n_1 p_2] \delta^{\alpha_1 \bbeta_1} \delta^{\beta_2 \balpha_2}  \tilde{\delta} (\lp_1 - p_1) e^{ip_2\cdot x} \,, \nn \\
\l q_{us-}^{\alpha_1}(p_1) \bar{q}_{-}^{\balpha_2}(p_2) |  (J^\dagger)_{2(us)0}^{\bbeta_1 \beta_2}(x) | 0 \r &= \sqrt{\frac{\omega_2}{2}} e^{i\Phi((J^\dagger)_{2(us)0})}  \l n_2 p_1\r  \delta^{\alpha_1 \bbeta_1} \delta^{\beta_2 \balpha_2}  \tilde{\delta} (\lp_2 - p_2)  e^{ip_1\cdot x} \,,   \nn \\ 
\l q_{us+}^{\alpha_1}(p_1) \bar{q}_{+}^{\balpha_2}(p_2) |  J_{2(\overline{us})0}^{\bbeta_1 \beta_2}(x) | 0 \r &= \sqrt{\frac{\omega_2}{2}} e^{i\Phi(J_{2(\overline{us})0})}  [n_2 p_1]  \delta^{\alpha_1 \bbeta_1} \delta^{\beta_2 \balpha_2}  \tilde{\delta} (\lp_2 - p_2) e^{ip_1\cdot x}\,,   \nn \\
\l q_{-}^{\alpha_1}(p_1) \bar{q}_{us-}^{\balpha_2}(p_2) |  (J^\dagger)_{1(\overline{us}) 0}^{\bbeta_1 \beta_2}(x) | 0 \r &= \sqrt{\frac{\omega_1}{2}}  e^{i\Phi((J^\dagger)_{1(\overline{us})0})}  \l n_1 p_2\r   \delta^{\alpha_1 \bbeta_1} \delta^{\beta_2 \balpha_2}  \tilde{\delta} (\lp_1 - p_1)  e^{ip_2\cdot x} \,.   \nn 
\end{align}
The phases appearing in the Feynman rules of \Eq{eq:feyn_col_us} are defined as
\begin{align}
& e^{i\Phi(J_{1(us)+})}=\frac{ [\bar n_1 p_1 ]}{\sqrt{2 \omega_1}} \,,
& e^{i\Phi(J_{1(us)-})}=\frac{ \l\bar n_1 p_1 \r}{\sqrt{2 \omega_1} } \,, \\
& e^{i\Phi(J_{2(\overline{us})+})}=\frac{ \l  \bar n_2 p_2 \r}{\sqrt{2 \omega_2} } \,,
& e^{i\Phi(J_{2(\overline{us})+})}=\frac{[  \bar n_2 p_2 ]}{\sqrt{2 \omega_2} } \,, \nn \\
& e^{i\Phi(J_{1(us)0})}=\frac{ \l n_1 \bn_1 \r [ \bn_1 p_1]}{2\sqrt{2 \omega_1} } \,,
& e^{i\Phi((J^\dagger)_{2(us)0})}=\frac{- [ n_2 \bn_2 ] \l \bn_2 p_2\r}{2\sqrt{2 \omega_2} } \,, \nn\\
& e^{i\Phi((J)_{2(\overline{us})0})}=\frac{- \l n_2 \bn_2 \r [ \bn_2 p_2]}{2\sqrt{2 \omega_2} } \,,
& e^{i\Phi((J^\dagger)_{1(\overline{us})0})}=\frac{ [ n_1 \bn_1 ] \l \bn_1 p_1\r}{2\sqrt{2 \omega_1} } \,. \nn
\end{align}
The phases involve only the momentum of the collinear field.
Additional $\cP_\perp^\pm$ insertions into the mixed ultrasoft-collinear currents are defined following the notation of \eq{p_perp_notation}, where we note that in \SCETi only the collinear fields carry label $\perp$ momentum. The Feynman rules for the currents involving two ultrasoft quark fields are
\begin{align}
\l q_{us+}^{\alpha_1}(p_1) \bar{q}_{us-}^{\balpha_2}(p_2) |  J_{(us)^2 12+}^{\bbeta_1 \beta_2} | 0 \r &=\frac{[n_2 p_1] \l n_1 p_2 \r}{\sqrt{2}\,  n_1 \cdot n_2} \, \delta^{\alpha_1 \bbeta_1} \delta^{\beta_2 \balpha_2}\,,\\
\l q_{us-}^{\alpha_1}(p_1) \bar{q}_{us+}^{\balpha_2}(p_2) |  J_{(us)^2 12-}^{\bbeta_1 \beta_2} | 0 \r &=\frac{ \l n_2 p_1\r [ n_1 p_2] }{\sqrt{2}\,  n_1 \cdot n_2}\, \delta^{\alpha_1 \bbeta_1} \delta^{\beta_2 \balpha_2}\,, \nn \\
\l q_{us+}^{\alpha_1}(p_1) \bar{q}_{us+}^{\balpha_2}(p_2) |  J_{(us)^2 12\, 0 }^{\bbeta_1 \beta_2} | 0 \r &=  [p_1 p_2] \, \delta^{\alpha_1 \bbeta_1} \delta^{\beta_2 \balpha_2}\,, \nn \\
\l q_{us-}^{\alpha_1}(p_1) \bar{q}_{us-}^{\balpha_2}(p_2) |  (J^\dagger)_{(us)^2 12\,  0}^{ \bbeta_1 \beta_2} | 0 \r &=\l p_1 p_2 \r\, \delta^{\alpha_1 \bbeta_1} \delta^{\beta_2 \balpha_2}\,. \nn
\end{align}
Due to the dependence on two ultrasoft momenta, there is no natural way to separate phases from these Feynman rules as was done for previous currents.


\subsection{Constructing Operator Bases}\label{sec:constr_subops}

In this section we describe how the building blocks of the previous sections can be combined to define bases of hard scattering operators. The hard scattering component of the SCET Lagrangian, \Eq{eq:SCETLagExpand}, has an explicit expansion in powers of $\lambda$,
\begin{equation} \label{eq:Leff_sub}
\cL_{\text{hard}} = \sum_{j\geq0} \cL^{(j)}_{\text{hard}}
\,.\end{equation}
Here $j$ denotes suppression by ${\cal O}(\lambda^j)$ with respect to the leading power hard scattering operators. The effective Lagrangian for hard scattering operators at each power is given by,
\begin{align} \label{eq:Leff_sub_explicit}
\cL^{(j)}_{\text{hard}} = \sum_{\{n_i\}} \sum_{A,\cdot\cdot} 
  \bigg[ \prod_{i=1}^{\ell_A} \int \! \! \df \omega_i \bigg] \,
& \vO^{(j)\dagger}_{A\,\lotsdots}\big(\{n_i\};
   \omega_1,\ldots,\omega_{\ell_A}\big) \nn\\
& \times
\vC^{(j)}_{A\,\lotsdots}\big(\{n_i\};\omega_1,\ldots,\omega_{\ell_A} \big)
\,.
\end{align}
The appropriate collinear sectors $\{n_i\}$ are determined by the hard process being considered, and the sum over $A,\cdot\cdot$ runs over the full basis of operators that appear at this order, which are specified by either explicit labels $A$ and/or helicity labels $\cdot\cdot$ on the operators and coefficients. The operators also satisfy momentum conservation for various momenta, including the ${\cal O}(\lambda^0)$ $\omega_i$'s.  Here the $\vC^{(j)}_{A}$ are ${\cal O}(\lambda^0)$ Wilson coefficients, and are also vectors in the color subspace in which the $\mathcal{O}(\lambda^j)$ hard scattering operators $\vec O_A^{(j)\dagger}$ are decomposed. Explicitly, in terms of color indices, we have
\begin{equation} \label{eq:Cpm_color}
C_{\lotsdots}^{a_1\dotsb\alpha_n}
= \sum_k C_{\lotsdots}^k T_k^{a_1\dotsb\alpha_n}
\equiv \vT^{ a_1\dotsb\alpha_n} \vC_{\lotsdots}
\,.\end{equation}
Note that the color bases used to decompose the hard scattering operators at each order in $\lambda$ are in general distinct, since at higher powers more building blocks appear which carry additional color indices. The Wilson coefficients depend only on the jet directions, $\{n_i\}$, and the large label components of the operators $\omega_i$. The number of $\omega_i$'s depends on the specific operator we are considering since at subleading power multiple collinear fields can appear in the same collinear sector and we must consider the inclusion of ultrasoft building blocks with no $\omega_i$ labels. For a given operator $A\cdot\cdot$, we label the number of $\omega_i$'s as $\ell_{A\cdot\cdot}$ in \eq{Leff_sub_explicit} and do the integral over each $\omega_i$. The operators $\vO^\dagger_{\lotsdots}$ are given by products of the quark and gluon helicity operators of \Tab{tab:helicityBB}.

At subleading power, this is complicated by the fact that the ultrasoft Wilson lines which appear in the ultrasoft gauge invariant building blocks only appear after the BPS field redefinition. There are two possible approaches to dealing with this issue, both of which give the same answer. We take the attitude that one can use whichever is most convenient. A priori, we might like to first know how the ultrasoft fields are color contracted in the operator in order to choose the appropriate collinear vector $n_i$ for defining each ultrasoft building block. In one approach we first determine the full color basis involving contractions for the ultrasoft fields' color indices, and then choose the building blocks. Alternatively, we can simply pick fixed $n_i$ vectors for the ultrasoft building blocks, so that we have common ultrasoft objects for all operators. In this case there will be products of ultrasoft Wilson lines in the operators to compensate for a choice that does not correspond with the color contraction.

In either approach a basis of operators can be constructed in the form
\begin{equation} \label{eq:Opm_BPScolor}
\vO^\dagger_{\lotsdots}
= O_{\lotsdots}^{a_1\dotsb \alpha_n}\, \vT_{\BPS}^{\, a_1\dotsb \bar \alpha_n}
\,.
\end{equation}
Here $O_{\lotsdots}^{a_1\dotsb \alpha_n}$ is formed from products of the collinear and ultrasoft gauge invariant helicity building blocks constructed in the previous sections, for which the complete list is given in \Tab{tab:helicityBB}. The meaning of $ \vT_{\BPS}^{\, a_1\dotsb \alpha_n}$ will be discussed below. When utilizing $\cP_{i\perp}^{\pm}$, $\partial_{us(i)\pm}$, and $\partial_{us(i)0}$ to construct operators, these derivatives can either act on a single building block object like $\cB^a_{i\pm}$ or $\cB^a_{us(i)\pm}$, or on a bilinear object like one of the currents $J$. For cases where they act on a bilinear object we use the curly bracket notation introduced in \eq{p_perp_notation} to indicated which of the two objects in the bilinear the derivative acts on. Note that $\cP_{i\perp}^{\pm}$ can only act on $n_i$-collinear building blocks, whereas $\partial_{us(i)\pm}$ and $\partial_{us(i)0}$ can act on any building block field. The convention for the subscripts used on $O^{a_1 \cdots \alpha_n}$ was defined below \eq{Opm_color}. A couple of more complicated examples of the use of this notation are
\begin{align}
  O_{A\:-0: +-(+\bar 0:-)}^{ a_1 a_2 a_3 a_4 \balpha \beta \bgamma \delta \bar\sigma \tau}
 &=  \cB_{us(1)-}^{a_1} \cB_{us(1)0}^{a_2} \cB_{1+}^{a_3} \cB_{2-}^{a_4} 
  J_{3+}^{\balpha\beta} J_{2\,\bar 0}^{\bgamma\delta}  J_{45-}^{\bar\sigma\tau} 
  \,, \\
 O_{B\: -:+-(\bar 0:-)[+:-]}^{ a_1 a_2 a_3 \balpha \beta \bgamma \delta}
 &= \cB_{us(1)-}^{a_1}  \big[ i\partial_{us(1)}^- \cB_{2+}^{a_2} \big] \cB_{2-}^{a_3}    J_{3\,\bar 0}^{\balpha \beta}  \big\{ P_{\perp}^+ J_{14-}^{\bgamma\delta} \big\}  \,.\nn
\end{align}
Besides illustrating the correspondence between the operator subscripts and the helicity building blocks, these examples also highlight some of the limits of the notation. In particular, specifying the collinear gluon building block helicities does not determine whether the corresponding building blocks are in the same or different collinear sectors, specifying the helicities for mixed collinear-ultrasoft currents does not fix the sector of the collinear field in this building block, and the notation for the derivatives does not fix which object they act on.  To distinguish these type of differences we adopt additional explicit labels on the operators, which here are $A$ and $B$. 

\begin{table}
\begin{center}
\begin{tabular}{|ccc|cccc|cccc|}
 \hline \phantom{x} & \phantom{x} & \phantom{x} 
   & \phantom{x} & \phantom{x} & \phantom{x} & \phantom{x} 
   & \phantom{x} & \phantom{x} & \phantom{x} & \phantom{x} 
 \\[-13pt]                      
 $\cB_{i\pm}^a$ & $J_{ij\pm}^{\balpha\beta}$ & $J_{ij0}^{\balpha\beta}$ & $\cP_{\perp}^{\pm}$ & $J_{i\pm}^{\balpha \beta}$ &
    $J_{i0}^{\balpha \beta}$ & $J_{i\bar 0}^{\balpha \beta}$  &
  $\cB^a_{us(i)\pm}$ & $\cB^a_{us(i)0}$ & 
   $\partial_{us(i)\pm}$ & $\partial_{us(i)0}$ 
 \\[3pt] 
 $\lambda$ &  $\lambda^2$ &  $\lambda^2$
   & $\lambda$ & $\lambda^2$& $\lambda^2$ & $\lambda^2$ 
   & $\lambda^2$ & $\lambda^2$ & $\lambda^2$ & $\lambda^2$ 
 \\
  \hline  
\end{tabular}\\
\vspace{0.3cm}
\begin{tabular}{|cccc|cc|}
  \hline \phantom{x} & \phantom{x} & \phantom{x} 
   & \phantom{x} & \phantom{x} & \phantom{x} 
 \\[-13pt]                        
 $J_{i(us)\pm}^{\balpha\beta}$  &
 $J_{i(\overline{us})\pm}^{\balpha\beta}$ &
 $J_{i(us)0}^{\balpha\beta}$ &
 $J_{i(\overline{us})0}^{\balpha\beta}$ &
 $J_{(us)^2ij\pm}$ & $J_{(us)^2ij0}$ 
\\[3pt] 
 $\lambda^4$ &  $\lambda^4$ &  $\lambda^4$
   & $\lambda^4$ & $\lambda^6$& $\lambda^6$
 \\ 
 \hline
\end{tabular}
\end{center}
\vspace{-0.3cm}
\caption{
Power counting for the complete set of helicity building block operators in $\text{SCET}_\text{I}$, where the definitions for these objects are given in Eqs.~(\ref{eq:cBpm_def},~\ref{eq:definitehelicityquarkdef},~\ref{eq:jpm_def},~\ref{eq:coll_subl},~\ref{eq:Pperppm},~\ref{eq:Bus},~\ref{eq:partialus},~\ref{eq:Jus}). The building blocks also include the conjugate currents $J^\dagger$ in cases where they are distinct from the ones shown.
}
\label{tab:helicityBB}
\end{table}

The $\vT_{\BPS}^{\, a_1\dotsb \alpha_n}$ in \eq{Opm_BPScolor} generalizes the color structure decomposition  of \Sec{sec:color}, because it is defined to include ultrasoft Wilson lines that arise from the BPS field redefinition. It is important to emphasize that products of ultrasoft Wilson lines, like $Y_{n_1}^\dagger Y_{n_2}$, should not be viewed as independent building blocks. The structure of ultrasoft Wilson lines is entirely determined by the form of the BPS field decoupling. Several examples will be given below to further clarify this point. The form in \eq{Opm_BPScolor} is convenient for factorization, since for operators involving only collinear field insertions it is already written in a factorized form. 

It is also useful to have an equivalent form of the operator basis where we leave the ultrasoft Wilson lines in the operator itself, which would maintain the exact color decompositions of \Sec{sec:color}. In order to do this, we can define an operator $\widetilde{O}$ such that after BPS field redefinitions we have
\begin{align}\label{eq:Otildedef}
\vO^\dagger_{\lotsdots}
= \widetilde{O}_{\lotsdots}^{a_1\dotsb \alpha_n}\, \vT^{\, a_1\dotsb\bar \alpha_n}.
\end{align}
As we will demonstrate below, converting between $O$ and $\widetilde{O}$ is a simple exercise in reorganizing where we place the ultrasoft Wilson lines, and either form is equally valid as a basis. The decomposition in \eq{Opm_BPScolor} provides a compact way to track both the ultrasoft Wilson lines and color structure in one object, and hence will be used for many of our later examples.

We can compare the complete basis of helicity operator building blocks given in \Tab{tab:helicityBB} with the traditional form of the building block basis given in Table~\ref{tab:PC}. While there are more building blocks with the helicity operators, there is a great benefit from the fact that each of them is a scalar. So, while constructing operators with \Tab{tab:PC} is a complex exercise in deducing all possible Lorentz structures, with the helicity operator approach we simply have to write down all possible products of the helicity building blocks at a given power.

In order to demonstrate how \eq{Opm_BPScolor} works in practice, we now give some simple examples. We begin by discussing cases involving only collinear fields. Here the BPS field redefinition is not necessary to define color decomposed helicity operators, but is essential in the proof of factorization.  Note that insertions of the $\cP_\perp$ operator have no effect, since they do not carry color and commute with the ultrasoft Wilson lines $[\cP_\perp, Y_n]=0$, so we will only consider examples without $\cP_\perp$ insertions.  

We begin with a simple leading power example with gluon and quark current building blocks,
\be \label{eq:349}
O^{a\bar \alpha \beta}_{+(\pm)}= \cB_{1 +}^a\,J_{23\pm}^{\balpha\bt}\,,     \qquad     O^{a\bar \alpha \beta}_{-(\pm)}= \cB_{1 -}^a\,J_{23\pm}^{\balpha\bt} \,.
\ee
In this case there is a unique color structure before BPS field redefinition, namely
\be \label{eq:gqqcolor}
\vT^{ a \al\bbeta} = \left ( T^a \right )_{\alpha \bbeta}\,.
\ee
After BPS field redefinition, we find the Wilson line structure,
\be
\vT_{\BPS}^{ a \al\bbeta} = \left ( Y^\dagger _{n_2} T^b \cY_{n_1}^{ba}   Y_{n_3}  \right )_{\alpha \bar\beta}\,.
\ee
The key point is that this structure is entirely determined by the form of the operator in \Eq{eq:349}, as well as the structure of the BPS field redefinition. The ultrasoft Wilson lines arise only from the BPS field redefinition of the collinear fields in the building blocks, and not from hard matching. We can also reorganize the ultrasoft Wilson lines to group them into the operator, with the form in \eq{Otildedef}, which gives
\begin{align}
\widetilde{O}^{a\bar \alpha \beta}_{+(\pm)}
 &= (\cY_{n_1} \cB_{1 +})^a (Y^\dagger_{n_2}\,J_{23 \pm} Y_{n_3})^{\balpha\bt}\,,     \qquad     \widetilde{O}^{a\bar \alpha \beta}_{-(\pm)}= (\cY_{n_1} \cB_{1 -})^a\,(Y^\dagger_{n_2} J_{23 \pm} Y_{n_3})^{\balpha\bt} \,.
\end{align}
with the color structure as given in \Eq{eq:gqqcolor}.

 Another illustrative example is the four quark operator discussed in \sec{color}
\begin{alignat}{2} \label{eq:qqQQ_basis}
&O_{(+;+)}^{\balpha\bt\bgamma\delta}
= J_{q12+}^{\balpha\bt}\, J_{q'34+}^{\bgamma\delta}
\, \qquad &
&O_{(+;-)}^{\balpha\bt\bgamma\delta}
= J_{q12+}^{\balpha\bt}\, J_{q'34-}^{\bgamma\delta}
\,,\\
&O_{(-;+)}^{\balpha\bt\bgamma\delta}
= J_{q12-}^{\balpha\bt}\, J_{q'34+}^{\bgamma\delta}
\,, \qquad&
&O_{(-;-)}^{\balpha\bt\bgamma\delta}
= J_{q12-}^{\balpha\bt}\, J_{q'34-}^{\bgamma\delta}
\,,\nn
\end{alignat}
with color basis
\begin{equation} \label{eq:qqqq_color_second}
 \vT^{\, \al\bbeta\ga\bdelta} =
\Bigl(
  \de_{\al\bdelta}\, \de_{\ga\bbeta}\,,\, \delta_{\al\bbeta}\, \de_{\ga\bdelta}
\Bigr)
\,.\end{equation}
After BPS field redefinition, the color basis becomes
\begin{equation}\label{eq:353}
\vT_{\BPS}^{\, \al\bbeta\ga\bdelta} =
\left( \left[ Y_{n_1}^\dagger Y_{n_4}  \right ]_{\al\bdelta}\,\left[ Y_{n_3}^\dagger Y_{n_2}  \right ]_{\ga\bbeta}\,,\, \left[ Y_{n_1}^\dagger Y_{n_2}  \right ]_{\al\bbeta}\, \left[ Y_{n_3}^\dagger Y_{n_4}  \right ]_{\ga\bdelta}
\right)
\,.\end{equation}
This demonstrates how the decomposition in \Eq{eq:Opm_BPScolor} provides a convenient way of organizing the ultrasoft Wilson lines for the collinear operators.  Once again, we could choose to simply organize the ultrasoft Wilson lines in the operators $\widetilde{O}^{\balpha \beta \bar{\gamma} \delta}$, which would then be multiplied by the color structure given in \Eq{eq:qqqq_color_second}.

Next we consider an example involving an ultrasoft building block. We again consider four quarks, but now using currents built from two $n_1$-collinear building blocks $\chi_{1\pm}^\beta$ and $\bar\chi_{1\pm}^\balpha$, one $n_2$-collinear building block $\bar\chi_{2\pm}^\bgamma$, and one ultrasoft building block $\psi_{us(i)\pm}^\delta$.  We choose to pair up these fields into the currents $J_{1\lambda}^{\balpha\beta}$ with $\lambda=\pm,0,\bar 0$, and $J_{2(us)\lambda'}^{\bgamma\delta}$ with $\lambda'=\pm,0$ or $(J^\dagger)_{2(\overline{us})0}^{\bgamma\delta}$ . This notation implies that we have made the choice of the $2$-sector for the ultrasoft building block field, $\psi_{us(2)\pm}^\delta$. The basis of operators is then 
\begin{align} \label{eq:usBBeg}
  O_{A(\lambda:\lambda')}^{\balpha\beta\bgamma\delta} &= J_{1\lambda}^{\balpha\beta}\: J_{2(us)\lambda'}^{\bgamma\delta}
  \,,\qquad O_{B(\lambda:0)}^{\balpha\beta\bgamma\delta} = J_{1\lambda}^{\balpha\beta}\: (J^\dagger)_{2(\overline{us})0}^{\bgamma\delta}
  \,,
\end{align}
where there are 16 distinct operators once we take into account the allowed helicity choices for $\lambda$, $\lambda'$. Again we have the color basis $\bar T^{\balpha\beta\bgamma\delta}$ in \eq{qqqq_color_second}. Due to the different structure of fields the color basis after BPS field redefinition for this example is
\begin{align} 
\vT_{\BPS}^{\, \al\bbeta\ga\bdelta} =
\left( \left[ Y_{n_1}^\dagger Y_{n_2}  \right ]_{\al\bdelta}\,\left[ Y_{n_2}^\dagger Y_{n_1}  \right ]_{\ga\bbeta}\,,\, \delta_{\al\bbeta}\, \delta_{\ga\bdelta}
\right)
\,.
\end{align}
In the second color structure we have no $Y_{n_1}$ Wilson lines because the $n_1$-collinear fields are color contracted, and the correct $Y_{n_2}^\dagger$ Wilson line is already contained in the $\psi_{us(2)\pm}^\delta$ building block inside $J_{2(us)\lambda'}^{\bgamma\delta}$.  In the first color structure we have $[Y_{n_2}^\dagger Y_{n_1}]$ determined by the collinear building blocks, and then need $[Y_{n_1}^\dagger Y_{n_2}]$ in order to swap the ultrasoft reference vector to the $1$-sector when this factor multiplies $J_{2(us)\lambda'}^{\bgamma\delta}$.  Writing out the $\widetilde O^{\balpha\beta\bgamma\delta} \vT^{\, \al\bbeta\ga\bdelta}$ form of the operators for this case, the two color structures give
\begin{align}
 &\widetilde O_{A(\lambda:\lambda')}^{\balpha\beta\bbeta\alpha} 
  = J_{1\lambda}^{\balpha\beta}\: 
    \big( Y_{n_1} Y_{n_2}^\dagger J_{2(us)\lambda'} Y_{n_2} Y_{n_1}^\dagger \big)^{\bbeta\alpha}
   \,, \qquad 
   \widetilde O_{A(\lambda:\lambda')}^{\balpha\alpha\bgamma\gamma} 
   = J_{1\lambda}^{\balpha\alpha}\: J_{2(us)\lambda'}^{\bgamma\gamma}
   \,, \\
   & \widetilde O_{B(\lambda:0)}^{\balpha\beta\bbeta\alpha} 
    = J_{1\lambda}^{\balpha\beta}\: 
    \big( Y_{n_1} Y_{n_2}^\dagger (J^\dagger)_{2(\overline{us})0} Y_{n_2} Y_{n_1}^\dagger \big)^{\bbeta\alpha}
  \,, \qquad  \widetilde O_{B(\lambda:0)}^{\balpha\alpha\bgamma\gamma} 
  = J_{1\lambda}^{\balpha\alpha}\: (J^\dagger)_{2(\overline{us})0}^{\bgamma\gamma}
  \,. \nn
\end{align}
From this example we see that the advantage of using $\vT_{\BPS}^{\, \al\bbeta\ga\bdelta}$ is that we do not need to be concerned with the color contractions when specifying $O_{(\lambda:\lambda')}^{\balpha\beta\bgamma\delta}$.  

Note that an equivalent way of specifying the basis in \eq{usBBeg} would be to have started with the other possible grouping of the fermion building blocks, using $J_{12+}^{\bgamma\beta}$, $J_{12-}^{\bgamma\beta}$, $J_{120}^{\bgamma\beta}$ or $(J^\dagger)_{12\,0}^{\bgamma\beta}$, together with $J_{1(us)+}$, $J_{1(us)-}$, $J_{1(us)0}$, or $(J^\dagger)_{1(\overline{us})0}$. This again gives 16 choices for the operator helicities, but now we are using the ultrasoft building block with the reference direction as the $1$-sector.  The final result is the same with relations between elements of the two bases.

As another example involving ultrasoft building blocks, we consider a dijet operator that in a traditional approach has an ultrasoft derivative insertion, $\bar \chi_{n_1+}^\balpha \big( \bar{n}_2 \cdot {D}_{us}  \big) \chi_{n_2 -}^\beta$. After BPS field redefinition, we can rewrite this operator as
\be
\bar \chi_{n_1+} \big( \bar{n}_2 \cdot {D}_{us}  \big) \chi_{n_2 -}\to \bar \chi_{n_1+} \left (    Y^\dagger _{n_1} Y_{n_2}  \right ) i\bar{n}_2 \cdot {\partial}_{us} \chi_{n_2 -}+ \chi_{n_1+} \left (    Y^\dagger _{n_1} Y_{n_2}  \right ) g {\cal B}^a_{us(2)0} T^a \chi_{n_2 -}\,.
\ee
Here we have chosen the $2$ sector to define the ultrasoft building block gluon field, ${\cal B}_{us(2)0}^a = \bn_2\cdot B_{us(2)}^a$. In terms of our basis of ultrasoft gauge invariant helicity operators, we can write the two operators as 
\be
O^{\bar \alpha \beta}=\frac{-2}{\sqrt{\vphantom{2} \omega_1 \,\omega_2}\,  [n_1 n_2] } \bar \chi^{\bar \alpha}_{1 +} i\partial_{us(2) 0} \chi^{\beta}_{2 -}=\big\{J_{120}^{\balpha \beta} (i\partial_{us(2) 0})^\dagger\big\}\,, \qquad
\vT_{\BPS}^{\alpha \bar \beta}=\left (    Y^\dagger _{n_1} Y_{n_2}  \right )_{\alpha \bar\beta}\,,
\ee
and 
\be
O^{\bar \alpha \beta a}= \frac{2}{\sqrt{\vphantom{2} \omega_1 \,\omega_2}\,  [n_1 n_2] } \bar \chi^{\bar \alpha}_{1 +} g {\cal B}^a_{us(2) 0} \chi^{\beta}_{2 -}=J_{120}^{\balpha \beta} g {\cal B}^a_{us(2) 0} \,, \qquad \vT_{\BPS}^{\alpha \bar \beta a}=\left (    Y^\dagger _{n_1} Y_{n_2}   T^a \right )^{\alpha \bar\beta}.
\ee

The decomposition can always be done in the form of \eq{Opm_BPScolor}. Using ultrasoft Wilson lines arising from the BPS field redifinition, gauge covariant derivatives can be converted into the ultrasoft partial derivative and the ultrasoft building block gluon field. Products of the building blocks will then be linked in color space by ultrasoft Wilson lines. In particular, the remaining ultrasoft derivatives do not act on the linking Wilson lines, they only act on the ultrasoft Wilson lines and other fields that appear in the definitions of the ultrasoft gauge invariant building blocks. In general we see that given a form of an operator $O$ with color structures $\bar T$, it is straightforward to determine $\bar T_{\rm BPS}$, and thus also $\widetilde O$.

\subsection{Extensions}\label{sec:extensions}

Throughout this paper we have discussed our operator basis in the language of the \SCETi theory with massless quarks and in $d=4$ dimensions. SCET is also used in its \SCETii (with soft rather than ultrasoft fields) and SCET$_+$ \cite{Bauer:2011uc,Procura:2014cba,Larkoski:2015zka,Pietrulewicz:2016nwo} incarnations,  with massive collinear particles, and with dimensional regularization in $d=4-2\epsilon$ dimensions, so we discuss the necessary extensions for each of these here.

\subsubsection{SCET$_{\text{II}}$}\label{sec:scet_II}

For a certain class of observables, including $p_T$ dependent measurements and exclusive decays, the theory \SCETii provides the appropriate effective field theory description~\cite{Bauer:2002aj}. In \SCETii the soft and collinear modes live on the same invariant mass hyperbola, and therefore modes mediating interactions between the soft and collinear modes are off-shell and can be integrated out of the theory, generating both collinear and soft Wilson lines. A convenient way of matching to the \SCETii theory is to first match QCD onto an \SCETi theory with a larger offshellness for the collinear fields \cite{Bauer:2002aj}.  The BPS field redefinition can then be used to decouple the ultrasoft and collinear modes, giving rise to Wilson lines in the operators, as discussed in \Sec{sec:BPS}. One can then match this decoupled theory to \SCETii by lowering the virtuality of the external collinear modes to the soft scale, and relabeling ultrasoft modes as soft.  This matching calculation will be trivial (1-to-1) in cases where there are not time-ordered products of two or more subleading operators or Lagrangians in the \SCETi theory  \cite{Bauer:2003mga}. Furthermore, in the matching procedure terms of a given order in $\lambda$ in the \SCETi theory will only contribute to terms at that same order or higher in the \SCETii theory.  The resulting Wilson coefficients in the \SCETii theory now also involve $n\cdot k_s$ momenta of soft building blocks, from integrating out hard-collinear momenta with offshellness of order $\bn\cdot p_n n\cdot k_s\sim Q^2\lambda$.  

\begin{table}
\begin{center}
\begin{tabular}{|ccc|cccc|ccccc|}
 \hline \phantom{x} & \phantom{x} & \phantom{x} 
   & \phantom{x} & \phantom{x} & \phantom{x} & \phantom{x} 
   & \phantom{x} & \phantom{x} & \phantom{x} & \phantom{x} 
 \\[-13pt]                      
 $\cB_{i\pm}^a$ & $J_{ij\pm}^{\balpha\beta}$ & $J_{ij0}^{\balpha\beta}$ & $\cP_{\perp}^{\pm}$ & $J_{i\pm}^{\balpha \beta}$ &
    $J_{i0}^{\balpha \beta}$ & $J_{i\bar 0}^{\balpha \beta}$  &
  $\cB^a_{s(i)\pm}$ & $\cB^a_{s(i)0}$ & 
   $\partial_{s(i)\pm}$ & $\partial_{s(i)0}$ & $\partial_{s(i)\bar 0}$ 
 \\[3pt] 
 $\lambda$ &  $\lambda^2$ &  $\lambda^2$
   & $\lambda$ & $\lambda^2$& $\lambda^2$ & $\lambda^2$ 
   & $\lambda$ & $\lambda$ & $\lambda$ & $\lambda$ & $\lambda$ 
 \\
  \hline  
\end{tabular}\\
\vspace{0.3cm}
\begin{tabular}{|cccc|cc|}
  \hline \phantom{x} & \phantom{x} & \phantom{x} 
   & \phantom{x} & \phantom{x} & \phantom{x} 
 \\[-13pt]                        
 $J_{i(s)\pm}^{\balpha\beta}$  &
 $J_{i(\overline{s})\pm}^{\balpha\beta}$ &
 $J_{i(s)0}^{\balpha\beta}$ &
 $J_{i(\overline{s})0}^{\balpha\beta}$ &
 $J_{(s)^2ij\pm}$ & $J_{(s)^2ij0}$ 
\\[3pt] 
 $\lambda^{5/2}$ &  $\lambda^{5/2}$ &  $\lambda^{5/2}$
   & $\lambda^{5/2}$ & $\lambda^3$& $\lambda^3$
 \\ 
 \hline
\end{tabular}
\end{center}
\vspace{-0.3cm}
\caption{
Power counting for the full set of helicity building block operators in $\text{SCET}_\text{II}$. Again we must add the conjugate currents $J^\dagger$ in cases where they are distinct from the ones shown.
}
\label{tab:helicityBB2}
\end{table}

Since the helicity operator building blocks listed in \Tab{tab:helicityBB} are defined after BPS field redefinition, they also directly apply to the description of the hard scattering operators in the \SCETii theory. We simply need to replace ultrasoft fields and Wilson lines by those involving soft fields, $q_{us}\to q_s$, $Y_{n_1}\to S_{n_1}$, etc. This matching should be done at the level of the $\widetilde O$ operators in the \SCETi theory, so that we do not have Wilson lines grouped with the color structures in the resulting \SCETii operators. There is a 1-to-1 correspondence between the appropriate building blocks in the two theories. We have the same building blocks for collinear fields, and operators are now built from the soft building blocks for gluons
\begin{align}
\cB^a_{s(i)\pm} = -\ve_{\mp\mu}(n_i, \bn_i)\,\cB^{a\mu}_{s(i)},\qquad  \cB^a_{s(i)0} =\bar n_\mu  \cB^{a \mu}_{s(i)}   
\,,
\end{align}
where $\cB_{s(i)}^{a\mu}$ is defined as in \eq{softgluondef} but with soft fields.  The soft quark building block $\psi_{s(i)}= S_{n_i}^\dagger q_s$ appears in currents that are directly analogs of those containing the ultrasoft quark building block, namely\footnote{The notation here is chosen to make the \SCETi to \SCETii matching simpler, which in some cases comes at the expense of using a different normalization for the soft and collinear currents in the \SCETii theory.}
\begin{align} \label{eq:Js_scetII}
 J_{i(s)\pm}^{\balpha\beta}
  &= \mp \frac{2}{\sqrt{\vphantom{2} \omega_{i}}} \:
  \frac{\ve_\mp^\mu(n_i, \bar n_i)}{ \l \bn_i \mp | n_i \pm \r}\: 
  \bar{\chi}^\balpha_{i\pm}\,  \gamma_\mu \psi^\beta_{s(i)\pm}
 \,,  \\
 J_{i(\overline{s})\pm}^{\balpha\beta}
  &=\mp \frac{2}{\sqrt{\vphantom{2} \omega_{i}}} 
  \frac{\ve_\mp^\mu(\bar{n}_i, n_i)}{ \l n_i \mp | \bar{n}_i \pm \r}\: \bar{\psi}_{s(i) \pm}^\balpha\, \gamma_\mu \chi^\beta_{i\pm} 
 \, , \nn \\
 J_{i(s)0}^{\balpha\beta}
  &= \sqrt{\frac{2}{\omega_{i}}}\:
  \bar \chi^\balpha_{i+}\psi^\beta_{s(i)-}
  \,, \qquad\qquad\qquad
  (J^\dagger)_{i(s)0}^{\balpha\beta}
  = \sqrt{\frac{2}{\omega_{i}}}\:
  \bar \psi^\balpha_{s(i)-} \chi^\beta_{i+}
  \,, \nn\\
 J_{i(\overline{s})0}^{\balpha\beta}
  &= \sqrt{\frac{2}{\omega_{i}}}\:
  \bar\psi^\balpha_{s(i)+} \chi^\beta_{i-}
  \,, \qquad\qquad\qquad
  (J^\dagger)_{i(\overline{s})0}^{\balpha\beta}
  = \sqrt{\frac{2}{\omega_{i}}}\:
  \bar \chi^\balpha_{i-}\psi^\beta_{s(i)+}
  \,, \nn\\
 J_{(s)^2 ij\pm}^{\balpha\beta}
  &= \mp\, \frac{\ve_\mp^\mu(n_i, n_j)}{\langle n_j\mp | n_i\pm\rangle}\: 
  \bar{\psi}^\balpha_{s(i)\pm} \gamma_\mu \psi^\beta_{s(j)\pm} 
  \,,\nn \\
 J_{(s)^2 ij 0}^{\balpha\beta}
  &=
  \bar \psi^\balpha_{s(i)+}\psi^\beta_{s(j)-}
  \,,\qquad\qquad\quad
 (J^\dagger)_{(s)^2 ij 0}^{\balpha\beta}
  =
  \bar \psi^\balpha_{s(i)-}\psi^\beta_{s(j)+}
  \,. \nn
\end{align}
The full set of \SCETii building blocks is listed in \Tab{tab:helicityBB2}. Here the soft derivatives $\partial_{s(i)\pm}$ and $\partial_{s(i)0}$ act only on soft building block fields. From these building blocks we see that the helicity formalism can also be used to greatly simplify the construction of operator bases for \SCETii processes.

\subsubsection{SCET$_+$}\label{sec:scet_plus}

In cases where multiple measurements are made on the same jet additional degrees of freedom must be added to SCET. A general class of effective theories to describe such situations is the class of SCET$_+$ theories \cite{Bauer:2011uc,Procura:2014cba,Larkoski:2015zka,Pietrulewicz:2016nwo}. In addition to collinear and soft modes, these effective field theories typically contain (multiple) collinear-soft modes, which exhibit both a collinear, and a soft scaling. Such effective field theories have been used, for example, for the calculation \cite{Larkoski:2015kga} of the $D_2$ \cite{Larkoski:2014gra} jet substructure observable.

While subleading power corrections to SCET$_+$ theories have not been studied, we wish to emphasize that our helicity operator approach extends also straightforwardly to such theories. In SCET$_+$ theories, subleading power hard scattering operators will involve not only collinear and (ultra)soft fields, but also collinear soft fields. Although we will not do it in this paper, it is then a simple exercise to write a basis of helicity operator building blocks, including also such collinear soft fields.  Indeed, the helicity operator formalism has already been used to simplify matching calculations in SCET$_+$ at leading power in \cite{Pietrulewicz:2016nwo}.

\subsubsection{SCET with Massive Collinear Quarks}\label{sec:scet_massive}

The effective field theory description of the dynamics of the ultrasoft and collinear modes, as discussed thus far, is appropriate for massless quarks. For certain cases of phenomenological relevance, including boosted top production, the quark mass is an IR scale with the same parametric scaling as the $\perp$ momenta of collinear particles. In this case, the quark mass must be included in $\cL_\dyn$ \cite{Leibovich:2003jd,Rothstein:2003wh} for collinear quarks, soft or ultrasoft quarks, or both. For example, the leading power collinear quark Lagrangian for massive quarks is given by 
\begin{align}\label{eq:collinear_massive}
\cL^{m(0)}_{n \xi} &= \bar{\xi}_n\left[ i n \cdot D_{ns} +\left( i \slashed{D}_{n \perp}-m \right) W_n \frac{1}{\overline{\cP}_n} W_n^\dagger \left( i \slashed{D}_{n \perp} +m \right) \right]  \frac{\slashed{\bar{n}}}{2} \xi_n\,.
\end{align}
where $i D_{ns} = in\cdot\partial_{us} + g n\cdot A_{us} + gn\cdot A_n$.

Since the mass appears as an IR scale in the effective theory, the hard scattering operators for the case of massive SCET are the same as for massless SCET and the helicity operator basis presented in this paper also applies. However, as compared to the leading power Lagrangian for massless collinear quarks, the mass terms in \Eq{eq:collinear_massive} imply that quark chirality is not conserved by the soft and collinear dynamics of the effective field theory. This means that symmetry arguments relying on the conservation of helicity no longer apply. (We will use such symmetry arguments to reduce the number of hard scattering operators that can contribute to the $e^+e^- \to $ dijets or constrained Drell-Yan cross-section at $\mathcal{O}(\lambda^2)$ for massless SCET.) Nevertheless, the helicity operator basis still provides a convenient way of organizing hard scattering SCET operators involving boosted massive quarks.

\subsubsection{Evanescent Operators}

One subtlety of the helicity operator basis is that it relies on massless quarks and gluons having two helicities, a feature which is specific to $4$ dimensions. In dimensional regularization, divergences are regularized by analytically continuing the particle momenta to $d=4-2\epsilon$ dimensions. In a general scheme, the helicities of quarks and gluons live in $d^g_s$, and $d^q_s$ dimensional spaces respectively, although in most commonly used schemes, only $d^g_s$ is analytically continued. Different schemes within dimensional regularization differ in their treatment of $d^g_s$ for internal (unobserved) and external (observed) particles.  Evanescent operators \cite{Buras:1989xd,Dugan:1990df,Herrlich:1994kh} are defined as those whose tree level matrix elements vanish as $\epsilon\to 0$. However in loop calculations these matrix elements can multiply $1/\epsilon$ poles and lead to contributions that must be included in matching and higher order anomalous dimension calculations. For explicit discussions within the context of SCET calculations, see \Refs{Hill:2004if,Becher:2004kk,Beneke:2005gs,Ali:2007sj}. Such evanescent operators can not be specified using only our helicity building block fields. 

In \Ref{Moult:2015aoa} a discussion of scheme dependence was given for leading power helicity operators, and it was shown that evanescent operators do not appear when using SCET helicity operators for leading power matching calculations in exclusive jet processes. However, evanescent operators could appear at loop level when working to subleading power. The required extension of our helicity operator basis depends in detail on the regularization scheme, but in general requires the inclusion of additional fields, for example an $\epsilon$ scalar gluon $\cB^a_{\epsilon}$ to encode the $(-2\epsilon)$ transverse degrees of freedom, and quark currents $J_\epsilon$ which involve Dirac structures that would vanish if $\epsilon=0$. Since we do not consider the explicit one-loop matching and evolution of the helicity operators in this paper, we postpone a detailed discussion of evanescent operators to future work. However, we expect that at each loop order, the possible evanescent operators can be easily identified and treated. We note that a calculation of the leading power inclusive jet and soft functions in different regularization schemes, including the treatment of $\epsilon$ scalar gluons was presented in \Ref{Broggio:2015dga}.

\subsection{Parity and Charge Conjugation Properties}\label{sec:PandC}

It may initially seem that having distinct operators for each external helicity configuration greatly increases the number of operators. However, as is known from the study of helicity amplitudes, this is not the case. Parity and charge conjugation relations allow one to relate operators with distinct helicity configurations. An understanding of these relations is therefore essential for minimizing the number of matching calculations. The use of parity and charge conjugation relations is not limited to theories which exhibit C or P symmetry. Amplitudes and Wilson coefficients can be decomposed into pieces each of which have definite properties under C or P.

The C/P properties for the helicity building blocks involving collinear fields  are as follows. Under parity, we have
\begin{align} \label{eq:subPproperties}
\P\, \cB^a_{i\pm}(n_i; \omega_i; x)\, \P
 &= \cB^a_{i \mp}(n_i^\P; \omega_i; x^\P)
\,,  \\
\P\, J^{\balpha\beta}_{i j\pm}(n_i,n_j; \omega_i,\omega_j; x)\, \P
 &= J^{\balpha\beta}_{i j\mp}(n_i^\P,n_j^\P; \omega_i,\omega_j; x^\P) 
 \nn \,,  \\
\P J_{i 0}^{\balpha \beta} (n_i; \omega_1,\omega_2;x) \P 
 &= J_{i  \bar 0 }^{\balpha \beta}(n_i^\P; \omega_1,\omega_2;x^\P) 
  \,, \nn \\
\P J_{i \bar 0}^{\balpha \beta} (n_i; \omega_1,\omega_2;x) \P 
 &= J_{i  0 }^{\balpha \beta}(n_i^\P; \omega_1,\omega_2;x^\P) 
\nn \,, \\
\P J_{i  \pm}^{\balpha \beta} (n_i; \omega_1,\omega_2;x) \P 
 &= J_{i  \mp }^{\balpha \beta}(n_i^\P; \omega_1,\omega_2;x^\P) 
  \nn\,,   
\end{align}
where we have made the dependence on $n_i$, $\omega_i$, and $x$ explicit, and the parity-transformed vectors are $x^\P_\mu = x^\mu$.
Under charge conjugation we have,  
\begin{align} \label{eq:subCproperties}
\C\, \cB^a_{i\pm}(n_i;\omega_i)\, T^a_{\alpha\bbeta}\,\C
 &= - \cB^a_{i \pm}(n_i;\omega_i) T^a_{\beta\balpha}
 \,,\\
\C\, J^{\balpha\beta}_{i j\pm}(n_i,n_j;\omega_i, \omega_j)\,\C 
 &= -J^{\bbeta\alpha}_{j i\mp} (n_j,n_i;\omega_j, \omega_i)
\,,\nn \\
\C J_{i 0}^{\balpha \beta}(n_i; \omega_1,\omega_2) \C
 &= -J_{i \bar 0 }^{\bbeta \alpha}(n_i; \omega_2,\omega_1)
  \,, \nn \\
\C J_{i \bar 0}^{\balpha \beta}(n_i; \omega_1,\omega_2) \C
 &= -J_{i 0}^{\bbeta \alpha}(n_i; \omega_2,\omega_1) 
 \nn \,, \\
\C J_{i\pm}^{\balpha \beta} (n_i; \omega_1,\omega_2)  \C 
 &= -J_{i  \pm }^{\bbeta \alpha} (n_i; \omega_2,\omega_1)
  \nn \,.
\end{align}
Under parity, the $\cP_{\perp}^{\pm}$ operators transform as
\begin{equation} \label{eq:Pperpparity}
\P \cP_{\perp}^{\pm} \P = -\cP_{\perp}^{\mp} \, .
\end{equation}
Since charge conjugation exchanges the order of fields within a quark current, we have
\begin{equation} \label{eq:Pperpcharge}
C \{ J_{i 0}^{\balpha \beta}  (\cP_{\perp}^{\pm})^\dagger\}    C =  -\{(\cP_{\perp}^{\pm}) J_{i \,  \,\bar 0}^{\bbeta \alpha}\} \,,
\end{equation} 
along with similar relations for the other operators that involve $\cP^{\pm}_\perp$ insertions.

Although for our main example in \sec{eeJets} we will not use the operators of \Eq{eq:Jus}, which involve ultrasoft quarks, for completeness we give their C/P properties. Under parity, the mixed ultrasoft collinear operators transform as
\begin{align}
\P J_{i(us) \pm}^{\balpha \beta}(n_i;\omega_i)\P&=     J_{i(us) \mp}^{\balpha \beta}(n_i^\P;\omega_i)
\,, \\
\P J_{i(\overline{us}) \pm}^{\balpha \beta}(n_i;\omega_i)\P&=     J_{i(\overline{us}) \mp}^{\balpha \beta}(n_i^\P;\omega_i)
\,, \nn \\
\P J_{i(us) 0}^{\balpha \beta}(n_i;\omega_i)\P&= (J^\dagger)_{i(us) 0}^{\balpha \beta}(n_i^\P;\omega_i)    
\,, \nn \\
\P J_{i(\overline{us}) 0}^{\balpha \beta}(n_i;\omega_i)\P&=     (J^\dagger)_{i(\overline{us}) 0}^{\balpha \beta}(n_i^\P;\omega_i)
\,, \nn \\
\P (J^\dagger)_{i(us) 0}^{\balpha \beta}(n_i;\omega_i)\P&=     J_{i(us) 0}^{\balpha \beta}(n_i^\P;\omega_i)
\,, \nn \\
\P (J^\dagger)_{i(\overline{us}) 0}^{\balpha \beta}(n_i;\omega_i)\P&=     J_{i(\overline{us}) 0}^{\balpha \beta}(n_i^\P;\omega_i)
\,, \nn 
\end{align}
while under charge conjugation, we have
\begin{align}
\C J_{i(us) \pm}^{\balpha \beta}(n_i;\omega_i)\C&=-J_{i(\overline{us}) \mp}^{\bbeta \alpha}(n_i;\omega_i)
\,, \\
\C J_{i(\overline{us}) \pm}^{\balpha \beta}(n_i;\omega_i)\C&=-J_{i(us) \mp}^{\bbeta \alpha}(n_i;\omega_i)
\,, \nn \\
\C J_{i(us) 0}^{\balpha \beta}(n_i;\omega_i)\C&=     -J_{i(\overline{us}) 0}^{\bbeta \alpha}(n_i;\omega_i)
\,, \nn \\
\C J_{i(\overline{us}) 0}^{\balpha \beta}(n_i;\omega_i)\C&=     -J_{i(us) 0}^{\bbeta \alpha}(n_i;\omega_i)
\,, \nn \\
\C (J^\dagger)_{i(us) 0}^{\balpha \beta}(n_i;\omega_i)\C&=  -(J^\dagger)_{i(\overline{us}) 0}^{\bbeta \alpha}(n_i;\omega_i)   
\,, \nn \\
\C (J^\dagger)_{i(\overline{us}) 0}^{\balpha \beta}(n_i;\omega_i)\C&=     -(J^\dagger)_{i(us) 0}^{\bbeta \alpha}(n_i;\omega_i)
\,. \nn 
\end{align}
The C/P properties of the currents involving two ultrasoft quarks are identical to those of standard quark bilinears, so we do not give them here. Finally, the C/P properties of the SCET$_\text{II}$ operators of \Eq{eq:Js_scetII}, which involve soft quarks, are easily obtained from the SCET$_\text{I}$ results above.

As a simple example to demonstrate the use of C/P relations, we consider $e^+e^- \to q\bar q$ through an off-shell photon at leading power in SCET, which we will consider in more depth in \sec{eeJets}. We will label the quark and antiquark by $1,2$ and the electron and positron by $3,4$.  It is well known that at leading power there is a single current using traditional SCET operators,
\begin{align}\label{eq:trad_operator}
J^{\mu \balpha \beta }= \bar{\chi}^\balpha_{1} \gamma^\mu \chi^\beta_{2}.
\end{align}
The free Lorentz index is contracted with the leptonic tensor to form a scalar. On the other hand, the helicity basis consists of four scalar operators, 
\begin{alignat}{2}\label{eq:ee_helop}
O^{(0)\balpha \beta}_{(+;+)}
&=   J^{\bar \alpha \beta}_{12+}\,J_{e+}
\,,& \qquad 
O^{(0)\balpha \beta}_{(+;-)}
&=  J^{\bar \alpha \beta}_{12+ }\,J_{e-}
\,,  \\
O^{(0)\balpha \beta}_{(-;+)}
&=  J^{\bar \alpha \beta}_{12- }\,J_{e+ }
\,, &\qquad
O^{(0)\balpha \beta}_{(-;-)}
&= J^{\bar \alpha \beta}_{12- }\,J_{e- }
\,,\nn
\end{alignat}
which already include the leptons through the lepton helicity current $J_{e\pm}$. The leptonic helicity currents are defined in an identical manner to the leading power QCD current  of \Eq{eq:jpm_def}, but without the corresponding Wilson lines or color indices
\begin{equation} 
J_{e\pm}\equiv J_{34\pm}
= \mp\, \sqrt{\frac{2}{\omega_3\, \omega_4}}\, \frac{   \ve_\mp^\mu(n_3, n_4) }{\langle n_4\mp | n_3\pm\rangle}   \, \bar{e}_{3\pm}\, \gamma_\mu e_{4\pm}
\,.\end{equation}
For the $e^+e^- \to q\bar q$ process, there is a unique color structure for either \eq{trad_operator} or \eq{ee_helop},
\begin{align}
\vT^{ \alpha \bar \beta}=(\delta_{\alpha \bar \beta})\,.
\end{align}

Invariance under parity implies that the Wilson coefficients for the helicity operators are related by
\begin{align} \label{eq:Cdijetparity}
C_{(+;+)}(n_1,n_2;\omega_1,\omega_2;\omega_3,\omega_4)
  &=C_{(-;-)}(n_1^\P,n_2^\P;\omega_1,\omega_2;\omega_3,\omega_4) 
 \,, \\
C_{(+;-)}(n_1,n_2;\omega_1,\omega_2;\omega_3,\omega_4)
 &=C_{(-;+)}(n_1^\P,n_2^\P;\omega_1,\omega_2;\omega_3,\omega_4) 
 \,. \nn
\end{align}
When doing the matching, we sum over the $n$'s as in \eq{Leff_sub_explicit}, so we are free to rewrite $n_i^\P \rightarrow n_i$. Since we consider the process to all orders in the strong interaction, but only leading order in the electromagnetic interaction, the leptons couple through the current $\langle 1\pm |\gamma^\mu | 2\pm \rangle=\langle 2\mp |\gamma^\mu | 1\mp \rangle$. This implies the further relation
\begin{align} \label{eq:CdijetC}
C_{(+;+)}(n_1,n_2;\omega_1,\omega_2;\omega_3,\omega_4)
  &=C_{(+;-)}(n_1,n_2;\omega_1,\omega_2;\omega_4,\omega_3) \,.
\end{align}
These relations can be easily checked by considering the tree level matching. At tree level, the Wilson coefficients are given by
\begin{align}\label{eq:ee_wilson}
C_{(+;+)}&=- e^2 q^2   
 \frac{2[13]\ang{24}}{s_{34}} \,,~~
 C_{(+;-)}=-e^2 q^2   
 \frac{2[14]\ang{23}}{s_{34}}  \,,~~ \\
C_{(-;+)}&=-e^2 q^2   
\frac{2[23]\ang{14}}{s_{34}}  \,,~~
C_{(-;-)}=-e^2 q^2
\frac{2[24]\ang{13}}{s_{34}}  \,. \nn
\end{align}
These satisfy the above relations, by noting from \app{helicity} that parity simply interchanges $[]\leftrightarrow \langle \rangle$.  Together the three relations in \eqs{Cdijetparity}{CdijetC} provide the necessary information to indicate that the matching onto the helicity operator basis came from a vector current.  Therefore, only one coefficient of the helicity operators needs to be computed in a matching calculation at any order in $\alpha_s$. Note that the basis constructed in \eq{ee_helop} also works for mediation through a Z-boson, where axial coupling also needs to be considered.

Further examples of the use of C and P to simplify helicity operator bases can be found in Ref.~\cite{Moult:2015aoa}, and below in \sec{eeJets}.

\subsection{Constraints from Angular Momentum Conservation}\label{sec:ang_cons}

The use of operators with definite helicities makes manifest symmetries related to rotational invariance. As discussed in detail in \cite{Kolodrubetz:2016uim}, constraints from conservation of angular momentum can greatly reduce the basis of hard scattering operators appearing at subleading powers, when multiple collinear fields can appear in each collinear sector. Conservation of angular momentum implies the general constraint \cite{Kolodrubetz:2016uim} 
\begin{align} \label{eq:Jmin}
J^{(i)}_\text{min}\leq \sum_{j \text{ with }  \hat n_j\neq \hat n_i} J_{\rm min}^{(j)} \,,
\end{align}
where $J^{(i)}_\text{min}$ is the minimum angular momentum carried by the $n_i$-collinear sector. This can be related to the helicities of the building blocks in a given sector by $J^{(i)}_\text{min}=|h_{n_i}^{\rm tot}|$, where the helicities in the $n_i$-collinear sector of some operator add up to $h_{n_i}^{\rm tot}$. From this, we immediately get the constraint,
\begin{align} \label{eq:hmin}
|h_{n_i}^\text{tot}| \leq \sum_{j \text{ with }  \hat n_j\neq \hat n_i}  | h_{n_j}^\text{tot}|  \,,
\end{align}
where it is important to count back-to-back collinear directions only once in this sum, considering the helicity about their common axis.

\begin{figure}[t!]
%
%
\begin{center}
\includegraphics[width=0.35\columnwidth]{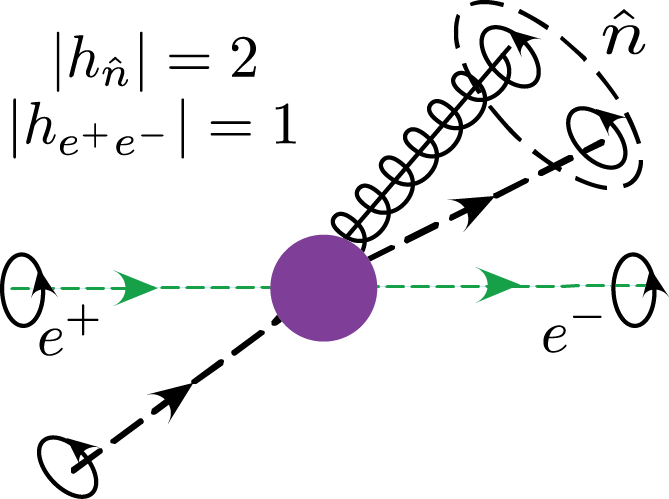} 
\hspace{1.4cm}
\includegraphics[width=0.35\columnwidth]{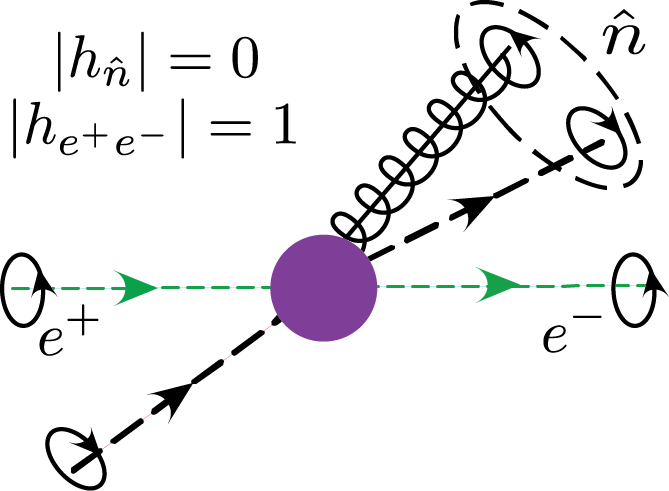} 
\raisebox{0cm}{ \hspace{-3.0cm} 
  $a$)\hspace{6.6cm}
  $b$)\hspace{2.4cm} }
\\[-25pt]
\end{center}
\vspace{-0.4cm}
\caption{ 
A illustration of the helicity selection rule for $e^+e^-\to $ dijets. In a) the collinear particles along the $n$ axis carry $|h=2|$, and have a vanishing projection onto the $J_{e\pm}$ current. In b), the collinear particles carry $|h=0|$ and therefore have a non-vanishing projection onto the $J_{e\pm}$ current. } 
\label{fig:helicity_constraint}
\end{figure}

These selection rules are particularly simple for the case of $e^+e^- \to$ dijets (or constrained Drell-Yan) which we study here. In this case, there are two axes along which particles move, namely the axis of the colliding $e^+e^-$ pairs, and the $n$ axis of the jets. In the helicity operator approach, the helicities of all operators are defined with respect to these axes. For the case of $e^+e^-\to$ dijets proceeding through an off-shell photon or $Z$ boson, the coupling to vector bosons guarantees that the electron pair has a combined helicity along the collision axis of $|h_{e^+e^-}|=1$, as shown in \Fig{fig:helicity_constraint}. To have a non-zero amplitude, the helicity state of the outgoing jets defined along the $n$ axis must have an overlap with this helicity $1$ state. In particular, we must have $h_{n}^\text{tot} = -1,0,1$. At subleading power, when there are multiple collinear fields in the $n$ and $\bar n$ sectors, this means that the helicities of the fields must be arranged in particular combinations, and considerably simplifies the basis. 

As an example, consider a subleading power operator involving an additional collinear gluon field in the $n$ collinear sector. Without imposing constraints from angular momentum conservation, a basis of allowed helicity operators is
\begin{alignat}{2} \label{eq:Z1_basis_cons}
&O_{+(+;\pm)}^{(1)a\,\balpha\bt}
=\cB_{n+}^a \, J_{n\bar n\,+}^{\balpha\bt}\,  J_{e\pm }
\,,\qquad &
&O_{+(-;\pm)}^{(1)a\,\balpha\bt}
= \cB_{n+}^a\, J_{n\bar n\,-}^{\balpha\bt}\, J_{e\pm }
\,,  \\
&O_{-(+;\pm)}^{(1)a\,\balpha\bt}
= \cB_{n-}^a\, J_{n\bar n\,+}^{\balpha\bt}\, J_{e\pm }
\,,\qquad &
&O_{-(-;\pm)}^{(1)a\,\balpha\bt}
= \cB_{n-}^a \, J_{n\bar n\, -}^{\balpha\bt}\,J_{e\pm }
\,. \nn
\end{alignat}
However, the first and fourth operators have $|h_{n}^\text{tot}| = 2$ and thus are not allowed, and can be eliminated from the basis. This configuration is shown schematically in \Fig{fig:helicity_constraint} a). Only the operators where the helicity of the $n$ collinear quark and gluon fields are opposite are allowed in the basis. The actual basis of allowed operators is therefore simpler, and is just given by
\begin{alignat}{2} \label{eq:Z1_basis_cons_reduced}
&O_{-(+;\pm)}^{(1)a\,\balpha\bt}
= \cB_{n-}^a\, J_{n\bar n\,+}^{\balpha\bt}\, J_{e\pm }
\,,\qquad &
&O_{+(-;\pm)}^{(1)a\,\balpha\bt}
= \cB_{n+}^a\, J_{n\bar n\,-}^{\balpha\bt}\, J_{e\pm }
\,,  
\end{alignat}
A schematic illustration of these configurations is shown in \Fig{fig:helicity_constraint} b). Using these restrictions, we have therefore eliminated half of the potential operators. Similar constraints will play an important role in simplifying the complete basis of subleading power operators given in \Sec{sec:eeJets}.

\section{Hard Scattering Operators with Two Collinear Directions}\label{sec:eeJets}

To demonstrate the simplicity of the helicity-operator approach, we construct the ${\cal O}(\lambda)$ and ${\cal O}(\lambda^2)$ basis of power suppressed hard scattering operators with two collinear directions. A summary of the complete set of operators is given in \Tab{tab:summary}. For concreteness, we take the color singlet particles involved in the hard scattering to be an $e^+e^-$ pair, with the interaction proceeding through an off-shell $\gamma$ or $Z$. The basis is valid to all orders in $\alpha_s$, and leading order in the electroweak couplings. Since the helicity operators from which our basis of hard scattering operators are composed are manifestly crossing symmetric, our operators can be used as a basis of hard scattering operators in factorization proofs for $e^+e^- \to$ dijet event shapes, constrained Drell-Yan, or DIS producing a single jet. 

The extension to $\cO(\lambda^2)$ is necessary, since when we square power suppressed jet amplitudes to calculate a cross section, the $\cO(\lambda)$ power corrections to the cross section tend to vanish \cite{Beneke:2003pa,Lee:2004ja,Freedman:2013vya}. We will explicitly show the vanishing of the contributions to the $\cO(\lambda)$ cross section from hard scattering contributions in \Sec{sec:ee_lambda}. We then present the hard scattering operators that contribute to the factorized cross section at $\cO(\lambda^2)$ in \Sec{sec:ee_lambda2}. A set of ${\cal O}(\lambda)$ and ${\cal O}(\lambda^2)$ operators for $e^+e^-\to$ dijets has been presented in \Ref{Freedman:2013vya}, although no claim of completeness was made, and a different formulation of SCET was used. In \Sec{sec:compare_earlier} we will briefly compare our all orders basis with the operators of \Ref{Freedman:2013vya}. 

{
\renewcommand{\arraystretch}{1.4}
\begin{table}[t!]
\hspace{-0.4cm} 
\scalebox{0.842}{
\begin{tabular}{| c | l | l | c |c| c |}
\hline 
$\!$Order$\!$ & $\!$Category$\!$ &  Operators (equation number) 
  & \#$\!$ helicity & \#$\!$ of
  & $\!\sigma_{2j}^{\cO(\lambda^2)}\!\! \ne\! 0\!$ 
\\[-8pt]
& & & configs & $\!$color$\!$ &
\\ \hline 
$\!\mathcal{O}(\lambda^0)\!$  
 & $\! e \bar e q \bar q$ & $O_{(\lambda_1;\pm)}^{(0)\balpha\bt}
  =J^{\balpha\bt}_{n \bar n \lambda_1}J_{e \pm }$ \,(\ref{eq:LP_basis})
  & 4 & 1 & $\checkmark$ \\
\hline
$\!\mathcal{O}(\lambda)\!$ 
 & $\! e \bar{e} q \bar{q} g$  
  & $O_{n\bar n1,2 \lambda_1 ( \lambda_2;\pm)}^{(1)a\,\balpha\bt}
  = \cB_{n,\bar n  \lambda_1}^a\, J_{n\bar n\, -\lambda_1}^{\balpha\bt}\,J_{e\pm }$ \,(\ref{eq:Z1_basis},\ref{eq:Z1_basis_flip}) 
  &8 & 1 & $\checkmark$  
  \\
 &   & $O_{\bar{n}\lambda_1 ( \lambda_2:\pm)}^{(1)a\,\balpha\bt}
  = \cB_{n \lambda_1}^a\, J_{\bar n  \lambda_2\, }^{\balpha\bt}\, J_{e\pm }$ \,(\ref{eq:Z1_basis_diff}) 
  &8  & 1 & $\checkmark$
  \\ \cline{2-6}
& $\! e \bar{e} ggg$  & $O_{\cB\lambda_1 \lambda_2 \lambda_3 (\pm)}^{(1)abc}
= S\, \, \cB_{n \lambda_1}^a\, \cB_{\bar n \lambda_2}^b\, \cB_{\bar n  \lambda_3}^c J_{e\pm }$ \,(\ref{eq:eeggg})
   &8 & 2 & $\checkmark$  
   \\ \hline
$\!\mathcal{O}(\lambda^2)\!$
 & $\! e \bar{e} q \bar q Q \bar Q$  &   
  $O_{qQ1(\lambda_1;\lambda_2:\pm)}^{(2)\balpha\bt\bgamma\delta}
  = J_{(q)n {\lambda_1}\, }^{\balpha\bt}\, J_{(Q) \bar n {\lambda_2}\, }^{\bgamma\delta}\,J_{e\pm } $ \,(\ref{eq:Z2_basis_qQ}) 
  & 8  & 2 & 
  \\
& & $O_{qQ2(\lambda_1;\lambda_1:\pm)}^{(2)\balpha\bt\bgamma\delta}
= J_{(q \bar{Q}) n  \lambda_1\, }^{\balpha\bt}\, J_{(Q \bar{q}) \bar{n}  \, \lambda_1\, }^{\bgamma\delta}\,J_{e\pm } $ \,(\ref{eq:Z2_basis_qQ_2})  
  &4 & 2 &    
  \\
& & $O_{qQ3(\lambda_1;-\lambda_1;\pm)}^{(2)\balpha\bt\bgamma\delta}
= J_{(q) n \bar n \lambda_1\, }^{\balpha\bt}\, J_{(Q) n\bar n -\lambda_1\, }^{\bgamma\delta}\,J_{e\pm } $ \,(\ref{eq:Z2_basis_qQ_3}) 
  &4 & 2 &    \\
& & $O_{qQ4(\lambda_1:\lambda_2;\pm)}^{(2)\balpha\bt\bgamma\delta}
= J_{(q)\bar n \lambda_1}^{\balpha\bt}\, J_{(Q) n \bar n\, {\lambda_2}\, }^{\bgamma\delta}\,J_{e\pm } $ \,(\ref{eq:Z3_basis_qQ}) 
  &8 & 2 & $\checkmark$      
  \\
& & $O_{qQ5(\lambda_1:\lambda_2;\pm)}^{(2)\balpha\bt\bgamma\delta}
=  J_{(q)\bar n \lambda_1\, }^{\balpha\bt}\, J_{(Q)\bar n n {\lambda_2}\, }^{\bgamma\delta}\,J_{e\pm } $ \,(\ref{eq:Z3_basis_qQ}) 
  &8 & 2 & $\checkmark$
  \\ \cline{2-6}
 & $\! e \bar{e} q \bar q q \bar q$  &   
  $O_{qq1(\lambda_1;\lambda_2:\pm)}^{(2)\balpha\bt\bgamma\delta}
  = J_{(q)n {\lambda_1}\, }^{\balpha\bt}\, J_{(q) \bar n {\lambda_2}\, }^{\bgamma\delta}\,J_{e\pm } $ \,(\ref{eq:Z2_basis_qq}) 
  & 8 & 2 & 
  \\
& & $O_{qq3(\lambda_1;-\lambda_1;\pm)}^{(2)\balpha\bt\bgamma\delta}
= J_{(q) n \bar n \lambda_1\, }^{\balpha\bt}\, J_{(q) n\bar n -\lambda_1\, }^{\bgamma\delta}\,J_{e\pm } $ \,(\ref{eq:Z2_basis_qq_3}) 
  &2 & 2 &    \\
& & $O_{qq4(\lambda_1:\lambda_2;\pm)}^{(2)\balpha\bt\bgamma\delta}
= J_{(q)\bar n \lambda_1}^{\balpha\bt}\, J_{(q) n \bar n\, {\lambda_2}\, }^{\bgamma\delta}\,J_{e\pm } $ \,(\ref{eq:Z3_basis_qq}) 
  &8 & 2 & $\checkmark$      
  \\
& & $O_{qq5(\lambda_1:\lambda_2;\pm)}^{(2)\balpha\bt\bgamma\delta}
=  J_{(q)\bar n \lambda_1\, }^{\balpha\bt}\, J_{(q)\bar n n {\lambda_2}\, }^{\bgamma\delta}\,J_{e\pm } $ \,(\ref{eq:Z3_basis_qq}) 
  &8 & 2 & $\checkmark$
  \\ \cline{2-6}
&  	$\! e \bar{e} q \bar q g g$  
  & $O_{\cB1\lambda_1 \lambda_2(\lambda_3;\pm)}^{(2)ab\, \balpha\bt}
  = S \cB_{n\lambda_1}^a \cB_{n \lambda_2}^b \, J_{n\bar n\, \lambda_3 }^{\balpha\bt}   \,J_{e\pm }$ \,(\ref{eq:eeqqgg_basis1})  
  &8 &3& $\checkmark$ 
  \\
& & $O_{\cB2\lambda_1 \lambda_2(\lambda_3;\pm)}^{(2)ab\, \balpha\bt}
  = S \cB_{n\lambda_1}^a \cB_{n \lambda_2}^b \, J_{\bar n n\, \lambda_3 }^{\balpha\bt}   \,J_{e\pm }$ \,(\ref{eq:eeqqgg_basis1a})  
  &8 &3& $\checkmark$ 
  \\
& & $O_{\cB3\lambda_1 \lambda_2(\lambda_3;\pm)}^{(2)ab\, \balpha\bt}
= \cB_{n\lambda_1}^a \cB_{\bar n \lambda_2}^b \, J_{n\bar n\, \lambda_3 }^{\balpha\bt}   \,J_{e\pm } $ \,(\ref{eq:eeqqgg_basis2}) 
  &12  &3& $\checkmark$  
  \\
& & $O_{\cB4\lambda_1 \lambda_2(\lambda_3:\pm)}^{(2)ab\, \balpha\bt}
=  \cB_{n \lambda_1}^a \cB_{\bar n \lambda_2}^b \, J_{n\,{\lambda_3} }^{\balpha\bt}   \,J_{e\pm } $ \,(\ref{eq:eeqqgg_basis3}) 
  & 8  &3&
  \\
& & $O_{\cB5\lambda_1 \lambda_2(\lambda_3:\pm)}^{(2)ab\, \balpha\bt}
=  \cB_{\bar n \lambda_1}^a \cB_{\bar n \lambda_2}^b \, J_{n\,{\lambda_3} }^{\balpha\bt}   \,J_{e\pm } $ \,(\ref{eq:eeqqgg_basis4}) 
  &  4  &3&
  \\ \cline{2-6}
&  	$\! e \bar{e} gggg$  & $O_{4g1\lambda_1 \lambda_2 \lambda_3 \lambda_4(\pm)}^{(2)a b c d}
= S  \cB^a_{n \lambda_1} \cB^b_{n \lambda_2} \cB^c_{\bn \lambda_3} \cB^d_{\bn \lambda_4} J_{e\pm } $ \,(\ref{eq:Z2_basis_gggg_1}) 
  & 6  &9&   
  \\
&   & $O_{4g2\lambda_1 \lambda_2 \lambda_3 \lambda_4(\pm)}^{(2)a b c d}
= S  \cB^a_{n \lambda_1} \cB^b_{\bn \lambda_2} \cB^c_{\bn \lambda_3} \cB^d_{\bn \lambda_4} J_{e\pm } $ \,(\ref{eq:Z2_basis_gggg_2}) 
  & 4 &9&
  \\ \cline{2-6}
&  	$\! \cP_\perp$   & $O_{\cP2 \lambda_1 (\lambda_2:\pm)[\lambda_{\cP}]}^{(2)a\,\balpha\bt}
= \cB_{n\lambda_1}^a \, \{J_{\bar n\, {\lambda_2}    }^{\balpha\bt}(\cP_{\perp}^{\lambda_{\cP}})^\dagger\}\,J_{e\pm }$  \,(\ref{eq:eeqqgpperp_basis_same})  
  &  8 & 1 &
  \\
& & $O_{\cP1n,\bar n \lambda_1 (\lambda_2;\pm)[\lambda_{\cP}]}^{(2)a\,\balpha\bt}
=\left [ \cP_{\perp}^{\lambda_{\cP}} \cB_{n,\bar n \lambda_1}^a \right ] \, J_{n\bar n\, \lambda_2}^{\balpha\bt}\,J_{e\pm } $ \,(\ref{eq:eeqqgpperp_basis},\ref{eq:eeqqgpperp_basis_flip}) 
  &24 & 1 & $\checkmark$   
  \\
& & $O_{\cP\cB \lambda_1 \lambda_2 \lambda_3(\pm)[\lambda_{\cP}]}^{(2)abc}
= S\, \cB_{n\lambda_1}^a\, \cB_{\bar n \lambda_2}^b\, \left [\cP_{\perp}^{\lambda_{\cP}} \cB_{\bar n \lambda_3}^c \right ]\, J_{e\pm } $  \,(\ref{eq:eegggpperp_basis}) 
  &8 & 2 &
  \\ \cline{2-6}
&  	$\!$Ultrasoft$\!\!\!$
 & $O_{\cB(us(i))\lambda_1:(\lambda_2;\pm) }^{(2)a\,\balpha\bt}
  =\cB_{us(i) \lambda_1}^a \, J_{n\bar n\, \lambda_2}^{\balpha\bt}\,J_{e\pm } $ \,(\ref{eq:soft_insert_basis},\ref{eq:soft_insert_basis2})
   &8 &1&    
   \\
& & $O_{\cB(us(i))0:(\lambda_1;\pm)}^{(2)a\,\balpha\bt}
  = \cB_{us(i)0}^a \, J_{n\bar n\,\lambda_1}^{\balpha\bt}\,J_{e\pm } $ \,(\ref{eq:soft_insert_basis},\ref{eq:soft_insert_basis2}) 
  & 8 &1&   $\checkmark$   
  \\
& & $O_{\partial(us(i))\lambda_1:(\lambda_2;\pm)}^{(2)\,\balpha\bt}
= \{\partial_{us(i)\lambda_1} \, J_{n\bar n\,\lambda_2}^{\balpha\bt}\}\,J_{e\pm } $ \,(\ref{eq:soft_derivative_basis}) 
  & 8 &1&   
  \\
& & $O_{\partial(us(i))0,\bar 0:(\lambda_1;\pm)}^{(2)\,\balpha\bt}
= \{\partial_{us(i)0,\bar 0} \, J_{n\bar n\,\lambda_1}^{\balpha\bt}\}\,J_{e\pm } $ \,(\ref{eq:soft_derivative_basis}) 
  & 8 &1&   $\checkmark$ 
  \\
& & $O_{(us(i))\lambda_1:\lambda_2 \lambda_3(\pm) }^{(2)abc}
=\cB_{us(i) \lambda_1}^a \,\cB_{n\, \lambda_2}^{b}\,\cB_{\bn\, \lambda_3}^{c}\,J_{e\pm } $ \,(\ref{eq:eegggus},\ref{eq:eegggus2}) 
  & 24  &2&   
  \\
& & $O_{\partial \cB (us(i))\lambda_1:\lambda_2 \lambda_3(\pm) }^{(2)ab}
=\left[ \partial_{us(i) \lambda_1} \cB_{n\, \lambda_2}\right] \cB_{\bn\, \lambda_3} J_{e\pm } $ (\ref{eq:eedggus},\ref{eq:eedggus2})\! 
  &  24   &2& 
  \\
\hline
\end{tabular}}
%
\vspace{-0.1cm}
\caption{Basis of hard scattering operators to ${\cal O}(\lambda^2)$ with an electron current $J_{e\pm}$ and two back-to-back collinear sectors. Here the $\lambda_i$ denote helicities, $S$ represents a symmetry factor, and $\cB_{n,\bar n \lambda_1}^a$ indicates $\cB_{n \lambda_1}^a$ or $\cB_{\bar n \lambda_1}^a$.  The allowed values for the $\lambda_i$ helicities are given in the indicated equation, and the total count is given in the indicated column. The last column indicates which operators can contribute to $e^+ e^- \to$ dijet event shapes and other two-direction processes up to $\mathcal{O}(\lambda^2)$ in the power expansion and at any order in $\alpha_s$, as discussed in detail in Sec. \ref{sec:ee_lambda2}. }
\label{tab:summary}
\vspace{-2.5cm}
\end{table}
}
	
One simplification that we make when constructing our basis is that we work in the center of mass frame, and only consider operators that are non-vanishing in this frame. This is natural for  Drell-Yan (CM frame), $e^+e^- \to$ dijet event shapes (the CM frame of the jets), and is also a convenient frame for theoretical studies of DIS (the Breit frame). Because of the conservation of label momentum in SCET, this choice of frame allows us to take the strongly interacting collinear sectors to be back-to-back. We therefore describe them by the back-to-back light-like vectors $n_1=n=(1,\hat n)$ and $n_2=\bn=(1,-\hat n)$. Due to this choice we will label the helicity currents with $n$ and $\bar n$, as in $J_{n\bn\pm}^{\balpha\beta}$, instead of with collinear sector numbers, $J_{12\pm}^{\balpha\beta}$. In \eq{Leff_sub_explicit} the hard Lagrangian in SCET is written as a sum over label momenta of the hard operators.  For the special case of two back-to-back collinear sectors this reduces to
\begin{align}\label{eq:sum_dir}
\cL_{\text{hard}}^{(j)} = \sum_{n} \sum_{A,\cdot\cdot}  
\bigg[ \prod_{i=1}^{\ell_A} \int \! \! \df \omega_i \bigg] \,
& \vO^{(j)\dagger}_{A\,\lotsdots}\big(n\bn;
   \omega_1,\ldots,\omega_{\ell_A}\big) \nn\\
& \times
\vC^{(j)}_{A\,\lotsdots}\big(n\bn;\omega_1,\ldots,\omega_{\ell_A} \big)
\,.
\end{align}
When constructing a complete basis, we therefore do not need to include operators which are identical up to the swap of $n\leftrightarrow \bar n$.  For a given operator, we can therefore choose the $n$ and $\bar n$ labels arbitrarily, and this can be done independently for each operator. When squaring matrix elements, all possible interferences are properly incorporated by accounting for the sum over directions in \Eq{eq:sum_dir}.


Furthermore, we choose to align the $n$ and $\nbar$ axes with the jets or protons, such that the overall label $\perp$ momentum of each collinear sector is zero.  As a consequence, operators involving $\cP_\perp$ acting on the complete set of fields in a  collinear sector vanish. Insertions of $\cP_\perp$ can still be non-vanishing when two or more collinear field building blocks appear in the same sector, and hence will first appear in our analysis at ${\cal O}(\lambda^2)$. The presence of ultrasoft degrees of freedom carrying residual momentum implies that, unlike the label perp momentum, the residual perp momentum of a collinear sector does not necessarily vanish, and can be exchanged with the ultrasoft sector.

All outgoing quark fields are taken to be massless. Chiral symmetry is violated in QCD by heavy quark masses, such as the top quark mass, and by non-perturbative effects. When matching QCD to SCET with only massless external fields we assume we are working below the scale where a top-quark can be produced, so top quarks appear only in closed loops and chirality is conserved at each order in $\alpha_s$ by the matching procedure (though not by the low energy non-perturbative dynamics, such as the chiral condensate). This remains true when considering QCD corrections to the Z exchange for any of the processes governed by the back-to-back collinear operators. All operators appearing in our basis must therefore preserve chirality. 

Throughout this section we will use $P_Z$ to denote the ratio of the $Z$ and photon propagators,
\begin{equation}
P_{Z}(s) = \frac{s}{s-M_{Z}^2 + i \Ga_{Z} M_{Z}}
\,,\end{equation}
and we will use $v^i_{L,R}$ for the coupling of particle $i$ to the Z boson, whose explicit expressions are given by
\begin{align}
 v_L^i = \frac{2 T_3^i - 2Q^i \sin^2 \theta_W}{\sin(2\theta_W)}
 \,, \quad
 v_R^i = - \frac{2Q^i \sin^2 \theta_W}{\sin(2\theta_W)}
\,,\end{align}
where $T_3^i$ is the third component of weak isospin. Since we use helicities to label the operators and Wilson coefficients, it is convenient to define the weak couplings in terms of helicities for both the quark and lepton currents,
\be
v_+^l=v_R^l, \qquad v_-^l=v_L^l, \qquad v_+^q=v_R^q,\qquad v_-^q=v_L^q\,.
\ee
For color, we use the normalization $\tr[T^a\,T^b] = 1/2\,\delta^{ab}$, i.e. $T_F=1/2$, and write the antisymmetric and symmetric structure constants of $SU(3)$, as $f^{abc}, \,d^{abc}$ respectively. In the case of the collinear operators, we present the color structure both before and after BPS field redefinition.

When labeling particles, the highest two subscripts will be used to refer to the electron and positron respectively, which will always appear in the current 
\begin{equation}
J_{e \pm } \equiv J_{e ij\pm }
= \mp \sqrt{\frac{2}{\omega_i\, \omega_j}}\, \ve_\mp^\mu(n_i, n_j)\,    \frac{\bar{e}_{i\pm} \gamma_\mu e_{j \pm}} {\langle n_j\mp | n_i\pm\rangle}
\,.
\end{equation}
Since this current appears in every operator, for notational convenience we will drop the explicit $ij$ label on the current, denoting it simply by $J_{e\pm}$.

Due to the relatively large number of operators present up to $\mathcal{O}(\lambda^2)$ we provide a summary of the complete set of operators in \Tab{tab:summary}, along with the number of helicity configurations for each operator. There are a total of $256$ helicity configurations in our operator basis, $128$ of which can contribute at $\cO(\lambda^2)$, of which $112$ have tree level contributions. This number does not include the different color configurations, which are also indicated in the table. The leading power basis can be found in \Sec{sec:LP}, the $\cO(\lambda)$ subleading basis in \Sec{sec:sub} and the $\cO(\lambda^2)$ subleading basis in  \Sec{sec:subsub}.

\subsection{Leading Power Operators}\label{sec:LP}

To begin this analysis, we review the leading power back-to-back operators and Wilson coefficients, which are combined according to \eq{Leff_sub_explicit} to give ${\cal L}_{\rm hard}^{(0)}$. By power counting, the leading power operators consist of either two collinear quark building blocks or two collinear gluon building blocks, one in the $n$-collinear and one in the $\bn$-collinear sector.  For the $eegg$ channel we have a process with one offshell spin-1 ($\gamma/Z$) particle and two onshell spin-1 particles ($gg$). The Wilson coefficients then all vanish by Yang's theorem~\cite{Landau:1948kw,PhysRev.77.242}, so we omit these operators. Therefore only the $ee q \bar q$ channel  contributes at leading power. While one should sum over the flavor of the outgoing quarks, this is trivial to implement, and therefore we do not make the flavor in the quark current explicit. The leading power helicity operators are given by
\begin{align}
& \boldsymbol{q \bar{q}:}{\vcenter{\includegraphics[width=0.18\columnwidth]{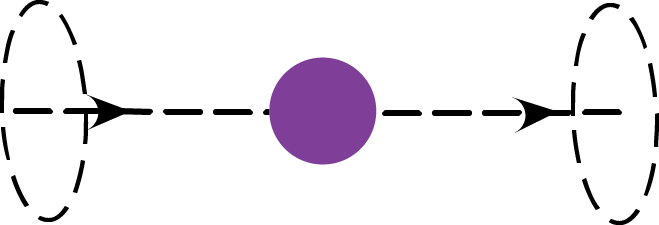}}}  \nn
\end{align}
\vspace{-0.4cm}
\begin{flalign}\label{eq:LP_basis}
O_{(+;\pm)}^{(0)\balpha\bt}
=J^{\balpha\bt}_{n \bar n+}J_{e \pm }\,, \qquad
O_{(-;\pm)}^{(0)\balpha\bt}
=J^{\balpha\bt}_{n \bar n-}  J_{e\pm } \,.
\end{flalign}
Here, and throughout this section, the bracketed superscript indicates the suppression in powers of $\lambda$ of the operator relative to the leading power operators. Since these are the leading power operators, it is zero in this case.

The color basis is one dimensional, and is given before and after BPS field redefinition by
\begin{align} \label{eq:leading_color}
 \vT^{\al\bbeta} = (\de_{\al\bbeta})\,, \qquad 
 \vT_{\BPS}^{\al\bbeta} =  \big[Y_n^\dagger Y_{\bar n} \big]_{\al\bbeta} 
\,.
\end{align}
Since the weak interactions break $C$ and $P$ symmetry, it is convenient to expand the Wilson coefficients into components with well defined $C/P$ properties. We use the decomposition
\begin{align} \label{eq:Z0_expand_Wilson}
\vec{C}^{(0)}_{(\lambda_q;\lambda_l)}(n,\bn; \omega_1, \omega_2; \omega_3,  \omega_4)
=  e^2  \,
\bigg\{ &\big[Q^\ell Q^q + v_{\lambda_l}^\ell v_{\lambda_q}^q P_Z(s_{34})\big] \vec{C}^{(0)}_{q(\lambda_q;\lambda_l)}(n,\bn; \omega_1, \omega_2; \omega_3,  \omega_4)
\nn\\ & 
\hspace{-2cm}+\sum_{j=1}^{n_f} \bigg[Q^\ell Q^j + \frac{v_{\lambda_l}^\ell}{2} (v_L^j +v_R^j) P_Zs_{34}) \bigg] \vec{C}^{(0)}_{v(\lambda_q;\lambda_l)}(n,\bn; \omega_1, \omega_2; \omega_3,  \omega_4) \nn \\
&\hspace{-2cm}+ \frac{v_{\lambda_l}^\ell}{\sin(2\theta_W)} P_Z(s_{34}) \vec{C}^{(0)}_{a(\lambda_q;\lambda_l)}(n,\bn; \omega_1, \omega_2; \omega_3,  \omega_4)  \bigg\}
\,, 
\end{align}
where $\lambda_q$ and $\lambda_l$ are the quark and lepton current helicities. Our notation follows that of \cite{Moult:2015aoa}. Here we have extracted only electroweak couplings from the Wilson coefficients, so that each of $\vec{C}_{q,v,a}$ is a power series in $\alpha_s$. In \eq{Z0_expand_Wilson} we have split the amplitude into a contribution, $\vec{C}_q$, arising from the matching contributions where the vector boson couples directly to the final-state quark line, and contributions $\vec{C}_v$, $\vec{C}_a$ where the vector boson couples to a quark loop through a vector or axial coupling, respectively. This decomposition is valid since we work only to leading order in the electroweak couplings. We have also made the assumption that all quarks, except for the top, are massless. This implies that only the $b,t$ isodoublet contributes to $\cA_a$. This assumption can trivially be lifted, but many helicity amplitudes are calculated assuming this approximation.

Charge and parity conjugation can be used to derive relations between the Wilson coefficients. Parity relates the Wilson coefficients by
\begin{align} \label{eq:leadingCP}
C^{(0)\alpha \bbeta}_{q\,(\lambda_{12};\lambda_{34})}(n,\bn; \omega_1, \omega_2; \omega_3,  \omega_4) &= C^{(0)\alpha \bbeta}_{q\,(-\lambda_{12};-\lambda_{34})}(n,\bn; \omega_1, \omega_2; \omega_3,  \omega_4) 
 \,,\\
C^{(0)\alpha \bbeta}_{v\,(\lambda_{12};\lambda_{34})}(n,\bn; \omega_1, \omega_2; \omega_3,  \omega_4) &= C^{(0)\alpha \bbeta}_{v\,(-\lambda_{12};-\lambda_{34})}(n,\bn; \omega_1, \omega_2; \omega_3,  \omega_4) 
 \,,\nn\\
C^{(0)\alpha \bbeta}_{a\,(\lambda_{12};\lambda_{34})}(n,\bn; \omega_1, \omega_2; \omega_3,  \omega_4) &= - C^{(0)\alpha \bbeta}_{a\,(-\lambda_{12};-\lambda_{34})}(n,\bn; \omega_1, \omega_2; \omega_3,  \omega_4) 
 \,. \nn
\end{align}
Here $\lambda_{12}=\pm$ denotes the helicity label of the helicity building block with momentum labels $\omega_1$ and $\omega_2$ describing the two collinear quark fields, and similarly for $\lambda_{34}$. This notation will be used throughout this section, with the additional allowance for $\lambda_i = 0$ or $\bar{0}$ when appropriate. For later applications, we also introduce the notation
\begin{align}
\text{if}\,\, \lambda = 0 \,\, \text{then}\, -\lambda = \bar{0}\,.
\end{align} 
Note that $n$ and $\bn$ are not swapped here since after applying parity we always make an additional swap $n\leftrightarrow \bn$ so that we get back the same form of operators. Since we always work to leading order in the weak and electromagnetic couplings, the leptons couple only through the vector and axial-vector currents which satisfy 
\begin{align}\label{eq:lo_weak}
\mae{3\pm}{\ga^\mu}{4\pm} = \mae{4\mp}{\ga^\mu}{3\mp}, \qquad
\mae{3\pm}{\ga^\mu\gamma_5}{4\pm} = - \mae{4\mp}{\ga^\mu\gamma_5}{3\mp} \,.
\end{align}
These relations will be used throughout our analysis. Here they imply
\begin{align} \label{eq:lep_reduce}
C_{q\,(\lambda_{12};\lambda_{34})}^{(0)\al\bbeta}(n,\bn; \omega_1, \omega_2; \omega_3,  \omega_4) &= C_{q\,(\lambda_{12};-\lambda_{34})}^{(0)\al\bbeta}(n,\bn; \omega_1, \omega_2; \omega_4,  \omega_3)
 \,,\\
C_{v\,(\lambda_{12};\lambda_{34})}^{(0)\al\bbeta}(n,\bn; \omega_1, \omega_2; \omega_3,  \omega_4) &= C_{v\,(\lambda_{12};-\lambda_{34})}^{(0)\al\bbeta}(n,\bn; \omega_1, \omega_2; \omega_4,  \omega_3)
 \,,\nn\\
C_{a\,(\lambda_{12};\lambda_{34})}^{(0)\al\bbeta}(n,\bn; \omega_1, \omega_2; \omega_3,  \omega_4) &= - C_{a\,(\lambda_{12};-\lambda_{34})}^{(0)\al\bbeta}(n,\bn; \omega_1, \omega_2; \omega_4,  \omega_3)
\,.\nn
\end{align}
The relations in \eqs{lo_weak}{lep_reduce} imply that only the three Wilson coefficients with the helicity label $(+;+)$ need to be calculated to get all twelve coefficients.

\subsection{Subleading Power Operators} \label{sec:sub}
From the power counting of the operators in \Tab{tab:PC}, we see that the $\cO(\lambda)$ suppressed operators have three ${\cal O}(\lambda)$ collinear building block fields, or two collinear building block fields and a single $\cP_\perp$  insertion. Our choice for $n$ and $\bn$ eliminates operators that have a $\cP_\perp$ acting on a complete collinear sector. Therefore, we only need to consider operators consisting of three collinear field building blocks at $\cO(\lambda)$. There are two possibilities for the field content of the operators: two collinear quarks and a collinear gluon, or three collinear gluons. We shall discuss each in turn.

The helicity operators involving two collinear quarks and a single collinear gluon consist of a single leptonic current, a quark current, and a collinear gluon building block field. For each helicity configuration, we must consider the cases where both collinear quarks are in the same sector, or in different sectors. The quarks are necessarily in a same chirality pair, which simplifies the operator basis. The basis of helicity operators was already constructed in \sec{ang_cons} and is given by
\begin{align}
&  \boldsymbol{(gq)_n (\bar q)_\bn :}{\vcenter{\includegraphics[width=0.18\columnwidth]{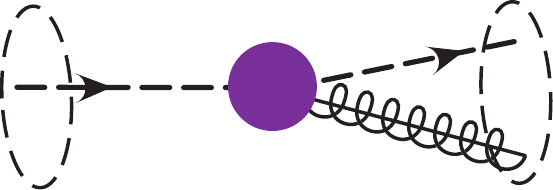}}}  \nn
\end{align}
\vspace{-0.4cm}
\begin{alignat}{2} \label{eq:Z1_basis}
&O_{n\bar n1+(-;\pm)}^{(1)a\,\balpha\bt}
= \cB_{n+}^a\, J_{n\bar n\,-}^{\balpha\bt} \,J_{e\pm }
\,,\qquad &
&O_{n\bar n1-(+;\pm)}^{(1)a\,\balpha\bt}
= \cB_{n-}^a \, J_{n\bar n\,+}^{\balpha\bt}\,J_{e\pm }
\,,
\end{alignat}
and
\begin{align}
&  \boldsymbol{(q)_n (g\bar q)_\bn :}{\vcenter{\includegraphics[width=0.18\columnwidth]{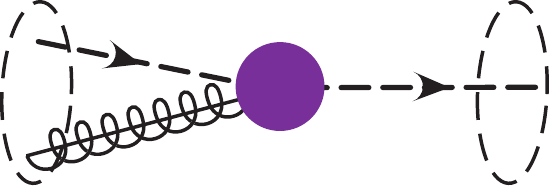}}}  \nn
\end{align}
\vspace{-0.4cm}
\begin{alignat}{2} \label{eq:Z1_basis_flip}
&O_{n\bar n2-(-;\pm)}^{(1)a\,\balpha\bt}
= \cB_{\bar n-}^a\, J_{n\bar n\,-}^{\balpha\bt} \,J_{e\pm }
\,,\qquad &
&O_{n\bar n2+(+;\pm)}^{(1)a\,\balpha\bt}
= \cB_{\bar n+}^a \, J_{n\bar n\,+}^{\balpha\bt}\,J_{e\pm }
\,,
\end{alignat}
for the case that the quarks are in different collinear sectors, and 
\vspace{0.3cm}
\begin{align}
& \boldsymbol{(g)_n (q \bar q)_\bn :}{\vcenter{\includegraphics[width=0.18\columnwidth]{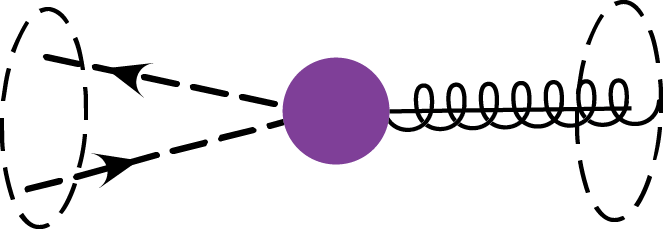}}}  \nn
\end{align}
\vspace{-0.4cm}
\begin{alignat}{2} \label{eq:Z1_basis_diff} 
 &O_{\bar{n}+(0:\pm)}^{(1)a\,\balpha\bt}
= \cB_{n+}^a\, J_{\bar n0\, }^{\balpha\bt}\,J_{e\pm }
\,,\qquad &
&O_{\bar{n}+(\bar 0:\pm)}^{(1)a\,\balpha\bt}
=  \cB_{n+}^a\, J_{\bar n\bar 0\, }^{\balpha\bt}\,J_{e\pm }
\,,\\
&O_{\bar{n}-(0:\pm)}^{(1)a\,\balpha\bt}
= \cB_{n-}^a\,  J_{\bar n0\, }^{\balpha\bt}\,J_{e\pm }
\,,\qquad &
&O_{\bar{n}-(\bar 0:\pm)}^{(1)a\,\balpha\bt}
=\cB_{n-}^a\, J_{\bar n\bar 0\, }^{\balpha\bt} \,J_{e\pm }
\,,\nn 
\end{alignat}
in the case that they are in the same sector. Note that the operators in \Eqs{eq:Z1_basis}{eq:Z1_basis_flip} are distinct, and therefore both need to be included in the basis, while in the case that both quarks are in the same sector, it is sufficient to chose both quark fields to be in the same sector, since the direction is summed over.  In \Eqs{eq:Z1_basis}{eq:Z1_basis_flip} we have eliminated two of the possible helicity combinations, as was discussed in detail in \Sec{sec:ang_cons}. 
The color basis is one dimensional, and is given by
\begin{equation} \label{eq:Z1q_color}
\vT^{a\, \al\bbeta} = T^a_{\al\bbeta}\,.
\end{equation}
After BPS field redefinition the structure of the ultrasoft Wilson lines is different depending on whether the quarks are in different or the same collinear sectors. We find
\begin{equation} \label{eq:BPSgqqcolor}
\vT_{\BPS}^{ a \al\bbeta} = \left (T^a Y_n^\dagger Y_{\bar n}  \right )_{\alpha \bar \beta}
\,,\qquad
\vT_{\BPS}^{ a \al\bbeta} = \left (Y_n^\dagger Y_{\bar n} T^a  \right )_{\alpha \bar \beta}
\,,\qquad
\vT_{\BPS}^{ a \al\bbeta} = \cY_n^{ba}   \cY_\bn^{bc}   T^c_{\alpha \bar \beta}
\,,
\end{equation}
for the operators in \Eqs{eq:Z1_basis}{eq:Z1_basis_flip} and \eq{Z1_basis_diff}, respectively. We have used \eq{adjointtofundamental} to simplify the equations in \Eq{eq:BPSgqqcolor} and we will continue to do so throughout this section when it simplifies the relevant Wilson line structures.

The Wilson coefficients of the operators in both \Eqs{eq:Z1_basis}{eq:Z1_basis_flip} and \eq{Z1_basis_diff} can be expanded as
\begin{align} \label{eq:Z1_qqg_expand}
  \vec{C}^{(1)}(n,\bn;\omega_1; \omega_2, \omega_3; \omega_4,\omega_5)
= &  e^2\,  \, 
\bigg\{ \big[Q^\ell Q^q + v_{\lambda_l}^\ell v_{\lambda_q}^q P_Z(s_{45})\big] \vec{C}^{(1)}_q(n,\bn;\omega_1; \omega_2, \omega_3; \omega_4,\omega_5) \nn 
\\ & 
+\sum_{j=1}^{n_f} \bigg[Q^\ell Q^j + \frac{v_{\lambda_l}^\ell}{2} (v_L^j +v_R^j) P_Zs_{45}) \bigg] \vec{C}^{(1)}_v(n,\bn;\omega_1; \omega_2, \omega_3; \omega_4,\omega_5) \nn \\
 & + \frac{v_{\lambda_l}^\ell}{\sin(2\theta_W)} P_Z(s_{45})\vec{C}^{(1)}_a(n,\bn;\omega_1; \omega_2, \omega_3; \omega_4,\omega_5)  \bigg\}
\,, 
\end{align}
where the components of the Wilson coefficient have the same meaning as in \eq{Z0_expand_Wilson}. C/P relations combined with the ability to flip the helicity of the electron current, as described in \Eq{eq:lo_weak}, give the following relations between Wilson coefficients
\begin{align} \label{eq:parityrel1}
\vec{C}^{(1)}_{v\,\lambda_1(\lambda_{23}:\lambda_{45})}(n,\bn;\{\omega_i\}) &= \vec{C}^{(1)}_{v\,-\lambda_1(-\lambda_{23}:-\lambda_{45})}(n,\bn;\{\omega_i\})\,,  \\ 
\vec{C}^{(1)}_{v\,\lambda_1(\lambda_{23}:\lambda_{45})}(n,\bn;\omega_1;\omega_2,\omega_3;\omega_4,\omega_5) &= -\vec{C}^{(1)}_{v\,\lambda_1(-\lambda_{23}:-\lambda_{45})}(n,\bn;\omega_1;\omega_3,\omega_2;\omega_5,\omega_4)
 \,,\nn \\
 \vec{C}^{(1)}_{v\,\lambda_1(\lambda_{23}:\lambda_{45})}(n,\bn;\omega_1;\omega_2,\omega_3;\omega_4,\omega_5) &= \vec{C}^{(1)}_{v\,\lambda_1(\lambda_{23}:-\lambda_{45})}(n,\bn;\omega_1;\omega_2,\omega_3;\omega_5,\omega_4)\,. \nn
\end{align}
Here the $\lambda_i$ denote generic helicity labels for the corresponding helicity building blocks, subject to the constraints of angular momentum conservation, as in \Eqs{eq:Z1_basis}{eq:Z1_basis_flip} and \eq{Z1_basis_diff}. $\vec C^{(1)}_q$ and $\vec C^{(1)}_a$ satisfy the same relations, but with an additional negative sign for each of the equations in the case of $\vec C^{(1)}_a$. We are using the notation where $\lambda = \pm 1, 0$ or $\bar{0}$. Additionally, if $\lambda = 0$, then we take $-\lambda = \bar{0}$. Combined, these relations imply that for each of the $\vec{C}^{(1)}_q$, $\vec{C}^{(1)}_v$, and $\vec{C}^{(1)}_a$, only a single Wilson coefficient needs to be calculated for the operators in \Eq{eq:Z1_basis} (for example, $\vec C_{+(- ; +)}$) and one for those in \Eq{eq:Z1_basis_diff} (for example, $\vec C_{+(0:+)}$). In particular, restricting to a mediating photon, and ignoring processes which proceed through a fermion loop, only a single Wilson coefficient is needed in each of the two cases.

Operators involving three collinear gluon fields do not appear in the matching until one-loop, and are therefore not of immediate phenomenological interest. However, we include them here both for completeness and to demonstrate the simplicity of the helicity operator approach. The basis of three gluon operators is
\begin{align}
& \boldsymbol{(g)_n (gg)_{\bn}:}{\vcenter{\includegraphics[width=0.18\columnwidth]{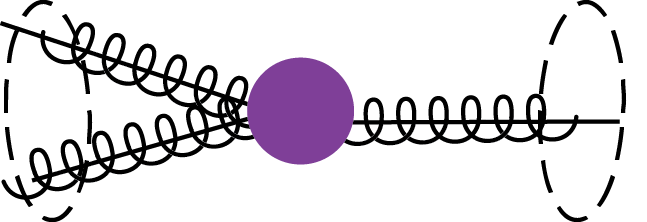}}}  \nn
\end{align}
\vspace{-0.4cm}
\begin{alignat}{2}\label{eq:eeggg}
 &O_{\cB+++(\pm)}^{(1)abc}
= \frac{1}{2}\, \cB_{n +}^a\, \cB_{\bar n+}^b\, \cB_{\bar n+}^c\,  J_{e\pm }
\,, \qquad &&O_{\cB---(\pm)}^{(1)abc}
= \frac{1}{2}\,   \cB_{n-}^a\, \cB_{ \bn -}^b\, \cB_{\bar n-}^c\, J_{e\pm }
\,, \nn\\ 
&O_{\cB++-(\pm)}^{(1)abc}
=   \cB_{n+}^a\, \cB_{ \bar n+}^b\, \cB_{ \bar n-}^c\, J_{e\pm }
\,, \qquad &&O_{\cB-+-(\pm)}^{(1)abc}
=  \cB_{n-}^a\, \cB_{ \bar n+}^b\, \cB_{ \bar n-}^c\, J_{e\pm }
\,,
\end{alignat}
where we have taken the two gluon fields to be in the $\bar n$ collinear sector.
Here the factors of $1/2$ are included for convenience as symmetry factors. Note that when writing this basis we have used the angular momentum constraints discussed in \Sec{sec:ang_cons} to eliminate the other two helicity combinations. These missing combinations have $h=\pm2$ about the $\hat n$ axes, and therefore vanish. 
The basis of color structures here is two dimensional,
\begin{equation} \label{eq:Z1g_color}
\vT^{abc} =
\begin{pmatrix}
  i  f^{abc} \\ d^{abc}
\end{pmatrix}
\,,
\qquad
\vT_{\BPS}^{abc} =
\begin{pmatrix}
  i  f^{a'b'c'}\, {\cal Y}_n^{a'a} {\cal Y}_\bn^{b'b} {\cal Y}_{\bn}^{c'c}  \\  
  d^{a'b'c'}\, {\cal Y}_n^{a'a} {\cal Y}_\bn^{b'b} {\cal Y}_{\bn}^{c'c} 
\end{pmatrix}
 =
\begin{pmatrix}
   i f^{bcd} {\cal Y}_{\bn}^{a'd} {\cal Y}_n^{a'a}   \\  
  d^{bcd}\, {\cal Y}_\bn^{a'd}  {\cal Y}_n^{a'a}  
\end{pmatrix}
\,.
\end{equation}
Once again we have simplified the Wilson line structures after the BPS field redefinition.

 For $e \bar e ggg $, the intermediate boson must couple to a fermion loop, so the Wilson coefficient can be expanded as
 \begin{align} \label{eq:Z1_ggg_expand}
 \vec{C}^{(1)}(n,\bn;\omega_1, \omega_2,\omega_3; \omega_4,\omega_5)
 &= e^2 \,
 \bigg\{\frac{v_{\lambda_l}^\ell}{\sin(2\theta_W)} P_Z(s_{45})  \vec{C}^{(1)}_a(n,\bn;\omega_1, \omega_2,\omega_3; \omega_4,\omega_5)  \\
 &+\sum_{j=1}^{n_f} \bigg[Q^\ell Q^j + \frac{v_{\lambda_l}^\ell}{2} (v_L^j +v_R^j) P_Zs_{45}) \bigg]  \vec{C}^{(1)}_v(n,\bn;\omega_1, \omega_2,\omega_3; \omega_4,\omega_5) \bigg\} \nn
 \,,
 \end{align}
 where, as in \eq{Z1_qqg_expand}, $\vec{C}^{(1)}_v$, $\vec{C}^{(1)}_a$ correspond to the contributions from the intermediate boson coupling through either the vector or axial couplings respectively, and we have suppressed the helicity labels on all coefficients. C/P relations combined with the ability to flip the helicity of the electron current, as described in \Eq{eq:lo_weak}, give the following relations between Wilson coefficients
\begin{align}  \label{eq:chargerel1}
\vec{C}^{(1)}_{v\,\lambda_1\lambda_2\lambda_3(\lambda_{45})}(n,\bn;\{\omega_i\}) &= \vec{C}^{(1)}_{ v\,-\lambda_1\,-\lambda_2\,-\lambda_3(-\lambda_{45})}(n,\bn;\{\omega_i\}) \,,  \\
\vec{C}^{(1)}_{v\,\lambda_1\lambda_2\lambda_3(\lambda_{45})}(n,\bn;\omega_1,\omega_2,\omega_3;\omega_4,\omega_5) &= \begin{pmatrix*}[r] -1 & 0 \\ 0 & 1\end{pmatrix*} \vec{C}^{(1)}_{v\,\lambda_1\lambda_2\lambda_3(-\lambda_{45})}(n,\bn;\omega_1,\omega_2,\omega_3;\omega_5,\omega_4) 
\nn \,, \\
\vec{C}^{(1)}_{v\,\lambda_1\lambda_2\lambda_3(\lambda_{45})}(n,\bn;\omega_1,\omega_2,\omega_3;\omega_4,\omega_5) &=  \vec{C}^{(1)}_{v\,\lambda_1\lambda_2\lambda_3(-\lambda_{45})}(n,\bn;\omega_1,\omega_2,\omega_3;\omega_5,\omega_4) \,, \nn
\end{align}
where the $\lambda_i=\pm$ denote generic helicity labels of the corresponding helicity building blocks, subject to the constraint of angular momentum conservation, as in \Eq{eq:eeggg}, and we have expressed the color structures in the bases of \Eq{eq:Z1g_color}. $\vec C^{(1)}_q$ and $\vec C^{(1)}_a$ satisfy the same relations but with an additional overall negative sign in both equations for $\vec C^{(1)}_a$.

The charge conjugation relations of \eq{chargerel1} imply that to all orders in $\alpha_s$ only the Wilson coefficients for the color structure $d^{abc}$ are non-zero for the vector current, whereas for the axial current, only the Wilson coefficients corresponding to the color structure $if^{abc}$ are non-zero. These statements remain true under renormalization group evolution. The helicity relations of \Eq{eq:chargerel1} then imply that only a single Wilson coefficient for a chosen helicity needs to be calculated for each of the color structures.

\subsection{Sub-subleading Power Operators} \label{sec:subsub}
The construction of the $\cO(\lambda^2)$ power suppressed operator basis is slightly more involved, so we divide the discussion into several subsections. The full list of operators can be found in Table \ref{tab:summary}. We separately discuss operators involving only collinear building block fields, operators involving $\cP_\perp$ insertions, and operators involving insertions of ultrasoft building blocks.

\subsubsection{Collinear Field Insertions}
Operators involving four collinear fields, corresponding to the partonic processes $e \bar e gggg$, $e \bar eq \bar q q\bar q$, $e \bar e q \bar q Q\bar Q $, $e \bar eggq\bar q$, appear at sub-subleading power. We will consider each in turn.

\vspace{0.4cm}
\noindent{\bf{Four Quark Operators:}}

We begin by considering the case of operators involving four collinear quark fields. When constructing the operator basis, we must consider the case that there are two collinear quarks in each collinear sector, or three collinear quarks in one sector, and one in the other. We must also treat separately the case of identical quark flavors $e \bar eq \bar q q\bar q$ and distinct massless quark flavors $e \bar e q \bar q Q\bar Q $. For the case of distinct quark flavors $e \bar e q \bar q Q\bar Q $ we will have a $q\leftrightarrow Q$ symmetry for the operators. Furthermore the two quarks of flavor $q$, and the two quarks of flavor $\bar Q$, are necessarily of the same chirality.  In the case that both quarks of the same flavor appear in the same current, the current will be labeled by the flavor. Otherwise, the current will be labeled with ($q \bar{Q}$) or ($Q\bar{q}$) appropriately. For all these cases, the color basis is
\begin{equation} \label{eq:qqqq_color}
\vT^{\, \al\bbeta\ga\bdelta} =
\Bigl(
\de_{\al\bdelta}\, \de_{\ga\bbeta}\,,\, \delta_{\al\bbeta}\, \de_{\ga\bdelta}
\Bigr)
\,.\end{equation}
We will give results for the corresponding $\bar T_{\rm BPS}^{\, \al\bbeta\ga\bdelta}$ basis as we consider each case below.

For the case of operators with distinct quark flavors $e \bar e q \bar q Q\bar Q $ and two collinear quarks in each of the $n$ and $\bn$ sectors there are three possibilities. There is either a quark anti-quark pair of the same flavor in each sector (e.g. $(q \bar q)_n(Q \bar Q)_\bn$), a quark and an antiquark of distinct flavors in the same sector (e.g. $(q \bar Q)_n(Q \bar q)_\bn$), or two quarks with distinct flavors in the same sector(e.g. $(q Q)_n(\bar q \bar Q)_\bn$). In the case that there is a quark anti-quark pair of the same flavor in each sector, the basis of helicity operators is
\begin{align}
 \boldsymbol{(q \bar q)_n(Q \bar Q)_\bn:}    {\vcenter{\includegraphics[width=0.18\columnwidth]{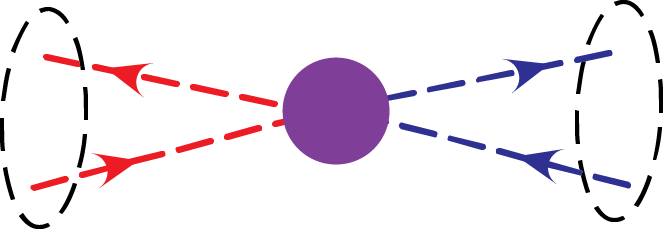}}}
\nn
\end{align}
\vspace{-0.4cm}
\begin{alignat}{2} \label{eq:Z2_basis_qQ}
&O_{qQ1(0;0:\pm)}^{(2)\balpha\bt\bgamma\delta}
= \, J_{(q)n 0\, }^{\balpha\bt}\, J_{(Q) \bar n 0\, }^{\bgamma\delta} J_{e\pm }
\,,\qquad &
&O_{qQ1(0;\bar 0:\pm)}^{(2)\balpha\bt\bgamma\delta}
= \, J_{(q) n  0\, }^{\balpha\bt}\, J_{(Q) \bar n \bar 0\,}^{\bgamma\delta} J_{e\pm }
\,,\\
&O_{qQ1(\bar 0;0:\pm)}^{(2)\balpha\bt\bgamma\delta}
= \,  J_{(q)n  \bar 0\, }^{\balpha\bt}\, J_{(Q) \bar n 0\, }^{\bgamma\delta} J_{e\pm }
\,,\qquad &
&O_{qQ1(\bar 0;\bar 0:\pm)}^{(2)\balpha\bt\bgamma\delta}
=\, J_{(q) n \bar 0\, }^{\balpha\bt}\, J_{(Q) \bar n \bar 0\, }^{\bgamma\delta} J_{e\pm }
\,,\nn
\end{alignat}
where we have chosen the $q$ quark to be in the $n$ sector. Since all the operators have total helicity $0$ along the $\hat n$ direction, there are only chirality constraints here and no constraints  from angular momentum conservation.  In the case that there is a quark anti-quark of distinct flavors in the same sector, chirality and angular momentum conservation constrains the basis to be 
\begin{align}
\boldsymbol{(q \bar Q)_n(Q \bar q)_\bn:}     {\vcenter{\includegraphics[width=0.18\columnwidth]{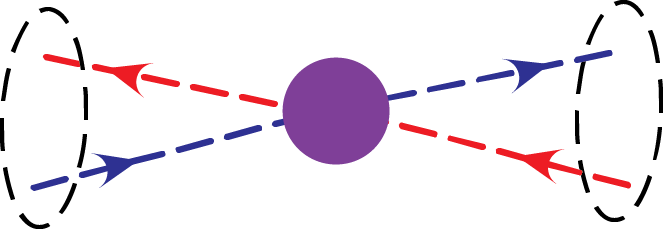}}}
\nn
\end{align}
\vspace{-0.4cm}
\begin{align} \label{eq:Z2_basis_qQ_2}
&O_{qQ2(0;0:\pm)}^{(2)\balpha\bt\bgamma\delta}
=\, J_{(q \bar{Q}) n 0\, }^{\balpha\bt}\, J_{(Q\bar{q} ) \bar{n} 0\, }^{\bgamma\delta} J_{e\pm } 
\,,\qquad 
O_{qQ2(\bar 0;\bar 0:\pm)}^{(2)\balpha\bt\bgamma\delta}
=\, J_{(q \bar{Q}) n \bar 0\, }^{\balpha\bt}\, J_{(Q\bar{q} )\bar{n} \bar 0\, }^{\bgamma\delta} J_{e\pm } 
\,.
\end{align}
For the operators in \eqs{Z2_basis_qQ}{Z2_basis_qQ_2} the color basis after BPS field redefinition is
\begin{align}  \label{eq:TBPS_OqQ12}
\vT_{\BPS}^{ \al\bbeta\ga\bdelta} &=
\left( \Big[ Y_{n}^\dagger Y_{\bar n}  \Big]_{\al\bdelta}\,\Big[ Y_{\bar n}^\dagger Y_{n}  \Big]_{\ga\bbeta}\,,\, \delta_{\al\bbeta}\, \delta_{\ga\bdelta}
\right)
\,.
\end{align}
When there are two quarks of distinct flavors in the same sector the basis of helicity operators is constrained by chirality and reduced further to just two operators by angular momentum conservation, giving 
\begin{align}
\boldsymbol{(q Q)_n(\bar q \bar Q)_\bn:}    {\vcenter{\includegraphics[width=0.18\columnwidth]{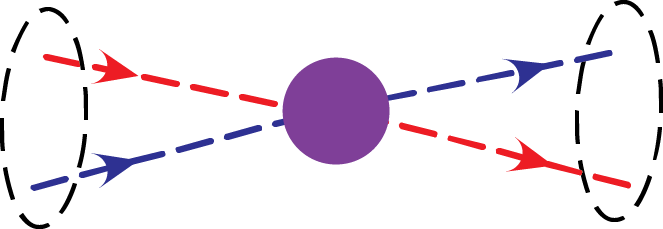}}}
\nn
\end{align}
\vspace{-0.4cm}
\begin{align}\label{eq:Z2_basis_qQ_3}
&O_{qQ3(+;-;\pm)}^{(2)\balpha\bt\bgamma\delta}
=\, J_{(q) n \bar n +\, }^{\balpha\bt}\, J_{(Q) n\bar n -\, }^{\bgamma\delta} J_{e\pm } 
\,,\qquad &
O_{qQ3(-;+;\pm)}^{(2)\balpha\bt\bgamma\delta}
= \, J_{(q)n \bar n -\, }^{\balpha\bt}\, J_{(Q) n\bar n +\, }^{\bgamma\delta} J_{e\pm }
\,.
\end{align}
For the operators in \eq{Z2_basis_qQ_3} the color basis after BPS field redefinition is
\begin{align} \label{eq:TBPS_OqQ3}
\vT_{\BPS}^{ \al\bbeta\ga\bdelta} &=
\left( \left[ Y_{n}^\dagger Y_{\bar n}  \right ]_{\al\bdelta}\,\left[ Y_{n}^\dagger Y_{\bar n}  \right ]_{\ga\bbeta}
 \,,\, 
 \left[ Y_{n}^\dagger Y_{\bar n}  \right ]_{\al\bbeta}\,\left[ Y_{n}^\dagger Y_{\bar n}  \right ]_{\ga\bdelta}
\right)
\,.
\end{align}

In the cases in \eqs{Z2_basis_qQ}{Z2_basis_qQ_2} where there is a quark and antiquark field in the same collinear sector, we have chosen to work in a basis using $J_{i0}^{\balpha\beta}$ and $J_{i\bar 0}^{\balpha\beta}$ which contain only fields in a single collinear sector. One could also construct an alternate form for the basis, for example using the currents $J_{n\bn\lambda}^{\balpha\beta}$. However, from the point of view of factorization, our basis is more convenient. The fields in the $n$ and $\bar n$ collinear sectors are only connected by color indices, which will simplify later steps of factorization proofs. In the following, we will make this choice for our basis whenever possible.

For identical quark flavors the operators have the same structure as in Eqs.~(\ref{eq:Z2_basis_qQ},\ref{eq:Z2_basis_qQ_2},\ref{eq:Z2_basis_qQ_3}), except the operators $O^{(2)}_{qQ1}$ and $O^{(2)}_{qQ2}$ are no longer distinct. A basis of operators is then given by
\begin{align}
  \boldsymbol{(q \bar q)_n (q \bar{q})_{\bn}:}     {\vcenter{\includegraphics[width=0.18\columnwidth]{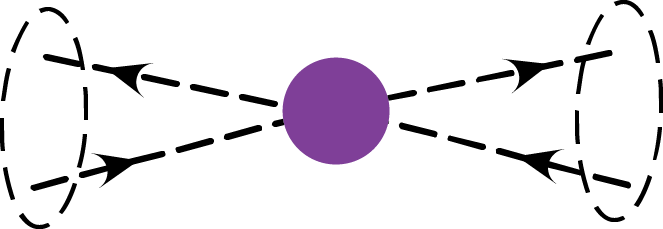}}}
\nn
\end{align}
\vspace{-0.4cm}
\begin{alignat}{2} \label{eq:Z2_basis_qq}
&O_{qq1(0;0:\pm)}^{(2)\balpha\bt\bgamma\delta}
= \, J_{(q)n 0\, }^{\balpha\bt}\, J_{(q) \bar n 0\, }^{\bgamma\delta} J_{e\pm }
\,,\qquad &
&O_{qq1(0;\bar 0:\pm)}^{(2)\balpha\bt\bgamma\delta}
= \, J_{(q) n  0\, }^{\balpha\bt}\, J_{(q) \bar n \bar 0\,}^{\bgamma\delta} J_{e\pm }
\,,\\
&O_{qq1(\bar 0;0:\pm)}^{(2)\balpha\bt\bgamma\delta}
= \,  J_{(q)n  \bar 0\, }^{\balpha\bt}\, J_{(q) \bar n 0\, }^{\bgamma\delta} J_{e\pm }
\,,\qquad &
&O_{qq1(\bar 0;\bar 0:\pm)}^{(2)\balpha\bt\bgamma\delta}
=\, J_{(q) n \bar 0\, }^{\balpha\bt}\, J_{(q) \bar n \bar 0\, }^{\bgamma\delta} J_{e\pm }
\,,\nn
\end{alignat}
and
\begin{align}
\boldsymbol{(q q)_n(\bar q \bar q)_\bn:}    {\vcenter{\includegraphics[width=0.18\columnwidth]{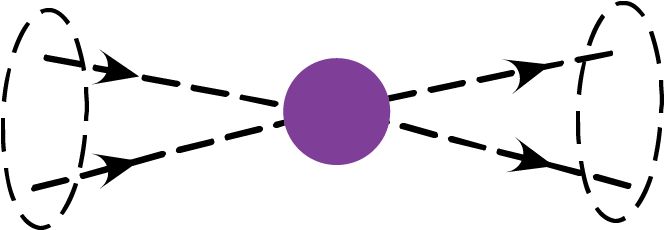}}}
\nn
\end{align}
\vspace{-0.4cm}
\begin{align}\label{eq:Z2_basis_qq_3}
&O_{qq3(+;-;\pm)}^{(2)\balpha\bt\bgamma\delta}
=\, J_{(q) n \bar n +\, }^{\balpha\bt}\, J_{(q) n\bar n -\, }^{\bgamma\delta} J_{e\pm } 
\,,\qquad &
O_{qq3(-;+;\pm)}^{(2)\balpha\bt\bgamma\delta}
= \, J_{(q)n \bar n -\, }^{\balpha\bt}\, J_{(q) n\bar n +\, }^{\bgamma\delta} J_{e\pm }
\,.
\end{align}
We also have the same color bases as in \eqs{TBPS_OqQ12}{TBPS_OqQ3} for $ O_{qq1}^{(2)}$ and $O_{qq3}^{(2)}$ respectively.

We must also consider the operators with three collinear quarks in one sector, and one quark in the other. To minimize the number of operators to display, we exploit the $q\leftrightarrow Q$ and $n\leftrightarrow \bar n$ symmetry to restrict ourselves to the case where the single quark (or antiquark) has flavor $Q$ and is in the $n$ collinear sector. The basis for the distinct flavor case with three quarks in the same collinear sector is then
\begin{align}
\boldsymbol{(Q)_n (\bar{Q} q \bar q)_{\bn}:}     {\vcenter{\includegraphics[width=0.18\columnwidth]{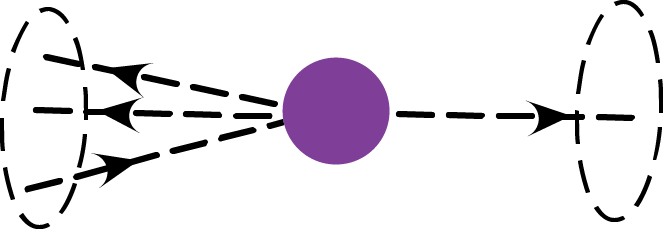}}}
\nn
\end{align}
\vspace{-0.4cm} 
\begin{align} \label{eq:Z3_basis_qQ}
& O_{qQ4(0:+;\pm)}^{(2)\balpha\bt\bgamma\delta}
=  J_{(q)\bar n 0\, }^{\balpha\bt}\, J_{(Q)n \bar n+\, }^{\bgamma\delta} \,J_{e\pm }
\,,\qquad
O_{qQ4(\bar 0:+;\pm)}^{(2)\balpha\bt\bgamma\delta}
=  J_{(q)\bar n \bar 0\, }^{\balpha\bt}\, J_{(Q) n \bar n +\,}^{\bgamma\delta} \,J_{e\pm }
\,,\\
& O_{qQ4(0:-;\pm)}^{(2)\balpha\bt\bgamma\delta}
= J_{(q)\bar n 0 \, }^{\balpha\bt}\, J_{(Q)n \bar n -\, }^{\bgamma\delta} \,J_{e\pm }
\,,\qquad
O_{qQ4(\bar 0:-;\pm)}^{(2)\balpha\bt\bgamma\delta}
= J_{(q)\bar n \bar 0\,}^{\balpha\bt}\, J_{(Q)n \bar n -\,  }^{\bgamma\delta} \,J_{e\pm }
\,, \nn \\
& O_{qQ5(0:+;\pm)}^{(2)\balpha\bt\bgamma\delta}
=  J_{(q)\bar n 0\,}^{\balpha\bt}\, J_{(Q)\bar n n +\,  }^{\bgamma\delta} \,J_{e\pm }
\,,\qquad
O_{qQ5(\bar 0:+;\pm)}^{(2)\balpha\bt\bgamma\delta}
= J_{(q)\bar n \bar 0\, }^{\balpha\bt}\, J_{(Q)\bar n n +\, }^{\bgamma\delta} \,J_{e\pm }
\,,\nn\\
& O_{qQ5(0:-;\pm)}^{(2)\balpha\bt\bgamma\delta}
= J_{(q)\bar n 0\, }^{\balpha\bt}\, J_{(Q)\bar n n -\, }^{\bgamma\delta} \,J_{e\pm }
\,,\qquad
O_{qQ5(\bar 0:-;\pm)}^{(2)\balpha\bt\bgamma\delta}
=J_{(q)\bar n \bar 0\,}^{\balpha\bt}\, J_{(Q)\bar n n -\,  }^{\bgamma\delta} \,J_{e\pm }
\,.\nn
\end{align}
Note that unlike the case with two quarks in each collinear sector in \Eq{eq:Z2_basis_qQ}, here angular momentum conservation does not impose constraints beyond those from chirality, and the flavor diagonal nature of QCD and tree level electroweak interactions. 
For the $O_{qQ4}^{(2)}$ operators the color basis after BPS field redefinition is
\begin{align}  \label{eq:TBPS_OqQ4}
\vT_{\BPS}^{ \al\bbeta\ga\bdelta} &=
\left( \,\delta_{\al\bdelta}\left[ Y_{n}^\dagger Y_{\bar n}  \right ]_{\ga\bbeta}
\,,\, 
\delta_{\al\bbeta}\left[ Y_{n}^\dagger Y_{\bar n}  \right ]_{\ga\bdelta}
\right)
\,,
\end{align}
while for the $O_{qQ5}^{(2)}$ operators the corresponding basis is
\begin{align}  \label{eq:TBPS_OqQ5}
\vT_{\BPS}^{ \al\bbeta\ga\bdelta} &=
\left(  \left[ Y_{\bar n}^\dagger Y_{n}  \right ]_{\al\bdelta}\,\delta_{\ga\bbeta}
\, \,,\, 
\delta_{\al\bbeta} \left[ Y_{\bar n}^\dagger Y_{n}  \right ]_{\ga\bdelta}
\right)
\,.
\end{align}
In the case of identical quark flavors, the same basis of eight terms as in \eq{Z3_basis_qQ} define $O_{qq4}^{(2)}$ and $O_{qq5}^{(2)}$, and the BPS color basis is as in \eq{TBPS_OqQ4} for $O_{qq4}^{(2)}$, and as in \eq{TBPS_OqQ5} for $O_{qq5}^{(2)}$. For convenience we add additional symmetry factors to the following operators,
\begin{align}
 \boldsymbol{(q)_n (\bar q q \bar{q})_{\bn}:}    {\vcenter{\includegraphics[width=0.18\columnwidth]{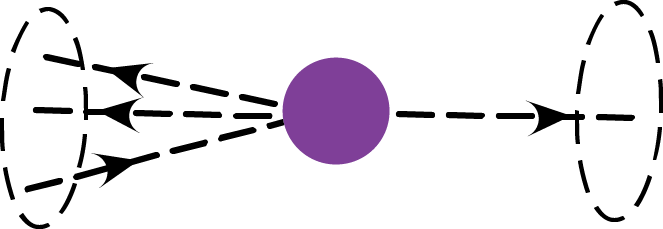}}}
\nn
\end{align}
\vspace{-0.4cm}
\begin{align} \label{eq:Z3_basis_qq}
& O_{qq4(0:+;\pm)}^{(2)\balpha\bt\bgamma\delta}
= \frac{1}{2}  J_{(q)\bar n 0\, }^{\balpha\bt}\, J_{(q)n \bar n+\, }^{\bgamma\delta} \,J_{e\pm }
\,,\qquad
O_{qq4(\bar 0:-;\pm)}^{(2)\balpha\bt\bgamma\delta}
=  \frac{1}{2} J_{(q)\bar n \bar 0\, }^{\balpha\bt}\, J_{(q) n \bar n +\,}^{\bgamma\delta} \,J_{e\pm }
\,,  \\
& O_{qq5(0:+;\pm)}^{(2)\balpha\bt\bgamma\delta}
=  \frac{1}{2} J_{(q)\bar n 0\,}^{\balpha\bt}\, J_{(q)\bar n n +\,  }^{\bgamma\delta} \,J_{e\pm }
\,,\qquad
O_{qq5(\bar 0:-;\pm)}^{(2)\balpha\bt\bgamma\delta}
=  \frac{1}{2} J_{(q)\bar n \bar 0\, }^{\balpha\bt}\, J_{(q)\bar n n +\, }^{\bgamma\delta} \,J_{e\pm }
\,.\nn
\end{align}

Having enumerated a complete basis for all types of four quark operators at ${\cal O}(\lambda^2)$, we now consider relations that follow from $C$ and $P$.
To make these relations explicit we expand the Wilson coefficients for $e\bar e q\bar qQ\bar Q$ as
\begin{align} \label{eq:Z2_expand}
  \vec{C}^{(2)}_{qQi}  (n,\bn;\omega_1, \omega_2; & \omega_3,  \omega_4;\omega_5,\omega_6)
 =  e^2  \,
 \bigg\{\big[Q^\ell Q^q + v_{\lambda_l}^\ell v_{\lambda_q}^q P_Z(s_{56})\big] \vec{C}^{(2)}_{q,qQi}(n,\bn;\omega_1, \omega_2; \omega_3, \omega_4;\omega_5,\omega_6)
 \nn \\ & 
 \hspace{-0.85cm}+ \big[Q^\ell Q^Q + v_{\lambda_l}^\ell v_{\lambda_Q}^Q P_Z(s_{56}) \big] \vec{C}^{(2)}_{Q,qQi}(n,\bn;\omega_1, \omega_2; \omega_3, \omega_4;  \omega_5,\omega_6) \nn \\
& \hspace{-0.85cm} +\frac{v_{\lambda_l}^\ell}{\sin(2\theta_W)} P_Z(s_{56}) \vec{C}^{(2)}_{a,qQi}(n,\bn;\omega_1, \omega_2; \omega_3, \omega_4;\omega_5,\omega_6)   \bigg\}
 \ \nn \\ & 
  \hspace{-0.85cm}+\sum_{j=1}^{n_f} \bigg[Q^\ell Q^j + \frac{v_{\lambda_l}^\ell}{2} (v_L^j +v_R^j) P_Zs_{56}) \bigg] \vec{C}^{(2)}_{v,qQi}(n,\bn;\omega_1, \omega_2; \omega_3, \omega_4;\omega_5,\omega_6)
 \,. 
\end{align}
Since we have accounted for symmetry factors explicitly in the the operators, for the case of identical quark flavors, $e\bar e q\bar qq\bar q$, we have the relation
\begin{align}
 & \vec{C}^{(2)}_{qqi}(n,\bn;\omega_1, \omega_2; \omega_3, \omega_4;\omega_5,\omega_6)=\nn \\
 &\hspace{3cm}\vec{C}^{(2)}_{qQi}(n,\bn;\omega_1, \omega_2; \omega_3, \omega_4;\omega_5,\omega_6)- \vec{C}^{(2)}_{qQi}(n,\bn;\omega_1, \omega_4; \omega_3,  \omega_2;\omega_5,\omega_6)
  \,.
\end{align}

We now discuss relations between different helicity operators due to symmetry constraints. C/P relations combined with the ability to flip the helicity of the electron current, as described in \Eq{eq:lo_weak}, give the following relations between Wilson coefficients
\begin{align} \label{eq:4quarkPrelation}
\vec{C}^{(2)}_{v,\,qQi(\lambda_{12}:\lambda_{34};\lambda_{56})}(n,\bn;\omega_1,\omega_2;\omega_3,\omega_4;\omega_5,\omega_6&) \hspace{-0.05cm}=\hspace{-0.05cm} \vec{C}^{(2)}_{v,\,qQi(-\lambda_{12}:-\lambda_{34};-\lambda_{56})}(n,\bn;\omega_1,\omega_2;\omega_3,\omega_4;\omega_5,\omega_6), \nn \\
\vec{C}^{(2)}_{v,\,qQj(\lambda_{12}:\lambda_{34};\lambda_{56})}(n,\bn;\omega_1,\omega_2;\omega_3,\omega_4;\omega_5,\omega_6&) \hspace{-0.05cm}=\hspace{-0.05cm} - \vec{C}^{(2)}_{v,\,qQj(-\lambda_{12}:-\lambda_{34};-\lambda_{56})}(n,\bn;\omega_2,\omega_1;\omega_4,\omega_3;\omega_6,\omega_5),
\nn \\
  \vec{C}^{(2)}_{v,\,qQi(\lambda_{12}:\lambda_{34};\lambda_{56})}(n,\bn;\omega_1,\omega_2;\omega_3,\omega_4;\omega_5,\omega_6&)\hspace{-0.05cm}=\hspace{-0.05cm} \vec{C}^{(2)}_{v,\,qQi(\lambda_{12}:\lambda_{34};-\lambda_{56})}(n,\bn;\omega_1,\omega_2;\omega_3,\omega_4;\omega_6,\omega_5). 
\end{align}
Here the $\lambda_i$ denote generic helicity labels of the corresponding helicity building blocks, subject to the constraints from angular momentum conservation discussed in this section. For the scalar currents, we are again using the convention that for $\lambda = 0$, $-\lambda = \bar{0}$. 
The same relations hold for $\vec{C}^{(2)}_{Q,qQi}$ and $\vec{C}^{(2)}_{q,qQi}$, and there is an additional overall minus sign in all these relations for $\vec{C}^{(2)}_{a,qQi}$.

\vspace{0.4cm}
\noindent{\bf{Two Quark-Two Gluon Operators:}}

We now consider the operators involving two collinear quark and two collinear gluon building blocks, corresponding to the partonic process $e \bar eq\bar qgg$. The quark and antiquark have the same chirality, but are not necessarily in the same collinear sector, as is also the case for the collinear gluons. The color basis for these channels is three dimensional, and we take our color basis to be
\begin{equation} \label{eq:ggqqll_color}
\vT^{\, ab \alpha\bbeta}
= \Bigl(
(T^a T^b)_{\alpha\bbeta}\,,\, (T^b T^a)_{\alpha\bbeta} \,,\, \tr[T^a T^b]\, \delta_{\alpha\bbeta}
\Bigr)
\,.\end{equation} 

We begin with operators that have the quarks in opposite collinear sectors, and two gluons in the same collinear sector. A basis for these operators is
\begin{align}
 \boldsymbol{(gg q)_n (\bar q)_{\bn}:}     {\vcenter{\includegraphics[width=0.18\columnwidth]{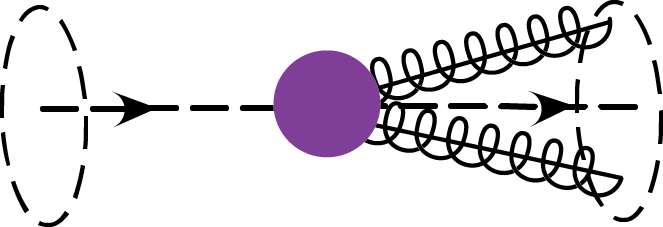}}}
\nn
\end{align}
\vspace{-0.4cm}
\begin{alignat}{2} \label{eq:eeqqgg_basis1}
&O_{\cB1++(-;\pm)}^{(2)ab\, \balpha\bt}
= \frac{1}{2}   \cB_{n+}^a\, \cB_{n+}^b\, J_{n\bar n\, - }^{\balpha\bt}   J_{e\pm }
\, , \qquad 
&&O_{\cB1--(+;\pm)}^{(2)ab\, \balpha\bt}
= \frac{1}{2}  \cB_{ n-}^a\, \cB_{ n-}^b \, J_{n\bar n\, + }^{\balpha\bt}   J_{e\pm }
\, ,  \nn\\
&O_{\cB1+-(+;\pm)}^{(2)ab\, \balpha\bt}
=  \cB_{n+}^a\, \cB_{n-}^b  \, J_{n\bar n\, +}^{\balpha\bt}   J_{e\pm }
\, , \qquad 
&&O_{\cB1+-(-;\pm)}^{(2)ab\, \balpha\bt}
=  \, \cB_{n+}^a\, \cB_{ n-}^b J_{n\bar n\, -}^{\balpha\bt}    J_{e\pm }
\, , 
\end{alignat}
\begin{align}
 \boldsymbol{(gg \bar q)_n (q)_{\bn}:}   {\vcenter{\includegraphics[width=0.18\columnwidth]{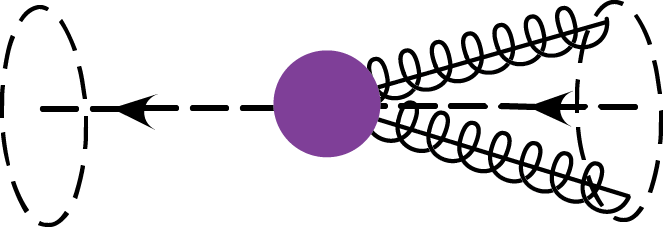}}}
\nn
\end{align}
\vspace{-0.4cm}
\begin{alignat}{2} \label{eq:eeqqgg_basis1a}
&O_{\cB2++(-;\pm)}^{(2)ab\, \balpha\bt}
= \frac{1}{2}   \cB_{n+}^a\, \cB_{n+}^b\, J_{\bar n n\, + }^{\balpha\bt}   J_{e\pm }
\, , \qquad 
&&O_{\cB2--(+;\pm)}^{(2)ab\, \balpha\bt}
= \frac{1}{2}  \cB_{ n-}^a\, \cB_{ n-}^b \, J_{\bar n n\, - }^{\balpha\bt}   J_{e\pm }
\, , \nn \\
&O_{\cB2+-(+;\pm)}^{(2)ab\, \balpha\bt}
=  \cB_{n+}^a\, \cB_{n-}^b  \, J_{\bar n n\, +}^{\balpha\bt}   J_{e\pm }
\, , \qquad 
&&O_{\cB2+-(-;\pm)}^{(2)ab\, \balpha\bt}
=  \, \cB_{n+}^a\, \cB_{ n-}^b J_{\bar n n\, -}^{\balpha\bt}    J_{e\pm }
\, . 
\end{alignat}
Here we have used constraints from angular momentum conservation to eliminate operators which do not have $h=0,\pm1$ along the $\hat n$ axis, and we have taken the two gluon fields to be in the $n$ collinear sector. For the operators in \eq{eeqqgg_basis1} the color basis after BPS field redefinition is
\begin{equation}
\vT_{\BPS}^{\, ab \alpha\bbeta}
= \Bigl(
(T^a T^bY^\dagger_n Y_{\bar n})_{\alpha\bbeta}\,,\, (T^b T^aY^\dagger_n Y_{\bar n})_{\alpha\bbeta} \,,\, \tr[T^a T^b]\, [Y^\dagger_n Y_{\bar n}]_{\alpha\bbeta}
\Bigr)
\,,
\end{equation}
while for the operators in \eq{eeqqgg_basis1a} it is 
\begin{equation}
\vT_{\BPS}^{\, ab \alpha\bbeta}
= \Bigl(
(Y^\dagger_\bn Y_{n} T^a T^b)_{\alpha\bbeta}\,,\, 
(Y^\dagger_\bn Y_{n}T^b T^a)_{\alpha\bbeta} \,,\, 
\tr[T^a T^b]\, [Y^\dagger_\bn Y_{n}]_{\alpha\bbeta}
\Bigr)
\,.
\end{equation}

Next we consider the operators with two gluon building blocks in distinct collinear sectors. When the quarks and gluons are both in distinct collinear sectors the basis of operators is 
\begin{align}
 \boldsymbol{(g q)_n (g \bar q)_{\bn}:}   {\vcenter{\includegraphics[width=0.18\columnwidth]{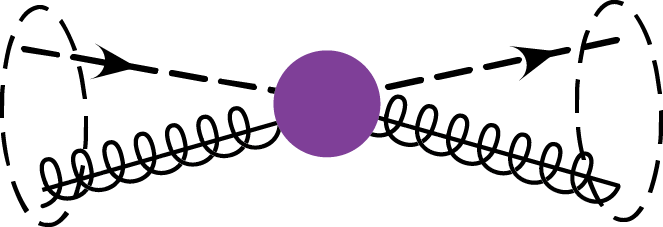}}}  \nn
\end{align}
\vspace{-0.45cm}
\begin{alignat}{2} \label{eq:eeqqgg_basis2}
&O_{\cB3++(+;\pm)}^{(2)ab\, \balpha\bt}
= \cB_{n+}^a\, \cB_{\bar n+}^b \, J_{n\bar n\, + }^{\balpha\bt}   J_{e\pm } \, , \qquad 
&&O_{\cB3--(-;\pm)}^{(2)ab\, \balpha\bt}
=    \cB_{n-}^a\, \cB_{\bar n-}^b \, J_{n\bar n\, -}^{\balpha\bt} J_{e\pm }\, ,  
\nn \\
&O_{\cB3++(-;\pm)}^{(2)ab\, \balpha\bt}
= \cB_{n+}^a\, \cB_{\bar n+}^b \, J_{n\bar n\, - }^{\balpha\bt}   J_{e\pm } \, , \qquad 
&&O_{\cB3--(+;\pm)}^{(2)ab\, \balpha\bt}
=    \cB_{n-}^a\, \cB_{\bar n-}^b \, J_{n\bar n\, +}^{\balpha\bt} J_{e\pm }\, , 
\\
&O_{\cB3+-(-;\pm)}^{(2)ab\, \balpha\bt}
=    \cB_{n+}^a\, \cB_{\bar n-}^b \, J_{n\bar n\, - }^{\balpha\bt}  J_{e\pm }\, , \qquad 
&&O_{\cB3-+(+;\pm)}^{(2)ab\, \balpha\bt}
=\cB_{n-}^a\, \cB_{\bar n+}^b  \, J_{n\bar n\, + }^{\balpha\bt} J_{e\pm }  \, , \nn
\end{alignat}
and the color basis after BPS field redefinition is
\begin{equation}
\vT_{\BPS}^{\, ab \alpha\bbeta}
= \Bigl(
(T^a  Y^\dagger_n Y_{\bar n} T^b)_{\alpha\bbeta}\,,\, (Y^\dagger_n T^d \cY_{\bn}^{db}  T^c \cY^{ca}_n Y_{\bar n})_{\alpha\bbeta} \,,\, \tr[ T^c \cY^{ca}_n T^d \cY_{\bn}^{db} ]\, [Y^\dagger_n Y_{\bar n}]_{\alpha\bbeta}
\Bigr)
 .
\end{equation}
Here operators with $J_{\bn n\lambda}^{\balpha\beta}$ are obtained from those in \eq{eeqqgg_basis2} by $n\leftrightarrow \bn$.  When the two quarks are in the same collinear sector the basis is given by
\begin{align}
 \boldsymbol{(g q\bar q)_n (g)_{\bn}:}   {\vcenter{\includegraphics[width=0.18\columnwidth]{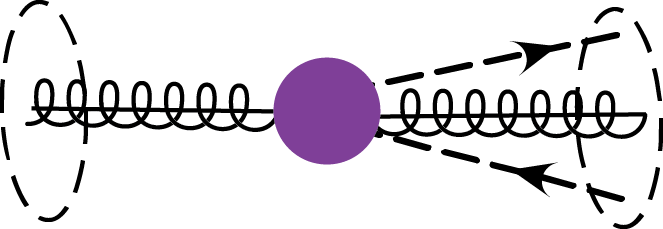}}} \nn
\end{align}
\vspace{-0.4cm}
\begin{alignat}{2} \label{eq:eeqqgg_basis3}
&O_{\cB4++(0:\pm)}^{(2)ab\, \balpha\bt}
=   \cB_{n+}^a\, \cB_{\bar n+}^b \, J_{n\,{0} }^{\balpha\bt}  J_{e\pm }\, , \qquad 
&&O_{\cB4++(\bar 0:\pm)}^{(2)ab\, \balpha\bt}
=\cB_{n+}^a\, \cB_{\bar n+}^b  \, J_{n\,{\bar 0} }^{\balpha\bt}   J_{e\pm }\, ,  \\
&O_{\cB4--(0:\pm)}^{(2)ab\, \balpha\bt}
=  \cB_{n-}^a\, \cB_{\bar n-}^b \, J_{n\,{0} }^{\balpha\bt}    J_{e\pm }\, , \qquad 
&&O_{\cB4--(\bar 0:\pm)}^{(2)ab\, \balpha\bt}
= \cB_{ n-}^a\, \cB_{\bar n-}^b  \, J_{n\,{\bar 0} }^{\balpha\bt}   J_{e\pm }\, . \nn
\end{alignat}
The color basis after BPS field redefinition is given by
\begin{equation}
\vT_{\BPS}^{\, ab \alpha\bbeta}
= \Bigl(
(\cY_n^T \cY_\bn)^{cb}  (T^a T^c)_{\alpha\bbeta} \,,\, 
(\cY_n^T \cY_\bn)^{cb}  (T^c T^a)_{\alpha\bbeta} \,,\, 
T_F (\cY_n^T \cY_\bn)^{ab} \, \delta_{\alpha\bbeta}
%
%
\Bigr)
\,.
\end{equation}
In writing \eq{eeqqgg_basis3} we have again used constraints of angular momentum conservation to restrict the allowed operators in the basis. 

Finally we consider the basis of operators with both quarks in the same collinear sector, and both gluons in the other collinear sector. Imposing angular momentum conservation reduces the basis to two distinct operators
\begin{align}
& \boldsymbol{(q \bar q)_n (gg)_{\bn}:}{\vcenter{\includegraphics[width=0.18\columnwidth]{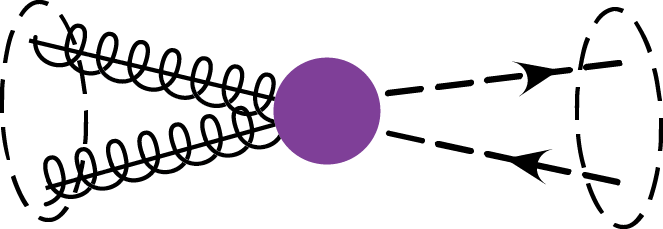}}}  \nn
\end{align}
\vspace{-0.4cm}
\begin{align}\label{eq:eeqqgg_basis4}
&O_{\cB5+-(0:\pm)}^{(2)ab\, \balpha\bt}
= \cB_{\bar n+}^a\, \cB_{ \bar n-}^b  \, J_{n\,{0} }^{\balpha\bt}   J_{e\pm }\,, \qquad 
&O_{\cB5+-(\bar 0:\pm)}^{(2)ab\, \balpha\bt}
= \cB_{\bar n+}^a\, \cB_{ \bar n-}^b  \, J_{n\,{\bar 0} }^{\balpha\bt}   J_{e\pm }\,.
\end{align}
The color basis after BPS field redefinition is
\begin{equation}
\vT_{\BPS}^{\, ab \alpha\bbeta}
= \Bigl(
(Y_n^\dagger Y_\bn T^a T^b Y^\dagger_{\bar n} Y_n )_{\alpha\bbeta}
  \,,\, 
(Y_n^\dagger Y_\bn T^b T^a Y^\dagger_{\bar n} Y_n )_{\alpha\bbeta}
   \,,\, 
\tr[T^a T^b]\, \delta_{\alpha\bbeta} \Bigr)
\,.
\end{equation}
We have chosen to write the operators with both quarks in the $n$ sector. 

For the $e\bar e q\bar q g g$ operators in Eqs.~(\ref{eq:eeqqgg_basis1},\ref{eq:eeqqgg_basis2},\ref{eq:eeqqgg_basis3},\ref{eq:eeqqgg_basis4}) we expand the Wilson coefficients as
 \begin{align} \label{eq:Z2_qqgg_expand}
& \vec{C}^{(2)}_{Bi}(n,\bn;\omega_1, \omega_2;  \omega_3,  \omega_4;\omega_5,\omega_6)
 = e^2\,  
 \bigg\{ \big[Q^\ell Q^q + v_{\lambda_l}^\ell v_{\lambda_q}^q P_Z(s_{56})\big] \vec{C}^{(2)}_{q,Bi}(n,\bn;\omega_1, \omega_2; \omega_3, \omega_4;\omega_5,\omega_6) \nn
 \\ & 
 \hspace{4cm}+\sum_{j=1}^{n_f} \bigg[Q^\ell Q^j + \frac{v_{\lambda_l}^\ell}{2} (v_L^j +v_R^j) P_Zs_{56}) \bigg] \vec{C}^{(2)}_{v,Bi}(n,\bn;\omega_1, \omega_2; \omega_3, \omega_4;\omega_5,\omega_6) \nn
 \\ &
\hspace{4cm} + \frac{v_{\lambda_l}^\ell}{\sin(2\theta_W)} P_Z(s_{56}) \vec{C}^{(2)}_{a,Bi}(n,\bn;\omega_1, \omega_2; \omega_3, \omega_4;\omega_5,\omega_6)  \bigg\}
 \,.
 \end{align}
C/P relations combined with the ability to flip the helicity of the electron current, as described in \Eq{eq:lo_weak}, give the following relations between Wilson coefficients
\begin{align} \label{eq:2q2gCPrelation}
&\vec{C}^{(2)}_{v,\,\cB\, i\lambda_1 \lambda_2(\lambda_{34}:\lambda_{56})}(n,\bn;\omega_1,\omega_2;\omega_3,\omega_4;\omega_5,\omega_6) \nn \\
& \hspace{4cm}= \vec{C}^{(2)}_{v,\,\cB\, i-\lambda_1 -\lambda_2(-\lambda_{34}:-\lambda_{56})}(n,\bn;\omega_1,\omega_2;\omega_3,\omega_4;\omega_5,\omega_6)\,,  \\
&\vec{C}^{(2)}_{v,\,\cB\, i\lambda_1 \lambda_2(\lambda_{34}:\lambda_{56})}(n,\bn;\omega_1,\omega_2;\omega_3,\omega_4;\omega_5,\omega_6) \nn  \\
 &\hspace{4cm}= \begin{pmatrix*}[r] 0 & 1 & 0 \\ 1 & 0 & 0 \\ 0 & 0 & 1\end{pmatrix*} \vec{C}^{(2)}_{v,\,\cB\, i\lambda_1 \lambda_2(-\lambda_{34}:-\lambda_{56})}(n,\bn;\omega_1,\omega_2;\omega_4,\omega_3;\omega_5,\omega_6)\,, 
\nn \\
&\vec{C}^{(2)}_{v,\,\cB\, i\lambda_1 \lambda_2(\lambda_{34}:\lambda_{56})}(n,\bn;\omega_1,\omega_2;\omega_3,\omega_4;\omega_5,\omega_6) \nn \\
& \hspace{4cm}= \vec{C}^{(2)}_{v,\,\cB\, i\lambda_1 \lambda_2(\lambda_{34}:-\lambda_{56})}(n,\bn;\omega_1,\omega_2;\omega_3,\omega_4;\omega_6,\omega_5)\,,  \nn
\end{align}
where the index $i$ runs from 1 to 4. Here the $\lambda_i$ denote generic helicity labels of the corresponding helicity building blocks, subject to the constraints from angular momentum conservation discussed in this section. The same two relations hold for $\vec{C}^{(2)}_{q}$, and hold with the addition of an overall minus sign for $\vec{C}^{(2)}_{a}$.

\vspace{0.4cm}
\noindent{\bf{Four Gluon Operators:}}

Finally, we consider $\mathcal{O}(\lambda^2)$ hard scattering operators involving four collinear gluons. The four gluon channel gives a highly suppressed contribution for $e^+e^- \to$ dijets and Drell-Yan, but we nevertheless present it here for completeness. It also provides a nice demonstration of the helicity basis approach, as the construction of a minimal basis of four gluon operators is quite difficult otherwise. The helicity operators that include four gluons were presented in the example of in \Eq{eq:helicitygggg} for the case of four well separated collinear sectors. To adapt these operators to the case of two collinear sectors, we need to restrict the sector labels to $n$ and $\bn$ and impose the angular momentum constraints of \sec{ang_cons}. The color basis for the four gluon operators will include more structures than were used in \Eq{eq:color_gggg}, as we now have to allow axial couplings and CP violation. Our choice for this basis is
\begin{equation} \label{eq:gggg_color}
\vT^{ abcd} =
\frac{1}{2}\begin{pmatrix}
\tr[abcd] + \tr[dcba] \\ \tr[acdb] + \tr[bdca] \\ \tr[adbc] + \tr[cbda] \\
\tr[abcd] - \tr[dcba] \\ \tr[acdb] - \tr[bdca] \\ \tr[adbc] - \tr[cbda] \\ 2\tr[ab]\, \tr[cd] \\ 2\tr[ac]\, \tr[db] \\ 2\tr[ad]\, \tr[bc]
\end{pmatrix}^{\!\!\!T}
\,.\end{equation}
For the specific case of SU($N_c$) with $N_c=3$ it is possible to further reduce the color basis by using relations of the form
\begin{align}
&\tr[abcd+dcba] + \tr[acdb+bdca] + \tr[adbc+cbda]
\nn\\ & \qquad
= \tr[ab]\tr[cd] + \tr[ac]\tr[db] + \tr[ad]\tr[bc]
\,.\end{align}
We prefer not to use this relation since it makes the structure more complicated, and does not hold for $N_c>3$, and hence one can not look at the large $N_c$ scaling of results if one uses such relations.

To construct a complete basis of four gluon operators with two collinear sectors, we need to consider two cases. First, when we have two gluons in each sector a basis of operators is
\begin{align}
& \boldsymbol{(gg)_n (gg)_{\bn}:}{\vcenter{\includegraphics[width=0.18\columnwidth]{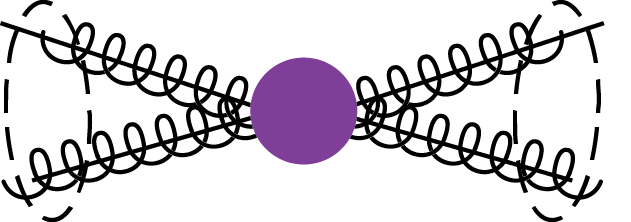}}}  \nn
\end{align}
\vspace{-0.4cm}
\begin{alignat}{2} \label{eq:Z2_basis_gggg_1}
&O_{4g1++++(\pm)}^{(2)a b c d}
= \,\frac{1}{4}  \cB^a_{n +} \cB^b_{n +} \cB^c_{\bn +} \cB^d_{\bn +} J_{e\pm }
\,,\qquad &
&O_{4g1+-+-(\pm)}^{(2)a b c d}
= \, \cB^a_{n +} \cB^b_{n -} \cB^c_{\bn +} \cB^d_{\bn -} J_{e\pm }
\,,\\
&O_{4g1----(\pm)}^{(2)a b c d}
= \,\frac{1}{4}  \cB^a_{n -} \cB^b_{n -} \cB^c_{\bn -} \cB^d_{\bn -} J_{e\pm }
\,,\qquad &\nn
\end{alignat}
where we have used angular momentum constraints to eliminate operators that contain only one $+$ or one $-$ helicity. The basis of color structures after BPS field redefinition is given by

\begin{equation}
\vT_{\text{BPS}}^{ abcd} =
\frac{1}{2}\begin{pmatrix}
(\tr[ T^{a'}   T^{b'}     T^{c'}   T^{d'}   ] + \tr[  T^{d'}   T^{c'}     T^{b'}   T^{a'}  ]) \cY^{a' a}_{n} \cY^{b' b}_{n} \cY^{c' c}_{\bn} \cY^{d' d}_{\bn}
\\
(\tr[ T^{a'}    T^{c'}   T^{d'}     T^{b'}  ] + \tr[ T^{b'}     T^{d'}   T^{c'}    T^{a'} ] ) \cY^{a' a}_{n} \cY^{b' b}_{n} \cY^{c' c}_{\bn} \cY^{d' d}_{\bn}
\\
(\tr[ T^{a'}    T^{d'}     T^{b'}     T^{c'}   ] + \tr[  T^{c'}     T^{b'}     T^{d'}    T^{a'}  ] ) \cY^{a' a}_{n} \cY^{b' b}_{n} \cY^{c' c}_{\bn} \cY^{d' d}_{\bn}
\\
(\tr[ T^{a'}   T^{b'}     T^{c'}   T^{d'}   ] - \tr[  T^{d'}   T^{c'}     T^{b'}   T^{a'}  ]) \cY^{a' a}_{n} \cY^{b' b}_{n} \cY^{c' c}_{\bn} \cY^{d' d}_{\bn}
\\
(\tr[ T^{a'}    T^{c'}   T^{d'}     T^{b'}  ] - \tr[ T^{b'}     T^{d'}   T^{c'}    T^{a'} ] ) \cY^{a' a}_{n} \cY^{b' b}_{n} \cY^{c' c}_{\bn} \cY^{d' d}_{\bn}
\\
(\tr[ T^{a'}    T^{d'}     T^{b'}     T^{c'}   ] - \tr[  T^{c'}     T^{b'}     T^{d'}    T^{a'}  ] ) \cY^{a' a}_{n} \cY^{b' b}_{n} \cY^{c' c}_{\bn} \cY^{d' d}_{\bn}
\\
\frac{1}{2}  \delta^{ab} \delta^{cd} 
\\
\frac{1}{2} (\cY^T_n \cY_\bn)^{ac} (\cY^T_n \cY_\bn)^{bd} 
\\
\frac{1}{2} (\cY^T_n \cY_\bn)^{ad} (\cY^T_n \cY_\bn)^{bc}
\end{pmatrix}^{\!\!\!T} 
.\end{equation}

The other relevant case has three gluons in one sector, and we can take advantage of the $n\leftrightarrow \bn$ symmetry to choose the three gluons to be in the $\bn$ collinear sector. The basis of operators is then given by
\begin{align}
& \boldsymbol{(g)_n (ggg)_{\bn}:}{\vcenter{\includegraphics[width=0.18\columnwidth]{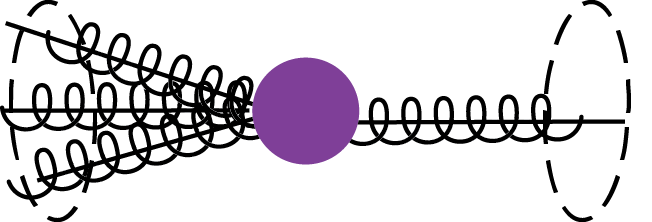}}}  \nn
\end{align}
\vspace{-0.4cm}
\begin{alignat}{2} \label{eq:Z2_basis_gggg_2}
&O_{4g2+++-(\pm)}^{(2)a b c d}
= \,\frac{1}{2}  \cB^a_{n +} \cB^b_{\bn +} \cB^c_{\bn +} \cB^d_{\bn -} J_{e\pm }
\,,\qquad &
&O_{4g2-+--(\pm)}^{(2)a b c d}
= \,\frac{1}{2}  \cB^a_{n -} \cB^b_{\bn +} \cB^c_{\bn -} \cB^d_{\bn -} J_{e\pm }
\,,
\end{alignat}
where we have once again used conservation of angular momentum to restrict to these particular helicity choices. In this case, we have 

\begin{equation}
\vT_{\text{BPS}}^{ abcd} =
\frac{1}{2}\begin{pmatrix}
(\tr[ T^{a'}   T^{b'}     T^{c'}   T^{d'}   ] + \tr[  T^{d'}   T^{c'}     T^{b'}   T^{a'}  ]) \cY^{a' a}_{n} \cY^{b' b}_{\bn} \cY^{c' c}_{\bn} \cY^{d' d}_{\bn}
\\
(\tr[ T^{a'}    T^{c'}   T^{d'}     T^{b'}  ] + \tr[ T^{b'}     T^{d'}   T^{c'}    T^{a'} ] ) \cY^{a' a}_{n} \cY^{b' b}_{\bn} \cY^{c' c}_{\bn} \cY^{d' d}_{\bn}
\\
(\tr[ T^{a'}    T^{d'}     T^{b'}     T^{c'}   ] + \tr[  T^{c'}     T^{b'}     T^{d'}    T^{a'}  ] ) \cY^{a' a}_{n} \cY^{b' b}_{\bn} \cY^{c' c}_{\bn} \cY^{d' d}_{\bn}
\\
(\tr[ T^{a'}   T^{b'}     T^{c'}   T^{d'}   ] - \tr[  T^{d'}   T^{c'}     T^{b'}   T^{a'}  ]) \cY^{a' a}_{n} \cY^{b' b}_{\bn} \cY^{c' c}_{\bn} \cY^{d' d}_{\bn}
\\
(\tr[ T^{a'}    T^{c'}   T^{d'}     T^{b'}  ] - \tr[ T^{b'}     T^{d'}   T^{c'}    T^{a'} ] ) \cY^{a' a}_{n} \cY^{b' b}_{\bn} \cY^{c' c}_{\bn} \cY^{d' d}_{\bn}
\\
(\tr[ T^{a'}    T^{d'}     T^{b'}     T^{c'}   ] - \tr[  T^{c'}     T^{b'}     T^{d'}    T^{a'}  ] ) \cY^{a' a}_{n} \cY^{b' b}_{\bn} \cY^{c' c}_{\bn} \cY^{d' d}_{\bn}
\\
\frac{1}{2}  (\cY^T_n \cY_\bn)^{ab} \delta^{cd} 
\\
\frac{1}{2} (\cY^T_n \cY_\bn)^{ac} \delta^{bd} 
\\
\frac{1}{2} (\cY^T_n \cY_\bn)^{ad} \delta^{bc}
\end{pmatrix}^{\!\!\!T} 
.\end{equation}
Once again for simplicity we have not used identities to simplify some of these Wilson line structures. Also we note that 

Just as in the case of three gluons, the $e \bar{e} gggg$ channel must proceed through a fermion loop, so we can decompose the Wilson coefficient as
\begin{align} \label{eq:Z2_gggg_expand}
\vec{C}^{(2)}(n,\bn;\omega_1, \omega_2,&\omega_3,\omega_4; \omega_5,\omega_6)
= e^2 \,
\bigg\{\frac{v_{\lambda_l}^\ell}{\sin(2\theta_W)} P_Z(s_{56})  \vec{C}^{(2)}_a(n,\bn;\omega_1, \omega_2,\omega_3, \omega_4;\omega_5,\omega_6) \nn \\
&\hspace{-1cm}+\sum_{j=1}^{n_f} \bigg[Q^\ell Q^j + \frac{v_{\lambda_l}^\ell}{2} (v_L^j +v_R^j) P_Zs_{56}) \bigg]  \vec{C}^{(2)}_v(n,\bn;\omega_1, \omega_2,\omega_3, \omega_4;\omega_5,\omega_6) \bigg\}  \,, 
\end{align}
where we have suppressed all of the helicity labels and $\vec{C}^{(2)}_a$ and $\vec{C}^{(2)}_v$ correspond to the axial or vector coupling contributions respectively. With this Wilson coefficient expansion and the color basis from \Eq{eq:gggg_color}, C/P relations combined with the ability to flip the helicity of the electron current, as described in \Eq{eq:lo_weak}, give the following relations between Wilson coefficients
\begin{align}  \label{eq:chargerelgggg}
\vec{C}^{(2)}_{v\,\lambda_1\lambda_2\lambda_3 \lambda_4(\lambda_{56})}(n,\bn;\{\omega_i\}) &= \vec{C}^{(2)}_{ v\,-\lambda_1\,-\lambda_2\,-\lambda_3\,-\lambda_4(-\lambda_{56})}(n,\bn;\{\omega_i\}) \,,  \\
\vec{C}^{(2)}_{v\,\lambda_1\lambda_2\lambda_3 \lambda_4(\lambda_{56})}(n,\bn;\omega_1,\omega_2,\omega_3,\omega_4;\omega_5,\omega_6) &= \hat{V}_{4g} \vec{C}^{(2)}_{v\,\lambda_1\lambda_2\lambda_3 \lambda_4(-\lambda_{56})}(n,\bn;\omega_1,\omega_2,\omega_3,\omega_4;\omega_6,\omega_5) \,, \nn \\
\vec{C}^{(2)}_{v\,\lambda_1\lambda_2\lambda_3 \lambda_4(\lambda_{56})}(n,\bn;\omega_1,\omega_2,\omega_3,\omega_4;\omega_5,\omega_6) &= \vec{C}^{(2)}_{v\,\lambda_1\lambda_2\lambda_3 \lambda_4(-\lambda_{56})}(n,\bn;\omega_1,\omega_2,\omega_3,\omega_4;\omega_6,\omega_5) 
\nn \,,
\end{align}
where $\hat{V}_{4g}$ is diagonal in the space defined by \Eq{eq:gggg_color} with +1 for the first three entries, -1 for the middle three entries and +1 for the final three entries. The $\lambda_i$ are generic helicity labels, but are restricted by the constraints from angular momentum conservation discussed earlier. $\vec{C}^{(2)}_a$ satisfies the same relations with an additional negative sign.

\subsubsection{$\cP_\perp^\pm$ Insertions}

For our choice of kinematics, hard scattering operators with explicit $\cP_\perp^\pm $ insertions first arise at $\cO(\lambda^2)$. Since operators involving a $\cP_\perp^\pm$ insertion that acts on an entire collinear sector vanish, the only non-vanishing $\cO(\lambda^2)$ operators involve single insertions of $\cP_\perp^\pm$ into subleading power operators with two collinear fields in the same sector. The contributing partonic processes are identical to those considered at subleading power, so we can decompose the Wilson coefficients following \eqs{Z1_qqg_expand}{Z1_ggg_expand}, and use the color bases of \Eqss{eq:Z1q_color}{eq:BPSgqqcolor}{eq:Z1g_color}.

For the insertions of $\cP_\perp^\pm$ into operators involving two quarks and a gluon, a basis of operators for the case that the quarks are in distinct collinear sectors is
\begin{align}
& \boldsymbol{(gq \cP_\perp)_n (\bar q)_{\bn}:}{\vcenter{\includegraphics[width=0.18\columnwidth]{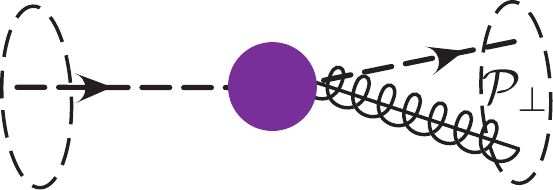}}}  \nn
\end{align}
\vspace{-0.4cm}
\begin{alignat}{2} \label{eq:eeqqgpperp_basis}
&O_{\cP1n + (+;\pm)[-]}^{(2)a\,\balpha\bt}
= \left [ \cP_{\perp}^{-} \cB_{n+}^a \right ]\, J_{n\bar n\, +}^{\balpha\bt}\, J_{e\pm }
\,,\qquad &
&O_{\cP1n - (-;\pm)[+]}^{(2)a\,\balpha\bt}
= \left [ \cP_{\perp}^{+} \cB_{n-}^a \right ]\, J_{n\bar n\, -}^{\balpha\bt}\, J_{e\pm }
\,,\\
&O_{\cP1n - (+;\pm)[+]}^{(2)a\,\balpha\bt}
=  \left [ \cP_{\perp}^{+} \cB_{n-}^a \right ]\, J_{n\bar n\, +}^{\balpha\bt}\, J_{e\pm }
\,,\qquad &
&O_{\cP1n - (+;\pm)[-]}^{(2)a\,\balpha\bt}
= \left [ \cP_{\perp}^{-} \cB_{n-}^a \right ]\, J_{n\bar n\, +}^{\balpha\bt}\, J_{e\pm }
\,,\nn\\
&O_{\cP1n + (-;\pm)[+]}^{(2)a\,\balpha\bt}
=  \left [ \cP_{\perp}^{+} \cB_{n+}^a \right ]\, J_{n\bar n\, -}^{\balpha\bt}\, J_{e\pm }
\,,\qquad &
&O_{\cP1n + (-;\pm)[-]}^{(2)a\,\balpha\bt}
= \left [ \cP_{\perp}^{-} \cB_{n+}^a \right ]\, J_{n\bar n\, -}^{\balpha\bt}\, J_{e\pm }
\,, \nn
\end{alignat}
and
\begin{align}
& \boldsymbol{(q)_n (g\bar q \cP_\perp)_{\bn}:}{\vcenter{\includegraphics[width=0.18\columnwidth]{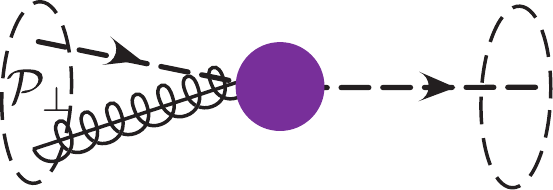}}}  \nn
\end{align}
\vspace{-0.4cm}
\begin{alignat}{2} \label{eq:eeqqgpperp_basis_flip}
&O_{\cP1\bar n + (+;\pm)[-]}^{(2)a\,\balpha\bt}
= \left [ \cP_{\perp}^{-} \cB_{\bn+}^a \right ]\, J_{n\bar n\, +}^{\balpha\bt}\, J_{e\pm }
\,,\qquad &
&O_{\cP1\bar n - (-;\pm)[+]}^{(2)a\,\balpha\bt}
= \left [ \cP_{\perp}^{+} \cB_{\bn-}^a \right ]\, J_{n\bar n\, -}^{\balpha\bt}\, J_{e\pm }
\,,\\
&O_{\cP1\bar n - (+;\pm)[+]}^{(2)a\,\balpha\bt}
=  \left [ \cP_{\perp}^{+} \cB_{\bn-}^a \right ]\, J_{n\bar n\, +}^{\balpha\bt}\, J_{e\pm }
\,,\qquad &
&O_{\cP1\bar n - (+;\pm)[-]}^{(2)a\,\balpha\bt}
= \left [ \cP_{\perp}^{-} \cB_{\bn-}^a \right ]\, J_{n\bar n\, -}^{\balpha\bt}\, J_{e\pm }
\,,\nn\\
&O_{\cP1\bar n + (-;\pm)[+]}^{(2)a\,\balpha\bt}
=  \left [ \cP_{\perp}^{+} \cB_{\bn+}^a \right ]\, J_{n\bar n\, +}^{\balpha\bt}\, J_{e\pm }
\,,\qquad &
&O_{\cP1\bar n + (-;\pm)[-]}^{(2)a\,\balpha\bt}
= \left [ \cP_{\perp}^{-} \cB_{\bn+}^a \right ]\, J_{n\bar n\, -}^{\balpha\bt}\, J_{e\pm }
\,, \nn
\end{alignat}
which we refer to as $\cP 1$ operators. In the case that they are in the same collinear sector the basis is,
\begin{align}
&   \boldsymbol{(g)_n (q\bar q\, \cP_\perp)_{\bn}:}{\vcenter{\includegraphics[width=0.18\columnwidth]{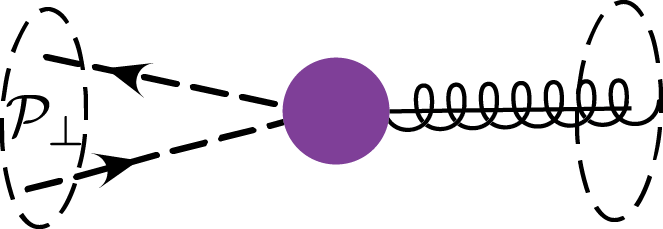}}}  \nn
\end{align}
\vspace{-0.4cm}
\begin{alignat}{2}\label{eq:eeqqgpperp_basis_same}
&O_{\cP2 + (0:\pm)[+]}^{(2)a\,\balpha\bt}
= \cB_{n+}^a\, \big\{ \cP_{\perp}^{+} J_{\bar n\, 0}^{\balpha\bt} \big\}\,  J_{e\pm }
\,,\qquad &
&O_{\cP2 - (0:\pm)[-]}^{(2)a\,\balpha\bt}
= \cB_{n-}^a\, \big\{ \cP_{\perp}^{-} J_{\bar n\, 0    }^{\balpha\bt} \big\} \, J_{e\pm } 
\,,\\
&O_{\cP2 + (\bar 0:\pm)[+]}^{(2)a\,\balpha\bt}
=  \cB_{n+}^a\,  \big\{ \cP_{\perp}^{+} J_{\bar n\, \bar0    }^{\balpha\bt} \big\}\,  J_{e\pm }
\,,\qquad &
&O_{\cP2 - (\bar 0:\pm)[-]}^{(2)a\,\balpha\bt}
= \cB_{n-}^a \, \big\{ \cP_{\perp}^{-} J_{\bar n\, \bar0    }^{\balpha\bt} \big\}  \, J_{e\pm }
\,,\nn
\end{alignat}
and we refer to these as $\cP2$ operators. If we integrate the $\cP_\perp^\pm$ by parts in \eq{eeqqgpperp_basis} then it gives the operators involving $[\cP_\perp^\pm J_{n\bn\lambda}^{\balpha\beta}]$, and doing this in \eq{eeqqgpperp_basis_same} gives the terms $\{ J_{\bn 0}^{\balpha\beta} (\cP_\perp^\pm)^\dagger\}$ and $\{ J_{\bn \bar 0}^{\balpha\beta} (\cP_\perp^\pm)^\dagger\}$, thus explaining why these structures do not appear as separate terms in the basis. In \eq{eeqqgpperp_basis} there is only only field in the $\bn$ direction, so any operators that contain $\{J_{n \bn \pm}^{\balpha \beta} (\cP_\perp^\lambda)^\dagger\}$ vanish. Similarly, in \eq{eeqqgpperp_basis_same}, any operators that contain $[\cP^\lambda_\perp \cB_{n}^\pm]$ are zero for our choice of kinematics. In both \eq{eeqqgpperp_basis} and \eq{eeqqgpperp_basis_same}  we have used angular momentum conservation of the hard scattering process to reduce the helicity combinations allowed in the basis. 

C/P relations combined with the ability to flip the helicity of the electron current, as described in \Eq{eq:lo_weak}, give the following relations between Wilson coefficients for the $\cP 1$ operators
\begin{align} \label{eq:Pparityrel1}
\vec{C}^{(2)}_{v,\,\cP 1 \lambda_1 (\lambda_{23}:\lambda_{45})[\lambda_{P}]}(n,\bn;\{\omega_i\})  &= - \vec{C}^{(2)}_{v,\,\cP 1 -\lambda_1 (-\lambda_{23};-\lambda_{45})[-\lambda_{P}]}(n^\P,\bn^\P;\{\omega_i\})\,,  \\ 
\vec{C}^{(2)}_{v,\,\cP 1 \lambda_1 (\lambda_{23}:\lambda_{45})[\lambda_{\cP}]}(n,\bn;\omega_1;\omega_2,\omega_3;\omega_4,\omega_5) &=\!-\vec{C}^{(2)}_{v,\,\cP 1 -\lambda_1 (-\lambda_{23}:-\lambda_{45})[-\lambda_{\cP}]}(n,\bn;\omega_1;\omega_3,\omega_2;\omega_5,\omega_4)\,, \nn \\
\vec{C}^{(2)}_{v,\,\cP 1 \lambda_1 (\lambda_{23}:\lambda_{45})[\lambda_{\cP}]}(n,\bn;\omega_1;\omega_2,\omega_3;\omega_4,\omega_5) &= \vec{C}^{(2)}_{v,\,\cP 1 \lambda_1  (\lambda_{23}:-\lambda_{45})[\lambda_{\cP}]}(n,\bn;\omega_1;\omega_2,\omega_3;\omega_5,\omega_4)\,,\nn
\end{align}
which hold for both the $n$ and $\bar n$ versions of the operators, 
and similarly for the $\cP 2$ operators
\begin{align}
\vec{C}^{(2)}_{v,\,\cP 2 \lambda_1 (\lambda_{23}:\lambda_{45})[\lambda_{\cP}]}(n,\bn;\{\omega_i\}) &= - \vec{C}^{(2)}_{v,\,\cP 2 -\lambda_1 (-\lambda_{23}:-\lambda_{45})[-\lambda_{\cP}]}(n^\P,\bn^\P;\{\omega_i\})\,,  \\ 
\vec{C}^{(2)}_{v,\,\cP 2 \lambda_1 (\lambda_{23}:\lambda_{45})[\lambda_{\cP}]}(n,\bn;\omega_1;\omega_2,\omega_3;\omega_4,\omega_5) &= \vec{C}^{(2)}_{v,\,\cP 2 \lambda_1  (-\lambda_{23}:-\lambda_{45})[\lambda_{\cP}]}(n,\bn;\omega_1;\omega_3,\omega_2;\omega_5,\omega_4)\,,\nn\\
\vec{C}^{(2)}_{v,\,\cP 2 \lambda_1 (\lambda_{23}:\lambda_{45})[\lambda_{\cP}]}(n,\bn;\omega_1;\omega_2,\omega_3;\omega_4,\omega_5) &= \vec{C}^{(2)}_{v,\,\cP 2 \lambda_1  (\lambda_{23}:-\lambda_{45})[\lambda_{\cP}]}(n,\bn;\omega_1;\omega_2,\omega_3;\omega_5,\omega_4)\,.\nn
\end{align}

A basis of operators involving three collinear gluon fields and a $\cP_\perp^\pm$ insertion is given by
\begin{align}
&  \boldsymbol{(g)_n (gg\, \cP_\perp)_{\bn}:}{\vcenter{\includegraphics[width=0.18\columnwidth]{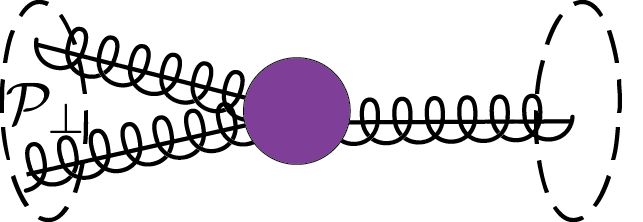}}}  \nn
\end{align}
\vspace{-0.4cm}
\begin{alignat}{2} \label{eq:eegggpperp_basis}
&O_{\cP\cB +++(\pm)[-]}^{(2)abc}
=  \cB_{n+}^a\, \cB_{\bar n+}^b\, \left [\cP_{\perp}^{-} \cB_{\bar n+}^c \right ] J_{e\pm }
\,,\qquad && O_{\cP\cB ---(\pm)[+]}^{(2)abc}
= \cB_{n-}^a\, \cB_{\bar n-}^b\, \left [\cP_{\perp}^{+} \cB_{\bar n-}^c \right ] J_{e\pm }\,,
 \nn \\
&O_{\cP\cB ++-(\pm)[+]}^{(2)abc}
= \, \cB_{n+}^a\, \cB_{ \bar n+}^b\, \left [\cP_{\perp}^{+} \cB_{\bar n-}^c \right ] J_{e\pm }
\,,\qquad 
&&O_{\cP\cB --+(\pm)[-]}^{(2)abc}
= \cB_{n-}^a\, \cB_{ \bn -}^b\, \left [\cP_{\perp}^{-} \cB_{\bar n +}^c \right ] J_{e \pm  }
\,.
\end{alignat}
We have used angular momentum conservation to eliminate certain helicity combinations. Note that the analogous operators with the helicities $O_{\cP\cB +-+(\pm)[+]}^{(2)abc}$ and $O_{\cP\cB -+-(\pm)[-]}^{(2)abc}$ are not eliminated, but instead are equivalent to those in the last row by integrating the $\cP_\perp^\pm$  by parts onto the other $\bn$-collinear field.

C/P relations combined with the ability to flip the helicity of the electron current, as described in \Eq{eq:lo_weak}, give the following relations between Wilson coefficients
\begin{align}  \label{eq:Pchargerel1}
\vec{C}^{(2)}_{v,\,\cP \cB \lambda_1 \lambda_2 \lambda_3 (\lambda_{45})[\lambda_{\cP}]}(n,\bn;\{\omega_i\}) &= -\vec{C}^{(2)}_{ v,\,\cP \cB  -\lambda_1 -\lambda_2 -\lambda_3 (-\lambda_{45}) [-\lambda_{\cP}]}(n^\P,\bn^\P;\{\omega_i\}) \,,  \\
&\hspace{-5.05cm}\vec{C}^{(2)}_{v,\,\cP \cB \lambda_1 \lambda_2 \lambda_3 (\lambda_{45}) [\lambda_{\cP}]}(n,\bn;\omega_1,\omega_2,\omega_3;\omega_4,\omega_5)  
\nn \\
&= \begin{pmatrix*}[r] -1 & 0 \\ 0 & 1\end{pmatrix*} \vec{C}^{(2)}_{v\,\cP \cB \lambda_1 \lambda_2 \lambda_3 (-\lambda_{45}) [\lambda_{\cP}]}(n,\bn;\omega_1,\omega_2,\omega_3;\omega_5,\omega_4) 
 \nn \,, \\
 &\hspace{-5.05cm}\vec{C}^{(2)}_{v,\,\cP \cB \lambda_1 \lambda_2 \lambda_3 (\lambda_{45}) [\lambda_{\cP}]}(n,\bn;\omega_1,\omega_2,\omega_3;\omega_4,\omega_5)   =\vec{C}^{(2)}_{v\,\cP \cB \lambda_1 \lambda_2 \lambda_3 (-\lambda_{45}) [\lambda_{\cP}]}(n,\bn;\omega_1,\omega_2,\omega_3;\omega_5,\omega_4) 
 \nn \,.
\end{align}
As was the case for the operators involving three collinear gluon fields discussed in \Sec{sec:sub}, the charge conjugation relations of \Eq{eq:Pchargerel1} imply that to all orders in $\alpha_s$ only the Wilson coefficients for the color structure $d^{abc}$ are non-zero for the vector current, whereas for the axial current, only the Wilson coefficients corresponding to the color structure $if^{abc}$ are non-zero. These statements remain true under renormalization group evolution.

\subsubsection{Ultrasoft Insertions}\label{sec:soft_basis}

 At $\cO(\lambda^2)$ operators first appear which involve a single  $B_{us(i)\lambda}^{a}$ with $\lambda=\pm,0$ (for example $B_{us(i)\lambda}^a J_{n\bn\lambda'}^{\balpha\beta}$) or an insertion of a ultrasoft derivative (for example, $\{ \partial_{us(i)0} J_{n\bn\lambda'}^{\balpha\beta} \}$).   There are no contributions involving the ultrasoft quark current building blocks, like $J_{i(us)\lambda}^{\balpha\beta}$. Even though these mixed ultrasoft-collinear currents have the correct power counting, they do not involve the collinear fields that are needed to conserve the large momentum flow in the hard scattering processes being considered.

Before listing the basis of operators, it is worth emphasizing the distinction between the treatment of label and residual $\perp$ momentum. In SCET$_\text{I}$ ultrasoft fields do not carry label momenta. Because only the collinear sectors carry label momentum, we are able to choose the collinear sectors back-to-back, with zero total $\perp$ momentum in each collinear sector. However, for the residual components of the momentum, it is inconsistent to simultaneously choose $\bar n=(1,-\vec n)$, and to set the $\perp$ component of the residual momentum in both sectors to zero. This is because the ultrasoft fields also carry $\cO(\lambda^2)$ residual momentum, which can cause the jet direction to recoil by this small amount. Furthermore, because ultrasoft fields carry residual momentum, we cannot, for example, say that the two collinear sectors carry equal and opposite residual momenta, and therefore we cannot in general relate ultrasoft derivatives acting on one sector to ultrasoft derivatives acting on another sector to reduce the basis. Ultrasoft derivatives acting on both sectors must therefore be included in the basis.

When constructing a basis of operators involving ultrasoft gluons, different choices can be made due to the fact that the ultrasoft gluons are not naturally associated with a given lightcone direction. This corresponds to a choice of which light like vector is used to define the $\cB_{us(n_i)}$ field of \Eq{eq:soft_gluon}. To guide our choice, we will always choose to work in a basis where ultrasoft derivatives acting on ultrasoft Wilson lines are absorbed into $\cB_{us}$ fields, and do not appear explicitly in the operator. As an example, consider the pre-BPS operators
\begin{align}
O^\mu_1=\bar \chi_{\bar n} \overrightarrow D_{us} \chi_n\,, \qquad O^\mu_2=\bar \chi_{\bar n} \overleftarrow D_{us} \chi_n\,,
\end{align}
where we have not made the contraction of the $\mu$ index explicit.
Performing the BPS field redefinition, we obtain 
\begin{align}
O^\mu_{1\text{BPS}}=\bar \chi_{\bar n}Y_{\bar n}^\dagger \overrightarrow D_{us} Y_n \chi_n\,, \qquad O^\mu_{2\text{BPS}}=\bar \chi_{\bar n} Y_{\bar n}^\dagger \overleftarrow D_{us} Y_n \chi_n
\end{align}
To absorb all ultrasoft derivatives acting on Wilson lines into $\cB_{us}$ fields, we can rearrange the Wilson lines in the operators as
\begin{align}
O^\mu_{1\text{BPS}}=\bar \chi_{\bar n}Y_{\bar n}^\dagger Y_n (Y_n^\dagger \overrightarrow D_{us} Y_n) \chi_n\,, \qquad O^\mu_{2\text{BPS}}=\bar \chi_{\bar n} (Y_{\bar n}^\dagger \overleftarrow D_{us}Y_{\bar n}) Y_{\bar n}^\dagger  Y_n \chi_n
\end{align}
Using the definition of the ultrasoft gluon field, \Eq{eq:soft_gluon}, we see that this can be written entirely in terms of $\partial_{us}$ operators acting on collinear fields, as well as the gauge invariant ultrasoft gluon fields $\cB_{us(n)}$ and $\cB_{us(\bar n)}$. In this way of organizing the basis, ultrasoft gluon fields defined using both $n$ and $\bar n$ are required. It should be clear from this example that it is also possible to work entirely with only $\cB_{us(n)}$ or $\cB_{us(\bar n)}$. However, in this case we see that we would have ultrasoft derivatives in the operators acting on dangling ultrasoft Wilson lines. To avoid this, and to make our basis more symmetric, we choose to work with both $\cB_{us(n)}$ or $\cB_{us(\bar n)}$.

For the operators involving one ultrasoft gluon and two collinear quarks, we have the basis
\begin{align}
&  \boldsymbol{g_{us}(q)_n (\bar q)_{\bn}:}{\vcenter{\includegraphics[width=0.18\columnwidth]{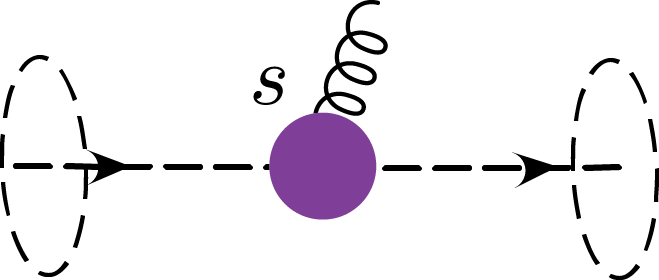}}}  \nn
\end{align}
\vspace{-0.4cm}
\begin{alignat}{2} \label{eq:soft_insert_basis}
&O_{\cB(us(n))-:(+;\pm)}^{(2)a\,\balpha\bt}
=  \cB_{us(n)-}^a\, J_{n\bar n\,+}^{\balpha\bt}\, J_{e\pm }
\,,\qquad &
&O_{\cB(us(n))+:(-;\pm)}^{(2)a\,\balpha\bt}
=\cB_{us(n)+}^a \, J_{n\bar n\, -}^{\balpha\bt}\, J_{e\pm }
\,,  \\
&O_{\cB(us(n))0:(+;\pm)}^{(2)a\,\balpha\bt}
=  \cB_{us(n)0}^a\, J_{n\bar n\,+}^{\balpha\bt}\, J_{e\pm }
\,,\qquad &
&O_{\cB(us(n))0:(-;\pm)}^{(2)a\,\balpha\bt}
= \cB_{us(n)0}^a \, J_{n\bar n\, -}^{\balpha\bt}\, J_{e\pm }
\,, \nn
\end{alignat}
with the unique color structure
\begin{equation} 
\vT_{\BPS}^{\,a\, \al\bbeta} = \left ( T^a Y^\dagger_{n} Y_{\bn} \right )_{\alpha \bar\beta}
\,,\end{equation}
and
\begin{alignat}{2} \label{eq:soft_insert_basis2}
&O_{\cB(us(\bar n))+:(+;\pm)}^{(2)a\,\balpha\bt}
=  \cB_{us(\bar n)+}^a\, J_{n\bar n\,+}^{\balpha\bt}\, J_{e\pm }
\,,\qquad &
&O_{\cB(us(\bar n))-:(-;\pm)}^{(2)a\,\balpha\bt}
=\cB_{us(\bar n)-}^a \, J_{n\bar n\, -}^{\balpha\bt}\, J_{e\pm }
\,,  \\
&O_{\cB(us(\bar n))0:(+;\pm)}^{(2)a\,\balpha\bt}
=  \cB_{us(\bar n)0}^a\, J_{n\bar n\,+}^{\balpha\bt}\, J_{e\pm }
\,,\qquad &
&O_{\cB(us(\bar n))0:(-;\pm)}^{(2)a\,\balpha\bt}
= \cB_{us(\bar n)0}^a \, J_{n\bar n\, -}^{\balpha\bt}\, J_{e\pm }
\,, \nn
\end{alignat}
with the unique color structure
\begin{equation} 
\vT_{\BPS}^{\,a\, \al\bbeta} = \left ( Y^\dagger_{n} Y_{\bn} T^a \right )_{\alpha \bar\beta}
\,.\end{equation}
The helicity selection rules act different for the two projections of the $\cB_{us}$ fields due to the different definition of helicity in the two cases.


The Wilson coefficients of the operators that include $\cB_{us(n)0}$ can be related to the Wilson coefficients of the leading power operators using RPI symmetry (see \cite{Larkoski:2014bxa}). In particular, we have 
\begin{align} \label{eq:usRPIrelation}
C^{(2)}_{\cB(us)0:(\lambda_1,\pm)}&=-\frac{\partial C^{(0)}_{(\lambda_1;\pm)} }{\partial \omega_1}  
\,, 
\end{align}
where $C^{(0)}_{(\lambda_1;\pm)}$ is the Wilson coefficient for the leading power dijet operator of \sec{LP}. As we will show in \Sec{sec:matching}, the leading power Wilson coefficients for the case of back to back jets are independent of $\omega_1$ and $\omega_2$, at tree level, so that this Wilson coefficient vanishes at tree level. We will also show explicitly that they do not arise in the tree level matching calculation in \Sec{sec:match_soft}. However, we will also show that the Wilson coefficient is non-vanishing at $\cO(\alpha_s)$.

The operators involving $\cB_{us(n)\pm}^a$ do not seem to be related to the leading power Wilson coefficient. In \cite{Larkoski:2014bxa} it was shown that in the general case of $N$ jets, certain subleading operators involving $\cB_{us(n)\pm}^a$ are generated by the RPI expansion of the leading power operator. However, these particular operators vanish for the case of back to back jets that we consider here. Interestingly, as we will show below, only the operators $O_{\cB(us)0:(+;\pm)}^{(2)a\,\balpha\bt}$ and $O_{\cB(us)\bar 0:(+;\pm)}^{(2)a\,\balpha\bt}$ can contribute to the dijet cross section at $\cO(\lambda^2)$, and therefore, the new Wilson coefficients (which are not related by RPI) do not contribute at this order.

We also have operators involving two collinear quark fields and a single ultrasoft derivative. In writing the basis, we can use the fact that quark equations of motion can be used to rewrite $in\cdot \partial \chi_n$ and $i\bar n \cdot \partial \chi_{\bar n}$ in terms of purely collinear operators. Therefore, these combinations of derivatives do not need to be included in our basis. A basis of derivative operators is then given by
\begin{align}
&  \boldsymbol{\partial_{us} (q)_n (\bar q)_{\bn}:}{\vcenter{\includegraphics[width=0.18\columnwidth]{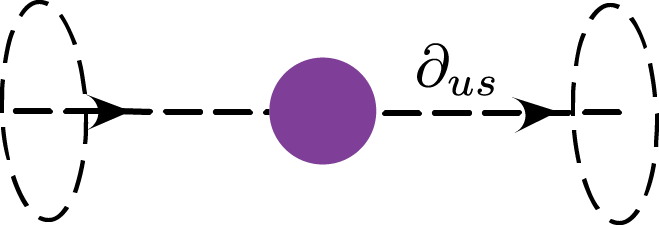}}}  \nn
\end{align}
\vspace{-0.4cm}
\begin{alignat}{2} \label{eq:soft_derivative_basis}
&O_{\partial(us(n))-:(+;\pm)}^{(2)\,\balpha\bt}
=  \{\partial_{us(n)-}\, J_{n\bar n\,+}^{\balpha\bt}\}\, J_{e\pm }
\,,\qquad &
&O_{\partial(us(n))+:(-;\pm)}^{(2)\,\balpha\bt}
=  \{\partial_{us(n)+} \, J_{n\bar n\, -}^{\balpha\bt}\}\, J_{e\pm }\,,
\nn \\
&O_{\partial(us(n))0:(+;\pm)}^{(2)\,\balpha\bt}
=  \{\partial_{us(n)0}\, J_{n\bar n\,+}^{\balpha\bt}\}\, J_{e\pm }
\,,\qquad &
&O_{\partial(us(n))0:(-;\pm)}^{(2)\,\balpha\bt}
=  \{\partial_{us(n)0} \, J_{n\bar n\, -}^{\balpha\bt}\}\, J_{e\pm }\,,
\nn \\
%
%
&O_{\partial^\dagger(us(\bar n))+:(+;\pm)}^{(2)\,\balpha\bt}
=  \{ J_{n\bar n\,+}^{\balpha\bt}\,     (i\partial_{us(\bar n)+})^\dagger\}\, J_{e\pm }
\,,\qquad &
&O_{\partial^\dagger(us(\bar n))-:(-;\pm)}^{(2)\,\balpha\bt}
=  \{  J_{n\bar n\, -}^{\balpha\bt}\, (i\partial_{us(\bar n)-})^\dagger\}\, J_{e\pm }\,,
\nn \\
%
%
&O_{\partial^\dagger(us(\bar n))\bar 0:(+;\pm)}^{(2)\,\balpha\bt}
=  \{ J_{n\bar n\,+}^{\balpha\bt}\, (i\partial_{us(\bar n)\bar 0})^\dagger\}\, J_{e\pm }
\,,\qquad &
&O_{\partial^\dagger(us(\bar n))\bar 0:(-;\pm)}^{(2)\,\balpha\bt}
=  \{  J_{n\bar n\, -}^{\balpha\bt}\,(i\partial_{us(\bar n)\bar 0})^\dagger\}\, J_{e\pm }
\,.
\end{alignat}
The helicity decomposition for these ultrasoft operators is more cumbersome due to the fact that not only the $\pm$ helicities appear, and for this particular case, it is perhaps simpler to use the more traditional operator basis, in contrast to the case with multiple collinear fields.

The color structure of these operators is exactly the same as for the leading power operator given in \eq{leading_color}.  The Wilson coefficients of the operators that include a $\partial_{us(n)0}$ or $\partial_{us(n)\bar 0}$ are related via RPI to the Wilson coefficients of the leading power operator by
\begin{align} \label{eq:uspartialRPIrelation}
C^{(2)}_{\partial(us)0:(\lambda_1,\pm)}=-\frac{\partial C^{(0)}_{(\lambda_1;\pm)} }{\partial \omega_1} \,, \qquad C^{(2)}_{\partial(us)\bar 0:(\lambda_1,\pm)}=-\frac{\partial C^{(0)}_{(\lambda_1;\pm)} }{\partial \omega_2}
\, .
\end{align}
This is true also of the operators where the derivatives act on the Wilson lines, as these arise only through the BPS field redefinition of the same operator.
As we will show in \Sec{sec:matching}, the leading power Wilson coefficients for the case of back to back jets are independent of $\omega_1$ and $\omega_2$ at tree level, so that these Wilson coefficients in fact vanish at the lowest order in the matching. We will also show this explicitly in the matching calculation in \Sec{sec:match_soft}. It is also interesting to mention the physical interpretation of the vanishing of these contributions. As was discussed in \cite{Larkoski:2014bxa} these derivative terms can be interpreted as the orbital angular momentum contribution to the tree level Low-Burnett-Kroll (LBK) theorem \cite{Low:1958sn,Burnett:1967km}, which vanishes for back to back jets.

In the case of ultrasoft derivative insertions, the operators that include a $\partial_{us(n)\pm}$  are related by RPI to $\cO(\lambda)$ operators that involve the insertion of $\cP_{\pm}$ into the leading power operators of \Eq{eq:LP_basis}. Indeed, RPI implies that the label momentum and derivative operator must always appear in the combination $\cP^\mu+i\partial^\mu$. Therefore the Wilson coefficients of these operators are equal. However, we have chosen to work in the center of mass frame, where insertions of $\cP_{\pm}$ into the leading power operator vanish, and so such operators do not appear explicitly in our basis. Therefore we include the Wilson coefficients of these operators in our basis.

We also have operators involving two collinear gluons and a single ultrasoft gluon field. Since their Wilson coefficients start at one-loop order these are of limited phenomenological relevance, but are included as a further example of our approach. The basis of such operators is given by
\begin{align}
& \boldsymbol{g_{us}(g)_n (g)_{\bn}:}{\vcenter{\includegraphics[width=0.18\columnwidth]{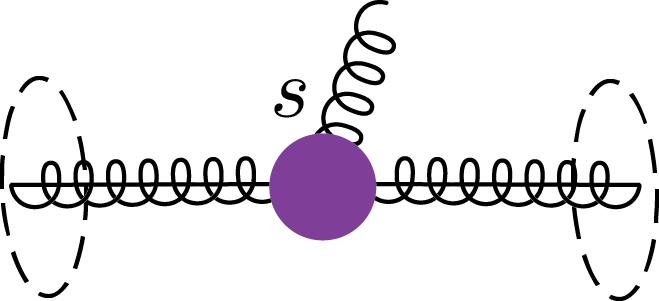}}}  \nn
\end{align}
\vspace{-0.4cm}
\begin{alignat}{2}\label{eq:eegggus}
 &O_{(us(n))+:++(\pm)}^{(2)abc}
=  \cB_{us(n)+}^a\, \cB_{n+}^b\, \cB_{\bar n+}^c\,  J_{e\pm }
\,, \qquad && O_{(us(n))+:--(\pm)}^{(2)abc} 
=   \cB_{us(n)+}^a\, \cB_{n-}^b\, \cB_{\bar n-}^c\,  J_{e\pm }
\,,  \\
 &O_{(us(n))-:++(\pm)}^{(2)abc}
 =  \cB_{us(n)-}^a\, \cB_{n+}^b\, \cB_{\bar n+}^c\,  J_{e\pm }
 \,, \qquad && O_{(us)-:--(\pm)}^{(2)abc} 
 =   \cB_{us(n)-}^a\, \cB_{n-}^b\, \cB_{\bar n-}^c\,  J_{e\pm }
 \,, \nn \\
  &O_{(us(n))+:-+(\pm)}^{(2)abc}
  =  \cB_{us(n)+}^a\, \cB_{n-}^b\, \cB_{\bar n+}^c\,  J_{e\pm }
  \,, \qquad && O_{(us(n))-:+-(\pm)}^{(2)abc} 
  =   \cB_{us(n)-}^a\, \cB_{n+}^b\, \cB_{\bar n-}^c\,  J_{e\pm }
\,,\nn
\end{alignat}
with the two dimensional basis of color structures,\footnote{In order to see how the Wilson line structure in \Eq{eq:Z2g_colorus} arises, we look at the object $D_{us}^{ab} \cB_{n}^c \cB_{\bn}^d$ pre-BPS field redefinitions. This object must be contracted with a tensor to make it a singlet under ultrasoft gauge transformations. Each of these resulting forms can be mapped onto the color structures of \Eq{eq:Z2g_colorus} after performing the BPS field redefinition} 
\begin{equation} \label{eq:Z2g_colorus}
\vT_{\BPS}^{abc} =
\begin{pmatrix}
i  f^{abd}\, \big({\cal Y}_n^T {\cal Y}_{\bar n}\big)^{dc} \\
 d^{abd}\, \big({\cal Y}_n^T {\cal Y}_{\bar n}\big)^{dc} 
\end{pmatrix}^T
\,,
\end{equation}
and
\begin{alignat}{2}\label{eq:eegggus2}
 &O_{(us(\bar n))-:++(\pm)}^{(2)abc}
=  \cB_{us(\bar n)-}^a\, \cB_{n+}^b\, \cB_{\bar n+}^c\,  J_{e\pm }
\,, \qquad && O_{(us(\bar n))-:--(\pm)}^{(2)abc} 
=   \cB_{us(\bar n)-}^a\, \cB_{n-}^b\, \cB_{\bar n-}^c\,  J_{e\pm }
\,,  \\
 &O_{(us(\bar n))+:++(\pm)}^{(2)abc}
 =  \cB_{us(\bar n)+}^a\, \cB_{n+}^b\, \cB_{\bar n+}^c\,  J_{e\pm }
 \,, \qquad && O_{(us(\bar n))+:--(\pm)}^{(2)abc} 
 =   \cB_{us(\bar n)+}^a\, \cB_{n-}^b\, \cB_{\bar n-}^c\,  J_{e\pm }
 \,, \nn \\
  &O_{(us(\bar n))-:-+(\pm)}^{(2)abc}
  =  \cB_{us(\bar n)-}^a\, \cB_{n-}^b\, \cB_{\bar n+}^c\,  J_{e\pm }
  \,, \qquad && O_{(us)+:+-(\pm)}^{(2)abc} 
  =   \cB_{us(\bar n)+}^a\, \cB_{n+}^b\, \cB_{\bar n-}^c\,  J_{e\pm }
\,,\nn
\end{alignat}
with the basis of color structures
\begin{equation} \label{eq:Z2g_colorus2}
\vT_{\BPS}^{abc} =
\begin{pmatrix}
i  f^{acd}\, \big({\cal Y}_{\bar n}^T {\cal Y}_{n}\big)^{db} \\
 d^{acd}\, \big({\cal Y}_{\bar n}^T {\cal Y}_{n}\big)^{db} 
\end{pmatrix}^T
\,.
\end{equation}
Here we have only included the $\vT_{\BPS}^{abc}$ version of the color structure here because the $\cB_{us(n)\lambda}^a$ are generated by BPS field redefiniton. When constructing this basis we have used the angular momentum constraints discussed in \Sec{sec:ang_cons} to eliminate the other two helicity combinations in each case.


In \eq{eegggus} we have not included the operators
\begin{align}
  &O_{(us(n))0:++(\pm)}^{(2)abc}
  =  \cB_{us(n)0}^a\, \cB_{n+}^b\, \cB_{\bar n+}^c\,  J_{e\pm }
  \,, \qquad && O_{(us(n))0:--(\pm)}^{(2)abc} 
  =   \cB_{us(n)0}^a\, \cB_{n-}^b\, \cB_{\bar n-}^c\,  J_{e\pm }\,,
  \nn \\
    &O_{(us(\bar n))0:++(\pm)}^{(2)abc}
  =  \cB_{us(\bar n)0}^a\, \cB_{n+}^b\, \cB_{\bar n+}^c\,  J_{e\pm }
  \,, \qquad && O_{(us(\bar n))0:--(\pm)}^{(2)abc} 
  =   \cB_{us(\bar n)0}^a\, \cB_{n-}^b\, \cB_{\bar n-}^c\,  J_{e\pm }
  \,. 
\end{align}
The coefficients of these operators are related by RPI to the derivative of the leading power operators for $e\bar e gg$, and therefore also vanish by Yang's theorem. The Wilson coefficients of the operators in \eq{eegggus} are not constrained by RPI considerations.

We can also consider operators with an insertion of $\partial_{us(n)}$ with two collinear gluons in different collinear sectors. As for the case of ultrasoft derivative insertions into the quark operators, the gluon equations of motion allow us to eliminate the operators $in\cdot \partial \cB_{n\perp}$ and $i\bar n\cdot \partial \cB_{\bar n\perp}$. However, these operators already vanish by Yang's theorem, as they are related by RPI to the leading power operators for $e\bar e gg$. A basis of helicity operators is then given by
\begin{align}
& \boldsymbol{\partial_{us}(g)_n (g)_{\bn}:}{\vcenter{\includegraphics[width=0.18\columnwidth]{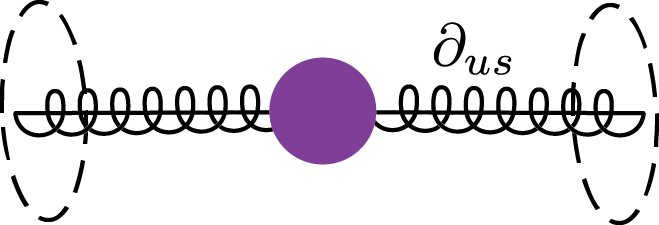}}}  \nn
\end{align}
\vspace{-0.4cm}
\begin{alignat}{2}\label{eq:eedggus}
&O_{\partial \cB  (us(n))+:++(\pm)}^{(2)ab}
=   \cB_{n+}^a\,\left[ \partial_{us(n)+} \cB_{\bar n+}^b  \right]\,  J_{e\pm }
\,, ~\, && O_{\partial \cB  (us(n))+:--(\pm)}^{(2)ab} 
=   \cB_{n-}^a \, \left[ \partial_{us(n)+} \cB_{\bar n-}^b\right]\,  J_{e\pm }
\,,
\nn  \\
&O_{\partial \cB  (us(n))-:++(\pm)}^{(2)ab}
=  \cB_{n+}^a  \, \left[ \partial_{us(n)-} \cB_{\bar n+}^b\right] \,  J_{e\pm }
\,, ~\, && O_{\partial \cB  (us(n))-:--(\pm)}^{(2)ab} 
=    \cB_{n-}^a \left[\partial_{us(n)-} \cB_{\bar n-}^b\right]\,  J_{e\pm }
\,,
 \nn \\
&O_{\partial \cB (us(n))+:-+(\pm)}^{(2)ab}
=   \cB_{n-}^a  \, \left[ \partial_{us(n)+} \cB_{\bar n+}^b \right]\,  J_{e\pm }
\,, \, ~&& O_{\partial \cB  (us(n))-:+-(\pm)}^{(2)ab} 
=    \cB_{n+}^a  \, \left [\partial_{us(n)-}\, \cB_{\bar n-}^b\right]\,  J_{e\pm }
\,,
\end{alignat}
with the basis of color structures
\begin{equation} \label{eq:Z2gd_colorus}
\vT_{\BPS}^{ab} =
\big({\cal Y}_n^T {\cal Y}_{\bar n}\big)^{ab} 
\,,
\end{equation}
and
\begin{alignat}{2}\label{eq:eedggus2}
&O_{\partial \cB  (us(\bar n))-:++(\pm)}^{(2)ab}
=  \left[ \partial_{us(\bar n)-}\, \cB_{n+}^a\right]\, \cB_{\bar n+}^b\,  J_{e\pm }
\,, \,~ && O_{\partial \cB (us(\bar n))-:--(\pm)}^{(2)ab} 
=  \left [ \partial_{us(\bar n)-}\, \cB_{n-}^a \right] \, \cB_{\bar n-}^b\,  J_{e\pm }
\,,
 \nn \\
&O_{\partial \cB  (us(\bar n))+:++(\pm)}^{(2)ab}
= \left [ \partial_{us(\bar n)+}\, \cB_{n+}^a \right] \, \cB_{\bar n+}^b\,  J_{e\pm }
\,, ~\, && O_{\partial \cB  (us(\bar n))+:--(\pm)}^{(2)ab} 
=  \left [ \partial_{us(\bar n)+}\, \cB_{n-}^a \right] \, \cB_{\bar n-}^b\,  J_{e\pm }
\,,
 \nn \\
&O_{\partial \cB  (us(\bar n))-:-+(\pm)}^{(2)ab}
=  \left [\partial_{us(\bar n)-}\, \cB_{n-}^a \right] \, \cB_{\bar n+}^b\,  J_{e\pm }
\,, ~\, && O_{\partial \cB  (us(\bar n))+:+-(\pm)}^{(2)ab} 
=  \left [ \partial_{us(\bar n)+}\, \cB_{n+}^a \right] \, \cB_{\bar n-}^b\,  J_{e\pm }
\,,
\end{alignat}
with the basis of color structures
\begin{equation} \label{eq:Z2gd_colorus2}
\vT_{\BPS}^{ab} =
\big({\cal Y}_{\bar n}^T {\cal Y}_{n}\big)^{ab} 
\,.
\end{equation}
We have not included any operator with $\partial_{us(n)0}$ or $\partial_{us(n)\bar 0}$  acting on two collinear gluons, as these will all have Wilson coefficients that are related to the coefficient of the two gluon operator, which vanishes due to Yang's theorem. 

\subsection{Cross Section Contributions}\label{sec:ee_discuss}

While the basis of hard scattering operators presented in this section is quite large, and we have focused on providing a complete basis to allow for an understanding of all possible contributions, many of these operators will not contribute to a calculation of a particular cross section. In this section we will consider the case of event shapes in $e^+e^-\to$ dijets, and discuss how symmetry arguments can be used to show which operators can contribute to the cross section up to $\cO(\lambda^2)$. In \Sec{sec:ee_lambda}, we begin by proving that the hard scattering operators do not generate a contribution to the cross section at $\cO(\lambda)$, and then in \Sec{sec:ee_lambda2}, we discuss which operators contribute to the cross section at $\cO(\lambda^2)$, and the particular form of their contribution, by listing the operator content of the resulting jet and soft functions. A summary of which operators contribute is given in \Tab{tab:summary}.

\subsubsection{Vanishing at $\cO(\lambda)$}\label{sec:ee_lambda}

For $e^+e^-\to$ dijets event shapes described by SCET$_\text{I}$, the leading $\cO(\lambda)$ power corrections vanish \cite{Beneke:2003pa,Lee:2004ja,Freedman:2013vya}. This is expected because fixed order calculations indicate the leading correction should scale as $e$, while an $\cO(\lambda)$ power correction would scale as $\sqrt{e}$ for our power counting. In this section, we use our formalism to show explicitly that this is the case for contributions from the hard scattering operators. Similar arguments can also be used to show that Lagrangian contributions vanish. The $\cO(\lambda)$ observable expansion terms also vanish, as discussed in \App{app:meas}.

While we will not discuss the factorization of the cross section in detail, the contribution of the hard scattering operators to the cross section at $\cO(\lambda)$ can be written
{\begin{small}
\begin{align}\label{eq:xsec_lam}
&\frac{d\sigma}{d\tau}^{(1)} \supset N \sum_{X,i}  \tilde \delta^{(4)}_q  \bra{0} C_i^{(1)} \tO_i^{(1)}(0) \ket{X}\bra{X} C^{(0)} \tO^{(0)}(0) \ket{0}   \delta\big( \tau - \tau^{(0)}(X) \big) +\text{h.c.} 
\,. 
\end{align}
\end{small}}
Here $N$ is a normalization factor.  We use the shorthand notation $\tilde \delta^{(4)}_q=(2\pi)^4\delta^4(q-p_X)$ for the momentum conserving delta function. The summation over all final states, $X$, includes phase space integrations. Here $L$ denotes the $e^+e^-$ leptonic initial state. The measurement of the observable is enforced by $ \delta\big( e - e(X) \big) $, where $e(X)$, returns the value of the observable $e$ as measured on the final state $X$. 

From the expression for the $\cO(\lambda)$ power correction to the cross section in \Eq{eq:xsec_lam}, we see that the only contributions from hard scattering operators arise from matrix elements of an  $\cO(\lambda)$ operator with an $\cO(\lambda^0)$ operator. When determining whether or not the insertion of a given operator vanishes, we can make arguments based on fermion number conservation or angular momentum conservation either before or after factorization. Before factorization, the matrix elements in the cross section given in \Eq{eq:Z1_basis} that contribute at $\cO(\lambda)$ with the insertion of an $\cO(\lambda)$ operator can be written as
\begin{align} \label{eq:subleadingmatrix_general}
\langle 0| O^{(0)}(x) \widehat \cM^{(0)} O^{(1)}(0) |0 \rangle\,.
\end{align}
Here the sum over the complete set of states $|X\r$ has been performed using the measurement operator $\widehat \cM^{(0)}$, where $\widehat \cM^{(0)} |X\r = \delta(e - e^{(0)}(X)) |X\r$. As we are taking a vacuum matrix element, we must have that $O^{(0)}(x) O^{(1)}(0)$ conserves fermion number and angular momentum. If this matrix element does not vanish, we can move to the factorized state, where we split the operators into components in the $n$, $\bn$ and ultrasoft sectors. This will give us factorized matrix elements that become our jet and soft functions,
\begin{align} \label{eq:subleadingmatrix_factorized}
\langle 0| O^{(0)}_n(x) \widehat \cM_n^{(0)} O_n^{(1)}(0) |0 \rangle \langle 0| O^{(0)}_{\bn}(x) \widehat \cM_n^{(0)} O_{\bn}^{(1)}(0) |0 \rangle \langle 0| O^{(0)}_{us}(x) \widehat \cM_{us}^{(0)} O_{us}^{(1)}(0) |0 \rangle\,.
\end{align}
In this form, it is clear that each sector must exhibit both fermion number and angular momentum conservation, so we can make these arguments at the level of the factorized matrix elements, providing an even stronger constraint. In other words, if we examine the field content in each sector, we must have the same number of quarks and anti-quarks and must conserve angular momentum. As shown in \Tab{tab:summary}, there are only two operator structures appearing at $\cO(\lambda)$, and a single operator at $\cO(\lambda^0)$ , so we can consider each of the possible contributions in turn. 

$ \boldsymbol{O^{(1)}_{ggg} O^{(0)}_{qq}}$ \textbf{ vanish:}

We first consider the $\cO(\lambda)$ hard scattering operators involving three collinear gluons, given in \Eq{eq:eeggg}. After factorization of this matrix element one obtains a vacuum matrix element involving a single quark field in each collinear sector, coming from the leading power operator. The leading order Lagrangian separately conserves fermion number in each collinear sector, and therefore this contribution vanishes.

$ \boldsymbol{O^{(1)}_{qqg} O^{(0)}_{qq}}$ \textbf{ vanish:}

Next we consider the contribution from the $\cO(\lambda)$ hard scattering operators involving two collinear quarks, and one collinear gluon. As for the three gluon operator, fermion number conservation immediately eliminates any possible contribution from the operators of \Eq{eq:Z1_basis_diff}, where the collinear quarks are both in the same sector. To eliminate the $\cO(\lambda)$ contribution from the operators of \Eq{eq:Z1_basis}, we can use symmetry arguments, similar to those in \Sec{sec:ang_cons}. For the operator of  \Eq{eq:Z1_basis}, the matrix elements entering the factorized expression for the cross section are of the form
\begin{align} \label{eq:subleadingmatrix}
\langle 0| (J^{\bar \alpha \beta}_{n\bar n \lambda_1}(x))^\dagger \widehat \cM^{(0)} J^{\bar \delta \gamma}_{n\bar n  \lambda_2}(0) \cB^a_{(n,\bar n) \lambda_3}(0) |0 \rangle\,.
\end{align}
 The $\lambda_i=\pm$ denote arbitrary helicities, and the $(n,\bar n)$ subscript on the gluon denotes that it can be associated with either collinear sector. Since RPI has been used to choose the axes of the collinear sectors as back-to-back, all helicities are defined with respect to a common $\hat{n}$ axis. The SCET Lagrangian preserves rotational invariance about the $\hat{n}$ axis, which implies that this matrix element vanishes since with an odd number of $\pm$ helicities it can not transform as a scalar. 
 An identical argument follows for the operator of \Eq{eq:Z1_basis_flip}.

\subsubsection{Relevant Hard Scattering Operators at $\cO(\lambda^2)$}\label{sec:ee_lambda2}

Having shown that there are no contributions to the dijet cross section at $\cO(\lambda)$, we now discuss contributions at $\cO(\lambda^2)$. To see which can contribute, we focus on contributions arising from our basis of hard scattering operators, although we will also briefly mention contributions from subleading Lagrangian insertions. Due to the power counting $\lambda \sim \sqrt{e}$,  the ${\cal O}(\lambda^2)$ power corrections correspond to ${\cal O}(e)$ power corrections, and will not in general vanish. While there are large number of $\cO(\lambda^2)$ hard scattering operators given in \Tab{tab:summary}, we can use similar arguments to those of \Sec{sec:ee_lambda} to severely restrict the number of operators that contribute to the dijet cross section at $\cO(\lambda^2)$. The resulting set of operators are indicated in \Tab{tab:summary}, and \Tab{tab:fact_func} shows which products of hard scattering operators contribute to the factorized cross section and the schematic form of the corresponding hard, jet and soft functions.

Despite the fact that we are working at subleading power, in many cases the jet and soft functions which appear in the factorization are identical to those at leading power, with only several new power suppressed functions appearing, as can be seen in \Tab{tab:fact_func}. For the case of the soft functions this simplification arises due to color coherence, allowing a simplification to the Wilson lines in the soft functions that appear. For quark-quark and gluon-gluon color channels the leading power soft functions are
\begin{align}
S_q^{(0)}=\frac{1}{N_c}  \tr \langle 0 | Y^\dagger_{\bar n} Y_n \widehat{\cM}^{(0)}Y_n^\dagger Y_{\bar n}|0\rangle\,, \qquad S_g^{(0)}=\frac{1}{(N_c^2-1)}  \tr \langle 0 | \cY^T_{\bar n} \cY_n \widehat{\cM}^{(0)}\cY_n^T \cY_{\bar n}|0\rangle\,,
\end{align}
and depend on the kinematic variables probed by the measurement operator $\widehat{\cM}^{(0)}$.
For the jet functions, this simplification occurs since the power correction is often restricted to a single collinear sector. The other collinear sector is then described by the leading power jet functions for quarks and gluons
\begin{align}
\delta^{\alpha\bar \beta} \Big( \frac{\Sl{n}}{2} \Big)^{\!ss'\!} J_q^{(0)}
  &= \int \!\!\frac{dx^-}{|\omega|}\: e^{\frac{i}{2} \ell^+ x^-} 
  \Big\langle 0\Big|\, \chi_{n}^{s\alpha} \big(x^- \text{\small $\frac{n}{2}$}\big) \,\hat{\delta}\, \bar \chi_{n,\omega}^{s'\bar\beta}(0) 
  \,\Big| 0 \Big\rangle
   \,, \\
\delta^{ab} g^{\mu\nu}_\perp J_g^{(0)}
  &=-\omega\!  \int \!\!\frac{dx^-}{|\omega|}\:  e^{\frac{i}{2} \ell^+ x^-}  \Big\langle 0 \Big|\, \cB^{\mu a}_{\perp} \big(x^- \text{\small $\frac{n}{2}$}\big)\, \hat{\delta}\, \cB^{\nu b}_{\perp,\omega}(0) \,\Big|0\Big\rangle
\,. \nn
\end{align}
The form of the leading power measurement function $\hat{\delta}$ appearing in these jet functions will depend on the precise factorization theorem being treated. Here we assume an SCET$_{\text{I}}$ type measurement that does not fix the $\cP_\perp$ of the measured particle. Often the jet functions are inclusive in which case $\hat{\delta}=(2\pi)^2 \delta^2(\omega_\perp+\cP_\perp^2)$, giving functions of a single invariant mass momentum variable, $J_{q}^{(0)}(\omega \ell^+-\vec \omega_\perp^{\,2})$ and $J_{g}^{(0)}(\omega \ell^+-\vec \omega_\perp^{\,2})$. 

Contributions to the cross section at $\cO(\lambda^2)$ whose power suppression arises solely from hard scattering operators, take the form either of a product of two $\cO(\lambda)$ operators or as a product of an $\cO(\lambda^2)$ operator and an $\cO(\lambda^0)$ operator:
{\begin{small}
\begin{align}\label{eq:xsec_lam2}
&\frac{d\sigma}{d\tau}^{(2)} \supset N \sum_{X,i}  \tilde \delta^{(4)}_q  \bra{0} C_i^{(2)} \tO_i^{(2)}(0) \ket{X}\bra{X} C^{(0)} \tO^{(0)}(0) \ket{0} \delta\big( \tau - \tau^{(0)}(X) \big) +\text{h.c.}\nn \\
&+ N \sum_{X,i,j}  \tilde \delta^{(4)}_q  \bra{0} C_i^{(1)} \tO_i^{(1)}(0) \ket{X}\bra{X} C_j^{(1)} \tO_j^{(1)}(0) \ket{0}   \delta\big( \tau - \tau^{(0)}(X) \big)+\text{h.c.} \,.
\end{align}
\end{small}}
We first consider the contributions from products of $\cO(\lambda)$ hard scattering operators, where we have several categories that could possibly contribute. 

\noindent \vbox{$ \boldsymbol{O^{(1)}_{ggg} O^{(1)}_{ggg}}$ \textbf{ contribute:}

 After factorization, the contribution to the $\cO(\lambda^2)$ cross section from the product of two $\cO(\lambda)$ three gluon operators of \Eq{eq:eeggg} gives rise to jet functions involving either two or four collinear gluon fields. The schematic factorization is given by $H_{q1}^{(0)} J_g^{(0)} J_{gg}^{(2)} S_g^{(0)}$, shown in \Tab{tab:fact_func}, where the jet functions $J_{gg}^{(2)}$ involving four $\cB$ fields have two different color contractions. Rotational invariance arguments, similar to those presented in \Sec{sec:ee_lambda}, but applied after factorization into separate matrix elements for the $n$ and $\bar n$ sectors, imply that jet functions involving three collinear gluon building block fields vanish. 
}

\noindent \vbox{
$ \boldsymbol{O^{(1)}_{qqg} O^{(1)}_{qqg}}$ \textbf{ contribute:}
	
Next consider the contribution to the $\cO(\lambda^2)$ cross section from the square of the operators involving two collinear quarks, and a collinear gluon, a basis of which were given in \Eqs{eq:Z1_basis}{eq:Z1_basis_flip} and \Eq{eq:Z1_basis_diff}. These give rise to the factorized contributions $H_{q2}^{(0)} J_g^{(0)} J_{qq}^{(2)} S_g^{(0)}$, and $H_{q3}^{(0)} J_q^{(0)} J_{qgg}^{(2)} S_{q}^{(0)}$ in \Tab{tab:fact_func}. Here the factorization theorems involve subleading jet functions $J_{qq}^{(2)}$ with four collinear quarks (one color contraction), or $J_{qgg}^{(2)}$ with two collinear quarks and two collinear gluons (one color contraction).  The exact color structure is not displayed, but is simple to obtain from the color structure of the hard scattering operators after BPS field redefinition. Rotational invariance arguments can be used to restrict the particular helicity configurations which give non-vanishing contributions. In both these cases the ultrasoft functions are leading power, and are actually given by the same product of Wilson lines that appear in leading power factorization theorems, as indicated in \Tab{tab:fact_func} with the notation $S_g^{(0)}$ and $S_{q}^{(0)}$.  
}

\noindent $ \boldsymbol{O^{(1)}_{ggg} O^{(1)}_{qqg}}$ \textbf{ vanish:}

The contribution from the product of an operator containing three collinear gluons and an operator containing a collinear gluon and two collinear quarks must vanish. If the two quarks are in different sectors, then each jet function will contain only one fermion and will vanish by fermion number conservation. With 2 quarks in the same sector, we have the product of $O_{\bar{n}\lambda_1 ( \lambda_2:\pm)}^{(1)a\,\balpha\bt}$ and $O_{\cB\lambda_3 \lambda_4 \lambda_5 (\pm)}^{(1)bcd}$, and we can see that one sector will contain three objects of helicity $\pm1$ and thus the jet function in that sector will vanish by the angular momentum arguments considered earlier. So, there is no nonvanishing contribution from an $\cO(\lambda)$ three gluon operator and an $\cO(\lambda)$ two quark, one gluon operator.

We now look at the contribution of $\cO(\lambda^2)$ hard scattering operators to the cross section. Since these operators are at the desired order, they must be combined with our leading operator which has a collinear quark building block in each of the $n$ and $\bn$ directions. Once again, there are several cases to consider.

\noindent $ \boldsymbol{O^{(2)}_{qqqq} O^{(0)}_{qq}}$ \textbf{ contribute:}

 First, we consider the inclusion of the four quark operators. Conservation of fermion number within each collinear sector ensures the the contributions to the cross section involving the four quark operators with two quarks within each collinear sector vanish. Therefore, only the four quark operators of \Eq{eq:Z3_basis_qQ}, involving three quarks in one collinear sector, and a single quark in the other collinear sector can give a nonzero result for the cross section at this order. They contribute to the factorized contribution $H_{q7}^{(0)} J_q^{(0)} J_{qq'}^{(2)} S_q^{(0)}$ shown in \Tab{tab:fact_func}, which involves  a jet function $J_{qq'}^{(2)}$ with $4$ $\chi_\bn$ (or 4 $\chi_n$)  fields (with two independent color contractions). Again, the corresponding soft function is simply the leading power $S_q^{(0)}$ in all cases.

\noindent $ \boldsymbol{O^{(2)}_{gggg} O^{(0)}_{qq}}$ \textbf{ vanish:}

The operators of \Eqs{eq:Z2_basis_gggg_1}{eq:Z2_basis_gggg_2} involving four collinear gluons do not contribute to the cross section at $\mathcal{O}(\lambda^2)$, since when multiplied with the leading power operator, the factorized matrix element would violate fermion number in the $n$ and $\bar n$ sectors. 

\noindent $ \boldsymbol{O^{(2)}_{ggg\cP} O^{(0)}_{qq}}$ \textbf{ vanish:}

 An identical argument also applies to the operators of \Eq{eq:eegggpperp_basis} involving three collinear gluons and a single $\cP_\perp$ insertion, which therefore do not contribute to the $\mathcal{O}(\lambda^2)$ cross-section.

{
\renewcommand{\arraystretch}{1.6}
\begin{table}[t!]
\begin{center}
\begin{tabular}{| c | l | c | c | c | c | }
	\hline
	& Operators & Factorization  & Jet $n$ &  Jet $\bar n$ & Soft  \\
	\hline 
	 $\!\!\mathbf{\mathcal{O}(\lambda^0)}\!\!$
	&$\! O^{(0)}  O^{(0)} $ 
	& $H_q^{(0)}  J_q^{(0)} J_q^{(0)} S_q^{(0)}$
	& $\bar \chi_n \,\hat\delta\, \chi_n$ 
	& $\bar \chi_\bn \,\hat\delta\,  \chi_\bn$
	& $Y_\bn^\dagger Y_n \widehat\cM^{(0)} Y_n^\dagger  Y_\bn$ 
	\\
	\hline
	$\!\!\mathbf{\mathcal{O}(\lambda^2)}\!\!$
	&$\! O_\cB^{(1)}  O_\cB^{(1)} $
	& $H_{q1}^{(0)} J_g^{(0)} J_{gg}^{(2)} S_g^{(0)}$  
	& $\cB_n \,\hat\delta\, \cB_n$  
	& $\!\!\cB_\bn \cB_\bn \hat\delta\, \cB_\bn \cB_\bn\!\!$  
	&  ${\cal Y}_n^T {\cal Y}_\bn \widehat\cM^{(0)} {\cal Y}_\bn^T {\cal Y}_n$  
	\\
	\cline{2-6}
	& $\! O_{\bar n}^{(1)}  O_{\bn}^{(1)} $
	& $H_{q2}^{(0)} J_g^{(0)} J_{qq}^{(2)} S_g^{(0)}$  
	& $\cB_n \,\hat\delta\, \cB_n$ 
	& $\bar \chi_\bn \chi_\bn \hat\delta\,  \bar \chi_\bn \chi_\bn $   
	& $\cY_\bn^T \cY_n \widehat\cM^{(0)}  \cY_n^T \cY_\bn $ 
	\\
	\cline{2-6}
	&$\! O_{n\bar n1}^{(1)}  O_{n\bar n1}^{(1)}$
	& $H_{q3}^{(0)} J_q^{(0)} J_{qgg}^{(2)}\, S_q^{(0)}$  
	& $\bar \chi_n \,\hat\delta\, \chi_n $ 
	& $\!\!\bar \chi_\bn  \cB_\bn \,\hat\delta\, \cB_\bn \chi_\bn\!\! $   
	& $Y_\bn^\dagger Y_n \widehat\cM^{(0)}  Y_n^\dagger Y_\bn$   
	\\
	\cline{2-6}
	&$\! O^{(0)}  O_{\cB1,\cB2}^{(2)} $
	& $H_{q4}^{(0)} J_q^{(0)} J_{qgg'}^{(2)} S_{q}^{(0)}$  
	& $\bar \chi_n \,\hat\delta\, \chi_n $ 
	& $\!\!\bar \chi_\bn \cB_\bn \cB_\bn \,\hat\delta\, \chi_\bn\!\!  $   
	& $Y_n^\dagger Y_\bn \widehat\cM^{(0)}   Y_\bn^\dagger Y_n$   
	\\
	\cline{2-6}
	&$\! O^{(0)}  O_{\cP1\bar n}^{(2)} $
	& $H_{q5}^{(0)} J_q^{(0)} J_{qgP}^{(2)} S_q^{(0)}$  
	& $\bar \chi_n \,\hat\delta\, \chi_n $ 
	& $\!\!\bar \chi_\bn [\cP_{\!\perp} \cB_\bn] \hat\delta\, \chi_\bn\!\!$   
	& $Y_n^\dagger Y_\bn \widehat\cM^{(0)} Y_\bn^\dagger Y_n$   
	\\
	\cline{2-6}
	&$\! O^{(0)} O_{\cB3}^{(2)} $
	& $\!\!H_{q6}^{(0)}\! J_{qg}^{(1)} J_{qg}^{(1)} \{ S_{qg}^{(0)}, S_q^{(0)} \}\!\!$  
	& $\!\!\bar \chi_n \cB_n \,\hat\delta\, \chi_n \!\!  $ 
	& $\bar \chi_\bn \,\hat\delta\, \cB_\bn  \chi_\bn  $   
	& $Y_\bn^\dagger Y_n \widehat\cM^{(0)}  Y_n^\dagger \cY_\bn \cY_n Y_\bn$   
	\\
	\cline{2-6}
	&$\! O^{(0)}  O_{qQ4,5}^{(2)}$
	& $H_{q7}^{(0)} J_q^{(0)} J_{qq'}^{(2)} S_q^{(0)}$  
	& $\bar \chi_n \,\hat\delta\, \chi_n   $ 
	& $\!\!\bar \chi_\bn \chi_\bn \bar \chi_\bn \,\hat\delta\, \chi_\bn\!\!$   
	& $Y_\bn^\dagger Y_n \widehat\cM^{(0)} Y_n^\dagger Y_\bn$   
	\\
	\cline{2-6}
	&$\! O^{(0)}  O_{\cB (us)0}^{(2)} \!\!$
	& $H_{q8}^{(0)} J_q^{(0)} J_{q}^{(0)} S_{q\cB}^{(2)}$  
	&  $\bar \chi_n\,\hat\delta\, \chi_n   $	 
	&$\bar \chi_\bn\,\hat\delta\, \chi_\bn   $  
	& $\!\!\cB_{us(n) 0}\, Y_n^\dagger Y_\bn \widehat\cM^{(0)} Y_\bn^\dagger Y_n\!\!$     
	\\
	\cline{2-6}
	& $\! O^{(0)}  O_{\partial(us)0}^{(2)} \!\! $
	& $H_{q9}^{(0)} J_q^{(0)} J_{q}^{(0)} S_{q\partial 0}^{(2)}$  
	&  $\bar \chi_n\,\hat\delta\, \chi_n   $	 
	&$\bar \chi_\bn\,\hat\delta\, \chi_\bn   $  
	& $\!\!\partial_{us(n) 0}\, Y_n^\dagger Y_\bn \widehat\cM^{(0)} Y_\bn^\dagger Y_n\!\!$       
	\\			
	\cline{2-6}
	& $\! O^{(0)}  O_{\partial(us)\bar 0}^{(2)}\!\!$
	& $H_{q10}^{(0)} J_q^{(0)} J_{q}^{(0)} S_{q\partial \bar 0}^{(2)}$  
	&  $\bar \chi_n\,\hat\delta\, \chi_n   $	 
	&$\bar \chi_\bn\,\hat\delta\, \chi_\bn   $  
	& $\!\!\partial_{us(n) \bar 0}\, Y_n^\dagger Y_\bn \widehat\cM^{(0)} Y_\bn^\dagger Y_n\!\!$       
	\\	
	\hline
\end{tabular}
\caption{Subleading jet and soft functions arising from products of hard scattering operators in the factorization of dijet event shapes and their field content. Helicity and color structures have been suppressed. We have not included products of operators whose jet and soft functions are identical to those given in the table by charge conjugation or $n\leftrightarrow \bar n$.}
\label{tab:fact_func}
\end{center}
\end{table}
}

\noindent $ \boldsymbol{O^{(2)}_{ggqq} O^{(0)}_{qq}}$ \textbf{ contribute:}

Another source of non-trivial contributions to the cross section at $\mathcal{O}(\lambda^2)$ comes from the hard scattering operators involving two collinear quarks, and two collinear gluons. To have a non-vanishing contribution, fermion number conservation within each collinear sector guarantees that the hard scattering operators must have a single quark building block field in each sector. This restricts us to the operators of \Eqs{eq:eeqqgg_basis1}{eq:eeqqgg_basis2}, as indicated by the check marks in \Tab{tab:summary}. These operators give rise to subleading jet functions involving two $\chi$ fields, and either one or two $\cB$ fields ($J_{qg}^{(1)}$ or $J_{qgg}^{(2)}$ respectively), as shown by the factorized contributions $H_{q6}^{(0)} J_{qg}^{(1)} J_{qg}^{(1)} \{S_{qg}^{(0)}, S_{q}^{(0)} \}$ and $H_{q4}^{(0)} J_q^{(0)} J_{qgg'}^{(2)} S_{q}^{(0)}$ in  \Tab{tab:fact_func}.  There is a unique color structure present in the definition of $J_{qg}^{(1)}$. The $J_{qgg'}^{(2)}$ here appear for two different color contractions and are distinct from the $J_{qgg}^{(2)}$ function appearing in the $H_{q3}^{(0)} J_q^{(0)} J_{qgg}^{(2)} S_{q}^{(0)}$ case due to the location of their measurement function.   The new soft function $S_{qg}^{(0)}$ has  the Wilson line structure shown by the Soft entry in \Tab{tab:fact_func}, and is generated with a unique color contraction. It only appears in the subleading power cross section even though it has leading power scaling.

\noindent $ \boldsymbol{O^{(2)}_{qqg\cP} O^{(0)}_{qq}}$ \textbf{ contribute:}

Similar arguments also apply to the subleading operators involving two collinear quarks, a collinear gluon, and a $\cP_\perp$ insertion. In particular, the operators of \Eq{eq:eeqqgpperp_basis_same} do not contribute due to fermion number conservation, while the operators of \Eqs{eq:eeqqgpperp_basis}{eq:eeqqgpperp_basis_flip} do contribute. These give rise to the factorized contribution $H_{q5}^{(0)} J_q^{(0)} J_{qgP}^{(2)} S_q^{(0)}$ in \Tab{tab:fact_func}, as well as an identical function with $n\leftrightarrow \bar n$. There is a unique color contraction for the jet function $J_{qgP}^{(2)}$. Rotational invariance arguments can be used to restrict the particular helicity configurations which give non-vanishing contributions.

\noindent $ \boldsymbol{O^{(2)}_{us} O^{(0)}_{qq}}$ \textbf{ contribute:}

As shown in \Tab{tab:fact_func}, the only operators involving an explicit ultrasoft field insertion which contribute to the dijet cross section at $\mathcal{O}(\lambda^2)$ are those involving the $0$ helicity component of the ultrasoft gluon, $\cB_{us(n) 0}$, or the ultrasoft partial operators $\partial_{us(n) 0}$, $\partial_{us(n) \bar 0}$. These operators give rise to the only subleading power soft functions arising from hard scattering operators, as seen in \Tab{tab:fact_func}. It is also interesting to note that for both these contributions, the hard function is fully determined by the RPI relations, as was argued in \Sec{sec:soft_basis}. The operators of \Eq{eq:soft_insert_basis} involving the $h=\pm$ components of the ultrasoft gluon field, along with the collinear quark current, vanish when multiplied with the leading power operator, due to angular momentum conservation. The contributions of the operators of \Eq{eq:eegggus} involving an ultrasoft gluon and two collinear gluons vanish when multiplied with the leading power operator due to fermion number conservation.

A complete analysis of all contributions in \Tab{tab:fact_func}, in particular of their detailed helicity and color structures, as well as their fixed order cross section contributions is beyond the scope of this paper, and is left for presentation in future work.

\subsection{Comparison with Earlier Literature}\label{sec:compare_earlier}

In this section we perform a brief comparison of our operator basis with the operators considered in \Ref{Freedman:2013vya}. These operators were used to study power corrections suppressed by ${\cal O}(\lambda^2)$ for the thrust event shape in $e^+e^-\to$ dijets. The goals of our two works are different. While we have focused on constructing a complete basis of operators valid at any order in $\alpha_s$, \Ref{Freedman:2013vya} instead derives only the set of operators that arise from tree level matching, and then uses them to explicitly calculate and confirm the ${\cal O}(\alpha_s\tau^0)$ terms in $d\sigma/d\tau$. 

Despite the difference in goals, we believe it is still interesting to perform a comparison between the forms of the operators in each case. Some care must be taken, since a different formulation of SCET (first presented in \Ref{Freedman:2011kj}) is used in \Ref{Freedman:2013vya}, as compared to this paper. In the formulation of SCET used in \Ref{Freedman:2013vya}, the dynamics of each collinear sector, as well as the ultrasoft sector, is described by a copy of the QCD Lagrangian, which does not have a power expansion. This implies that the operator basis of \Ref{Freedman:2013vya}  must also include terms which incorporate what would be termed subleading Lagrangian corrections in the standard formulation of SCET. Additionally, while the standard formulation of SCET used in this paper implements separate momentum conservation of residual and label momentum using the multipole expansion with labels, in the formulation of \Ref{Freedman:2013vya}, momentum is not conserved, and additional operators must be included to compensate for this. These terms must be distinguished to perform the comparison. At leading power there are no subleading Lagrangian insertions, and therefore the organization of the two SCET frameworks is equivalent.

At $\mathcal{O}(\lambda)$ in the power expansion, the basis of \Ref{Freedman:2013vya} does not consider operators involving three collinear gluons, as given in \Eq{eq:eeggg}. These operators have vanishing Wilson coefficient at tree level, with the first non-zero correction appearing through diagrams involving a quark loop, but can contribute to event shape cross sections at $\mathcal{O}(\lambda^2)$ and higher orders in the $\alpha_s$ expansion, as shown in \Tab{tab:summary}. Operators with two collinear quarks in different sectors, and a collinear gluon are also given in \Ref{Freedman:2013vya} in their operators $O_2^{(1_{a_n})}$, $O_2^{(1_{b_n})}$ and $O_2^{(1_{a_{\bar n}})}$, $O_2^{(1_{b_{\bar n}})}$. Operators with two collinear quarks in the same sector, and a collinear gluon in the other sector, corresponding to  \Eq{eq:Z1_basis_diff} are given in \Ref{Freedman:2013vya} by $O_2^{(1_{e_n})}$, $O_2^{(1_{f_n})}$. It can be shown using the conservation of the QCD current that to all orders in $\alpha_s$ in the matching, these operators appear only in the combinations $O_2^{(1_{a_n})}-O_2^{(1_{b_n})}$, $O_2^{(1_{a_{\bar n}})}-O_2^{(1_{b_{\bar n}})}$, and $O_2^{(1_{e_n})}-O_2^{(1_{f_n})}$, as discussed in \Sec{sec:ang_cons}. This was first shown in \cite{Freedman:2014uta}, and used to simplify the basis of \Ref{Freedman:2013vya}. In this case, upon setting the total $\cP_\perp$ in each sector to be zero, we find agreement with the operators in 
\Eqs{eq:Z1_basis}{eq:Z1_basis_diff}. In \Ref{Freedman:2013vya}, they also include $\mathcal{O}(\lambda)$ corrections to the ultrasoft sector in (A7). In the SCET framework used in this paper, these do not appear in our operator basis due to power counting. Instead, corrections to the ultrasoft dynamics are incorporated through subleading Lagrangian insertions.

At $\mathcal{O}(\lambda^2)$ in the power expansion, the basis of \Ref{Freedman:2013vya} neglects operators involving four collinear quark fields, as included in our basis in Eqs. (\ref{eq:Z2_basis_qQ})-(\ref{eq:Z3_basis_qq}). These operators appear in the tree level matching, and satisfy all necessary symmetry relations to contribute to event shape cross sections at $\mathcal{O}(\lambda^2)$ and $\alpha_s^2$. It would be interesting to compute their explicit contribution to the cross-section, or to present an argument showing that they do not contribute.  The operators involving four collinear gluons of \Eqs{eq:Z2_basis_gggg_1}{eq:Z2_basis_gggg_2} are also not considered in \Ref{Freedman:2013vya}. These operators are of limited phenomenological relevance, as they first appear at loop level in the matching, and furthermore, do not contribute to event shape cross sections at $\mathcal{O}(\lambda^2)$, as explained. In \Ref{Freedman:2013vya} they also neglect operators involving two quarks in the same sector combined with two gluons or one gluon and one $\cP_\perp$ insertion, as in \Eqs{eq:eeqqgg_basis3}{eq:eeqqgpperp_basis_same}. Again, we have shown for the particular case of dijet event shapes, that such operators do not contribute at $\cO(\lambda^2)$. Operators involving two quarks in opposite collinear sectors with two collinear gluons are included in \Ref{Freedman:2013vya} in operators $O_2^{(2 {b_n})}$ and $O_2^{(2 {b_{\bar n }})}$. In this case, our basis is quite different, as we have used the gluon equations of motion in the effective theory to eliminate the $n\cdot \cB_n$ field, as is commonly done in the SCET literature \cite{Marcantonini:2008qn}. The operators $O_2^{(2 {a_n})}$, $O_2^{(2 {A_n})}$, $O_2^{(2 {\delta_n})}$, $O_2^{(2 {a_{\bar n}})}$, $O_2^{(2 {A_{\bar n }})}$, and $O_2^{(2 {\delta_{\bar n}})}$ from \Ref{Freedman:2013vya} will contribute to this case after simplification with the equations of motion. This elimination is important for the construction of our helicity basis, as it  allows us to work only in terms of the physical polarizations of the $\cB_n$ field, namely $\cB_\perp^\pm$, and thus simplifies our basis. For the case of two gluons and two quarks, \Ref{Freedman:2013vya} does not include operators where the two quarks are in the same sector or the two gluons are in separate sectors, which can first contribute at $\alpha_s^2$. As was the case at $\mathcal{O}(\lambda)$, \Ref{Freedman:2013vya} also includes operators which in our language arise from subleading Lagrangian insertions. Counting only operators that arise from two quarks in separate sectors and two gluons in the same sector, we see that we have 4 operators while  \Ref{Freedman:2013vya} has 8, which implies that the simplifications from working in the center of mass frame and using the equations of motion are useful for reducing the number of operators.

Finally, \Ref{Freedman:2013vya} also has operators in (A11) involving the expansion of momentum conserving delta functions, denoted $\mathcal{O}^{(\delta)}$, which are required in their formulation of SCET. In the standard formulation of SCET, both label, and residual momenta are conserved, so that these operators are not required in our basis. This distinction also modifies the measurement functions in the two approaches, and the subleading operators $\mathcal{O}^{(\delta)}$ are required to maintain consistency between the approaches. Indeed, a particularly convenient feature of our construction is that \Tab{tab:helicityBB} gives the full list of building blocks required to construct a complete basis of hard scattering operators for an arbitrary number of collinear directions and to arbitrary power in the expansion parameter $\lambda$. The completeness of our operator basis ensures that it is closed under renormalization group evolution.

\section{Matching at Subleading Power}
\label{sec:matching}

Having identified the relevant operators which contribute to the cross section in \Tab{tab:summary}, in this section we carry out the matching to determine the lowest order Wilson coefficients for these operators. At ${\cal O}(\lambda^2)$ operators with up to four collinear fields are present in the basis. The amplitudes are known for $e^+e^- \to 3$ partons (and related crossings) at $2$ loops \cite{Garland:2001tf,Garland:2002ak}, and $e^+e^- \to 4$ partons (and related crossings) at $1$ loop \cite{Bern:1996ka,Bern:1997sc}. As the focus of the present paper is on the structure of the operators, we will content ourselves with performing the tree level matching, leaving the higher loop matching for future work. This implies in particular that we do not match to operators involving only gluon fields, which necessarily first appear at loop level. Although we have emphasized the utility of the helicity basis for counting operators, in this section we will perform the calculation using free Lorentz indices, and then projecting onto definite helicities to obtain the Wilson coefficients for the helicity operators. This  allows us to carry out the matching for a starting current with a general Dirac structure $\Gamma$. We will give the results for the operators both in terms of standard Lorentz and Dirac structures, and then projected onto our helicity operator basis. 

We begin in \Sec{sec:matching_general} with a general discussion of matching at subleading power. We then consider explicit matching calculations.
From \Tab{tab:summary}, we see that we must therefore match only to operators involving one additional collinear gluon, which are considered in \Sec{sec:matching_NLP}, two additional collinear gluons, which are considered in \Sec{sec:matching_NNLP_qqgg}, or an additional $q\bar q$ pair, which are considered in \Sec{sec:matching_NNLP_4q}. Although the operators involving insertions of ultrasoft operators which contribute to the $\cO(\lambda^2)$ cross section have their Wilson coefficients determined by RPI, we also perform the matching calculation with a single ultrasoft gluon in \Sec{sec:match_soft} to explicitly verify this.


\subsection{General Formalism}\label{sec:matching_general}

In this section we briefly describe matching to the subleading helicity operators in \SCETi. 
As in the leading power case, QCD is matched to an effective SCET hard scattering Lagrangian, which governs the interactions at the hard scale. As has been discussed, this hard scattering Lagrangian has an explicit expansion in powers of $\lambda$,
\begin{equation} \label{eq:Leff_sub}
\cL_{\text{hard}} = \sum_{j\geq0} \cL^{(j)}_{\text{hard}}
\,,\end{equation}
where $j$ denotes suppression by ${\cal O}(\lambda^j)$ with respect to the leading power hard scattering operators. The effective Lagrangian for hard scattering operators at each power is given by,
\begin{align} \label{eq:Leff_sub_explicit_later}
\cL^{(j)}_{\text{hard}} = \sum_{\{n_i\}} \sum_{A,\cdot\cdot} 
  \bigg[ \prod_{i=1}^{\ell_A} \int \! \! \df \omega_i \bigg] \,
& \vO^{(j)\dagger}_{A\,\lotsdots}\big(\{n_i\};
   \omega_1,\ldots,\omega_{\ell_A}\big) \nn\\
& \times
\vC^{(j)}_{A\,\lotsdots}\big(\{n_i\};\omega_1,\ldots,\omega_{\ell_A} \big)
\,.
\end{align}
The Wilson coefficients, $\vC^{(j)}_{A}$, are determined by performing a matching calculation from QCD to SCET. When matching at subleading powers, one must take into account subleading Lagrangian insertions with lower power hard scattering operators, arising from the fact that the Lagrangian describing the ultrasoft and collinear dynamics is also a power expansion in $\lambda$,
\begin{align}
\cL_\dyn=\cL^{(0)}+\cL^{(1)}+\cdots \,.
\end{align}
To any given power we will only need to consider a finite number of insertions of $\cL^{(i)}$ for $i \geq1$, as these are constrained by the power counting. However, we must consider arbitrary $\cL^{(0)}$ insertions, constrained only by the order in $\alpha_s$ to which we are working.  

Consider the tree level matching at order $\lambda^p$. We assume that the Wilson coefficients at all lower powers, $\vec C^{(q)}_A$ with $q<p$, have already been determined. The matching can be performed with an arbitrary external state $\bra{X}$, as long as it is chosen to have non-zero matrix elements for the specific color and helicity structure of the operator. (If the basis is over complete with regards to color, then the organization of the matching results will also depend on the convention adopted, but we do not encounter this issue in this paper.) In general these specifications will not pick out a particular operator from among those with the same color and helicity structures, so there would still be a sum on $A$. The remaining distinguishing feature used is the dependence on momenta (for example, looking at the $p_\perp$-momenta that appear to determine where a $\cP_\perp^\pm$ is acting in the operator). This then enables us to write down a matching equation for a fixed $A$ and fixed helicities, and color channel. For notational simplicity we assume all these specifications to be made with the state $\bra{X_A}$. 
Such an external state can arise at tree level only from $\vO^{(p)\dagger}_A$, or from subleading Lagrangian insertions into lower power operators. One then has the following matching equation
\begin{align} \label{eq:lamp_match}
-i(\cA^{\tree})^{(\lambda^p)} 
 =C^{(p)}  \big\langle X_A\big|  \vO^{(p)\dagger}_A   \big|0 \big\rangle^{\tree}_{\cL^{(0)}}
 + \sum_{n=1} \big\langle X_A\big| [\mbox{\small$\prod$} \cL^{(k)}]^{(n)} \cL^{(p-n)}_{\text{hard}} \big|0 \big\rangle^{\tree}_{\cL^{(0)}}\,,
\end{align}
where $[\mbox{\small$\prod$} \cL^{(k)}]^{(n)}$ represents all combinations of subleading SCET Lagrangian insertions $\cL^{(k\ge 1)}$ whose powers sum to $n$. The ${\cal L}^{(0)}$ subscript on the matrix elements indicates that they are evaluated with any number of insertions of the leading power SCET Lagrangian. Here $(\cA^{\tree})^{(\lambda^p)}$ is the corresponding amplitude in full QCD expanded to pick out the order $\lambda^p$ component. The momentum of external particles in this amplitude are assigned a power counting corresponding to the building blocks in $\vO^{(p)\dagger}_A$. This means that if two external particles are in the same collinear sector then we take their collinear limit when expanding the matrix element, and that we take the ultrasoft limit for momenta of ultrasoft particles. The matrix element, $\bra{X_A}  \vO^{(p)\dagger}_A   |0 \rangle^{\tree}_{\cL^{(0)}}$, is evaluated using the Feynman rules for the helicity operators. For the case with $N$ distinct collinear particles at leading power, the sum on the right side of \eq{lamp_match} does not contribute, as there is nothing to expand when each of the particles are separated to distinct sectors. However, at subleading power either collinear particles in the same sector or ultrasoft particles will give contributions in this expansion. Most modern fixed order computations are performed using spinor helicity and color decomposition techniques, which give compact results. Using our basis of helicity operators, the helicity amplitudes can be directly used, since we match to external states with definite helicities. For the particularly simple cases in this paper, we find it is easier to simply project the different helicities from a tensor, however, the helicity formalism still provides a powerful way of organizing the operators.

Beyond tree level, the matching is still conceptually straightforward, but technically more demanding due to the need to evaluate loop diagrams in both the full and effective theories. \eq{lamp_match} still holds, but now we must consider each of the Wilson coefficients and matrix elements as an expansion in $\alpha_s$, and go beyond the leading term.  At leading power it is often possible to arrange the choice of IR and UV regulators such that the SCET loop diagrams are scaleless in the matching calculation. At subleading power we in general need more than one particle in a given sector in order to have nonzero overlap with the operator, and this can introduce non-trivial momentum scales in the SCET loop integrals such that  loop calculations in SCET can not be avoided in this manner.

\subsection{Setup and Leading Power Matching}
\label{sec:matching_LP}

In the following sections we carry out the matching starting from a full QCD quark current with an arbitrary spin structure $\Gamma$,
\begin{align}
J^{\Gamma}=\bar \psi \Gamma \psi\,.
\end{align}
For vector and axial-vector $\Gamma$ this current is (partially)-conserved in QCD, while for the scalar, pseudo-scalar and tensor cases $J^{\Gamma}$ has an anomalous dimension in QCD. We will denote the full theory vertex with an $\otimes$ symbol, and hard scattering operators in the effective theory will be denoted with a purple circle.

The leading power Wilson coefficient is of course well known, however, we reproduce it here for completeness. The unique leading power operator, written in the form of a current is
\begin{align}
\cO^{(0)\Gamma}=\bar \chi_{n} \Gamma \chi_{\bar n}\,,
\end{align}
It's Wilson coefficient for any spin structure $\Gamma$ is given to $\cO(\alpha_s)$ by 
\begin{align}\label{eq:LP_wilson}
C^{(0)}=1+\frac{\alpha_s(\mu) C_F}{4\pi} \left( -\log^2 \left[  \frac{-\omega_1 \omega_2-i0}{\mu^2} \right] +3\log \left[  \frac{-\omega_1 \omega_2-i0}{\mu^2} \right] -8 +\frac{\pi^2}{6} \right) \,.
\end{align}
Throughout this section, we will restrict ourselves to the tree level matching, however, we have given the Wilson coefficient of \Eq{eq:LP_wilson} to one loop, since it will be used to demonstrate the RPI relations of \Sec{sec:soft_basis} for the operators involving ultrasoft insertions, which are first non-trivial at this order.

\subsection{Subleading Matching with a Single Collinear Gluon}
\label{sec:matching_NLP}

We begin by considering the matching to operators involving two collinear quark fields and a collinear gluon field, along with a possible $\cP_\perp$ insertion. A basis of such operators was given in \Eqs{eq:Z1_basis}{eq:Z1_basis_diff}, as well as \Eq{eq:eeqqgpperp_basis}.

The matching coefficients for these operators can be derived by considering matrix elements with a $q\bar q g$ final state. At $\mathcal{O}(\alpha_s)$, the QCD diagrams for the production of a single gluon from a quark current with spin structure $\Gamma$ are
\begin{align}
\fd{3cm}{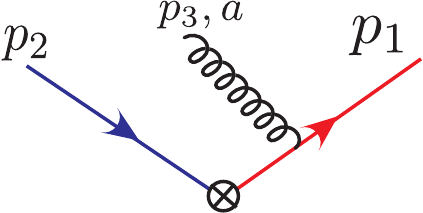} &= \bar{u}(p_1) (i g T^a \Sl{\epsilon}^*) \frac{ i (\Sl{p}_1 + \Sl{p}_3)}{ (p_1+p_3)^2} \Gamma v(p_2) \,, \nn\\
\fd{3cm}{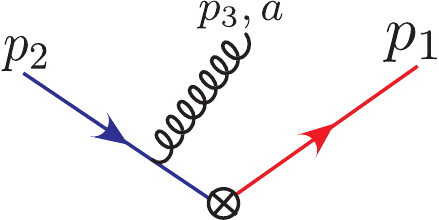}&= \bar{u}(p_1) \Gamma \frac{- i (\Sl{p}_2 + \Sl{p}_3)}{ (p_2+p_3)^2}  (i g T^a \Sl{\epsilon}^*)v(p_2) \,.
\end{align}
The required matching coefficients is obtained by expanding these amplitudes in the required kinematic limits, namely as the gluon becomes collinear with either the quark or antiquark, or when the quark and antiquark become collinear. We will consider each of these cases in turn. 

We note that while we will restrict our attention to tree level matching in this section, it would be particularly interesting to extend the matching to one-loop. Indeed, the one-loop matching to the operators involving a single additional collinear gluon field is the only ingredient related to the hard scattering operators, beyond the matching coefficients presented in this paper, that would be required to perform a full analytic $\cO(\alpha_s^2)$ fixed order calculation of the subleading cross section for an $e^+e^-\to$ dijets observable. We leave this to future work.

\vspace{0.2cm} \noindent 
\underline{
$\boldsymbol{(q)_{ n} (\bar q g)_{\bar n}:}$
}

We first consider the case where the gluon and antiquark are $\bar n$ collinear, and the quark is $n$ collinear. In this case we have hard scattering operators at both $\mathcal{O}(\lambda)$, which are independent of $\cP_\perp$ and operators at $\mathcal{O}(\lambda^2)$, which depend on $\cP_\perp$. To extract the matching coefficients for both sets of operators in a single calculation, we expand the QCD amplitudes with a non-zero $\cP_\perp$ between the gluon and antiquark. We take the momenta of the particles as
\begin{align}
p_1^\mu=\frac{\omega_1}{2}n^\mu \,, \qquad
p_2^\mu=\frac{\omega_2}{2} \bar n^\mu+p_\perp^\mu +\frac{ p_{2r}}{2}  n^\mu\,, \qquad
p_3^\mu=\frac{\omega_3}{2}\bar n^\mu -p_\perp^\mu +\frac{ p_{3r}}{2}  n^\mu\,, 
\end{align}
where the light cone components are assigned a collinear scaling. The $\omega_i$ and $p_\perp$ are taken to be purely label momenta. Following the notation of \Sec{sec:coll}, throughout this section, we will use the notation
\begin{align}
u_n(i)=P_n u(p_i)\,, \qquad \text{and} \qquad v_n(i)=P_n v(i)\,,
\end{align}
for the projected SCET spinors, where the momentum $p_i$ is $n$ collinear, and similarly for the case that it is $\bar n$ collinear. They obey
\begin{align}
u(p_i) = \Big( 1+ \frac{\Sl{p}_{i\perp}}{\bar{n} \cdot p_i} \frac{\Sl{\bar{n}}}{2} \Big)  u_n(i) \,, 
 \qquad
u(p_i) = \Big( 1+ \frac{\Sl{p}_{i\perp}}{n \cdot p_i} \frac{\Sl{n}}{2} \Big) u_{\bar n}(i) \,,
\end{align}
for the $n$ collinear and $\bar n$ collinear cases respectively, with direct analogs for the $v(p_i)$ spinors.

Expanding the first diagram in $\lambda$, we find
\begin{align}\label{eq:ggqn_match1}
\fd{3cm}{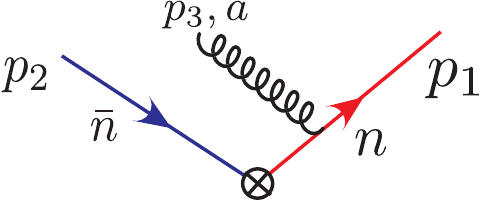}
  &=-  g \frac{n\cdot \epsilon^*}{ \omega_3} \bar{u}_{ n}(1) T^a \Gamma  v_{\bar n}(2) \\
&\hspace{-2cm} 
- \frac{g}{\omega_1 \omega_3}\bar{u}_{n}(1)  T^a \left(  \omega_3 \Sl{\epsilon}_\perp^*\frac{\Sl{\bar n}}{2}\Gamma- n\cdot\epsilon^*   \frac{\Sl{\bar n}}{2}  \Sl{p}_\perp \Gamma   +\frac{ n\cdot \epsilon^* \, \omega_1}{\omega_2} \Gamma \Sl{p}_\perp \frac{\Sl{n}}{2} \right) v_{\bar n}(2) \nn \\
 &\hspace{-2cm} - \frac{g}{\omega_1 \omega_3}\bar{u}_{n}(1)  T^a 
 \biggl( n\cdot \epsilon^*\,  p_{3r}  \Gamma - \Sl{\epsilon}_\perp^* \Sl{p}_\perp \Gamma + \frac{\omega_3}{\omega_2}\Sl{\epsilon}_\perp^*\frac{\Sl{\bar n}}{2} \Gamma \Sl{p}_\perp \frac{\Sl{n}}{2}-\frac{n\cdot\epsilon^*}{\omega_2} \frac{\Sl{\bar n}}{2} \Sl{p}_\perp \Gamma \Sl{p}_\perp \frac{ \Sl{n}}{2}    \biggr) v_{\bar n}(2)
\nn \\
&\hspace{-2cm}+\mathcal{O}(\lambda^3)
\,, \nn
\end{align}
where terms in the first line scale like $\mathcal{O}(\lambda^0)$, terms in the second line scale like $\mathcal{O}(\lambda^1)$, and so on. Similarly, we find
\begin{align}\label{eq:ggqn_match2}
\fd{3cm}{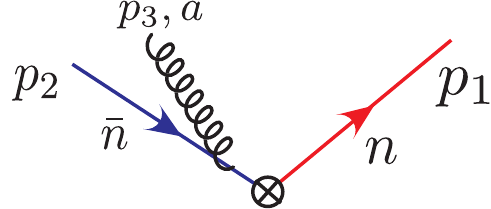}
 &=\frac{g}{(p_{2r}+p_{3r})}\bar u_{n}(1) T^a \Gamma 
 \Bigl(  \epsilon^*\cdot \bar n +\frac{ \Sl{\epsilon}_\perp^*  \Sl{p}_\perp   }{\omega_2}     \Bigr)v_{\bar n}(2)\nn \\
& +\frac{g}{2(\omega_{2}+\omega_{3})}\bar u_{n}(1) T^a \Gamma \Bigl(    \frac{\epsilon^*\cdot n }{\omega_2} \Sl{p}_\perp \Sl{n}  +  \Sl{n} \Sl{\epsilon}^*_\perp     \Bigr) v_{\bar n}(2)\,.
\end{align} 
The first line is $\mathcal{O}(\lambda^0)$ and second line is $\mathcal{O}(\lambda^1)$. Interestingly, this result is exact with no corrections beyond $\cO(\lambda)$.

The $\mathcal{O}(\lambda^0)$ terms are reproduced by $T$-products of the leading SCET Lagrangian.
Since we have performed the matching with no residual momenta for the large or $\perp$ components, all subleading Lagrangian contributions from the $T$-products in the second term on the right hand side of \Eq{eq:lamp_match} vanish. This includes those with insertions of the ultrasoft derivative operators of  \Eqs{eq:soft_derivative_basis}{eq:eedggus} and the leading order SCET Lagrangian.  Since we have shown by combining the RPI relation of \Eq{eq:uspartialRPIrelation} with the matching to the leading power operator in \Eq{eq:LP_wilson} that these operators first appear at $\cO(\alpha_s)$, such terms will not contribute to the matching. With this setup the QCD result at $\mathcal{O}(\lambda^1)$ and $\mathcal{O}(\lambda^2)$ must therefore be exactly reproduced by our basis of hard scattering operators.

Starting at $\mathcal{O}(\lambda)$, summing the results of \Eqs{eq:ggqn_match1}{eq:ggqn_match2} and rearranging, we find
\begin{align}
\left( \fd{3cm}{figures/matching_subleadingvertex_casea1_low}+\fd{3cm}{figures/matching_subleadingvertex_casea2_low}\right)\left. \vphantom{\fd{3cm}{figures/matching_subleadingvertex_casea2_low}} \right|_{\mathcal{O}(\lambda)}& \\
&\hspace{-8cm} 
= - \frac{g}{\omega_1}    \bar u_{n} (1) T^a \Bigl(  \Sl{\epsilon}^*_\perp+\frac{\epsilon^*\cdot n \Sl{p}_\perp}{\omega_3}  \Bigr)         \frac{\Sl{\bar n}}{2}  \Gamma v_{\bar n}(2)
+  \frac{g}{(\omega_2 +\omega_3)}   \bar u_{n} (1) T^a \Gamma \frac{\Sl{n}}{2} \Bigl(  \Sl{\epsilon}^*_\perp+\frac{\epsilon^*\cdot n \Sl{p}_\perp}{\omega_3}  \Bigr) v_{\bar n}(2)\,.\nn
\end{align}
This can be recognized as the one gluon matrix element of the operator
\begin{align}\label{eq:matched_1gluon_barn_general}
\mathcal{O}^{(1)\Gamma}_{\cB \bar n}&=\frac{g}{\omega_2+\omega_3} \bar \chi_{n,\omega_1} \Gamma \frac{\Sl{n}}{2}\Sl{\cB}_{\perp \bar n,\omega_3} \chi_{\bar n,-\omega_2}-\frac{g}{\omega_1}\bar \chi_{n,\omega_1}\Sl{\cB}_{\perp \bar n,\omega_3}  \frac{\Sl{\bar n}}{2}    \Gamma   \chi_{\bar n,-\omega_2}\,.
\end{align}
In the particular case that $ \Gamma=\gamma^\mu$, this simplifies to
\begin{align}\label{eq:matched_1gluon_barn}
\mathcal{O}^{(1)\gamma^\mu}_{\cB \bar n}&= g\left(  \frac{\omega_1 n^\mu-(\omega_2 +\omega_3) \bar n^\mu }{\omega_1 (\omega_2 +\omega_3)}  \right) \bar \chi_{n, \omega_1} \Sl{\cB}_{\perp \bar n,\omega_3} \chi_{\bar n,-\omega_2}\,,
\end{align}
which we note for later use.
In the center-of-momentum frame conservation of momentum implies that $\omega_1=\omega_2+\omega_3=Q$, further simplifying the structure of the operator. However, throughout this section, we will not perform such a simplification, and will write the result for the Wilson coefficient for generic values of the label momenta.

At $\mathcal{O}(\lambda^2)$, only the diagram in \Eq{eq:ggqn_match1} contributes. The result can be simplified using the on-shell condition for the collinear gluon,
\begin{align}
p_{3r}\omega_3+p_\perp^2=0\,,
\end{align}
after which one finds
\begin{align}
&\left( \fd{3cm}{figures/matching_subleadingvertex_casea1_low}+\fd{3cm}{figures/matching_subleadingvertex_casea2_low}\right)\left. \vphantom{\fd{3cm}{figures/matching_subleadingvertex_casea2_low}}\right|_{\mathcal{O}(\lambda^2)}\\
&=\frac{g}{\omega_1 \omega_2}   \bar u_{n} (1) T^a\Bigl( \Sl{\epsilon}^*_\perp +\frac{\epsilon^*\cdot  n \Sl{p}_\perp}{\omega_3}  \Bigr) \frac{\Sl{\bar n}}{2}  \Gamma \frac{\Sl{n}}{2} \Sl{p}_\perp v_{\bar n}(2)
+\frac{g}{\omega_1 \omega_3}   \bar u_{n} (1)T^a \Bigl( \Sl{\epsilon}^*_\perp +\frac{\epsilon^*\cdot  n \Sl{p}_\perp}{\omega_3}  \Bigr) \Sl{p}_\perp  \Gamma  v_{\bar n}(2)\,.\nn
\end{align}
This can be recognized as the one gluon matrix element of the two SCET hard scattering operators
\begin{align}\label{eq:matched_nbarP}
\mathcal{O}^{(2)\mu}_{\cP \bar n1}
  &=- \frac{g}{\omega_1 \omega_3}  \bar \chi_{ n,\omega_1} 
  \big[\Sl\cB_{\perp \bar n, \omega_3} \Sl{\cP}^\dagger_{\perp}\big]  
  \Gamma \chi_{\bar n,-\omega_2} \,, \nn \\
\mathcal{O}^{(2)\mu}_{\cP \bar n2}
  &= - \frac{g}{\omega_1 \omega_2}  \bar \chi_{n,\omega_1} 
  \Big[\Sl\cB_{\perp \bar n, \omega_3} \frac{\Sl{\bar n}}{2} \Gamma \frac{\Sl{n}}{2} \Sl{\cP}^\dagger_{\perp}\Big]  \chi_{\bar n,-\omega_2} 
 \,.
\end{align}
Of particular interest is the fact that the Wilson coefficients exhibit singularities as the momentum fraction of the gluon or quark vanishes, which will be associated with logarithms in the subleading power cross section. 

\vspace{0.2cm} \noindent 
\underline{
$\boldsymbol{(gq)_{n} (\bar q )_{\bar n}:}$
}

In the case that the gluon is collinear with the quark, we can immediately obtain the operators by charge conjugation, and a relabeling. To be clear on the particle labeling, we consider
\begin{align}
p_1^\mu=\frac{\omega_1}{2}n^\mu \,, \qquad
p_2^\mu=\frac{\omega_2}{2} \bar n^\mu+p_\perp^\mu +\frac{ p_{2r}}{2}  n^\mu\,, \qquad
p_3^\mu=\frac{\omega_3}{2}n^\mu -p_\perp^\mu +\frac{ p_{3r}}{2}  \bar n^\mu\,. 
\end{align}
Analogously to the case that the gluon is in the same collinear sector as the antiquark, we find the $\mathcal{O}(\lambda)$ operator
\begin{align}\label{eq:matched_1gluon_n_general}
\mathcal{O}^{(1)\mu}_{\cB n}&=-\frac{g}{\omega_1+\omega_3} \bar \chi_{n,\omega_1} \Sl{\cB}_{\perp n,\omega_3} \frac{\Sl{\bar n}}{2}    \Gamma  \chi_{\bar n,-\omega_2}+\frac{g}{\omega_2}\bar \chi_{n,\omega_1}       \Gamma   \frac{\Sl{n}}{2}  \Sl{\cB}_{\perp  n,\omega_3}    \chi_{\bar n,-\omega_2}\,.
\end{align}
Again, in the case that we take $ \Gamma=\gamma^\mu$, this simplifies to 
\begin{align}\label{eq:matched_1gluon_n}
\mathcal{O}^{(1)\mu}_{\cB n}&=g\left(  \frac{(\omega_1+\omega_3) n^\mu-\omega_2 \bar n^\mu }{\omega_2 (\omega_1 +\omega_3)}  \right) \bar \chi_{n, \omega_1} \Sl{\cB}_{\perp n,\omega_3} \chi_{\bar n,-\omega_2}\,.
\end{align}
The two $\mathcal{O}(\lambda^2)$ operators are
\begin{align}\label{eq:matched_nP}
\mathcal{O}^{(2)\mu}_{\cP n1}
  &=-\frac{g}{\omega_1 \omega_2} \bar \chi_{n,\omega_1}  
  \Bigl[ \Sl{\cP}_\perp \frac{\Sl{\bar n}}{2} 
  \Gamma \frac{\Sl{n}}{2} \Sl\cB_{\perp n, \omega_3}  \Bigr] 
   \chi_{\bar n,-\omega_2}
  \,, \nn \\
\mathcal{O}^{(2)\mu}_{\cP n2}
  &= -  \frac{g}{\omega_2 \omega_3}  
  \bar \chi_{n,\omega_1}   \Gamma 
  \bigl[ \Sl{\cP}_\perp  \Sl\cB_{\perp n, \omega_3} \bigr] 
  \chi_{\bar n,-\omega_2}
  \,.
\end{align}


\vspace{0.2cm} \noindent 
\underline{
$\boldsymbol{(q \bar q)_{\bar n} ( g)_{n}:}$
}

We now consider the case where both the quark and anti-quark are in the same ($\bar n$) collinear sector and the gluon is in the $\bar n$ collinear sector. As discussed in \Sec{sec:eeJets}, in this case, the operators of this form only contribute to the cross section in the form of a matrix element with themselves. We therefore only need the matching to $\mathcal{O}(\lambda)$, if we are interested in $\mathcal{O}(\lambda^2)$ contributions to the cross section. We can therefore set the $\perp$ momentum to zero when performing the matching.  We take the kinematics as
\begin{align}
p_1^\mu=\frac{\omega_1}{2}\bar n^\mu \,, \qquad
p_2^\mu=\frac{\omega_2}{2}\bar n^\mu\,, \qquad
p_3^\mu=\frac{\omega_3}{2}n^\mu\,.
\end{align}
Expanding the QCD results the ${\cal O}(\lambda^0)$ terms cancel between diagrams, and at ${\cal O}(\lambda)$ we find
\begin{align}\label{eq:matching_qqsame1}
\left( \fd{3cm}{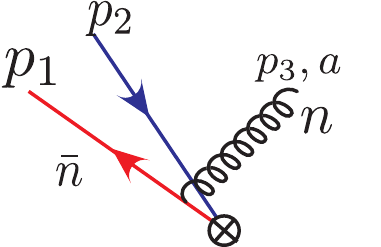} \right)  \left. \vphantom{\fd{3cm}{figures/matching_subleadingvertex_casec1_low}}   \right|_{\mathcal{O}(\lambda)}
 &= - \frac{g}{\omega_1}\bar u_{\bar n}(1) T^a \Sl{\epsilon}^*_\perp \frac{\Sl{n}}{2}  \Gamma     v_{\bar n}(2)\,,
 \\
\left( \fd{3cm}{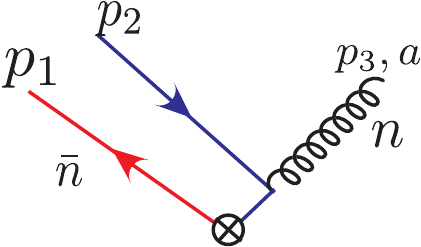} \right)  \left. \vphantom{\fd{3cm}{figures/matching_subleadingvertex_casec2_low}}   \right|_{\mathcal{O}(\lambda)}
 &= 
 + \frac{g}{\omega_2 }  \bar u_{\bar n}(1) T^a \Gamma  \frac{\Sl{n}}{2} \Sl{\epsilon}^*_\perp  v_{\bar n}(2)
 \,. \nn
 \end{align}
We therefore see that for this configuration, both QCD diagrams contribute.
These contributions can be recognized as the one gluon matrix element of the two SCET operators
\begin{align}\label{eq:match_2qsame}
\cO^{(1)\mu}_{\chi\chi\bar n1}&=-\frac{g}{\omega_1}  \bar \chi_{\bar n, \omega_1} \Sl{\cB}_{\perp  n,\omega_3} \frac{\Sl{n}}{2}  \Gamma  \chi_{\bar n,-\omega_2}\,, \nn \\
\cO^{(1)\mu}_{\chi\chi\bar n2}&=\frac{g}{\omega_2}  \bar \chi_{\bar n, \omega_1}   \Gamma \frac{\Sl{n}}{2} \Sl{\cB}_{\perp n,\omega_3}   \chi_{\bar n,-\omega_2}\,.
\end{align}
Both Wilson coefficients exhibit a singularity as the energy fraction as one or the other quark goes to zero. It is interesting to note that this structure is dictated by the RPI-III symmetry of the theory. In particular, due to the presence of the $\Sl{n}$ projector which necessarily appears between the two $\bar n$ collinear quark fields, the Wilson coefficient must behave like $1/\omega_i$, where $i=2$, or $3$. This hints at the possible universality of this structure in subleading power collinear limits.

\subsection{Subleading Matching with Two Collinear Gluons}
\label{sec:matching_NNLP_qqgg}

We now consider the matching to operators involving two collinear gluons. As discussed in \Sec{sec:eeJets}, there are two relevant classes of such operators that contribute to the cross section at $\mathcal{O}(\lambda^2)$: those involving two collinear gluons in the same collinear sector, and those involving two gluons in opposite sectors, both of which have have the quark anti-quark pair in opposite sectors. We consider each case in turn. 

\vspace{0.2cm} \noindent 
\underline{
$\boldsymbol{(\bar q)_{\bar n} ( qgg)_{n}:}$
}

A basis of $\mathcal{O}(\lambda^2)$ operators with two collinear gluons in the same sector was given in \Eq{eq:eeqqgg_basis1}. Since these operators get their power suppression from the explicit collinear gluon fields, we can simplify the matching by taking the particle momenta as
\begin{align}
p_1^\mu&=\frac{\omega_1}{2}n^\mu \,, \qquad
p_2^\mu=\frac{\omega_2}{2}\bar n^\mu\,, \nn \\
p_3^\mu&=\frac{\omega_3}{2}n^\mu+p_\perp^\mu +\frac{p_{3r}}{2} \bar n^\mu\,, \qquad
p_4^\mu=\frac{\omega_4}{2}n^\mu-p_\perp^\mu +\frac{p_{4r}}{2} \bar n^\mu\,, 
\end{align}
where the particle labeling is the same as before, but with $p_4$ labeling the additional gluon. This choice of momenta also removes any contributions from subleading Lagrangian insertions. We do however, get contributions from the subleading hard scattering operators, $\mathcal{O}^{(2)\mu}_{\cP  n1,2}$, which must be disentangled from the operator coefficients we want to determine. For these operators we find
\begin{align}\label{eq:2q2gsame_SCET}
\left( \fd{3cm}{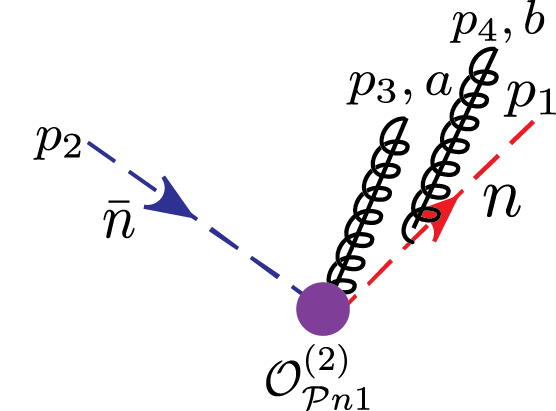}+\fd{3cm}{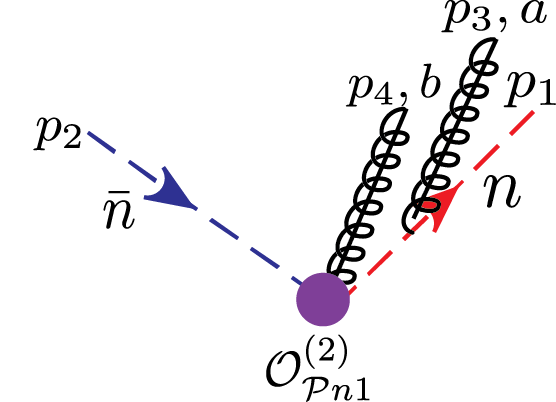}\right)  \left. \vphantom{\fd{3cm}{figures/matching_2q2g_same_SCET_contribution_3_b}} \right |_{\mathcal{O}(\lambda^2)}&  \\
&\hspace{-4.5cm} =    \frac{-g^2\omega_4}{\omega_1 \omega_2 (\omega_1+\omega_4)} \bar u_n (1) T^b T^a  \Sl{\epsilon}^*_{4\perp} \frac{\Sl{\bar n}}{2}  \Gamma  \frac{\Sl{n}}{2}  \Sl{\epsilon}^*_{3\perp} v_{\bar n} (2) +\left[ (3,a)\leftrightarrow (4,b)  \right] ,   \nn \\
\left( \fd{3cm}{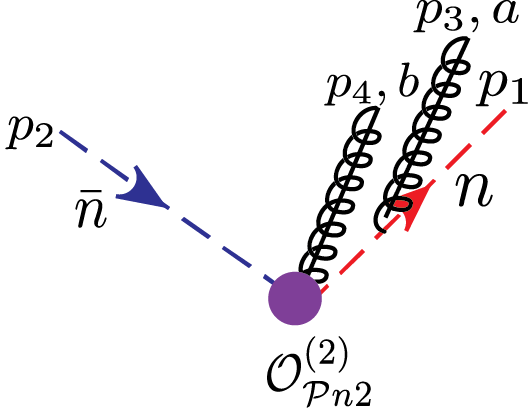}+\fd{3cm}{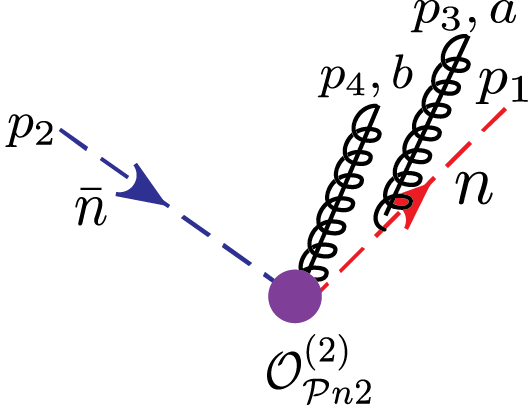} \right)\left. \vphantom{\fd{3cm}{figures/matching_2q2g_same_SCET_contribution_4_b}} \right |_{\mathcal{O}(\lambda^2)}&\nn \\
&\hspace{-4.5cm} =   \frac{g^2}{\omega_1 \omega_2 \omega_3 p_{4r}}   \bar u_n (1) T^b T^a \Sl{\epsilon}^*_{4\perp}   \Sl{p}_\perp  \Gamma  \Sl{p}_\perp    \Sl{\epsilon}^*_{3\perp}  v_{\bar n} (2) +\left[ (3,a)\leftrightarrow (4,b)  \right]\,. \nn
\end{align}
Interestingly, we see that with our choice of momentum, this gives rise to one term which localizes, with an $\cO(\lambda^0)$ denominator, and a term which does not, with a $1/p_{4r}$. The corresponding non-abelian graphs vanish with our choice of momentum,
\begin{align}
\fd{3cm}{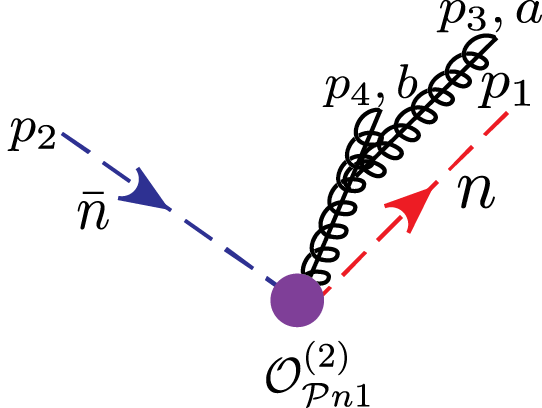}=0\,, \qquad
\fd{3cm}{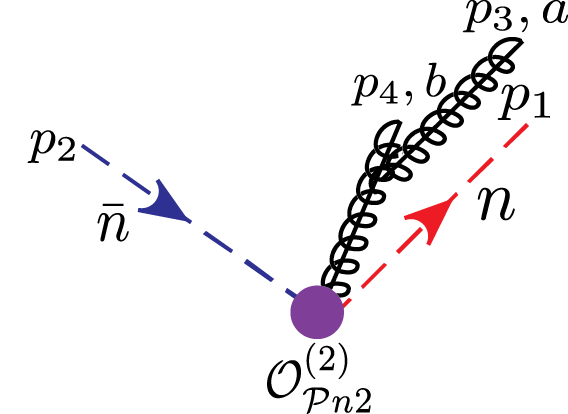}=0\,.
\end{align}
Note that we do not have to consider the operators involving ultrasoft derivatives from \Eqs{eq:soft_derivative_basis}{eq:eedggus} in the matching, since we have chosen to only use a residual momentum component for the small momentum component, and we have shown by RPI that this component of the derivative first appears at $\cO(\alpha_s)$.

We now consider the expansion of the full theory diagrams.
We begin by expanding each of the QCD diagrams for the production of two gluons off of the $q\bar q$ pair to this order. For the independent emission diagrams where both gluons with $n$-collinear scaling are emitted from the quark with $\bar n$-collinear scaling, we have
\begin{align}\label{eq:qqggmatch_1}
&\fd{3cm}{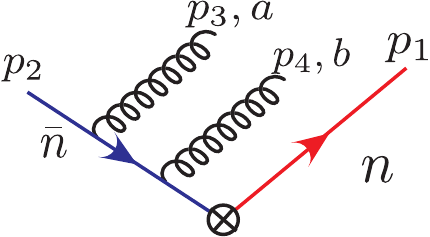}+\fd{3cm}{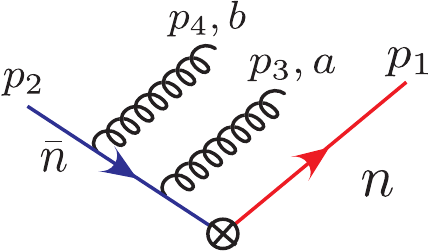}  \\
&= \bar{u}(p_1)   \Gamma  \frac{-i (\Sl{p}_2 + \Sl{p}_3 + \Sl{p}_4)}{(p_2 + p_3 + p_4)^2} (i g T^b) \Sl{\epsilon}^*_{4} \frac{ -i (\Sl{p}_2 + \Sl{p}_3)}{(p_2 + p_3)^2} (i g T^a) \Sl{\epsilon}_{3}^* v(p_2) +\left[ (3,a)\leftrightarrow (4,b)  \right]\,. \nn
\end{align}
Expanding, we can pick out the $\mathcal{O}(\lambda^2)$ contribution of these diagrams, which is given by
\begin{align}
\left(  \fd{3cm}{figures/matching_2q2g_diagram1_low}+\fd{3cm}{figures/matching_2q2g_diagram2_low}  \right)\left. \vphantom{\fd{3cm}{figures/matching_2q2g_diagram2_low}}  \right|_{\mathcal{O}(\lambda^2)}&\\
&\hspace{-3cm}= -\frac{g^2}{\omega_2(\omega_3+\omega_4)}\bar u_{n}(1)  T^b T^a  \Gamma  \Sl{\epsilon}^*_{4\perp} \Sl{\epsilon}^*_{3\perp}  v_{\bar n}(2)+\left[ (3,a)\leftrightarrow (4,b)  \right]\,,\nn
\end{align}
which is local, having only $\cO(\lambda^0)$ momenta in the denominator.
Similarly, for the independent emission diagrams where a single gluon is emitted from the $n$ collinear quark, we have
\begin{align}\label{eq:qqggmatch_2}
&\fd{3cm}{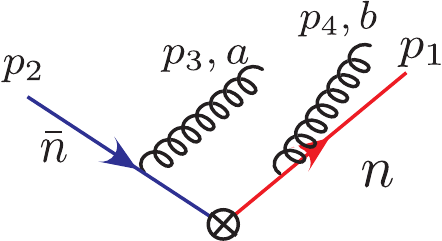}+\fd{3cm}{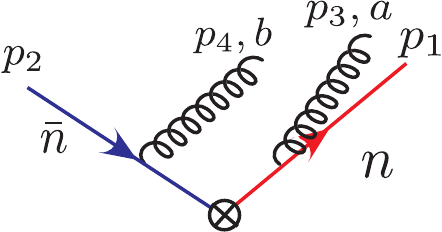} \\
&= \bar{u}(p_1) (i g T^b) \Sl{\epsilon}^*_{4} \frac{i (\Sl{p}_1 + \Sl{p}_4)}{(p_1 + p_4)^2}  \Gamma \frac{ -i (\Sl{p}_2 + \Sl{p}_3)}{(p_2 + p_3)^2} (i g T^a) \Sl{\epsilon}_{3}^* v(p_2) +\left[ (3,a)\leftrightarrow (4,b)  \right]
\,. \nn
\end{align}
The $\mathcal{O}(\lambda^2)$ contribution of these diagrams is given by
\begin{align}
\left( \fd{3cm}{figures/matching_2q2g_diagram3_low}+\fd{3cm}{figures/matching_2q2g_diagram4_low}  \right)\left. \vphantom{\fd{3cm}{figures/matching_2q2g_diagram4_low}}\right |_{\mathcal{O}(\lambda^2)}&  \\
&\hspace{-6cm}= -g^2  \frac{1}{\omega_1 \omega_2} \bar u_n (1)T^b T^a \Sl{\epsilon}^*_{4\perp} \frac{\Sl{\bar n}}{2}  \Gamma \frac{\Sl{n}}{2} \Sl{\epsilon}^*_{3\perp} v_{\bar n}(2)    +\left[ (3,a)\leftrightarrow (4,b)  \right]\,  \nn\\
&\hspace{-6cm} +g^2 \frac{1}{\omega_1 \omega_2 \omega_3 p_{4r}}  \bar u_n (1)T^b T^a  \Sl{\epsilon}^*_{4\perp} \Sl{p}_\perp  \Gamma  \Sl{p}_\perp  \Sl{\epsilon}^*_{3\perp} v_{\bar n} (2)   +\left[ (3,a)\leftrightarrow (4,b)  \right]\,. \nn
\end{align}
Here we recognize both a local term, as well as a $T$-product like term, which is reproduced by the SCET diagrams in \Eq{eq:2q2gsame_SCET}. This is expected from the topology of the diagram.

There are also non-abelian diagrams. In the case that the gluon is emitted from the $\bar n$ collinear quark, we find
\begin{align}\label{eq:qqggmatch_3}
\fd{3cm}{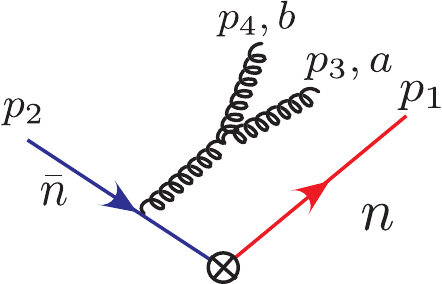} &= -i g^2 f^{abc} \bar u(p_1)  \Gamma T^c \frac{(\Sl{p}_2+\Sl{p}_3+\Sl{p}_4 )}{(p_2+p_3+p_4)^2(p_3+p_4)^2}\gamma^\rho v(p_2)\nn \\
&\cdot  \left[  g^{\mu \nu}(p_4-p_3)^\rho-g^{\nu \rho}(2p_4+p_3)^\mu +g^{\rho \mu}(p_4+2p_3)^\nu  \right]\epsilon^*_{4\nu} \epsilon^*_{3\mu} \,.
\end{align}
The $\mathcal{O}(\lambda^2)$ contribution of this diagram vanishes
\begin{align}
\fd{3cm}{figures/matching_2q2g_diagram7_low}\left. \vphantom{\fd{3cm}{figures/matching_2q2g_diagram7_low}}\right |_{\mathcal{O}(\lambda^2)}&=0 \,.
\end{align}

Finally, there are three diagrams in which both gluons are emitted from the $n$ collinear quark 
\begin{align}
&\left(   \fd{3cm}{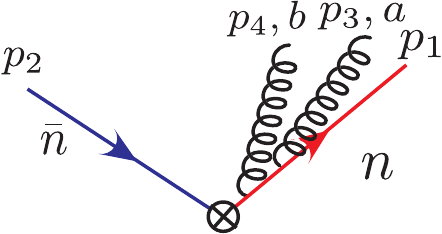}+\fd{3cm}{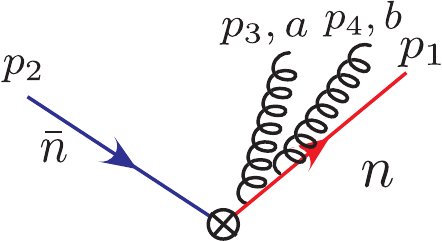}+\fd{3cm}{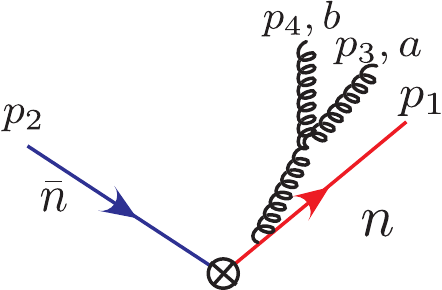}\right) \left. \vphantom{\fd{3cm}{figures/matching_2q2g_diagram8_low}} \right |_{\mathcal{O}(\lambda^2)}=0\,.
\end{align}
Each of these diagrams individually gives a vanishing contribution at $\cO(\lambda^2)$, as might naively be expected due to the presence of on-shell propagators.

Subtracting the SCET matrix elements from the full theory results, we find that the result is given by the tree level matrix element of the two operators
\begin{align}\label{eq:matched_twogluon_samesector}
\cO_{\cB\cB1}^{(2)\mu}&=\frac{-g^2 }{ \omega_2 (\omega_1+\omega_3)}    \bar \chi_{n,\omega_1} \Sl{\cB}_{n\perp,\omega_3} \frac{\Sl{\bar n}}{2}    \Gamma   \frac{\Sl{n}}{2}  \Sl{\cB}_{n\perp,\omega_4} \chi_{\bar n,-\omega_2}\,, \nn \\
\cO_{\cB\cB2}^{(2)\mu}&=\frac{-g^2}{\omega_2(\omega_3+\omega_4)}   \bar \chi_{n,\omega_1}  \Gamma  \Sl{\cB}_{n\perp, \omega_3}  \Sl{\cB}_{n\perp, \omega_4}\chi_{\bar n,-\omega_2}\,.
\end{align}
The behavior of these Wilson coefficients is interesting, in that they exhibit a singularities as a pair of collinear particles in the $n$ direction simultaneously have their energy approach zero.

\vspace{0.2cm} \noindent 
\underline{
$\boldsymbol{(q)_{\bar n} (\bar qgg)_{n}:}$
}

To obtain the matching to the operators involving two collinear gluons and an antiquark in the same collinear sector, we can simply apply charge conjugation to the operators in \Eq{eq:matched_twogluon_samesector} and then relabel to obtain
\begin{align}\label{eq:matched_twogluon_samesector_conj}
\cO_{\cB\cB1'}^{(2)\mu}&=\frac{-g^2 }{ \omega_1 (\omega_2+\omega_4)}    \bar \chi_{\bn,\omega_1} \Sl{\cB}_{n\perp,\omega_3} \frac{\Sl{\bar n}}{2}    \Gamma   \frac{\Sl{n}}{2}  \Sl{\cB}_{n\perp,\omega_4} \chi_{n,-\omega_2}\,, \nn \\
\cO_{\cB\cB2'}^{(2)\mu}&=\frac{-g^2}{\omega_1(\omega_3+\omega_4)}   \bar \chi_{\bn,\omega_1}    \Sl{\cB}_{n\perp, \omega_3}  \Sl{\cB}_{n\perp, \omega_4} \Gamma\chi_{n,-\omega_2}\,.
\end{align}

\vspace{0.2cm} \noindent 
\underline{
$\boldsymbol{(\bar qg)_{\bar n} ( qg)_{n}:}$
}

A basis of $\mathcal{O}(\lambda^2)$ operators with two collinear gluons in the opposite sectors was given in \Eq{eq:eeqqgg_basis2}. These operators get their power suppression from the explicit collinear gluon fields, and to simplify the matching we decompose the particle momenta as
\begin{align}
p_1^\mu&=\frac{\omega_1}{2}n^\mu+p_{1\perp}+\frac{p_{1r}}{2}\bar n^\mu \,, \qquad
p_2^\mu=\frac{\omega_2}{2}\bar n^\mu+p_{2\perp} +\frac{p_{2r}}{2}n^\mu\,, \nn \\
p_3^\mu&=\frac{\omega_3}{2}\bar n^\mu-p_{2\perp} +\frac{p_{3r}}{2}n^\mu\,, \qquad
p_4^\mu=\frac{\omega_4}{2} n^\mu-p_{1\perp}+\frac{p_{4r}}{2}\bar n^\mu\,, 
\end{align}
where the particle labeling is the same as before, but with $p_4$ labeling the additional gluon. Furthermore, we choose the gluon polarizations to be perp in the matching. This choice of momenta also removes any contributions from subleading Lagrangian insertions. However, there are also SCET contributions coming from the $\cO(\lambda^2)$ operators $\mathcal{O}^{(2)\mu}_{\cP \bar n1,2}$, and $\mathcal{O}^{(2)\mu}_{\cP  n1,2}$. The four contributions from these operators are given by
\begin{align}
\fd{3cm}{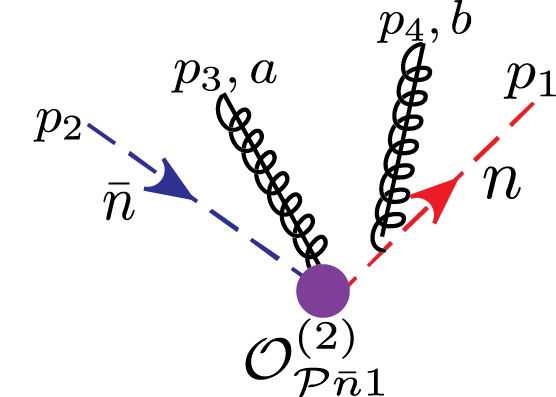}
\hspace{0.4cm}
&=- \frac{g^2\, \bar u_n (1) T^b T^a \Sl{p}_{1\perp} \Sl{\epsilon}^*_{4\perp}  \Sl{\epsilon}^*_{3\perp}  \Sl{p}_{2\perp}   \Gamma   v_{\bar n}(2)}{\omega_1 \omega_3(\omega_1+\omega_4)(p_{1r}+p_{4r})} 
\,,   \nn \\
\fd{3cm}{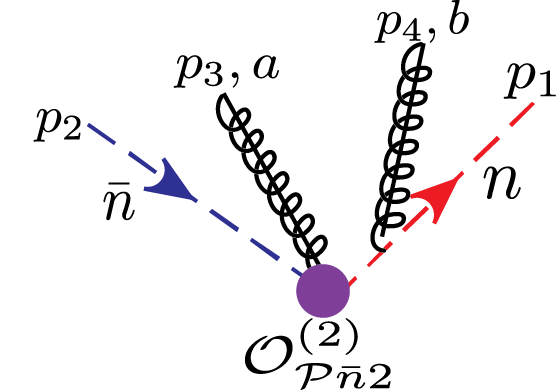}
\hspace{0.4cm}
&= - \frac{g^2\, \bar u_n (1) T^b T^a \Sl{p}_{1\perp} \Sl{\epsilon}^*_{4\perp} \Sl{\epsilon}^*_{3\perp} \frac{\Sl{\bar n}}{2}  \Gamma \frac{\Sl{n}}{2} \Sl{p}_{2\perp}  v_{\bar n}(2)}{\omega_1 \omega_2(\omega_1+\omega_4)(p_{1r}+p_{4r})} 
\,, \nn \\
\fd{3cm}{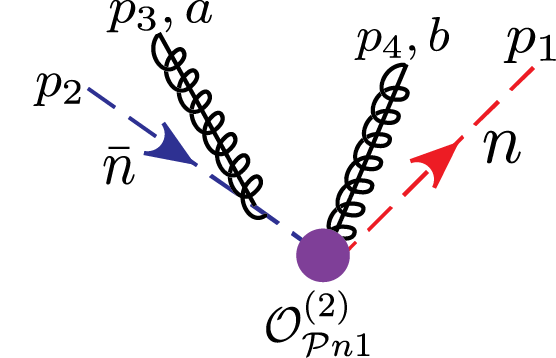} 
\hspace{0.4cm}
&= - \frac{g^2\, \bar u_n (1) T^b T^a \Sl{p}_{1\perp} \frac{\Sl{\bar n}}{2} \Gamma \frac{\Sl{n}}{2} \Sl{\epsilon}^*_{4\perp}\Sl{\epsilon}^*_{3\perp}\Sl{p}_{2\perp}   v_{\bar n}(2)}{(\omega_2+\omega_3)\omega_1 \omega_2 (p_{2r}+p_{3r})} 
\,, \nn \\
\fd{3cm}{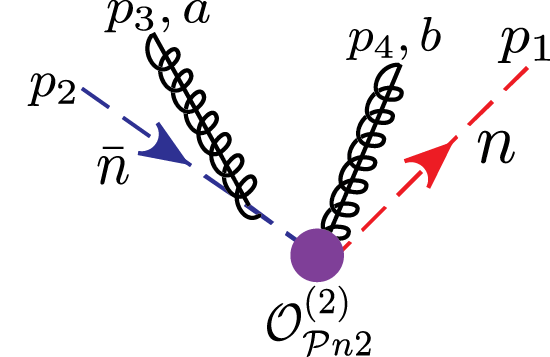}
\hspace{0.4cm}
&=- \frac{g^2\, \bar u_n (1)  T^b T^a   \Gamma  \Sl{p}_{1\perp} \Sl{\epsilon}^*_{4\perp}\Sl{\epsilon}^*_{3\perp} \Sl{p}_{2\perp}   v_{\bar n}(2)}{(\omega_2+\omega_3)\omega_2 \omega_4 (p_{2r}+p_{3r})}
\,.
\end{align}
Note that we do not have to consider the operators involving ultrasoft derivatives from \Eqs{eq:soft_derivative_basis}{eq:eedggus} in the matching, since we have chosen to only have a residual momentum for the small momentum component, and we have shown by the RPI relation of \Eq{eq:uspartialRPIrelation} combined with the matching to the leading power operator in \Eq{eq:LP_wilson} that this component of the derivative first appears at $\cO(\alpha_s)$.

Since the matrix elements in full QCD are identical to the case when the gluons are in different collinear sectors, as was just considered, here we just give the results of the diagrams evaluated to $\mathcal{O}(\lambda^2)$. Unlike the previous case where both gluons were in the same collinear sector, in this case all possible diagrams contribute at $\mathcal{O}(\lambda^2)$. Using also the direction of the gluon to indicate whether it is taken to have $n$-collinear or $\bar n$-collinear momentum, we find
\begin{align}\label{eq:qqggmatch_alldiagrams}
&\fd{3cm}{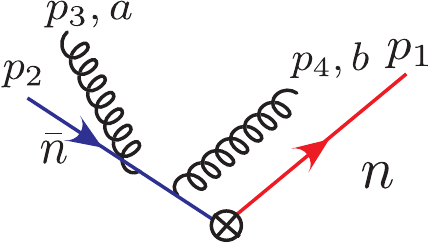} \left. \vphantom{\fd{3cm}{figures/matching_2q2g_diagram1_opp_low} }    \right |_{\mathcal{O}(\lambda^2)}  
 = -\frac{g^2\,  \bar u_n (1) T^b T^a  \Gamma  \Sl{\epsilon}^*_{4\perp}   \Sl{\epsilon}^*_{3\perp}    v_{\bar n}(2) }{(\omega_2+\omega_3)\omega_4 }   
\nn \\
&\hspace{4.5cm}
- \frac{g^2\,\bar u_n (1)  T^b T^a \Gamma  \Sl{p}_{1\perp} \Sl{\epsilon}^*_{4\perp}\Sl{\epsilon}^*_{3\perp} \Sl{p}_{2\perp}   v_{\bar n}(2) }{(\omega_2+\omega_3)\omega_2 \omega_4 (p_{2r}+p_{3r})}  \nn \\
&\hspace{4.5cm}
- \frac{g^2\,\bar u_n (1) T^b T^a \Sl{p}_{1\perp}\frac{\Sl{\bar n}}{2}  \Gamma \frac{\Sl{n}}{2} \Sl{\epsilon}^*_{4\perp}\Sl{\epsilon}^*_{3\perp}\Sl{p}_{2\perp}   v_{\bar n}(2) }{(\omega_2+\omega_3)\omega_1 \omega_2 (p_{2r}+p_{3r})} 
 \,,\nn \\
&\fd{3cm}{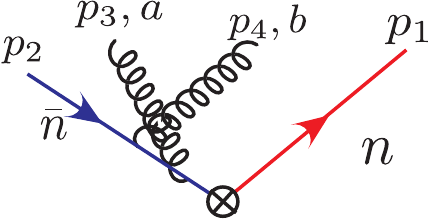}   \left. \vphantom{\fd{3cm}{figures/matching_2q2g_diagram2_opp_low}}     \right |_{\mathcal{O}(\lambda^2)}   
=  -\frac{g^2\, \bar u_n (1) T^a T^b  \Gamma  \Sl{\epsilon}^*_{3\perp}   \Sl{\epsilon}^*_{4\perp}    v_{\bar n}(2)}{\omega_2\omega_4 }   
 \,, 
\nn \\
&\fd{3cm}{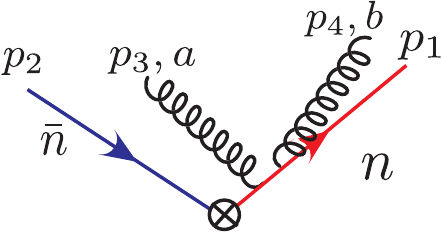}   \left. \vphantom{\fd{3cm}{figures/matching_2q2g_diagram5_opp_low}}     \right|_{\mathcal{O}(\lambda^2)}   
=-\frac{g^2\,\bar u_n (1)  T^b T^a  \Sl{\epsilon}^*_{4\perp}   \Sl{\epsilon}^*_{3\perp}  \Gamma   v_{\bar n}(2)  }{(\omega_1+\omega_4)\omega_3 }    
 \nn\\
&\hspace{4.5cm}
-\frac{g^2\, \bar u_n (1) T^b T^a\Sl{p}_{1\perp} \Sl{\epsilon}^*_{4\perp} \Sl{\epsilon}^*_{3\perp} \frac{\Sl{\bar n}}{2} \Gamma \frac{\Sl{n}}{2}  \Sl{p}_{2\perp} v_{\bar n}(2)}{\omega_1 \omega_2(\omega_1+\omega_4)(p_{1r}+p_{4r})} 
\nn \\
&\hspace{4.5cm}
- \frac{g^2\, \bar u_n (1) T^b T^a\Sl{p}_{1\perp} \Sl{\epsilon}^*_{4\perp}  \Sl{\epsilon}^*_{3\perp} \Sl{p}_{2\perp}    \Gamma   v_{\bar n}(2)}{\omega_1 \omega_3(\omega_1+\omega_4)(p_{1r}+p_{4r})} 
\,, \nn \\
&\fd{3cm}{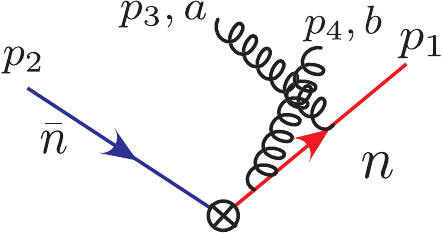}   \left. \vphantom{\fd{3cm}{figures/matching_2q2g_diagram6_opp_low}}       \right|_{\mathcal{O}(\lambda^2)}  
 =-\frac{g^2\,\bar u_n (1)  T^a T^b  \Sl{\epsilon}^*_{3\perp}   \Sl{\epsilon}^*_{4\perp}  \Gamma   v_{\bar n}(2) }{\omega_1\omega_3 }    \,, \nn\\
&\fd{3cm}{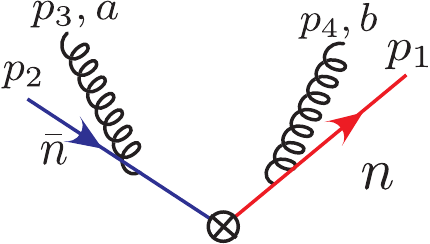}   \left. \vphantom{\fd{3cm}{figures/matching_2q2g_diagram3_opp_low}}         \right|_{\mathcal{O}(\lambda^2)}   
= -\frac{g^2\,\bar u_n (1)  T^b T^a  \Sl{\epsilon}^*_{4\perp}  \frac{\Sl{\bar n}}{2} \Gamma \frac{\Sl{n}}{2}  \Sl{\epsilon}^*_{3\perp}    v_{\bar n}(2)  }{(\omega_2+\omega_3)(\omega_1+\omega_4) }   
  \,, \nn\\
&\fd{3cm}{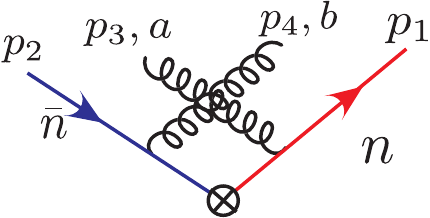}   \left. \vphantom{\fd{3cm}{figures/matching_2q2g_diagram4_opp_low}}       \right|_{\mathcal{O}(\lambda^2)}  
 = \frac{-g^2\,\bar u_n (1) T^a T^b\Sl{\epsilon}^*_{3\perp} \frac{\Sl{\bar n}}{2} \Gamma \frac{\Sl{n}}{2}  \Sl{\epsilon}^*_{4\perp}    v_{\bar n}(2) }{\omega_1 \omega_2}  
 \,, \nn \\
&\fd{3cm}{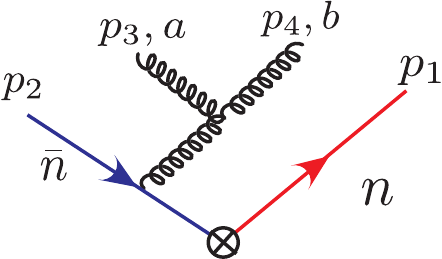}   \left. \vphantom{\fd{3cm}{figures/matching_2q2g_diagram7_opp_low}}       \right|_{\mathcal{O}(\lambda^2)}  
 =-\frac{ig^2 f^{abc} \epsilon^*_{3\perp} \cdot \epsilon^*_{4\perp}\,
  \bar u_n (1) T^c   \Gamma  v_{\bar n}(2) }{\omega_3 \omega_4}    \,, \nn\\
&\fd{3cm}{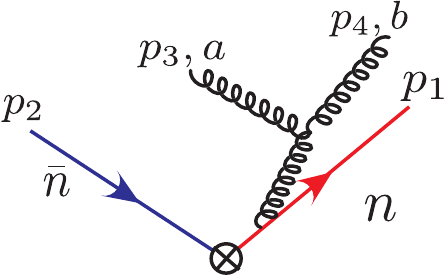}    \left. \vphantom{\fd{3cm}{figures/matching_2q2g_diagram8_opp_low}}        \right|_{\mathcal{O}(\lambda^2)}  
 =-\frac{ig^2 f^{abc}\epsilon^*_{3\perp} \cdot \epsilon^*_{4\perp} \,
  \bar u_n (1) T^c   \Gamma  v_{\bar n}(2) }{\omega_3 \omega_4}    \,.
\end{align}

The sum of all these contributions must be reproduced by hard scatting operators in SCET, after the non-local terms, which are easily recognizable, have been subtracted. The simplest approach is just to associate to each of the above diagrams an operator with a different Lorentz and color structures. This makes the structure the most transparent, and simplifies the projection to helicity operators. It also allows us to easily treat a completely general Dirac structure, $ \Gamma$, inserted at the vertex. We can therefore write the operators generated by the tree level matching as
\begin{align}\label{eq:matched_twogluon_diffsector}
\cO^{(2)\mu}_{\cB\bar\cB1}
 &= \frac{-g^2}{(\omega_2+\omega_3) \omega_4}  \,
  \bar \chi_{n,\omega_1}  \Gamma \Sl{\cB}_{\perp n, \omega_4}  \Sl{\cB}_{\perp \bar n, \omega_3} \chi_{\bar n, -\omega_2}
\,, \quad
\cO^{(2)\mu}_{\cB\bar\cB2}
 = \frac{-g^2}{\omega_2 \omega_4} \,
  \bar \chi_{n,\omega_1}  \Gamma   \Sl{\cB}_{\perp \bar n, \omega_3}  \Sl{\cB}_{\perp n, \omega_4}  \chi_{\bar n, -\omega_2}
\,, \nn \\
\cO^{(2)\mu}_{\cB\bar\cB3}
  &= \frac{-g^2}{(\omega_1+\omega_4) \omega_3}  \,
 \bar \chi_{n,\omega_1}  \Sl{\cB}_{\perp n, \omega_4}  \Sl{\cB}_{\perp \bar n, \omega_3}  \Gamma    \chi_{\bar n, -\omega_2}
\,, \quad
\cO^{(2)\mu}_{\cB\bar\cB4}
   = \frac{-g^2}{\omega_1 \omega_3} \,
 \bar \chi_{n,\omega_1}    \Sl{\cB}_{\perp \bar n, \omega_3}  
 \Sl{\cB}_{\perp n, \omega_4}  \Gamma    \chi_{\bar n, -\omega_2}
\,, \nn \\
\cO^{(2)\mu}_{\cB\bar\cB5}
  &= \frac{-g^2}{(\omega_2+\omega_3)(\omega_1+\omega_4)}  \,
  \bar \chi_{n,\omega_1}  \Sl{\cB}_{\perp n, \omega_4}  \frac{\Sl{\bar n}}{2}  \Gamma \frac{\Sl{n}}{2}  \Sl{\cB}_{\perp \bar n, \omega_3}    \chi_{\bar n, -\omega_2}
\,, \nn \\
\cO^{(2)\mu}_{\cB\bar\cB6}
  &=\frac{-g^2}{\omega_1 \omega_2} \,
  \bar \chi_{n,\omega_1}     \Sl{\cB}_{\perp \bar n, \omega_3} \frac{\Sl{\bar n}}{2}  \Gamma \frac{\Sl{n}}{2}  \Sl{\cB}_{\perp n, \omega_4}  \chi_{\bar n, -\omega_2}
\,,\nn \\
\cO^{(2)\mu}_{\cB\bar\cB7}
 &= \frac{-2g^2}{\omega_3 \omega_4}  \,
  \bar \chi_{n,\omega_1}   \bigl(  \cB_{\perp \bar n, \omega_3} \cdot  \cB_{\perp n, \omega_4} -  \cB_{\perp n, \omega_4}  \cdot    \cB_{\perp \bar n, \omega_3}  \bigr)    \Gamma   \chi_{\bar n, -\omega_2}  
 \,.
\end{align}
A number of these operators exhibit singularities in the Wilson coefficients as a single one of the gluons or quarks becomes soft. This behavior is distinct from that of the Wilson coefficients of the operators involving two gluons in a single collinear sector, given in \Eq{eq:matched_twogluon_samesector}, which only have a singularity when a pair of particles becomes soft. Of particular interest, are the operators $\cO^{(2)\mu}_{\cB\bar\cB2}$, $\cO^{(2)\mu}_{\cB\bar\cB4}$, $\cO^{(2)\mu}_{\cB\bar\cB6}$, $\cO^{(2)\mu}_{\cB\bar\cB7}$, which have a separate divergence in each collinear sector.

In \Sec{sec:proj_hel_nnlp} these operators will be projected to a helicity basis, which simplifies their structure.

\subsection{Subleading Matching with Four Collinear Quarks}\label{sec:matching_NNLP_4q}

Finally, we consider the matching to the operators involving four collinear quark fields. As discussed in \Sec{sec:eeJets}, the only relevant such operators for the cross section at $\mathcal{O}(\lambda^2)$ involve a single collinear quark operator in one sector, and three collinear quark operators in the other sector. A basis of such operators was given in \Eq{eq:Z3_basis_qQ}.
Since the operators get their power suppression from the explicit collinear quark fields, we can simplify the matching by taking the particle momenta as
\begin{align}
p_1^\mu&=\frac{\omega_1}{2} n^\mu \,, \qquad
p_2^\mu=\frac{\omega_2}{2}\bar n^\mu\,, \nn \\
p_3^\mu&=\frac{\omega_3}{2}\bar n^\mu+p_\perp^\mu+\frac{p_{3r}}{2} n^\mu\,, \qquad
p_4^\mu=\frac{\omega_4}{2}\bar n^\mu -p_\perp^\mu+\frac{p_{4r}}{2}n^\mu\,.
\end{align}
The momentum labeling corresponds to that shown in the figure in \Eq{eq:qqqqmatch_1}. This choice of momentum also removes contributions from subleading Lagrangian insertions, as well as from the $\cO(\lambda^2)$ single gluon operators
\begin{align}
\fd{3cm}{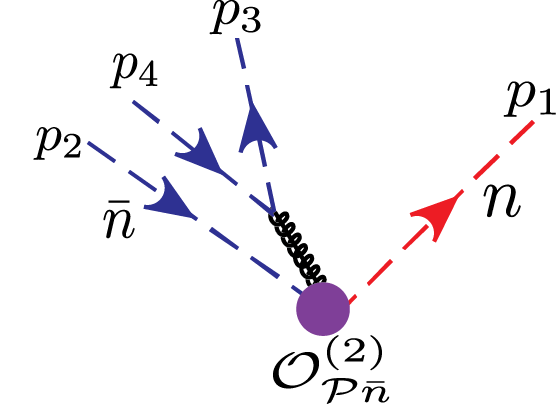}=0\,.
\end{align}
The QCD diagrams at $\mathcal{O}(\lambda^2)$ must therefore be exactly reproduced by the hard scattering operators. As before, the matching coefficients can be calculated by expanding the QCD diagrams. In the case where a gluon splits into a quark anti-quark pair, with one in each sector, we find
\begin{align}\label{eq:qqqqmatch_1}
\fd{3cm}{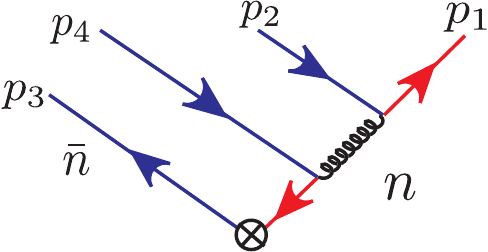} &= \bar{u}(p_1) (i g T^a \gamma_\nu) v(p_2) \frac{-i}{(p_1+p_2)^2} \bar{u}(p_3)  \Gamma \frac{-i (\Sl{p}_1 + \Sl{p}_2 + \Sl{p}_4)}{(p_1 + p_2 + p_4)^2} (ig T^a \gamma^\nu) v(p_2) \,, \nn \\
%
\fd{3cm}{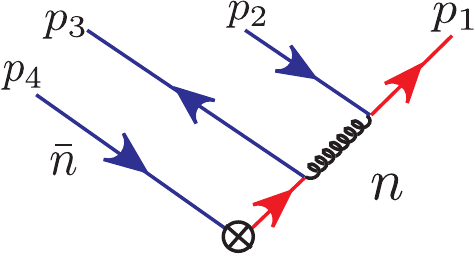}    &=  \bar{u}(p_1) (i g T^a \gamma_\nu) v(p_2) \frac{-i}{(p_1+p_2)^2} \bar{u}(p_3) (ig T^a \gamma^\nu) \frac{i (\Sl{p}_1 + \Sl{p}_2 + \Sl{p}_3)}{(p_1 + p_2 + p_3)^2}  \Gamma  v(p_2) \,.
\end{align}
The $\mathcal{O}(\lambda^2)$ contribution of these diagrams is given by
\begin{align}
\fd{3cm}{figures/matching_2q2Q_diagram1_low} \left. \vphantom{\fd{3cm}{figures/matching_2q2Q_diagram1_low}}\right |_{\mathcal{O}(\lambda^2)}&= \frac{g^2}{\omega_1\omega_2 (\omega_2 + \omega_4) } \bar{u}_{n} (1) T^a \gamma_\nu v_{\bar n} (2) \bar{u}_{\bar n} (3)  \Gamma  \frac{\Sl {n}}{2}  \gamma^\nu T^a v_{\bar n}(4) \,, \nn\\
\fd{3cm}{figures/matching_2q2Q_diagram1a_low} \left. \vphantom{\fd{3cm}{figures/matching_2q2Q_diagram1a_low}   }\right |_{\mathcal{O}(\lambda^2)}&=-\frac{g^2}{\omega_1\omega_2 (\omega_2 + \omega_3) } \bar{u}_{n} (1) T^a \gamma_\nu v_{\bar n} (2) \bar{u}_{\bar n} (3) \gamma^\nu T^a   \frac{\Sl {n}}{2}   \Gamma  v_{\bar n}(4) \,.
\end{align}
We could also have contributions from a gluon emitted from the $n$ collinear quark, which splits into two quarks, both in the $\bar n$ collinear sector. The full theory diagram is given by
\begin{align}
\fd{3cm}{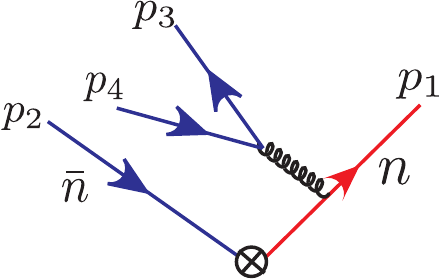}=\bar u (p_3) (igT^a) \gamma^\rho v(p_4)  \bar u(p_1) (igT^a) \gamma^\rho i \frac{\Sl{p}_1+\Sl{p}_3+\Sl{p}_4}{(p_1+p_3+p_4)^2}  \Gamma v(p_2) \frac{(-i)}{(p_3+p_4)^2}\,.
\end{align}
Expanding this, we find that its $\cO(\lambda^2)$ term vanishes
\begin{align}
\left.\fd{3cm}{figures/matching_2q2Q_diagram2_low}\right |_{\mathcal{O}(\lambda^2)}&= 0\,.
\end{align}
Finally, there is a contribution from a gluon emitted from the $\bar n$ collinear quark, which splits into two quarks. The full theory diagram is given by
\begin{align}
\fd{3cm}{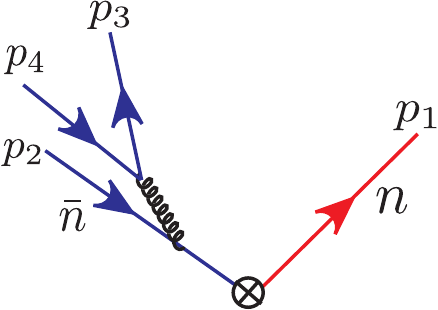}=\bar u (p_3) (igT^a) \gamma^\rho v(p_4)  \bar u(p_1)  \Gamma    (-i) \frac{\Sl{p}_2+\Sl{p}_3+\Sl{p}_4}{(p_2+p_3+p_4)^2} (igT^a) \gamma^\rho  v(p_2) \frac{(-i)}{(p_3+p_4)^2}\,.
\end{align}
Expanding this to $\cO(\lambda^2)$, we find that its contribution at this order also vanishes
\begin{align}
\left.\fd{3cm}{figures/matching_2q2Q_diagram3_low}\right |_{\mathcal{O}(\lambda^2)}&=0\,.
\end{align}
We therefore find that the QCD result is reproduced by two SCET operators
\begin{align}\label{eq:matched_4qn}
\cO^{(2)\Gamma}_{4\chi1a}
  &=\frac{g^2}{\omega_1 \omega_2 (\omega_2 + \omega_4)} \, 
\Big[ \bar \chi_{n, \omega_1} T^a \gamma_\nu \chi_{\bar n,-\omega_2}\Big]
\Big[ \bar \chi_{\bar n, \omega_3}  \Gamma  \frac{\Sl n}{2}  \gamma^\nu T^a \chi_{\bar n,-\omega_4} \Big]
 \,, \nn \\
\cO^{(2)\Gamma}_{4\chi1b}
 &=-\frac{g^2}{\omega_1 \omega_2 (\omega_2 + \omega_3)}\, \Big[ \bar \chi_{n, \omega_1} T^a \gamma_\nu \chi_{\bar n,-\omega_2}\Big]
\Big[ \bar \chi_{\bar n, \omega_3} \gamma^\nu T^a  \frac{\Sl n}{2}    \Gamma \chi_{\bar n,-\omega_4} \Big]
\,.
\end{align}
Similarly, for the case where there is an antiquark in the $n$ collinear sector instead of a quark, the SCET operators are given by
\begin{align}\label{eq:matched_4qbarn}
\cO^{(2)\Gamma}_{4\bar \chi1a}
  &=\frac{g^2}{\omega_1 \omega_2 (\omega_2 + \omega_4) } \, 
\Big[ \bar \chi_{\bar n, \omega_2} T^a \gamma_\nu \chi_{n,-\omega_1}\Big]
\Big[ \bar \chi_{\bar n, \omega_3}  \Gamma  \frac{\Sl n}{2}  \gamma^\nu T^a \chi_{\bar n,-\omega_4} \Big]
\,, \nn \\
\cO^{(2)\Gamma}_{4\bar \chi1b}
  &=-\frac{g^2}{\omega_1 \omega_2 (\omega_2 + \omega_3) } \,
\Big[ \bar \chi_{\bar n, \omega_2} T^a \gamma_\nu \chi_{n,-\omega_1} \Big]  \Big[ \bar \chi_{\bar n, \omega_3} \gamma^\nu T^a \frac{\Sl n}{2} \Gamma   \chi_{\bar n,-\omega_4} \Big] 
\,.
\end{align}
Both operators exhibit singularities as certain quarks go soft.

\subsection{Subleading Matching with a Single Ultrasoft Gluon}\label{sec:match_soft}

In this section we consider the matching with a the emission of a single ultrasoft gluon, which allows us to probe the operators involving ultrasoft insertions. These operators are determined by RPI, so the determination of their Wilson coefficients does not require a new calculation, nevertheless it is instructive to see how this matching works.\footnote{Note that we have also shown that they do not contribute to the dijet cross section at $\mathcal{O}(\alpha_s \lambda^2)$.} For the corresponding tree level matching computation, the necessary SCET diagram calculations were carried out explicitly in Ref.~\cite{Larkoski:2014bxa} for an arbitrary current. For completeness we give the full theory and SCET results needed for the matching calculation in the case studied here. This illustrates how contributions from the SCET Lagrangian insertions are treated in the matching, as discussed in general in \Sec{sec:matching_general}, which can not be avoided in this case.  We carry out this matching prior to making the BPS field redefinition.

To perform the matching calculation, we route the residual momenta such that it is only carried by the external ultrasoft particle,  any intermediate collinear fields, and then out through the hard current. We take
\begin{align}
p_1 =\frac{\omega_1}{2} n^\mu\,, \qquad p_2 =\frac{\omega_2}{2} \bar n^\mu\,, \qquad p_s=\frac{n \cdot p_s}{2}\bar n^\mu+p_{s\perp} +\frac{\bar n \cdot p_s}{2}n^\mu\,,
\end{align}
where the momentum $p_s$ is ultrasoft and hence purely residual.
Expanding the QCD diagrams, we find 
\begin{align}
\fd{3cm}{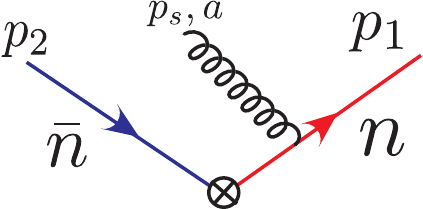} 
&=\frac{-g \epsilon_\mu^* \bar u_n(1) T^a 
\Bigl[ (\omega_1 \!+\! \bn\cdot p_s) n^\mu + \gamma_\perp^\mu \Sl{p}_{s\perp}
  \! + (\gamma_\perp^\mu n\cdot p_s \!-\! \Sl{p}_{s\perp}\! n^\mu)
  \frac{\bnslash}{2}
 \Bigr] \Gamma v_{\bar n}^*(2)
  }{\omega_1\, n\cdot p_s}
 ,\nn\\
\fd{3cm}{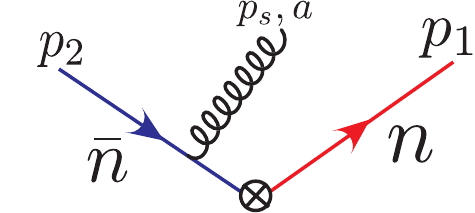}  
 &= \frac{g \epsilon_\mu^* \bar u_n(1) T^a \Gamma 
 \Bigl[ (\omega_2 \!+\! n\cdot p_s)\bn^\mu + \Sl{p}_{s\perp}\gamma_\perp^\mu
 \!+ \frac{\nslash}{2} (\gamma_\perp^\mu \bn\cdot p_s 
    \!-\! \Sl{p}_{s\perp}\! \bn^\mu) 
  \Bigr] v_{\bar n}^*(2)
  }{\omega_2\, \bn\cdot p_s} .
\end{align}
Next we drop the standard eikonal ${\cal O}(\lambda^0)$ terms, and using the equations of motion $n\cdot p_s \bn\cdot p_s = -p_{s\perp}^2$ to write the result in a form which is clearly gauge invariant graph by graph for the ${\cal O}(\lambda^2)$ terms.  This gives
\begin{align}  \label{eq:fullsoft1gluonmatch}
\fd{3cm}{figures/matching_subleadingvertex_2_soft_low} \left. \vphantom{\fd{3cm}{figures/matching_subleadingvertex_2_soft_low}}\right |_{\mathcal{O}(\lambda^2)}
& \!\! = \frac{-g\epsilon_\mu^*  \bar u_n(1) T^a
   }{\omega_1\, n\cdot p_s}
  \biggl\{
  \frac{p_{s\perp}^\mu n\mcdot p_s \!-\! p_{s\perp}^2\! n^\mu}{n\mcdot p_s}
  + \frac{[\gamma_\perp^\mu,\Sl{p}_{s\perp}]}{2}
  + \frac{(\gamma_\perp^\mu n\mcdot p_s \!-\! \Sl{p}_{s\perp}\! n^\mu)\bnslash}{2}
  \biggr\} 
 \nn\\ 
& \quad \times \Gamma v_{\bar n}^*(2)  
 \,, \nn\\[5pt]
%
\fd{3cm}{figures/matching_subleadingvertex_1_soft_low} \left. \vphantom{\fd{3cm}{figures/matching_subleadingvertex_1_soft_low}}\right |_{\mathcal{O}(\lambda^2)}
& \!\! = \frac{g\epsilon_\mu^*  \bar u_n(1) \Gamma
   }{\omega_2\, \bn\cdot p_s}
  \biggl\{
  \frac{p_{s\perp}^\mu \bn\mcdot p_s \!-\! p_{s\perp}^2\! \bn^\mu}
   {\bn\mcdot p_s}
  - \frac{[\gamma_\perp^\mu,\Sl{p}_{s\perp}]}{2}
  + \frac{\nslash(\gamma_\perp^\mu \bn\mcdot p_s \!-\! \Sl{p}_{s\perp}\! \bn^\mu)}{2}
  \biggr\}
 \nn\\ 
& \quad \times T^a v_{\bar n}^*(2)  
\,.
\end{align}
These results are written in the same form given in Ref.~\cite{Larkoski:2014bxa}, which is also the form predicted by the LBK relation~\cite{Low:1958sn,Burnett:1967km}.

Unlike our earlier analysis of the purely collinear graphs, with a ultrasoft particle present there are residual momenta in collinear propagators.  Therefore, in the matching with \Eq{eq:lamp_match} there are non-vanishing contributions from insertions of the subleading power SCET Lagrangians. Insertions of $\cL^{(1)}$ vanish as they involve label $\perp$ momentum, which we chose to be zero. The Feynman rules for the corrections from $\cL^{(2)}$ to an $n$-collinear quark propagator, and to the emission of a ultrasoft gluon are given by
\begin{align}
&\fd{3cm}{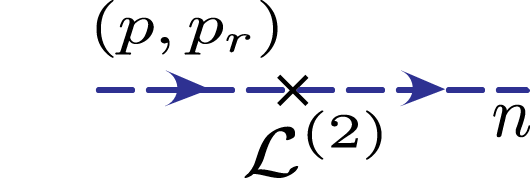}
  \: =i \frac{\Sl{\bar n}}{2}\frac{p_{r\perp}^2}{\bar n\cdot p}
\,,
& 
\raisebox{0.3cm}{\fd{2.7cm}{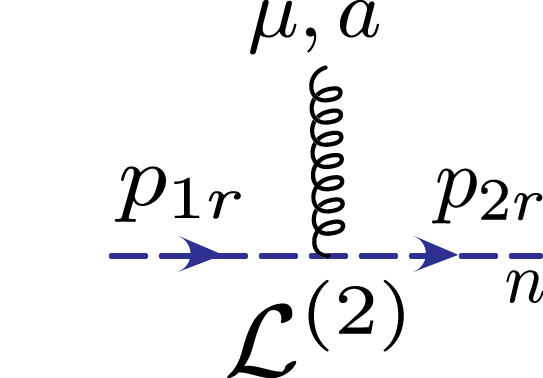}}
& \: =igT^a   \biggl(   \frac{\gamma^\mu_\perp\Sl{p}_{1r\perp}}{\bar n \cdot p}   +   \frac{\Sl{p}_{2r\perp}\gamma^\mu_\perp}{\bar n \cdot p}        \biggr) \frac{\bnslash}{2}.
\end{align}
Computing the SCET diagrams that contribute to the matrix element, we find that the propagator insertions of $\cL^{(2)}$ and corrections to the gluon emission from $\cL^{(2)}$ give
\begin{align}
& \fd{3cm}{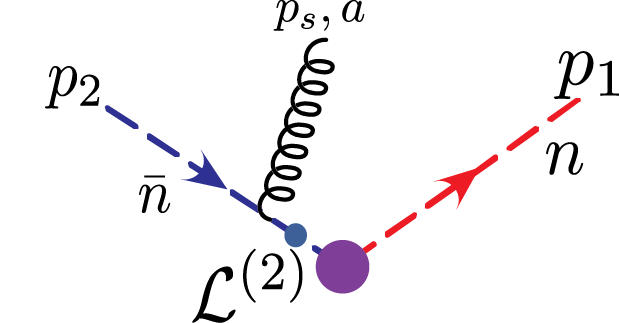} +\fd{3cm}{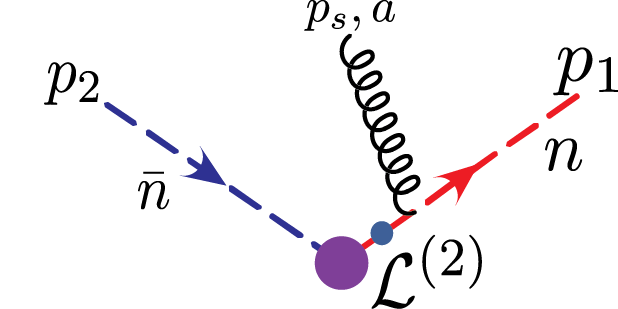} 
=g\biggl( 
 -\frac{n \cdot \epsilon^*\, \bar n \cdot p_s}{\omega_1\,n\cdot p_s}    
 +\frac{\bar n \cdot \epsilon^*\, n \cdot p_s}{\omega_2\,\bar n\cdot p_s}
 \biggr) \bar u_n (1) T^a  \Gamma  v_{\bar n}^* (2)
\,, \nn\\
& \fd{3cm}{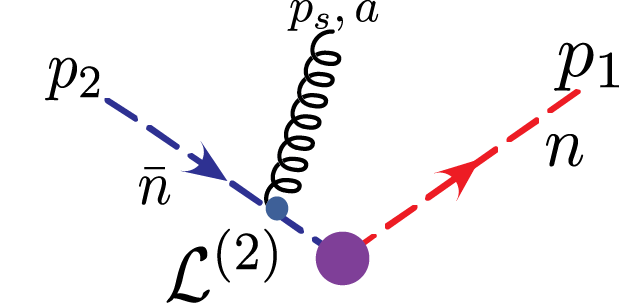} +\fd{3cm}{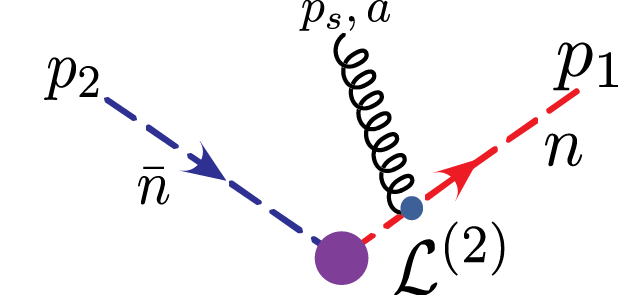} 
=g\, \bar u_n (1) T^a \biggl( 
  - \frac{\Sl{\epsilon}^*_\perp \Sl{p}_{s\perp} \Gamma}
      {\omega_1\,n\cdot p_s}           
  + \frac{\Gamma \Sl{p}_{s\perp} \Sl{\epsilon}^*_\perp}
      {\omega_2\,  \bn \cdot p_s}   
 \biggr) v_{\bar n}^* (2) 
\,,\nn\\
& \text{Sum} 
  = \frac{ g \bar u_n(1) T^a \Gamma}{\omega_2\, \bn\cdot p_s}
   \biggl\{ \frac{p_{s\perp}^\mu \bn\mcdot p_s \!-\! p_{s\perp}^2\! \bn^\mu}
   {\bn\mcdot p_s} - \frac{[\gamma_\perp^\mu,\Sl{p}_{s\perp}]}{2} 
   \biggr\} v_\bn^*(2) 
\nn\\
&\qquad\ 
  - \frac{ g \bar u_n(1) T^a}{\omega_1\, n\cdot p_s}
   \biggl\{ \frac{p_{s\perp}^\mu n\mcdot p_s \!-\! p_{s\perp}^2\! n^\mu}{n\mcdot p_s}  + \frac{[\gamma_\perp^\mu,\Sl{p}_{s\perp}]}{2}
   \biggr\} \Gamma v_\bn^*(2)
  \,.
\end{align}
Note that the sum is gauge invariant for each leg.  Subtracting off these SCET matrix elements leaves only the third term in each line of the full theory results in \eq{fullsoft1gluonmatch}. These terms are reproduced by the SCET operators
\begin{align}\label{eq:soft_current_perp}
\cO^{(2)\Gamma}_{s\perp1}&
  = \bar \chi_{n,\omega_1}  \Gamma\, \frac{i\Sl{D}_{\perp us}}{(-\omega_2)} 
     \frac{\Sl{n}}{2}  \chi_{\bar n, -\omega_2}
\,, 
& \cO^{(2)\Gamma}_{s\perp2}&
  =  \bar \chi_{n,\omega_1}  \frac{\Sl{\bar n}}{2}  
  \frac{(-i \overleftarrow{\Sl{D}}_{\perp us})}{\omega_1} 
  \Gamma \chi_{\bar n, -\omega_2}\,. 
\end{align}
Each of these operators gives two contributions, one where the $D_\perp$ produces a ultrasoft gluon at the vertex, and one where the $D_\perp$ acts as a derivative, and the ultrasoft gluon is produced by an insertion of $\cL^{(0)}$:
\begin{align}
 \Bigl\langle \cO^{(2)\Gamma}_{s\perp1} \Bigr\rangle
  &= g\, \bar u_n(1) \Gamma T^a \frac{\nslash}{2} \biggl( \frac{\Sl{\epsilon}_\perp}{\omega_2}
   -\frac{\Sl{p}_{s\perp}\bn\cdot\epsilon^*}{\omega_2\,\bn\cdot p_s} \biggr)
  v_\bn^*(2)
  \,,\nn\\[5pt]
 \Bigl\langle \cO^{(2)\Gamma}_{s\perp2} \Bigr\rangle
  &= - g\, \bar u_n(1) T^a \biggl( 
   \frac{\Sl{\epsilon}_\perp}{\omega_1}
   -\frac{\Sl{p}_{s\perp} n\cdot\epsilon^*}{\omega_1\, n\cdot p_s} \biggr)
  \frac{\bnslash}{2} \Gamma v_\bn^*(2)
  \,.
\end{align}
The Wilson coefficients of these operators are both fixed by RPI to all orders in perturbation theory, and our tree level matching result agrees with this relation.

We see that only the $D_\perp$, or correspondingly in terms of helicity operators, the $\partial_{\pm}$ and $\cB_{\pm}$, are required to reproduce the tree level QCD result.  Operators of the form $\bar \chi_{\bar n, \omega_1} i n\cdot D_{us}  \Gamma \chi_{n, \omega_2}$ or $\bar \chi_{\bar n, \omega_1} i \bar n\cdot \overleftarrow{D}_{us}  \Gamma \chi_{n, \omega_2}$ are turned into purely collinear operators by the leading power SCET quark equations of motion, and therefore are not included in the basis, as was discussed for the helicity operators in \Sec{sec:soft_basis}.
Prior to the BPS field redefinition, the operators involving the other components of $D^\mu_{us}$, which could appear at the one gluon level are
\begin{align}
\cO_{s\bn}^{(2)\Gamma} &= C_{s\bn}^{(2)}\:
 \bar \chi_{n, \omega_1} (-i \bn\cdot \overleftarrow{D}_{us})
 \Gamma \chi_{\bn, -\omega_2}
  \,, 
&\cO_{sn}^{(2)\Gamma} &= C_{s n}^{(2)}\:
 \bar \chi_{n, \omega_1} \Gamma  i n\cdot D_{us}  
  \chi_{\bn, -\omega_2}
  \,, 
\end{align}
but these two operators do not show up in the tree level matching since their coefficients $C_{sn}^{(2)}$ and $C_{s\bn}^{(2)}$ are ${\cal O}(\alpha_s)$.  As noted in \Sec{sec:soft_basis}, the Wilson coefficients of these two operators are related to derivatives of the leading power operator. Since the Wilson coefficient of the leading power operator is $1$ at tree level, the derivative RPI relation implies that the Wilson coefficient of these operators vanish, which is reproduced by our explicit tree level matching. Using the RPI relations of \Sec{sec:soft_basis}, we can determine the Wilson coefficients of these operators at one-loop. Taking derivatives of the leading power Wilson coefficient of \Eq{eq:LP_wilson} gives
\begin{align}\label{eq:wilson_coeff_soft_mu}
C_{sn}^{(2)}=\frac{\partial C^{(0)}}{d\omega_1} 
 &=-\frac{\alpha_s C_F}{4\pi} \frac{1}{\omega_1}
 \biggl[  -2 \ln \Bigl( \frac{-\omega_1\omega_2-i0}{\mu^2} \Bigr) + 3 \biggr] 
 + {\cal O}(\alpha_s^2)
  \,, \\
C_{s\bar n}^{(2)}=\frac{\partial C^{(0)}}{d\omega_2}
 &=-\frac{\alpha_s C_F}{4\pi} \frac{1}{\omega_2}
  \biggl[ -2 \ln \Bigl( \frac{-\omega_1\omega_2-i0}{\mu^2} \Bigr) +3 \biggr] 
 + {\cal O}(\alpha_s^2)
 \,.  \nn
\end{align}
In the calculation of the cross section, conservation of momentum will enforce $\omega_1=\omega_2$, so that the Wilson coefficients of these two operators are identical. This will hold to all orders in perturbation theory. Combined with the arguments of \Sec{sec:ee_lambda2}, showing that the $\perp$ ultrasoft operators don't contribute to the $e^+e^-\to$ dijets cross section at $\cO(\lambda^2)$, this shows that ultrasoft operators will first contribute at $\cO(\alpha_s^2)$. The $\cO(\alpha_s)$ ultrasoft contributions therefore arise only from the subleading SCET Lagrangian. 

\subsection{Projection onto Helicities}\label{sec:proj_hel}

Given the general result for the Wilson coefficients in SCET, in this section we project the operators onto helicities to show how they appear in the helicity operator basis. For simplicity, we will do this projection for the specific case of a vector current, namely $ \Gamma=\gamma^\mu$, as relevant for $e^+e^-\to $ dijets proceeding through an off shell photon. We then compare with the operators of the helicity basis given in \Sec{sec:eeJets}.

In setting up the decomposition of the Wilson coefficients, as given, for example, for the leading power operator in \Eq{eq:Z0_expand_Wilson}, the couplings of the photon to both the electrons and the quarks has been extracted. To extract the Wilson coefficient and helicity operators from one of the currents, $\cO_j^{(i)\nu}$ we can simply expand the contraction 
\begin{align}\label{eq:contraction_master}
\bar v_e \gamma^\mu u_e  \left(  \frac{-ig_{\mu \nu}}{Q^2} \right) \cO_j^{(i)\nu}\,,
\end{align}
using the completeness relation for polarization vectors
\begin{align}\label{eq:completeness2}
\sum\limits_{\lambda=\pm} \epsilon^\lambda_\mu(n_i,\bar n_i) \left (\epsilon^\lambda_\nu(n_i,\bar n_i) \right )^*=-g_{\mu \nu}+\frac{n_{i\mu} \bar n_{i\nu}+n_{i\nu} \bar n_{i\mu}}{n_i\cdot \bar n_i} = - g^\perp_{\mu \nu} ( n_i, \bn_i)\,.
\end{align}
Here we have taken $s=Q^2$, so that a large light cone momentum $Q$ is deposited in each hemisphere.

The helicity operator basis of \Sec{sec:eeJets} involves helicity operators defined with respect to the axes of the jets, namely $n$ and $\bar n$, as well as with respect to the axes of the incoming electrons, namely $n_e$ and $\bar n_e$. In performing the projection in \Eq{eq:contraction_master}, the non-trivial angular dependence of the Wilson coefficient arises from inner products of the polarization vectors with respect to the axis of the electrons and those defined with respect to the axis of the jets. While we will not in general expand these inner products of polarization vectors, as they allow for an intuitive interpretation of each term, we note that they can straightforwardly be written in terms of spinor products as
\begin{align}\label{eq:polar_innerprod}
\epsilon_+(n,\bar n)\cdot \epsilon_+ (n_e, \bar n_e) &=\frac{[n n_e]\langle \bar n_e \bar n\rangle}{\langle \bar n n \rangle \langle \bar n_e n_e\rangle}\,, \qquad
\epsilon_+(n,\bar n)\cdot \epsilon_- (n_e, \bar n_e) =-\frac{[n \bar n_e]\langle n_e \bar n\rangle}{\langle \bar n n \rangle [ \bar n_e n_e ]}\,, \nn\\
\epsilon_-(n,\bar n)\cdot \epsilon_+ (n_e, \bar n_e) &=-\frac{[\bar n n_e]\langle \bar n_e n \rangle}{[\bar n n] \langle\bar n_e n_e \rangle}\,, \qquad
\epsilon_-(n,\bar n)\cdot \epsilon_- (n_e, \bar n_e) =\frac{[\bar n \bar n_e]\langle n_e n \rangle}{[ \bar n n ] [\bar n_e n_e ]}\,, 
\end{align}
Since these factors will appear whenever two spin $1$ currents are projected onto each other, we will define a shorthand notation
\begin{align}
\epsilon_+(n,\bar n)\cdot \epsilon_+ (n_e, \bar n_e) =\dpp\,,  \qquad   \epsilon_+(n,\bar n)\cdot \epsilon_- (n_e, \bar n_e) = \dpm\,, \nn \\
\epsilon_-(n,\bar n)\cdot \epsilon_+ (n_e, \bar n_e) =\dmp\,,  \qquad   \epsilon_-(n,\bar n)\cdot \epsilon_- (n_e, \bar n_e) = \dmm\,.
\end{align}
Similarly, inner products of the $n$ and $\bar n$ vectors with the electron polarization vectors can be expanded as
\begin{align}
n\cdot \epsilon_+ (n_e, \bar n_e) &= \frac{[n_e n] \langle n \bar n_e \rangle}{\sqrt{2} \langle \bar n_e n_e \rangle  }\,, \qquad    \bar n\cdot \epsilon_+ (n_e, \bar n_e) = \frac{[n_e \bar n] \langle \bar n \bar n_e \rangle}{\sqrt{2} \langle \bar n_e n_e \rangle  }\,, \nn \\
n\cdot \epsilon_- (n_e, \bar n_e) &= -\frac{\langle n_e n\rangle [ n \bar n_e ]}{\sqrt{2} [ \bar n_e n_e ]  }\,, \qquad    \bar n\cdot \epsilon_- (n_e, \bar n_e) = -\frac{\langle n_e \bar n\rangle [ \bar n \bar n_e ]}{\sqrt{2} [ \bar n_e n_e ]  }\,.
\end{align}   
We will occasionally perform this expansion if it simplifies the result.

The spinor products in general give a phase 
\begin{align}
\langle i j \rangle =\sqrt{|s_{ij}|}e^{i\phi_{ij}}\,, \qquad  [ij]=\sqrt{|s_{ij}|}e^{-i(\phi_{ij}+\pi)}\,,
\end{align}
see \App{app:helicity} for details. To slightly simplify the expressions, we will choose to define our spinors with respect to to the jet axis $n$, which we take to be in the $z$ direction, $n^\mu=(1,0,0,1)$. In this case we simplify the spinor products between $n$ and $\bar n$
\begin{align}
[n \bar n]=-\sqrt{2}\,, \qquad \langle n \bar n \rangle=\sqrt{2} \,.
\end{align}
We also note that depending on the collinear sector of the operator, the helicities are defined either with respect to the $n$ or $\bar n$ axes. The relation
\begin{align}
\epsilon^\mu_\pm(n, \bar n)=\epsilon^\mu_\mp(\bar n, n)\,,
\end{align}
allows for the trivial exchange of the corresponding decompositions.

Finally, we comment on the organization of this section. Instead of providing tables of the Wilson coefficients, we find that it is more transparent to show the decomposition of each of the currents generated in the tree level matching onto the helicity operators. The way we have performed the matching, we have typically associated a single Feynman diagram in the matching with an SCET current. Therefore, each of these currents typically projects onto a single helicity operator. The Wilson coefficients for a particular helicity operator can then trivially be read off from the corresponding projection.

\subsubsection{Leading Power}\label{sec:proj_hel_lp}

We begin by considering the projection onto the leading power helicity basis of \Sec{sec:LP}. Since this is the first operator we are considering, we go through the projection onto the helicity basis in slightly more detail. Using the completeness relation for the polarization vectors of \Eq{eq:completeness2},
as well as the projection relations for the SCET spinors, we find
\begin{align}
&\hspace{-1cm}\bar v_e \gamma^\mu u_e  \left(  \frac{-ig_{\mu \nu}}{Q^2} \right) \cO^{(0)\nu}=\nn \\
&\left(  \frac{-i}{Q^2} \right) \epsilon_-(n_e, \bar n_e)\cdot \epsilon_+(n, \bar n)   \left( \epsilon_+^\mu(n_e, \bar n_e) \bar v_e \gamma^\mu u_e \right)   \left( \epsilon_-^\rho (n,\bar n)  \cO^{(0)}_\rho  \right)\nn \\
+&\left(  \frac{-i}{Q^2} \right) \epsilon_+(n_e, \bar n_e)\cdot \epsilon_+(n, \bar n)   \left( \epsilon_-^\mu(n_e, \bar n_e)   \bar v_e \gamma^\mu u_e \right)   \left( \epsilon_-^\rho (n,\bar n)  \cO^{(0)}_\rho  \right)\nn \\
+&\left(  \frac{-i}{Q^2} \right) \epsilon_-(n_e, \bar n_e)\cdot \epsilon_-(n, \bar n)   \left( \epsilon_+^\mu (n_e, \bar n_e)  \bar v_e \gamma^\mu u_e \right)   \left( \epsilon_+^\rho (n,\bar n)  \cO^{(0)}_\rho  \right)\nn \\
+&\left(  \frac{-i}{Q^2} \right) \epsilon_+(n_e, \bar n_e)\cdot \epsilon_-(n, \bar n)   \left( \epsilon_-^\mu (n_e, \bar n_e)  \bar v_e \gamma^\mu u_e \right)   \left( \epsilon_+^\rho (n,\bar n)  \cO^{(0)}_\rho  \right)\,.
\end{align}
Here we see explicitly the four different helicity combinations which arise from the single operator, and that the Wilson coefficient arises as the inner product of the helicity vectors defined with the different axes, as expected form the point of view of spin projection.
Written in terms of the helicity currents, we have
\begin{align}\label{eq:LP_helicity_matched}
\bar v_e \gamma^\mu u_e  \left(  \frac{-ig_{\mu \nu}}{Q^2} \right) \cO^{(0)\nu}&= \delta_{\alpha \bbeta}\left(  \frac{-i}{\sqrt{2}} \right)   \left (\dpm   \left.   [\bar n_e n_e]  J_{\bar e e-} J_{\bar q q +}^{\balpha \beta}     \right. -  \dpp      \left.  \langle \bar n_e n_e\rangle  J_{\bar e e+} J_{\bar q q +}^{\balpha \beta}     \right.  \right.  \left. \right. \nn \\
&\hspace{1cm}\left. +\dmm       \left.   [\bar n_e n_e]  J_{\bar e e-} J_{\bar q q -}^{\balpha \beta}     \right.  -\dmp          \left.   \langle \bar n_e n_e\rangle J_{\bar e e+} J_{\bar q q -}^{\balpha \beta}     \right. \right) \,. 
\end{align}
We therefore see that each of the helicity combinations in the leading power operator basis of \Eq{eq:LP_basis} is reproduced, as required from the $C$ and $P$ relations. The Wilson coefficients can then immediately be read off of \Eq{eq:LP_helicity_matched}. Note that the spinor products appearing in this expression can be simplified given a particular choice of reference axes, but we have chosen to leave them in this general form.

\subsubsection{Subleading Power}\label{sec:proj_hel_nlp}

We now consider the projection onto the $\cO(\lambda)$ subleading power  helicity operator basis of \Sec{sec:sub}.

\vspace{0.2cm} \noindent 
\underline{
$\boldsymbol{(\bar q )_{\bar n} ( qg)_{n}}$ and $\boldsymbol{(\bar qg )_{\bar n} ( q)_{n}:}$
}

We begin by considering the case where the quark and antiquark are in different collinear sectors. The operators found in \Eqs{eq:matched_1gluon_barn}{eq:matched_1gluon_n} can be expanded in terms of the helicity currents. We have
\begin{align}
\bar \chi_{n,\omega_1} \Sl{\cB}_{\perp n} \chi_{\bar n,-\omega_2} &=\sqrt{\omega_1 \omega_2}  J_{n\bar n -}\cB_{n+}  +   \sqrt{\omega_1 \omega_2}  J_{n\bar n +}\cB_{n-}\,, \nn \\
\bar \chi_{n,\omega_1} \Sl{\cB}_{\perp \bar n,\omega_3} \chi_{\bar n,-\omega_2} &=\sqrt{\omega_1 \omega_2} J_{n\bar n -}\cB_{\bar n-}  +   \sqrt{\omega_1 \omega_2}  J_{n\bar n +}\cB_{\bar n +}\,,
\end{align}
where the helicity of the gluon fields is opposite in the two cases due to the fact that it is defined with respect to different axes.
From this, we immediately obtain the helicity expansions of the two operator. In the case that the gluon is in the $n$ collinear sector, we have
\begin{align}\label{eq:proj_1Bn}
&\bar v_e \gamma^\mu u_e  \left(  \frac{-ig_{\mu \nu}}{Q^2} \right) \cO^{(1)\nu}_{\cB  n}=gT^a_{\alpha \bbeta}\left(\frac{-i \sqrt{\omega_1 \omega_2}}{Q}\right)\\
&\left [ [\bar n_e n_e] \left(  \frac{(\omega_1+\omega_3) n\cdot \epsilon^-(n_e, \bar n_e) -\omega_2 \bar n\cdot \epsilon^-(n_e, \bar n_e)    }{\omega_2 (\omega_1+\omega_3)}  \right)  \left(J_{\bar e e -} J^{\balpha \beta}_{n\bar n -}\cB^a_{n+}   +  J_{\bar e e -} J^{\balpha \beta}_{n\bar n +}\cB^a_{n-}    \right) \right.   \nn \\
&\left .+ \langle \bar n_e n_e \rangle  \left(  \frac{(\omega_1+\omega_3) \cdot \epsilon^+(n_e, \bar n_e) -\omega_2 \bar n\cdot \epsilon^+(n_e, \bar n_e)   }{\omega_2 (\omega_1+\omega_3)}   \right) \left(J_{\bar e e +} J^{\balpha \beta}_{n\bar n -}\cB^a_{n+}   +  J_{\bar e e +} J^{\balpha \beta}_{n\bar n +}\cB^a_{n-}    \right) \right ]\nn\,,
\end{align}
and in the case that the gluon is in the $\bar n$ collinear sector, we find
\begin{align}\label{eq:proj_1Bnbar}
&\bar v_e \gamma^\mu u_e  \left(  \frac{-ig_{\mu \nu}}{Q^2} \right) \cO^{(1)\nu}_{\cB  \bar n}=gT^a_{\alpha \bbeta}\left(\frac{-i \sqrt{\omega_1 \omega_2}}{Q}\right) \\
&\left [ [\bar n_e n_e] \left(  \frac{\omega_1 n\cdot \epsilon^-(n_e, \bar n_e) -(\omega_2+\omega_3)  \bar n\cdot \epsilon^-(n_e, \bar n_e)    }{\omega_1 (\omega_2+\omega_3)}  \right)  \left(J_{\bar e e -} J^{\balpha \beta}_{n\bar n -}\cB^a_{\bar n-}   + J_{\bar e e -} J^{\balpha \beta}_{n\bar n +}\cB^a_{\bar n+}    \right) \right .  \nn \\
&\left . +  \langle \bar n_e n_e \rangle  \left(  \frac{\omega_1 \cdot \epsilon^+(n_e, \bar n_e) -(\omega_2+\omega_3) \bar n\cdot \epsilon^+(n_e, \bar n_e)   }{\omega_1 (\omega_2+\omega_3)}   \right) \left( J_{\bar e e +} J^{\balpha \beta}_{n\bar n -}\cB^a_{\bar n-}   +  J_{\bar e e +} J^{\balpha \beta}_{n\bar n +}\cB^a_{\bar n+}    \right) \right ] \,.\nn
\end{align}
Note that while the Wilson coefficients of the individual helicity operators are more complicated, and explicitly involve spinor products which incorporate the angular dependence of the scattering process, the helicity structure is very simple. Indeed, comparing the operators generated, with those in the basis of
 \Eq{eq:Z1_basis}, we explicitly see that the helicity selection rules are respected by the tree level matching. This will continue to hold to all orders in perturbation theory. 

\vspace{0.2cm} \noindent 
\underline{
$\boldsymbol{(\bar q q )_{\bar n} ( g)_{n}:}$
}

We now consider the operator with a single collinear gluon, where both the quarks are in the same sector. In performing the matching, we found the two operators of \Eq{eq:match_2qsame}.
Projecting these onto helicities, we find
\begin{align}
\bar v_e \gamma^\mu u_e  \left(  \frac{-ig_{\mu \nu}}{Q^2} \right) \cO^{(1)\nu}_{\chi\chi\bar n1}&= T^a_{\alpha \bbeta}\frac{-g}{\omega_1} \left(  \frac{-i\sqrt{2 \omega_1 \omega_2}}{Q} \right)\,  \\
&\left [  \dpm    \left.   [\bar n_e n_e]  J_{\bar e e-} \cB^a_{n+} J^{\balpha \beta}_{\bar n 0}    \right.   \right .-  \dpp       \left.  \langle \bar n_e n_e\rangle  J_{\bar e e+}  \cB^a_{n+} J^{\balpha \beta}_{\bar n 0}  \right.\nn \\
&+ \dmm       \left.   [\bar n_e n_e]  J_{\bar e e-}  \cB^a_{n-} J^{\balpha \beta}_{\bar n \bar 0}   \right. \left .-  \dmp           \left.   \langle \bar n_e n_e\rangle J_{\bar e e+}   \cB^a_{n-} J^{\balpha \beta}_{\bar n \bar 0}  \right. \right ] \,,\nn
\end{align}
and
\begin{align}
\bar v_e \gamma^\mu u_e  \left(  \frac{-ig_{\mu \nu}}{Q^2} \right) \cO^{(1)\nu}_{\chi\chi\bar n2}&=T^a_{\alpha \bbeta} \frac{g}{\omega_2} \left(  \frac{-i\sqrt{2 \omega_1 \omega_2}}{Q} \right)  \\
&\left [  \dpm     \left.   [\bar n_e n_e]  J_{\bar e e-} \cB^a_{n+} J^{\balpha \beta}_{\bar n \bar 0}    \right. \right.  -  \dpp   \left.  \langle \bar n_e n_e\rangle  J_{\bar e e+}  \cB^a_{n+} J^{\balpha \beta}_{\bar n \bar 0}  \right.\nn \\
&+  \dmm       \left.   [\bar n_e n_e] J_{\bar e e-}  \cB^a_{n-} J^{\balpha \beta}_{\bar n 0}   \right. \left.- \dmp           \left.   \langle \bar n_e n_e\rangle J_{\bar e e+}   \cB^a_{n-} J^{\balpha \beta}_{\bar n  0}  \right. \right ]\,. \nn
\end{align}
Note that all the operators expected in the basis of \Eq{eq:Z1_basis_diff} are generated, although half come from the first Lorentz structure generated in the tree level matching, while half come from the second Lorentz structure. Again, the expected helicity selection rules, namely that the two quarks have net helicity zero is respected.

\subsubsection{Subsubleading Power}\label{sec:proj_hel_nnlp}

We now consider the projection of the operators found in the tree level matching onto the $\cO(\lambda^2)$ helicity operator basis of \Sec{sec:subsub}.

\vspace{0.2cm} \noindent 
\underline{
$\boldsymbol{(\bar q g \cP_\perp)_{\bar n} ( q)_{n}:}$
}

We first consider the operators involving an insertion of the  $\cP_\perp$ operator, which contribute to the cross section at $\cO(\lambda^2)$. The basis of helicity operators was given in  \Eq{eq:eeqqgpperp_basis}, and they involve a single collinear gluon field, and a single insertion of the $\cP_\perp$ operator.
The two currents found in the tree level matching were given in \Eq{eq:matched_nbarP}. We find
\begin{align}
\hspace{0cm}\bar v_e \gamma^\mu u_e  \left(  \frac{-ig_{\mu \nu}}{Q^2} \right) \cO^{(2)\nu}_{\cP\bar n1}&=-T^a_{\alpha \bbeta} \frac{g}{\omega_1 \omega_3}\left(  \frac{-i\sqrt{2 \omega_1 \omega_2} }{Q} \right)  \\
&\left[  \dpm     \left.   [\bar n_e n_e]  J_{\bar e e-}  J^{\balpha \beta}_{n \bar n+}  [ \cB^a_{\bar n-} \cP_{+}^\dagger]   \right.      \right. - \dpp      \left.  \langle \bar n_e n_e\rangle J_{\bar e e+} J^{\balpha \beta}_{n \bar n+}    [ \cB^a_{\bar n-} \cP_{+}^\dagger]   \right.\nn \\
&+\dmm        \left.   [\bar n_e n_e]  J_{\bar e e-} J^{\balpha \beta}_{n \bar n -}  [\cB^a_{\bar n+}  \cP_{-}^\dagger]   \right.  \left. - \dmp           \left.   \langle \bar n_e n_e\rangle  J_{\bar e e+} J^{\balpha \beta}_{n \bar n -} [ \cB^a_{\bar n+} \cP_{-}^\dagger]    \right.   \right] \,,\nn
\end{align}
and
\begin{align}
\hspace{0cm}\bar v_e \gamma^\mu u_e  \left(  \frac{-ig_{\mu \nu}}{Q^2} \right) \cO^{(2)\nu}_{\cP\bar n2}&= T^a_{\alpha \bbeta}\frac{g}{\omega_1 \omega_2}\left(  \frac{-i\sqrt{2 \omega_1 \omega_2}}{Q} \right)   \\
&\left[ \dpm    \left.   [\bar n_e n_e]  J_{\bar e e-} J^{\balpha \beta}_{n \bar n +} [ \cB^a_{\bar n-} \cP^\dagger_{-} ]   \right.      \right.- \dpp      \left.  \langle \bar n_e n_e\rangle  J_{\bar e e+}  J^{\balpha \beta}_{n \bar n +}[ \cB^a_{\bar n-} \cP^\dagger_{-}]   \right.\nn \\
&+  \dmm       \left.   [\bar n_e n_e] J_{\bar e e-}  J^{\balpha \beta}_{n \bar n -}  [\cB^a_{\bar n+}  \cP^\dagger_{+} ]     \right. \left. - \dmp           \left.   \langle \bar n_e n_e\rangle J_{\bar e e+}  J^{\balpha \beta}_{n \bar n -}  [\cB^a_{\bar n+}  \cP^\dagger_{+}]     \right.   \right] \,. \nn
\end{align}
We see that these project down into two different classes of helicity operators. Of the operators in the helicity basis of \Eq{eq:eeqqgpperp_basis} the first four are absent, while the last eight obtain non-zero matching coefficients at tree level. As expected all expected helicity selection rules are respected.

\vspace{0.2cm} \noindent 
\underline{
$\boldsymbol{(\bar q )_{\bar n} (gg q)_{n}:}$
}

We now consider the projection onto the helicity operators for the case of two collinear gluons in the same collinear sector. A basis of helicity operators for this case was given in \Eq{eq:eeqqgg_basis1}.
The operators arising in the tree level matching were given in \Eq{eq:matched_twogluon_samesector}. Projecting onto helicities, we find
\begin{align}
\bar v_e \gamma^\mu u_e  \left(  \frac{-ig_{\mu \nu}}{Q^2} \right) \cO^{(2)\nu}_{\cB \cB1}&=\frac{1}{2} \left( (T^a T^b)_{\alpha \bbeta} +(T^b T^a)_{\alpha \bbeta}  \right) \frac{g^2 \sqrt{ \omega_1 \omega_2}}{\omega_2(\omega_1+\omega_3)} \left(  \frac{-i\sqrt{2}}{Q} \right)   \\
&\left[  \dpm     \left.   [\bar n_e n_e]  J_{\bar e e-} J^{\balpha \beta}_{n\bar n+} \cB^a_{n-} \cB^b_{n-}    \right.      \right.-  \dpp     \left.  \langle \bar n_e n_e\rangle  J_{\bar e e+} J^{\balpha \beta}_{n\bar n +}  \cB^a_{n-} \cB^b_{n-}    \right.\nn \\
&+ \dmm        \left.   [\bar n_e n_e]  J_{\bar e e-} J^{\balpha \beta}_{n\bar n -} \cB^a_{n+} \cB^b_{+}    \right. \left. -  \dmp           \left.   \langle \bar n_e n_e\rangle  J_{\bar e e+} J^{\balpha \beta}_{n\bar n -}  \cB^a_{n+} \cB^b_{n+}     \right.   \right] \,, \nn
\end{align}
where we have included the appropriate symmetry factor. And
\begin{align}
\bar v_e \gamma^\mu u_e  \left(  \frac{-ig_{\mu \nu}}{Q^2} \right) \cO^{(2)\nu}_{\cB\cB2}&=\frac{1}{2} \left( (T^a T^b)_{\alpha \bbeta} +(T^b T^a)_{\alpha \bbeta}  \right) \frac{-g^2 \sqrt{ \omega_1 \omega_2}}{\omega_2(\omega_3+\omega_4)} \left(  \frac{-i\sqrt{2}}{Q} \right)   \\
&\left[  \dpm     \left.   [\bar n_e n_e]  J_{\bar e e-} J^{\balpha \beta}_{n\bar n-} \cB^a_{n-} \cB^b_{n+}    \right.      \right.-  \dpp     \left.  \langle \bar n_e n_e\rangle  J_{\bar e e+} J^{\balpha \beta}_{n\bar n -}  \cB^a_{n-} \cB^b_{n+}    \right.\nn \\
&+ \dmm        \left.   [\bar n_e n_e]  J_{\bar e e-} J^{\balpha \beta}_{n\bar n +} \cB^a_{n+} \cB^b_{-}    \right. \left. -  \dmp           \left.   \langle \bar n_e n_e\rangle  J_{\bar e e+} J^{\balpha \beta}_{n\bar n +}  \cB^a_{n+} \cB^b_{n-}     \right.   \right] \,. \nn
\end{align}
Again we see that of the helicity combinations present in the basis of \Eq{eq:eeqqgg_basis1}, four are not present, while the other eight are generated in the tree level matching.

\vspace{0.2cm} \noindent 
\underline{
$\boldsymbol{(\bar q g )_{\bar n} ( qg)_{n}:}$
}

Similar projections apply to the case when the two gluons are in different collinear sectors. A basis of helicity operators for this case was given in \Eq{eq:eeqqgg_basis2}. The operators arising in the tree level matching were given in \Eq{eq:matched_twogluon_diffsector}. Because of the way we have organized the operators, namely each corresponding to a particular Feynman diagram, the helicity structure of the Feynman diagram is reflected in particular helicity correlations in the projected helicity operator. However, summing over all the Feynman diagrams, we will generate all the different helicity combinations in our helicity basis.
We will write the color structure of the operators in terms of the color basis of \Eq{eq:ggqqll_color}. The organization of the operators in \Eq{eq:matched_twogluon_diffsector} corresponds quite closely to this color decomposition, and therefore the projection of each operator will typically contribute to only one of the elements of the color basis.

We begin by projecting onto helicity operators the operators $\cO^{(2)\nu}_{\cB\bar \cB1}$ and $\cO^{(2)\nu}_{\cB\bar \cB2}$ of  \Eq{eq:matched_twogluon_diffsector}. We find
\begin{align}\label{eq:2B_proj_1}
\bar v_e \gamma^\mu u_e  \left(  \frac{-ig_{\mu \nu}}{Q^2} \right) \cO^{(2)\nu}_{\cB\bar \cB1}&= \frac{-g^2}{(\omega_2+\omega_3) \omega_4} (T^b T^a)_{\alpha \bbeta} \left(  \frac{-i\sqrt{2\omega_1 \omega_2}}{Q} \right)  \\
&\left[   \dpm     \left.   [\bar n_e n_e]  J_{\bar e e-} J^{\balpha \beta}_{n\bar n +} \cB^b_{n+} \cB^a_{\bar n+}   \right.      \right.-  \dpp      \left.  \langle \bar n_e n_e\rangle  J_{\bar e e+} J^{\balpha \beta}_{n\bar n +}  \cB^b_{n+} \cB^a_{\bar n+}   \right.\nn \\
&+  \dmm         \left.   [\bar n_e n_e]  J_{\bar e e-} J^{\balpha \beta}_{n\bar n -} \cB^b_{n-} \cB^a_{\bar n-}     \right. \left. - \dmp           \left.   \langle \bar n_e n_e\rangle  J_{\bar e e+} J^{\balpha \beta}_{n\bar n -}  \cB^b_{n-} \cB^a_{\bar n-}   \right.   \right] \,,\nn
\end{align}
and
\begin{align}\label{eq:2B_proj_2}
\bar v_e \gamma^\mu u_e  \left(  \frac{-ig_{\mu \nu}}{Q^2} \right) \cO^{(2)\nu}_{\cB\bar \cB2}&= \frac{-g^2}{\omega_2 \omega_4} (T^a T^b)_{\alpha \bbeta} \left(  \frac{-i\sqrt{ 2\omega_1 \omega_2}}{Q} \right)  \\
&\left[  \dpm     \left.   [\bar n_e n_e]  J_{\bar e e-} J^{\balpha \beta}_{n\bar n +} \cB^a_{\bar n-} \cB^b_{n-}   \right.      \right.-   \dpp       \left.  \langle \bar n_e n_e\rangle J_{\bar e e+} J^{\balpha \beta}_{n\bar n+}  \cB^a_{\bar n-} \cB^b_{n-}    \right.\nn \\
&+ \dmm        \left.   [\bar n_e n_e]  J_{\bar e e-} J^{\balpha \beta}_{n\bar n -} \cB^a_{\bar n+} \cB^b_{n+}    \right. \left. -  \dmp            \left.   \langle \bar n_e n_e\rangle  J_{\bar e e+} J^{\balpha \beta}_{n\bar n -}  \cB^a_{\bar n+} \cB^b_{n+}    \right.   \right] \,, \nn
\end{align}
which is identical up to a relabeling of $n \leftrightarrow \bar n$, and flipping the helicities correspondingly for the gluon fields. Note that because the two gluon helicity fields are in the $\bar n$ and $n$ directions, the helicities of the $\cB$ fields are defined with respect to different axes.

We can similarly project the operators $\cO^{(2)\nu}_{\cB\bar \cB3}$ and $\cO^{(2)\nu}_{\cB\bar \cB4}$ onto the helicity operator basis. We find
\begin{align}\label{eq:2B_proj_3}
\bar v_e \gamma^\mu u_e  \left(  \frac{-ig_{\mu \nu}}{Q^2} \right) \cO^{(2)\nu}_{\cB\bar \cB3}&= \frac{-g^2}{(\omega_1+\omega_4) \omega_3} (T^b T^a)_{\alpha \bbeta} \left(  \frac{-i\sqrt{ 2\omega_1 \omega_2}}{Q} \right)  \\
&\left[  \dpm    \left.   [\bar n_e n_e]  J_{\bar e e-} J^{\balpha \beta}_{n\bar n +} \cB^b_{n-} \cB^a_{\bar n-}    \right.      \right.-  \dpp      \left.  \langle \bar n_e n_e\rangle  J_{\bar e e+} J^{\balpha \beta}_{n\bar n +}  \cB^b_{n-} \cB^a_{\bar n-}    \right.\nn \\
&+ \dmm         \left.   [\bar n_e n_e]  J_{\bar e e-} J^{\balpha \beta}_{n\bar n -} \cB^b_{n+} \cB^a_{\bar n+}    \right. \left. - \dmp            \left.   \langle \bar n_e n_e\rangle  J_{\bar e e+} J^{\balpha \beta}_{n\bar n -}  \cB^b_{n+} \cB^a_{\bar n+}     \right.   \right] \,,\nn
\end{align}
and
\begin{align}\label{eq:2B_proj_4}
\bar v_e \gamma^\mu u_e  \left(  \frac{-ig_{\mu \nu}}{Q^2} \right) \cO^{(2)\nu}_{\cB\bar \cB4}&= \frac{-g^2}{\omega_1 \omega_3} (T^a T^b)_{\alpha \bbeta} \left(  \frac{-i\sqrt{ 2\omega_1 \omega_2}}{Q} \right)  \\
&\left[   \dpm     \left.   [\bar n_e n_e]  J_{\bar e e-} J^{\balpha \beta}_{n\bar n +} \cB^a_{\bar n+} \cB^b_{n+}    \right.      \right.-  \dpp      \left.  \langle \bar n_e n_e\rangle  J_{\bar e e+} J^{\balpha \beta}_{n\bar n +}  \cB^a_{\bar n+} \cB^b_{n+}    \right.\nn \\
&+\dmm         \left.   [\bar n_e n_e]  J_{\bar e e-} J^{\balpha \beta}_{n\bar n -} \cB^a_{\bar n-} \cB^b_{n-}    \right. \left. -  \dmp           \left.   \langle \bar n_e n_e\rangle  J_{\bar e e+} J^{\balpha \beta}_{n\bar n -}  \cB^a_{\bar n-} \cB^b_{n-}     \right.   \right] \,. \nn
\end{align}
In this case both the $\cB$ fields are again in the same helicity, but their correlation with the helicity of the quark current is opposite to that of \Eqs{eq:2B_proj_1}{eq:2B_proj_2}.

The operators  $\cO^{(2)\nu}_{\cB\bar \cB5}$ and $\cO^{(2)\nu}_{\cB\bar \cB6}$ give a different helicity contribution, where the gluon fields have opposite helicity. We have 
\begin{align}\label{eq:2B_proj_5}
\bar v_e \gamma^\mu u_e  \left(  \frac{-ig_{\mu \nu}}{Q^2} \right) \cO^{(2)\nu}_{\cB \bar \cB5}&= \frac{g^2}{(\omega_2+\omega_3)(\omega_1+\omega_4)} (T^b T^a)_{\alpha \bbeta}\left(  \frac{-i\sqrt{ 2\omega_1 \omega_2}}{Q} \right)   \\
&\left[   \dpm     \left.   [\bar n_e n_e]  J_{\bar e e-} J^{\balpha \beta}_{n\bar n +} \cB^b_{n-} \cB^a_{\bar n+}    \right.      \right.-  \dpp       \left.  \langle \bar n_e n_e\rangle  J_{\bar e e+} J^{\balpha \beta}_{n\bar n +}  \cB^b_{n-} \cB^a_{\bar n+}   \right.\nn \\
&+ \dmm        \left.   [\bar n_e n_e]  J_{\bar e e-} J^{\balpha \beta}_{n\bar n -} \cB^b_{n+} \cB^a_{\bar n-}    \right. \left. -  \dmp           \left.   \langle \bar n_e n_e\rangle  J_{\bar e e+} J^{\balpha \beta}_{n\bar n -}  \cB^b_{n+} \cB^a_{\bar n-}    \right.   \right] \,, \nn
\end{align}
and 
\begin{align}\label{eq:2B_proj_6}
\bar v_e \gamma^\mu u_e \left(  \frac{-ig_{\mu \nu}}{Q^2} \right) \cO^{(2)\nu}_{\cB \bar \cB6}&= \frac{g^2 }{\omega_1 \omega_2} (T^a T^b)_{\alpha \bbeta} \left(  \frac{-i \sqrt{2\omega_1 \omega_2}}{Q} \right)  \\
&\left[  \dpm     \left.   [\bar n_e n_e]  J_{\bar e e-} J^{\balpha \beta}_{n\bar n +} \cB^a_{\bar n+} \cB^b_{ n-}   \right.      \right.-  \dpp     \left.  \langle \bar n_e n_e\rangle  J_{\bar e e+} J^{\balpha \beta}_{n\bar n+}  \cB^a_{\bar n+} \cB^b_{ n-}    \right.\nn \\
&+ \dmm        \left.   [\bar n_e n_e]  J_{\bar e e-} J^{\balpha \beta}_{n\bar n -} \cB^a_{\bar n-} \cB^b_{n+}     \right. \left. -  \dmp           \left.   \langle \bar n_e n_e\rangle J_{\bar e e+} J^{\balpha \beta}_{n\bar n -}  \cB^a_{\bar n-} \cB^b_{n+}    \right.   \right] \,. \nn
\end{align}

Finally, the operator $\cO^{(2)\nu}_{\cB\bar \cB7}$ gives
\begin{align}\label{eq:2B_proj_7}
&\bar v_e \gamma^\mu u_e  \left(  \frac{-ig_{\mu \nu}}{Q^2} \right) \cO^{(2)\nu}_{\cB \bar \cB7}=-\frac{2g^2}{\omega_3 \omega_4} \left(  \frac{-i\sqrt{ \omega_1 \omega_2}}{\sqrt{2}Q} \right) \\
&  \left(  \cB^a_{\omega_3+}\cB^b_{\omega_4+}(T^aT^b)_{\alpha \bbeta}+  \cB^a_{\omega_3-}\cB^b_{\omega_4-}  (T^aT^b)_{\alpha \bbeta}  -    \cB^b_{\omega_4+}\cB^a_{\omega_3+}(T^bT^a)_{\alpha \bbeta}-  \cB^b_{\omega_4-}\cB^a_{\omega_3-} (T^bT^a)_{\alpha \bbeta}   \right) \nn \\
&\left[  \dpm   \left.   [\bar n_e n_e]  J_{\bar e e-} J^{\balpha \beta}_{n\bar n +}     \right.      \right.-   \dpp      \left.  \langle \bar n_e n_e\rangle  J_{\bar e e+} J^{\balpha \beta}_{n\bar n +}     \right.+\dmm         \left.   [\bar n_e n_e]  J_{\bar e e-} J^{\balpha \beta}_{n\bar n -}     \right. \left. - \dmp           \left.   \langle \bar n_e n_e\rangle  J_{\bar e e+} J^{\balpha \beta}_{n\bar n -}     \right.   \right] \,. \nn
\end{align}
Therefore we see that all the different helicity combinations present in the basis of \Eq{eq:eeqqgg_basis2} are generated in the tree level matching.

\vspace{0.2cm} \noindent 
\underline{
$\boldsymbol{(\bar q Q\bar Q)_{\bar n} ( q)_{n}:}$
}

The four quark operators generated in the matching were given in \Eq{eq:matched_4qn}. Projecting onto the basis of helicity operators, and using the color basis of \Eq{eq:qqqq_color}, we find
\begin{align}
\bar v_e \gamma^\mu u_e \left(  \frac{-ig_{\mu \nu}}{Q^2} \right) \cO^{(2)\nu}_{4\chi1a}&= \frac{g^2 \sqrt{ \omega_1 \omega_2 \omega_3 \omega_4} }{\omega_1 \omega_2 (\omega_2 + \omega_4)}  \frac{1}{2}(\delta_{\gamma \bbeta} \delta_{\bdelta \alpha} -\frac{1}{N_c} \delta_{\gamma \bdelta} \delta_{\alpha \bbeta}) \left(  \frac{-i}{\sqrt{2}Q} \right) \\
&\left[ \dpm    \left.   [\bar n_e n_e]  J_{\bar e e-}  J^{\bgamma \delta}_{(q) n\bar n+} J^{\balpha \beta}_{(Q)\bar n \bar 0}   \right.      \right.-   \dpp      \left.  \langle \bar n_e n_e\rangle \ J_{\bar e e+}  J^{\bgamma \delta}_{(q) n\bar n+} J^{(Q)\balpha \beta}_{\bar n \bar 0}   \right.\nn \\
&+ \dmm        \left.   [\bar n_e n_e]  J_{\bar e e-} J^{\bgamma \delta}_{(q) n\bar n-} J^{\balpha \beta}_{(Q)\bar n  0}    \right. \left. - \dmp           \left.   \langle \bar n_e n_e\rangle  J_{\bar e e+}  J^{\bgamma \delta}_{(q) n\bar n-} J^{\balpha \beta}_{(Q)\bar n  0}    \right.   \right] \,,\nn
\end{align}
and
\begin{align}
\bar v_e \gamma^\mu u_e  \left(  \frac{-ig_{\mu \nu}}{Q^2} \right) \cO^{(2)\nu}_{4\chi1b}&=\frac{g^2 \sqrt{ \omega_1 \omega_2 \omega_3 \omega_4} }{\omega_1 \omega_2 (\omega_2 + \omega_3) }  \frac{1}{2}(\delta_{\gamma \bbeta} \delta_{\bdelta \alpha} -\frac{1}{N_c} \delta_{\gamma \bdelta} \delta_{\alpha \bbeta}) \left(  \frac{-i}{\sqrt{2}Q} \right) \\
&\left[ \dpm    \left.   [\bar n_e n_e]  J_{\bar e e-}  J^{\bgamma \delta}_{(q) n\bar n+} J^{\balpha \beta}_{(Q)\bar n 0}   \right.      \right.-   \dpp      \left.  \langle \bar n_e n_e\rangle \ J_{\bar e e+}  J^{\bgamma \delta}_{(q) n\bar n+} J^{(Q)\balpha \beta}_{\bar n  0}   \right.\nn \\
&+ \dmm        \left.   [\bar n_e n_e]  J_{\bar e e-} J^{\bgamma \delta}_{(q) n\bar n-} J^{\balpha \beta}_{(Q)\bar n  \bar 0}    \right. \left. - \dmp           \left.   \langle \bar n_e n_e\rangle  J_{\bar e e+}  J^{\bgamma \delta}_{(q) n\bar n-} J^{\balpha \beta}_{(Q)\bar n  \bar 0}    \right.   \right] \,.\nn
\end{align}
We see that all the helicity combinations of \Eq{eq:Z3_basis_qQ} are generated in the tree level matching, as are both possible color structures. Furthermore, all the expected helicity selection rules are respected.
The projection of the operator \Eq{eq:matched_4qn} with an antiquark instead of a quark in the $n$ collinear sector proceeds identically, so we will not discuss it explicitly.

\subsection{Discussion}\label{sec:discuss}
Having performed the tree level matching both onto operators formed of standard Dirac and Lorentz structures, as well as projecting these onto operators in the helicity basis of \Sec{sec:eeJets}, here we briefly comment on the two approaches, and their advantages and disadvantages. The major advantage of the helicity operator approach is in enumerating a complete basis, and in making symmetry arguments about which operators will contribute to the factorization. This is greatly simplified by the use of helicity operators. Enumerating the basis is straightforward, and non-trivial relationships between different operators, for example using spin Fierz relations are absent. On the other hand, enumerating a complete and minimal basis of operators in the standard approach (which we did not do in this paper) is significantly more complicated. Furthermore, many of the symmetry arguments that were used to show that particular matrix elements of operators do not contribute to the cross section are obscured. Therefore for performing the formal factorization to all orders in perturbation theory, where a complete basis is essential, the helicity operator approach offers a clear advantage. This will become even more essential if one were to consider the case of three jets, for example. As noted in the case of dijets, a large number of the operators don't contribute. This can be seen easily in the helicity operator approach where many symmetries are manifest.

However, as has been seen in the matching, since each of the helicity building blocks is a scalar object, in the helicity approach all angular correlations are contained in the Wilson coefficients. This means that the Wilson coefficients are slightly more complicated objects. For example, in the case of the $e^+e^-\to$ dijets that we have considered here, the Wilson coefficients contain the inner products of polarization vectors, which carry all angular dependence. Furthermore, for relatively simple final states, such as those considered here, it is arguably more efficient to compute first a general current, and then project to helicity amplitudes. This may no longer be true if more jet directions are involved, since for higher multiplicity states, particularly involving gluons, it is well known that the calculation of individual helicity amplitudes can be significantly simpler.
Regardless of the exact techniques used to perform the matching, the helicity approach offers significant formal advantages for understanding the all orders structure of the operators, and for generating complete operator basis. The exact procedure which is then used to match to these operators can then be whatever is most convenient.

In a forthcoming paper we have performed a similar analysis, using the helicity operators introduced in this paper to study a complete subleading operator basis for $gg\to H$, as well as performing the tree level matching to those operators which contribute to the cross section at $\cO(\lambda^2)$ \cite{usggh:2016}. In that case we also found the helicity operator approach to be convenient. In particular, the spin zero nature of the Higgs leads to even more stringent helicity selection rules than for the $\bar q \Gamma q$ current considered here. The structure of the Wilson coefficients in the case of $gg\to H$ is also simpler due to the fact that there are no angular correlations between the initial and final state, and therefore the Wilson coefficients for the helicity operators remain simple. We therefore believe that the use of helicity operators is a particularly powerful approach for simplifying subleading power operator bases in SCET, much more generally than the $\bar q \Gamma q$ current considered here.

\section{Conclusions}\label{sec:conclusions}

In this paper we have presented a basis of SCET helicity operator building blocks valid to all orders in the power expansion. This involved the use of helicity operator building blocks with multiple collinear fields in the same collinear sector, as well as ultrasoft gauge invariant helicity fields describing ultrasoft degrees of freedom, as summarized in \Tab{tab:helicityBB}. These operators allow for efficient organization of both helicity and color information. At subleading power interesting selection rules arise from the conservation of angular momentum \cite{Kolodrubetz:2016uim}, which constrain the allowed hard scattering operators in a basis. The use of helicity operators, color organization, and ultrasoft gauge invariant building blocks greatly simplifies the construction of subleading power operator bases in SCET. 

To demonstrate the efficiency of the helicity-operator approach, we explicitly constructed a complete basis of hard scattering operators from a spin-1 current with two back-to-back collinear sectors up to $\cO(\lambda^2)$. Due to the manifest crossing symmetry of our operator basis, this basis is applicable to studying power corrections for a number of phenomenologically relevant processes, including $e^+e^-\to$ dijets, $e^- p\to e^-$ jet, and constrained Drell-Yan. As an example, we discussed in some detail the structure of the factorization in SCET for $e^+e^-\to$ dijet event shapes at subleading power, and detailed the different sources of power corrections. Symmetry relations, which are manifest in the helicity operator basis, were used to show the vanishing of hard scattering contributions to the dijet cross section at $\mathcal{O}(\lambda)$. Using our basis of hard scattering operators we enumerated and studied the field content of the subleading jet and soft functions which arise from the subleading hard scattering operators at $\mathcal{O}(\lambda^2)$ in the expansion of the cross section. We then performed a tree level matching calculation, showing the operators which arise at tree level, both in a more standard notation in terms of Dirac and Lorentz structures, as well as projected into the helicity basis. We contrasted the different forms of the operators and their utility for different purposes. The explicit results for the matching of the subleading power operators will be useful for further studies of power corrections both in fixed order and resummed perturbation theory. 

Since relatively little is known about the structure of factorization theorems at subleading power a number of directions exist for future study. It would be interesting to study in more detail RPI relations between operators for the subleading dijet operators to understand if relations beyond those given in this paper could be derived.  The renormalization group evolution of the subleading power helicity operators in SCET is of considerable interest. The anomalous dimensions of leading power operators are well understood, and exhibit many universal features due to their connections with the soft limits of gauge theories, and it would interesting to determine to what extent such features persist to subleading power, and what new properties emerge.  A study of the RG structure of $\cO(\lambda)$ operators was considered in \Ref{Freedman:2014uta}. The RG evolution of higher twist operators has also been well studied \cite{Balitsky:1987bk,Ratcliffe:1985mp,Ji:1990br,Ali:1991em,Kodaira:1996md,Balitsky:1996uh,Mueller:1997yk,Belitsky:1997zw,Belitsky:1999ru,Belitsky:1999bf,Vogelsang:2009pj,Braun:2009mi}, and should exhibit similar structures. Finally, an understanding of the numerical impact of the subleading corrections for $e^+e^-\to $ dijets event shapes would be of considerable interest, for example, for improving extractions of $\alpha_s$ performed in \cite{Hoang:2015hka, Abbate:2010xh, Becher:2008cf, Chien:2010kc}.

Another potentially interesting application of subleading factorization theorems is to improving subtraction schemes for higher order perturbative calculations. Subtraction schemes based on factorization theorems include the recently introduced $N$-jettiness subtraction scheme \cite{Boughezal:2015aha,Boughezal:2015dva,Gaunt:2015pea}, based on the $N$-jettiness event shape \cite{Stewart:2010tn}, as well as the SCET based subtraction scheme for NNLO semileptonic top quark decays of \Ref{Gao:2012ja}. Subleading factorization theorems would allow for the subtraction of the next-to-singular terms, potentially improving the numerical accuracy and speed of the techniques. This was emphasized in \cite{Gaunt:2015pea}, and was first studied numerically in \cite{Boughezal:2016wmq}. The feasibility of extending these schemes to subleading power will rely on a convenient organization of the subleading factorization theorem, which should be aided by the simplicity of the helicity operator approach. A detailed study of power corrections for $N$-jettiness subtractions using the operators in this paper was presented in \cite{Moult:2016fqy}, where the terms of $\cO(\alpha_s^2 \log^3 \tau)$ were computed explicitly. See also \cite{Boughezal:2016zws} for a calculation of the power corrections using alternative methods.

More broadly, we also envision that the helicity operator approach could be useful for constructing subleading operators in other processes, including $B$ physics and higher twist DIS where power corrections have been more thoroughly studied.  Although power corrections have yet to begin to play an important phenomenological role in jet physics, we have demonstrated that a particular part of the factorization at subleading power, namely the construction of a basis of hard scattering operators, can be greatly simplified by the use of helicity operators, which we hope will prove useful in the future study of factorization theorems at subleading power.

\begin{acknowledgments}
	
We thank Matthew Schwartz, Simon Freedman, Raymond Goerke, Andrew Larkoski, Duff Neill, Gherardo Vita and Hua Xing Zhu for helpful discussions.  We thank the Erwin Schroedinger Institute and the organizers of the ``Jets and Quantum Fields for LHC and Future Colliders''  and ``Challenges and Concepts for Field Theory and Applications in the Era of LHC Run-2''  workshops for hospitality and support while portions of this work were completed. This work was supported in part by the Office of Nuclear Physics of the U.S. Department of Energy under the Grant No. DE-SCD011090, by the Office of High Energy Physics of the U.S. Department of Energy under Contract No. DE-AC02-05CH11231, NSERC of Canada, the Center for the Fundamental Laws of Nature at Harvard University, and the LDRD Program of LBNL. I.S. was also supported by the Simons Foundation through the Investigator grant 327942.

\end{acknowledgments}

\appendix

\section{Spinor Helicity Identities and Conventions}
\label{app:helicity}

The four-component spinor $u(p)$ of a massless Dirac particle with momentum $p$, satisfies the massless Dirac equation,
\begin{equation} \label{eq:Dirac}
\Sl p\, u(p)=0
\,, \qquad
p^2 = 0
\,,\end{equation}
as does the charge conjugate (antiparticle) spinor $v(p)$. We can therefore choose a representation such that $v(p) = u(p)$. We denote the spinors and conjugate spinors for the two helicity states by
\begin{align} \label{eq:braket_def}
\ket{p\pm} = \frac{1 \pm \ga_5}{2}\, u(p)
\,,\qquad
\bra{p\pm} = \mathrm{sgn}(p^0)\, \bar{u}(p)\,\frac{1 \mp \ga_5}{2}\,.
\end{align}
Here the $\mathrm{sgn}(p^0)$ is included in the definition to simplify relations under crossing symmetry.

The spinors $\ket{p \pm}$ have an overall phase that is left free by the Dirac equation. Using the Dirac representation,
\begin{equation}
\ga^0 = \begin{pmatrix} 1 & 0 \\ 0 & -1 \end{pmatrix}
\,,\quad
\ga^i = \begin{pmatrix} 0 & \sigma^i \\ -\sigma^i & 0 \end{pmatrix}
\,,\quad
\ga_5 = \begin{pmatrix} 0 & 1 \\ 1 & 0 \end{pmatrix}
\,.\end{equation}
If we take $n_i^\mu = (1,0,0,1)$, we get the standard solutions~\cite{Dixon:1996wi}
\begin{equation} \label{eq:ket_explicit}
\ket{p+} = \frac{1}{\sqrt{2}}
\begin{pmatrix}
\sqrt{p^-} \\
\sqrt{p^+} e^{i  \phi_p} \\
\sqrt{p^-} \\
\sqrt{p^+} e^{i  \phi_p}
\end{pmatrix}
,\quad
\ket{p-} = \frac{1}{\sqrt{2}}
\begin{pmatrix}
\sqrt{p^+} e^{-i \phi_p} \\
-\sqrt{p^-} \\
-\sqrt{p^+} e^{-i \phi_p} \\
\sqrt{p^-}
\end{pmatrix}
,\end{equation}
where
\begin{equation}
p^\pm = p^0 \mp p^3
\,,\qquad
\exp(\pm i  \phi_p) = \frac{p^1 \pm i  p^2}{\sqrt{p^+ p^-}}
\,.\end{equation}
Using these conventions, we have
\begin{align}
\langle i j \rangle =\sqrt{|s_{ij}|}e^{i\phi_{ij}}\,, \qquad  [ij]=\sqrt{|s_{ij}|}e^{-i(\phi_{ij}+\pi)}\,,
\end{align}
where
\begin{align}
\cos \phi_{ij}= \frac{k_i^1k_j^+-k_j^1k_i^+}{\sqrt{|s_{ij}| k_i^+ k_j^+}}\,, \qquad \sin \phi_{ij}= \frac{k_i^2k_j^+-k_j^2 k_i^+}{\sqrt{|s_{ij}| k_i^+ k_j^+}}   \,.
\end{align}
For negative $p^0$ and $p^\pm$ we use the usual branch of the square root, such that for $p^0 > 0$
\begin{equation}
\ket{(-p)\pm} = i  \ket{p\pm}
\,.\end{equation}
We also define, $\bra{p\pm}$, the conjugate spinors, as
\begin{equation}
\bra{p\pm} = \mathrm{sgn}(p^0)\, \overline{\ket{p\pm}}
\,.\end{equation}
We include the additional minus sign for negative $p^0$ as we want to use the same branch of the square root for both types of spinors. We see for $p^0 > 0$
\begin{equation}
\bra{(-p)\pm} = - \overline{\ket{(-p)\pm}} = -(-i ) \bra{p\pm} = i \bra{p\pm}
\,.\end{equation}
This makes all spinor identities correct for momenta of both signs, allowing easier utilization of crossing symmetry. These signs will appear in expressions with explicit complex conjugation, including the most important example,
\begin{equation} \label{eq:spin_conj}
\l p-|q+\r^* = \mathrm{sgn}(p^0 q^0)\, \l q+|p-\r
\,.\end{equation}
The spinor products are denoted by
\begin{equation}
\langle p q \rangle = \langle p-|q+\rangle
\,,\qquad
[p q] = \langle p+|q-\rangle
\,.\end{equation}
Several useful identities satisfied by the spinor products are
\begin{align}
&\ang{pq} = - \ang{qp}
\,,\qquad
[pq] = - [qp]
\,, \qquad [p|\ga^\mu \ket{p}=\bra{p} \ga^\mu |p] = 2p^\mu\,, \\
&\ket{p\pm}\bra{p\pm} = \frac{1 \pm \ga_5}{2}\,\pslash
\,, \qquad
\pslash=|p]\bra{p}+\ket{p}[p|,\nn\\
&\langle pq \rangle [qp] = \frac{1}{2}\,\tr\bigl\{(1 - \ga_5) \pslash \qslash \bigr\} = 2 p\cdot q \,, \qquad
|\ang{pq}| = |[pq]| = \sqrt{|2p \cdot q|}\,, \nn\\
&\bra{p}\gamma^\mu |q]  = [q|\gamma^\mu \ket{p}
\,, \qquad
[p|\gamma_\mu \ket{q} [k|\gamma^\mu \ket{l}= 2[pk]\ang{lq}
\,, \nn\\
&\ang{pq} \ang{kl} = \ang{pk} \ang{ql} + \ang{pl} \ang{kq}
\,.\nn
\end{align}
Momentum conservation $\sum_{i=1}^n p_i = 0$ also implies the relation
\begin{equation} \label{eq:spinormomcons}
\sum_{i=1}^n [ji] \ang{ik} = 0
\,.\end{equation}
Under parity the spinors transform as
\begin{align} \label{eq:spinorparity}
\ket{p^\P\pm} = \pm e^{\pm i \phi_p}\gamma^0\,\ket{p\mp}\,, \quad
\ang{p^\P q^\P} = -e^{i (\phi_p + \phi_q)} [pq]
\,, \quad
[p^\P q^\P] = -e^{-i (\phi_p + \phi_q)} \ang{pq}
\,.\end{align}
The polarization vector satisfies the completeness relation
\begin{align}
\sum\limits_{\lambda=\pm} \epsilon^\lambda_\mu(p,q) \left (\epsilon^\lambda_\nu(p,q) \right )^*=-g_{\mu \nu}+\frac{p_\mu q_\nu+p_\nu q_\mu}{p\cdot q}.
\end{align}
In SCET the projected collinear quark fields 
\begin{align}
\ket{p\pm}_{n}=\frac{\nslash\bnslash}{4}\ket{p\pm}\,,
\end{align}
satisfy the relation
\begin{equation}
\nslash\,\Bigl(\frac{\nslash\bnslash}{4}\ket{p\pm}\Bigr) = 0
\,,\end{equation}
and are therefore proportional to $\ket{n\pm}$. Working in the basis in \eq{ket_explicit}, we find
\begin{align}\label{eq:ketn}
\frac{\nslash\bnslash}{4} \ket{p}=& \sqrt{p^0}\left[ \cos\left (\frac{\theta_n}{2}\right )\cos \left(\frac{\theta_p}{2}\right )
+e^{i(\phi_p-\phi_n)} \sin \left(\frac{\theta_n}{2}\right )   \sin \left(\frac{\theta_p}{2} \right )   \right ] \ket{n}\,,  \\
\frac{\nslash\bnslash}{4} |p]=& \sqrt{p^0}\left [ e^{i \left(\phi _p-\phi _n\right)} \cos \left(\frac{\theta _n}{2}\right)  \cos \left(\frac{\theta _p}{2}\right) 
 +\sin \left(\frac{\theta _n}{2}\right) \sin \left(\frac{\theta _p}{2}\right)   \right ] |n]\,.\nn
\end{align}
Here $\theta_n, \phi_n$, and $\theta_p, \phi_p$, are the polar and azimuthal angle of the $n$ and $p$ vectors, respectively.
Choosing $n_i^\mu=p_i^\mu/p_i^0$, we have
\begin{align}
\frac{\nslash\bnslash}{4} \ket{p\pm}=\sqrt{\frac{\bn_i\cdot p}{2}}\: \ket{n_i\pm}\,.
\end{align}

\section{Subleading Measurement Function}
\label{app:meas}

In this section we detail the factorization of the subleading measurement function, to illustrate its structure, and how the expansion can be systematically performed. We restrict ourselves to what we will refer to as ``pseudo-additive observables'' which we define as those observables with measurement functions that can be factorized into contributions from collinear and ultrasoft modes at each order in the power expansion in the form
\begin{align}\label{eq:fact_meas}
 e^{(i)}(X)= e_n^{(i)}(X_n,G_\bn,G_s)+ e_\bn^{(i)}(X_\bn,G_n,G_s) + e_s^{(i)}(X_s,G_n,G_\bn)\,.
\end{align}
Here $G_n$, $G_\bn$, $G_s$ refer to global properties of the corresponding sectors that can be defined independent of the order in perturbation theory to which one is working.\footnote{An example of a factorization theorem for which non-trivial factors of $G_n, G_\bn, G_s$ appear is the so called ``soft haze'' factorization theorem of \Ref{Larkoski:2015kga}.}  The sum over intermediate states in each sector can be performed by introducing measurement functions $\widehat \cM_n^{(i)}$, $\widehat \cM_\bn^{(i)}$, and $\widehat \cM_s^{(i)}$. The measurement functions act as $\widehat \cM_n^{(i)} |X\r = \delta(e - e_n^{(i)}(X))|X\r$, and similarly for the other sectors \cite{Lee:2006nr,Sveshnikov:1995vi,Korchemsky:1997sy,Bauer:2008dt,Belitsky:2001ij}. The contributions to the cross section at each order in the power expansion can then be expressed as a sum of vacuum matrix elements, involving a measurement function insertion, and each containing only collinear $n$, collinear $\bar n$, or ultrasoft fields. 

We consider explicitly the factorization for the thrust observable \cite{Farhi:1977sg}
\begin{align}\label{eq:thrust_defn}
T= \text{max}_{\hat t} \frac{\sum_i |\hat t \cdot \vec p_i|}{\sum_i |\vec p_i|}\,,
\end{align}
assuming massless particles.
It is convenient to work with
\begin{align}
\tau=1-T\,,
\end{align}
which vanishes in the dijet limit, and can be described by an SCET$_\text{I}$ factorization theorem with $\lambda \sim \sqrt{\tau}$ \cite{Fleming:2007qr,Schwartz:2007ib,Fleming:2007xt,Becher:2008cf,Abbate:2010xh}. We will explicitly construct the subleading power corrections to the thrust measurement function, performing the expansion to $\cO(\lambda^2)$
\begin{align}
\tau=\tau^{(0)}+\tau^{(1)}+\tau^{(2)}\,,
\end{align}
using the same SCET formalism used to enumerate our operator basis.

We consider an $e^+e^-$ event at center of mass energy $Q$, and we work in the center of mass frame. We label our lightlike vectors in the effective theory as $n^\mu=(1,\vec n)$, and $\bar n=(1,-\vec n)$. The axis defined by the unit vector $\vec t$, which satisfies $\vec t \cdot \vec t=1$, is referred to as the thrust axis. Thrust is maximized when the thrust axis is aligned with the hemisphere whose  particles have the largest three-momentum along $\vec t$. We therefore partition our event into hemispheres based on this criteria, which we call $J_1$, and $J_2$. For particles $i\in J_1$, we have $\vec p_i \cdot \vec t >0$, while for particles  $i\in J_2$, we have $\vec p_i \cdot \vec t <0$. We also define the thrust light-cone vectors, $ t^\mu = (1,\vec t\;)$ and $\bar{t}^\mu =  (1,-\vec t\;)$. As shorthand, we will use the notation
\begin{align}
P^\mu_{J_1} = \sum_{i\in J_1} p_i^\mu\,, \qquad P^\mu_{J_2} = \sum_{i\in J_2} p_i^\mu\,.
\end{align}

We have chosen to define $\vec n$ so that we are in a frame where the total label $\perp$ component of each collinear sector vanishes. This choice still allows an infinite family of possible frame choices as it does not restrict the residual momenta.  In particular, the residual momentum of the collinear sectors need not vanish when one experiences recoil at $\mathcal{O}(\lambda^2)$. To make this frame restriction manifest, we split the momenta in each hemisphere as
\begin{align}
 P^\mu_{J_1} =\frac{\overline{\cP}_{J_1}}{2} n^\mu +(p^{r\mu}_n +k_1^{(2)\mu})\,, \qquad   
 P^\mu_{J_2} =\frac{ \cP_{J_2}}{2} \bn^\mu +(p^{r\mu}_\bn +k_2^{(2)\mu})\,.
\end{align}
Here $\overline{\cP}_{J_1}$ and   $\cP_{J_2}$ denote the large component of the label momenta, $p^r_n$ and $p^r_\bn$ denote the residual momenta of the collinear sectors and
\begin{align}
k_1^{(2)\mu} &= \sum_{\text{soft}, i} k_i^\mu \, \theta(\vec n\cdot \vec k_i)
\,,\qquad
k_2^{(2)\mu} = \sum_{\text{soft}, i} k_i^\mu \, \theta(-\vec n\cdot \vec k_i)
  \,, 
\end{align}
which we will use to express our results.

We will not present a derivation, but simply state the results for the measurement functions.
The familiar result for the leading measurement function is
\begin{align}
\tau^{(0)}
  &= \frac{n\cdot \bigl(p_n^r + k_1^{(2)}\bigr)}{Q} + \frac{\bn\cdot \bigl(p_\bn^r+k_2^{(2)}\bigr)}{Q}
 \,. \nn
\end{align}
At the next order in the power expansion we have no ${\cal O}(\lambda)$ terms, 
\begin{align} \label{eq:app-lambda-thrust-zero}
\tau^{(1)}= & 0\,,
\end{align}
as expected. After some algebra, at $\cO(\lambda^2)$ we find the rather simple final result
\begin{align} \label{eq:tau4full}
\tau^{(2)}&=-\frac{2}{Q^2} \Bigl(\vec{p}^{\,r}_{n\perp} +\vec k_{1\perp}^{(2)} \Bigr)^2 
  \\
&=-\frac{2}{Q^2} \Bigl[
  \bigl(\vec{p}^{\,r}_{n\perp} \bigr)^2 +2\vec k_{1\perp}^{(2)} \cdot \vec{p}^{\,r}_{n\perp} +\bigl(\vec k_{1\perp}^{(2)} \bigr)^2  \Bigr]
 \,. \nn
\end{align}
Thus the subleading thrust measurement function depends on only the squared total ${\cal O}(\lambda^2)$ perpendicular momentum in a hemisphere.
This can also be written in a form that is symmetric in $\bar n$ and $n$,
\begin{align} \label{eq:tau4full_sym}
\tau^{(2)}&=-\frac{1}{Q^2} \Bigl(\vec{p}^{\,r}_{n\perp} +\vec k_{1\perp}^{(2)} \Bigr)^2 -\frac{1}{Q^2} \Bigl(\vec{p}^{\,r}_{\bar n\perp} +\vec k_{2\perp}^{(2)} \Bigr)^2 
  \\
&=-\frac{1}{Q^2} \Bigl[
  \bigl(\vec{p}^{\,r}_{n\perp} \bigr)^2 +2\vec k_{1\perp}^{(2)} \cdot \vec{p}^{\,r}_{n\perp} +\bigl(\vec k_{1\perp}^{(2)} \bigr)^2  \Bigr]   -\frac{1}{Q^2} \Bigl[
  \bigl(\vec{p}^{\,r}_{\bar n\perp} \bigr)^2 +2\vec k_{2\perp}^{(2)} \cdot \vec{p}^{\,r}_{\bar n\perp} +\bigl(\vec k_{2\perp}^{(2)} \bigr)^2  \Bigr]
 \,. \nn
\end{align}
In either form the $\tau^{(2)}$ power correction to the thrust observable is always negative. Note that there is a cross term in the measurement function at $\cO(\lambda^2)$. However, from the point of view of the ultrasoft (collinear) sector, this term depends only on a global property of the collinear (ultrasoft) sector, and therefore satisfies the general form of \Eq{eq:fact_meas}. 

A derivation of the subleading power measurement function for thrust was first presented in \cite{Freedman:2013vya} using the SCET framework of \cite{Freedman:2011kj}. Due to the fact that residual momentum is not conserved in their framework, their $\tau^{(1)}$ measurement function does not vanish, and a more complicated form of the $\tau^{(1)}$ measurement function is obtained.  Nevertheless, the strategy for computing the subleading measurement functions in the two setups is common, and hence was not repeated here.

\section{Generalized Basis with $\cP_{\perp n}$, $\cP_{\perp \bar n}\ne 0$}

Throughout the main text, we have restricted ourselves to back to back axes and made the assumption that the total label perp momentum in each collinear sector vanishes. When working at subleading power it is convenient to keep the axes back to back, and in certain cases, it is therefore necessary to generalize the basis to the case that there is a non-zero total perp momentum in either collinear sector. In this appendix, we provide the additional operators that appear in this situation, as well as their tree level matching. Since there are only a few additional operators, we distinguish them simply by their field content. Furthermore, for conciseness, we give only the operators that can interfere with the leading power operator. In particular, we do not give the operators involving two collinear quark fields in the same sector, along with an insertion of the $\cP_\perp$ operator in the general frame. We also perform the tree level matching onto these operators.

We first consider the operators involving $\cP_\perp$ insertions with just a single collinear quark field in each collinear sector. These appear at $\cO(\lambda)$ with a single $\cP_\perp$ insertion, and at  $\cO(\lambda^2)$ with two $\cP_\perp$ insertions. At $\cO(\lambda)$, we have the operators 
\begin{align}
& \boldsymbol{(\cP_\perp q)_n \bar{q}_{\bar n}:}{\vcenter{\includegraphics[width=0.18\columnwidth]{figures/Leading_qq}}}  \nn
\end{align}
\vspace{-0.4cm}
\begin{flalign}\label{eq:app_Pperp_basis_NLP}
O_{\cP n(+;\pm)[-]}^{(1)\balpha\bt}
=\left\{ \cP_\perp^- J^{\balpha\bt}_{n \bar n+} \right\} J_{e \pm }\,, \qquad
O_{\cP n(-;\pm)[+]}^{(1)\balpha\bt}
=\left\{ \cP_\perp^+ J^{\balpha\bt}_{n \bar n-} \right\} J_{e\pm } \,,
\end{flalign}
where the $\cP_\perp$ operator acts on the outgoing quark, and 
\begin{align}
& \boldsymbol{q_n (\cP_\perp \bar{q}_{\bar n}):}{\vcenter{\includegraphics[width=0.18\columnwidth]{figures/Leading_qq}}}  \nn
\end{align}
\vspace{-0.4cm}
\begin{flalign}\label{eq:app_Pperp_basis_NLP_bar}
O_{\cP \bar n(+;\pm)[-]}^{(1)\balpha\bt}
=\left\{ J^{\balpha\bt}_{n \bar n+} \cP_\perp^{-\dagger} \right\}J_{e \pm }\,, \qquad
O_{\cP \bar n(-;\pm)[+]}^{(1)\balpha\bt}
=\left\{J^{\balpha\bt}_{n \bar n-} \cP_\perp^{+\dagger} \right\} J_{e\pm } \,.
\end{flalign}
where the $\cP_\perp$ operator acts on the outgoing antiquark. Note that we do not allow the $\cP_\perp$ operator to act on the electron current, as this can always be removed using integration by parts.

There are also $\cO(\lambda^2)$ operators involving two insertions of the $\cP_\perp$ operators. A basis of such operators is given by
\begin{align}
& \boldsymbol{(\cP_\perp\cP_\perp q)_n \bar{q}_{\bar n}:}{\vcenter{\includegraphics[width=0.18\columnwidth]{figures/Leading_qq}}}  \nn
\end{align}
\vspace{-0.4cm}
\begin{flalign}\label{eq:app_Pperp_basis_NNLP}
O_{\cP \cP n(+;\pm)[--]}^{(2)\balpha\bt}
=\left\{ \cP_\perp^- \cP_\perp^- J^{\balpha\bt}_{n \bar n+} \right\} J_{e \pm }\,, \qquad
O_{\cP \cP n(-;\pm)[++]}^{(2)\balpha\bt}
=\left\{ \cP_\perp^+ \cP_\perp^+ J^{\balpha\bt}_{n \bar n-} \right\} J_{e\pm } \,, \nn \\
O_{\cP \cP n(+;\pm)[+-]}^{(2)\balpha\bt}
=\left\{ \cP_\perp^+ \cP_\perp^- J^{\balpha\bt}_{n \bar n+} \right\} J_{e \pm }\,, \qquad
O_{\cP \cP n(-;\pm)[+-]}^{(2)\balpha\bt}
=\left\{ \cP_\perp^+ \cP_\perp^- J^{\balpha\bt}_{n \bar n-} \right\} J_{e\pm } \,,
\end{flalign}
when both $\cP_\perp$ operators act on the quark sector,
\begin{align}
& \boldsymbol{q_n (\cP_\perp\cP_\perp \bar{q}_{\bar n}):}{\vcenter{\includegraphics[width=0.18\columnwidth]{figures/Leading_qq}}}  \nn
\end{align}
\vspace{-0.4cm}
\begin{flalign}\label{eq:app_Pperp_basis_NNLP_bar}
O_{\cP \cP \bar n(+;\pm)[--]}^{(2)\balpha\bt}
=\left\{ J^{\balpha\bt}_{n \bar n+} \cP_\perp^{-\dagger} \cP_\perp^{-\dagger} \right\}J_{e \pm }\,, \qquad
O_{\cP \cP \bar n(-;\pm)[++]}^{(2)\balpha\bt}
=\left\{J^{\balpha\bt}_{n \bar n-} \cP_\perp^{+\dagger} \cP_\perp^{+\dagger} \right\} J_{e\pm } \,, \nn \\
O_{\cP \cP \bar n(+;\pm)[-+]}^{(2)\balpha\bt}
=\left\{ J^{\balpha\bt}_{n \bar n+} \cP_\perp^{-\dagger} \cP_\perp^{+\dagger} \right\}J_{e \pm }\,, \qquad
O_{\cP \cP \bar n(-;\pm)[-+]}^{(2)\balpha\bt}
=\left\{J^{\balpha\bt}_{n \bar n-} \cP_\perp^{-\dagger} \cP_\perp^{+\dagger} \right\} J_{e\pm } \,,
\end{flalign}
when both $\cP_\perp$ operators act on the antiquark sector, and
\begin{align}
& \boldsymbol{(\cP_\perp q_n) (\cP_\perp \bar{q}_{\bar n}):}{\vcenter{\includegraphics[width=0.18\columnwidth]{figures/Leading_qq}}}  \nn
\end{align}
\vspace{-0.4cm}
\begin{flalign}\label{eq:app_Pperp_basis_NNLP_bar}
O_{\cP \cP(+;\pm)[--]}^{(2)\balpha\bt}
=\left\{ \cP_\perp^{-}  J^{\balpha\bt}_{n \bar n+}  \cP_\perp^{-\dagger} \right\}J_{e \pm }\,, \qquad
O_{\cP \cP(-;\pm)[++]}^{(2)\balpha\bt}
=\left\{ \cP_\perp^{+}  J^{\balpha\bt}_{n \bar n-}  \cP_\perp^{+\dagger} \right\} J_{e\pm } \,, \nn \\
O_{\cP \cP(+;\pm)[-+]}^{(2)\balpha\bt}
=\left\{ \cP_\perp^{-} J^{\balpha\bt}_{n \bar n+}  \cP_\perp^{+\dagger} \right\}J_{e \pm }\,, \qquad
O_{\cP \cP(-;\pm)[-+]}^{(2)\balpha\bt}
=\left\{  \cP_\perp^{-} J^{\balpha\bt}_{n \bar n-}  \cP_\perp^{+\dagger} \right\} J_{e\pm } \,, \nn \\
O_{\cP \cP(+;\pm)[+-]}^{(2)\balpha\bt}
=\left\{ \cP_\perp^{+} J^{\balpha\bt}_{n \bar n+}  \cP_\perp^{-\dagger} \right\}J_{e \pm }\,, \qquad
O_{\cP \cP(-;\pm)[+-]}^{(2)\balpha\bt}
=\left\{  \cP_\perp^{+} J^{\balpha\bt}_{n \bar n-}  \cP_\perp^{-\dagger} \right\} J_{e\pm } \,.
\end{flalign}
when one operator acts on either collinear sector. In all these cases, the basis of color structures is identical to that of the leading power operator in \Eq{eq:leading_color}.

Finally, we must consider the $\cO(\lambda^2)$ operators involving an insertion of the $\cP_\perp$ operator along with a $\cB_\perp$ insertion. In the case that the $\cP_\perp$ in each sector is non-vanishing, we must consider the possibility that the $\cP_\perp$ operator acts on any field, thus making the basis more cumbersome than in the center of mass frame.  In the general case, the basis is given by
\begin{align}
& \boldsymbol{(gq \cP_\perp)_n (\bar q)_{\bn}:}{\vcenter{\includegraphics[width=0.18\columnwidth]{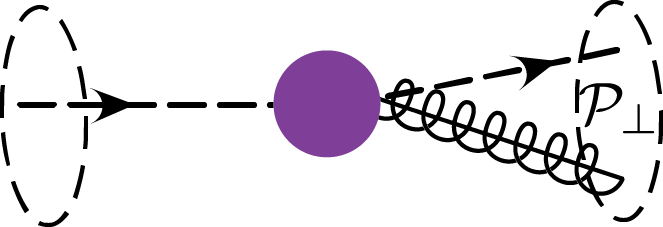}}}  \nn
\end{align}
\vspace{-0.4cm}
\begin{alignat}{2} \label{eq:app_perp_gluon}
&O_{\cP\cB + (+;\pm)[-]}^{(2)a\,\balpha\bt}
= \left [ \cP_{\perp}^{-} \cB_{n+}^a \right ]\, J_{n\bar n\, +}^{\balpha\bt}\, J_{e\pm }
\,,\qquad &
&O_{\cP\cB - (-;\pm)[+]}^{(2)a\,\balpha\bt}
= \left [ \cP_{\perp}^{+} \cB_{n-}^a \right ]\, J_{n\bar n\, -}^{\balpha\bt}\, J_{e\pm }
\,,\\
&O_{\cP\cB - (+;\pm)[+]}^{(2)a\,\balpha\bt}
=  \left [ \cP_{\perp}^{+} \cB_{n-}^a \right ]\, J_{n\bar n\, +}^{\balpha\bt}\, J_{e\pm }
\,,\qquad &
&O_{\cP\cB - (+;\pm)[-]}^{(2)a\,\balpha\bt}
= \left [ \cP_{\perp}^{-} \cB_{n-}^a \right ]\, J_{n\bar n\, +}^{\balpha\bt}\, J_{e\pm }
\,,\nn\\
&O_{\cP\cB + (-;\pm)[+]}^{(2)a\,\balpha\bt}
=  \left [ \cP_{\perp}^{+} \cB_{n+}^a \right ]\, J_{n\bar n\, -}^{\balpha\bt}\, J_{e\pm }
\,,\qquad &
&O_{\cP\cB + (-;\pm)[-]}^{(2)a\,\balpha\bt}
= \left [ \cP_{\perp}^{-} \cB_{n+}^a \right ]\, J_{n\bar n\, -}^{\balpha\bt}\, J_{e\pm }
\,, \nn
\end{alignat}
when the $\cP_\perp$ operator acts on the $\cB_\perp$ field,
\begin{align}
& \boldsymbol{(gq \cP_\perp)_n (\bar q)_{\bn}:}{\vcenter{\includegraphics[width=0.18\columnwidth]{figures/Subleading_qg_orderlam_perp}}}  \nn
\end{align}
\vspace{-0.4cm}
\begin{alignat}{2} \label{eq:app_perp_quark}
&O_{\cP\chi + (+;\pm)[-]}^{(2)a\,\balpha\bt}
= \cB_{n+}^a \left\{ \cP_{\perp}^{-} J_{n\bar n\, +}^{\balpha\bt} \right\}\, J_{e\pm }
\,,\qquad &
&O_{\cP\chi - (-;\pm)[+]}^{(2)a\,\balpha\bt}
=  \cB_{n-}^a \,\left\{  \cP_{\perp}^{+} J_{n\bar n\, -}^{\balpha\bt}\right\}\, J_{e\pm }
\,,\\
&O_{\cP\chi - (+;\pm)[+]}^{(2)a\,\balpha\bt}
=   \cB_{n-}^a \, \left\{  \cP_{\perp}^{+}J_{n\bar n\, +}^{\balpha\bt}\right\}\, J_{e\pm }
\,,\qquad &
&O_{\cP\chi - (+;\pm)[-]}^{(2)a\,\balpha\bt}
=  \cB_{n-}^a \, \left\{ \cP_{\perp}^{-} J_{n\bar n\, +}^{\balpha\bt}\right\}\, J_{e\pm }
\,,\nn\\
&O_{\cP\chi + (-;\pm)[+]}^{(2)a\,\balpha\bt}
=   \cB_{n+}^a \, \left\{  \cP_{\perp}^{+}J_{n\bar n\, -}^{\balpha\bt}\right\}\, J_{e\pm }
\,,\qquad &
&O_{\cP\chi + (-;\pm)[-]}^{(2)a\,\balpha\bt}
=  \cB_{n+}^a \, \left\{  \cP_{\perp}^{-} J_{n\bar n\, -}^{\balpha\bt}\right\}\, J_{e\pm }
\,, \nn
\end{alignat}
when it acts on the quark field, and
\begin{align}
& \boldsymbol{(gq \cP_\perp)_n (\bar q)_{\bn}:}{\vcenter{\includegraphics[width=0.18\columnwidth]{figures/Subleading_qg_orderlam_perp}}}  \nn
\end{align}
\vspace{-0.4cm}
\begin{alignat}{2} \label{eq:app_perp_antiquark}
&O_{\cP\bar \chi + (+;\pm)[-]}^{(2)a\,\balpha\bt}
= \cB_{n+}^a \left\{ J_{n\bar n\, +}^{\balpha\bt}  \cP_{\perp}^{\dagger-}  \right\}\, J_{e\pm }
\,,\qquad &
&O_{\cP\bar \chi - (-;\pm)[+]}^{(2)a\,\balpha\bt}
=  \cB_{n-}^a \,\left\{   J_{n\bar n\, -}^{\balpha\bt}  \cP_{\perp}^{\dagger+}   \right\}\, J_{e\pm }
\,,\\
&O_{\cP\bar \chi - (+;\pm)[+]}^{(2)a\,\balpha\bt}
=   \cB_{n-}^a \, \left\{ J_{n\bar n\, +}^{\balpha\bt}   \cP_{\perp}^{\dagger+} \right\}\, J_{e\pm }
\,,\qquad &
&O_{\cP\bar \chi - (+;\pm)[-]}^{(2)a\,\balpha\bt}
=  \cB_{n-}^a \, \left\{ J_{n\bar n\, +}^{\balpha\bt} \cP_{\perp}^{\dagger-}    \right\}\, J_{e\pm }
\,,\nn\\
&O_{\cP\bar \chi + (-;\pm)[+]}^{(2)a\,\balpha\bt}
=   \cB_{n+}^a \, \left\{  \cP_{\perp}^{+}J_{n\bar n\, -}^{\balpha\bt}  \cP_{\perp}^{\dagger+}   \right\}\, J_{e\pm }
\,,\qquad &
&O_{\cP\bar \chi + (-;\pm)[-]}^{(2)a\,\balpha\bt}
=  \cB_{n+}^a \, \left\{  \cP_{\perp}^{-} J_{n\bar n\, -}^{\balpha\bt}  \cP_{\perp}^{\dagger-}  \right\}\, J_{e\pm }
\,, \nn
\end{alignat}
when it acts on the antiquark field. 

We now perform the tree level matching to these operators. The first two groups of operators can be matched to using a $q\bar q$ external state. We take the momenta as
\begin{align}
p_1^\mu=\omega_1 \frac{n^\mu}{2}+p_{1\perp}+p_{1r}\frac{\bar n^\mu}{2}\,, \qquad p_2^\mu=\omega_2 \frac{\bar n^\mu}{2}+p_{2\perp}+p_{2r}\frac{n^\mu}{2}\,,
\end{align}
where $p_1$ denotes the momentum of the quark and $p_2$ the momentum of the anti-quark.
Expanding the tree level result for generic kinematics, we find
\begin{align}
\bar u_n \Gamma u_{\bar n}    +\bar u_n \frac{\Sl{\bar n}}{2}   \frac{\Sl{p}_{1\perp}}{\omega_1} \Gamma u_{\bar n}   +\bar u_n \Gamma \frac{\Sl{p}_{2\perp}}{\omega_2}   \frac{\Sl{n}}{2}  u_{\bar n}+ \bar u_n  \frac{\Sl{p}_{1\perp}}{\omega_1}\frac{\Sl{\bar n}}{2}  \Gamma  \frac{\Sl{n}}{2}  \frac{\Sl{p}_{2\perp}}{\omega_2}  u_{\bar n}\,,
\end{align}
From these we can immediately read off the Wilson coefficients and structure of the operators appearing in the matching,
\begin{align}
\cO^{(1)}_{\cP n}=-\frac{1}{\omega_1}  \bar \chi_{n,\omega_1} \Sl{\cP}_\perp^\dagger \frac{\Sl{\bar n}}{2} \Gamma \chi_{\bar n,-\omega_2}\,, \qquad \cO^{(1)}_{\cP \bar n}=-\frac{1}{\omega_2}  \bar \chi_{n,\omega_1} \Gamma  \frac{\Sl{\bar n}}{2}  \Sl{\cP}_\perp \chi_{\bar n,-\omega_2}\,,
\end{align}
and
\begin{align}\label{eq:double_pperp}
\cO^{(2)}_{\cP\cP}=\frac{1}{\omega_1 \omega_2}  \bar \chi_{n,\omega_1} \Sl{\cP}_\perp^\dagger \frac{\Sl{\bar n}}{2} \Gamma  \frac{\Sl{n}}{2}  \Sl{\cP}_\perp \chi_{\bar n,-\omega_2}\,.
\end{align}
Note that the Wilson coefficients of the operators with a single $\cP_\perp$ insertion are equal by RPI to the operators of \Eq{eq:soft_current_perp} involving an ultrasoft perp derivative. Note also that the operators involving two insertions of the $\cP_\perp$ operator in the same collinear sector do not appear in the tree level matching.

We now consider the matching to the operators involving one additional collinear gluon field, and an additional $\cP_\perp$ insertion, in the case that we do not restrict to the center of mass frame. Recall that the relevant QCD matrix elements are given by
\begin{align}
\fd{3cm}{figures/matching_subleadingvertex_2_low} &= \bar{u}(p_1) (i g T^a \Sl{\epsilon}^*) \frac{ i (\Sl{p}_1 + \Sl{p}_3)}{ (p_1+p_3)^2} \Gamma v(p_2) \,, \nn\\
\fd{3cm}{figures/matching_subleadingvertex_1_low}&= \bar{u}(p_1) \Gamma \frac{- i (\Sl{p}_2 + \Sl{p}_3)}{ (p_2+p_3)^2}  (i g T^a \Sl{\epsilon}^*)v(p_2) \,.
\end{align}
We will explicitly consider the case where the gluon is in the $\bar n$ sector. We can then simplify the matching by performing it in two steps. First, we can set the total $\perp$ momentum of the $\bar n$ collinear sector to zero, and extract the operator involving the $\cP_\perp$ acting on the $n$ collinear sector. Having extracted that Wilson coefficient, we can then set the perp in the $n$-collinear sector to zero, and then extract the other two Wilson coefficients. This suffices since the operators are linear in $\cP_\perp$. Note that for general kinematics, there are also $T$-product contributions arising from the $\cO(\lambda^2)$ operator of \Eq{eq:double_pperp}, and an insertion of the leading power SCET Lagrangian, or an emission from the Wilson line in the vertex. However, for the particular choice of kinematics, both these $T$-product contributions vanish.

We begin by extracting the Wilson coefficient for the case that the $\cP_\perp$ acts on the $n$ collinear sector. We take the momenta as
\begin{align}
p_1^\mu=\frac{\omega_1}{2}n^\mu+p_{1perp}^\mu+\frac{p_{1r}}{2}\bar n^\mu \,, \qquad
p_2^\mu=\frac{\omega_2}{2} \bar n^\mu+p_\perp^\mu +\frac{ p_{2r}}{2}  n^\mu\,, \qquad
p_3^\mu=\frac{\omega_3}{2}\bar n^\mu -p_\perp^\mu +\frac{ p_{3r}}{2}  n^\mu\,, 
\end{align}
Expanding the QCD amplitudes, and keeping only terms involving $\cP_{\perp}$ acting on the $n$ collinear sector, we find
\begin{align}
\left. \fd{3cm}{figures/matching_subleadingvertex_2_low}\right |_{\cO(\lambda^2)} &= -\frac{g}{\omega_1(\omega_2+\omega_3)} \bar u_n(p_1) \Sl{p}_{1\perp} T^a \frac{\Sl{\bar n}}{2} \Gamma \frac{\Sl{n}}{2}\left(  \Sl{\epsilon}_\perp-n\cdot \epsilon \frac{\Sl{p}_{2\perp}}{\omega_2}  \right) v_{\bar n}(p_2)\,,
\end{align}
and
\begin{align}
\left. \fd{3cm}{figures/matching_subleadingvertex_1_low}\right |_{\cO(\lambda^2)}&=-\frac{g}{\omega_1 \omega_3}  \bar u_n(p_1) \Sl{p}_{1\perp} T^a  \left(  \Sl{\epsilon}_\perp-n\cdot \epsilon \frac{\Sl{p}_{2\perp}}{\omega_2}  \right)   v_{\bar n}(p_2)\,.
\end{align}
At tree level, we therefore find the operators
\begin{align}
\cO^{(2)}_{\cP \chi 1}&= -\frac{g}{\omega_1(\omega_2+\omega_3)} \bar \chi_{n,\omega_1} \Sl{\cP}_\perp^\dagger   \frac{\Sl{\bar n}}{2} \Gamma \frac{\Sl{n}}{2}  \Sl{\cB}_{\perp \bar n,\omega_3} \chi_{\bar n,-\omega_2}\,, \nn \\
\cO^{(2)}_{\cP \chi 2}&=-\frac{g}{\omega_1 \omega_3} \bar \chi_{n,\omega_1} \Sl{\cP}_\perp^\dagger \Sl{\cB}_{\perp \bar n, \omega_3} \Gamma \chi_{\bar n,-\omega_2} \,.
\end{align}

Now, to extract the Wilson coefficients of the other operators, where the $\cP_\perp$ acts in the $\bar n$ sector, we choose the kinematics as
\begin{align}
p_1^\mu=\frac{\omega_1}{2}n^\mu \,, \qquad
p_2^\mu=\frac{\omega_2}{2} \bar n^\mu+p_{2\perp}^\mu +\frac{ p_{2r}}{2}  n^\mu\,, \qquad
p_3^\mu=\frac{\omega_3}{2}\bar n^\mu +p_{3\perp}^\mu +\frac{ p_{3r}}{2}  n^\mu\,. 
\end{align}
Expanding the QCD diagrams, we find
\begin{align}
\left. \fd{3cm}{figures/matching_subleadingvertex_2_low}\right |_{\cO(\lambda^2)} &=-\frac{g}{\omega_1 \omega_3} \bar u_n (p_1) \Sl{\epsilon}_{3\perp} \Sl{p}_{3\perp} T^a  \Gamma v_{\bar n}(p_2)  +  \frac{g}{\omega_1 \omega_2} \bar u_n (p_1) \frac{\Sl{\bar n}}{2} \Gamma \frac{\Sl{n}}{2} \Sl{p}_{2\perp} v_{\bar n}(p_2)\,,
\end{align}
and
\begin{align}
\left. \fd{3cm}{figures/matching_subleadingvertex_1_low}\right |_{\cO(\lambda^2)} &=0\,.
\end{align}
We therefore find the operators
\begin{align}
\mathcal{O}^{(2)\mu}_{\cP \bar n1}
  &=- \frac{g}{\omega_1 \omega_3}  \bar \chi_{ n,\omega_1} 
  \big[\Sl\cB_{\perp \bar n, \omega_3} \Sl{\cP}^\dagger_{\perp}\big]  
  \Gamma \chi_{\bar n,-\omega_2} \,, \nn \\
\mathcal{O}^{(2)\mu}_{\cP \bar n2}
  &= - \frac{g}{\omega_1 \omega_2}  \bar \chi_{n,\omega_1} 
  \Sl\cB_{\perp \bar n, \omega_3} \frac{\Sl{\bar n}}{2} \Gamma \frac{\Sl{n}}{2}  \Big[\Sl{\cP}_{\perp} \chi_{\bar n,-\omega_2} \Big]
 \,.
\end{align}
Note, that restricting to zero total $\perp$ momentum in the $\bar n$ collinear sector, i.e. setting $p_{3\perp}=-p_{2\perp}$, we recover the result of \Eq{eq:matched_nbarP}.

In the case that the gluon operator is in the $n$ collinear sector, we have the corresponding operators
\begin{align}
\cO_{\cP \chi 1}^{(2)}&= -\frac{g}{\omega_2(\omega_1+\omega_3)} \bar \chi_{n,\omega_1} \Sl{\cB}_{\perp  n,\omega_3}    \frac{\Sl{\bar n}}{2} \Gamma \frac{\Sl{n}}{2} \Bigl[\Sl{\cP}_\perp \chi_{\bar n,-\omega_2} \Bigr]\,, \nn \\
\cO_{\cP \chi 2}^{(2)}&=-\frac{g}{\omega_2 \omega_3} \bar \chi_{n,\omega_1} \Gamma \Sl{\cB}_{\perp n,\omega_3}   \Bigl[ \Sl{\cP}_\perp  \chi_{\bar n,-\omega_2}\Bigr]\,,
\end{align}
and
\begin{align}
\mathcal{O}^{(2)\mu}_{\cP n1}
  &=-\frac{g}{\omega_1 \omega_2} \Bigl[\bar \chi_{n,\omega_1}  
   \Sl{\cP}_\perp^\dagger \Bigr] \frac{\Sl{\bar n}}{2} 
  \Gamma \frac{\Sl{n}}{2} \Sl\cB_{\perp n, \omega_3} 
   \chi_{\bar n,-\omega_2}
  \,, \nn \\
\mathcal{O}^{(2)\mu}_{\cP n2}
  &= -  \frac{g}{\omega_2 \omega_3}  
  \bar \chi_{n,\omega_1}   \Gamma 
  \bigl[ \Sl{\cP}_\perp  \Sl\cB_{\perp n, \omega_3} \bigr] 
  \chi_{\bar n,-\omega_2}
  \,.
\end{align}
The additional operators given in this appendix demonstrate the simplifications that can be achieved by working in the center of mass frame. There are however cases of interest, for example, beam thrust, which involves hadronic radiation in both the final and initial state, where such a convenient frame cannot be chosen, and this extended operator basis must be used.

\bibliography{../overallbib}{}
\bibliographystyle{jhep}

\end{document}